\documentclass[11pt]{article}
\pdfoutput=1
\usepackage{jheppub}
\usepackage{amsfonts}
\usepackage{amscd}
\usepackage{amssymb}
\usepackage{amsmath,bbm}
\usepackage{epsfig}
\usepackage{latexsym}
\usepackage{mathtools}
\usepackage{arydshln}
\usepackage{tikz-cd}
\usepackage{hyperref}
\usepackage{color}
\usepackage[T1]{fontenc}
\usepackage{slashed}
\usepackage{euscript, mathrsfs}
\usepackage[vcentermath]{youngtab}
\usepackage{marginnote}
\usepackage{subcaption}
\usepackage{bbm}

\def\be{\begin{equation}}
\def\ee{\end{equation}}
\def\bea{\begin{eqnarray}}
\def\eea{\end{eqnarray}}
\def \beaa {\begin{equation}\begin{aligned}}
\def \eeaa {\end{aligned}\end{equation}}
\def \ba  {\begin{eqnarray}}
\def \ea  {\end{eqnarray}}

\newcommand{\nn}{\nonumber}

\newcommand\diff{\mathrm{d}}
\newcommand{\ii}{\mathrm{i}}

\newcommand\cG{\mathcal{G}}
\newcommand\cH{\mathcal{H}}
\newcommand\cI{\mathcal{I}}

\newcommand\cN{\mathcal{N}}

\newcommand\cT{\mathcal{T}}

\def\arg{\operatorname{arg}}

\def\rel{{\rm rel}\,}
\def\Tr{{\rm Tr}\,}

\def\CT{{\mathcal T}}

\def\ICP{{\mathbb{CP}}}
\def\CW{{\mathcal W}}
\def\CB{{\mathcal B}}

\def\CT{{\mathcal T}}

\def\ba{{\mathbf a}}

\def\CI{{\mathcal I}}
\def\CX{{\mathcal X}}

\def\CS{{\mathcal S}}

\def\IR{{\mathbb{R}}}
\def\IZ{{\mathbb{Z}}}
\def\IN{{\mathbb{N}}}

\def\IC{{\mathbb{C}}}

\def\tCW{\widetilde{\mathcal{W}}}

\def\nn{\nonumber}

\def\e{{\rm e}}

\def\CG{{\cal G}}

\def\CT{{\cal T}}

\def\ba{{\mathfrak a}}

\def\fh{{\mathfrak h}}

\def\hTr {{\widehat{\Tr}}}

\renewcommand{\(}{\left(}
\renewcommand{\)}{\right)}
\newcommand{\CN}{{\mathcal{N}}}
\def\tm{\widetilde{m}}

\def\bS{\mathcal{S}}
\def\bM{{M}}

\title{
An infrared bootstrap of the Schur index with surface defects
}

\author[a,b]{Martin Fluder}
\author[c]{Pietro Longhi}

\affiliation[a]{Kavli IPMU (WPI), UTIAS, The University of Tokyo, Kashiwa, Chiba 277-8583, Japan}
\affiliation[b]{Walter Burke Institute for Theoretical Physics, California Institute of Technology,\\ Pasadena, CA 91125, USA}
\affiliation[c]{Institute for Theoretical Physics, ETH Zurich, CH - 8093, Zurich, Switzerland}

\emailAdd{martin.fluder@ipmu.jp, longhip@phys.ethz.ch}

\preprint{CALT-TH-2019-015, IPMU19-0071}

\abstract{
The infrared formula relates the Schur index of a 4d $\mathcal{N}=2$ theory to its wall-crossing invariant, \emph{a.k.a.} BPS monodromy. A further extension of this formula, proposed by C\'ordova, Gaiotto and Shao, includes contributions by various types of line and surface defects. We study BPS monodromies in the presence of vortex surface defects of \emph{arbitrary} vorticity for general class~$\mathcal{S}$ theories of type $A_1$ engineered by UV curves with at least one regular puncture. The trace of these defect BPS monodromies is shown to coincide with the action of certain $q$-difference operators acting on the trace of the (pure) 4d BPS monodromy. We use these operators to develop a ``bootstrap'' (of traces) of BPS monodromies, relying only on their infrared properties, thereby reproducing the very general ultraviolet characterization of the Schur index. 
}

\begin{document}
\maketitle
\bibliographystyle{JHEP}

\noindent
\pagebreak
\section{Introduction}

Supersymmetric quantum field theories provide a fertile playground for understanding various aspects of quantum field theories, including their relation to deep mathematical structures. Theories with extended ($\CN>1$) supersymmetry typically contain  ``protected'' sectors, whose study allows probing various regimes of the system, while still retaining some control. 
A prototypical class of examples are four-dimensional $\CN=2$ supersymmetric field theories. 
The infrared (IR) description of such theories on the Coulomb branch has long been understood in terms of a so-called Seiberg-Witten curve~\cite{Seiberg:1994rs,Seiberg:1994aj}. This geometric picture emerged from the study of a protected sector of these theories, and its range of validity extends from weakly coupled regimes all the way to strong coupling.

More recently, a set of 4d $\CN=2$ theories, dubbed ``class $\CS$ theories''~\cite{Gaiotto:2009we,Gaiotto:2009hg}, has been the focus of intense research. 
These theories are defined by compactifying 6d (maximally-supersymmetric) $\cN=(2,0)$ superconformal field theories (SCFTs) on a Riemann surface $C$ -- the ``UV curve'' -- with punctures. 
Since their inception, it has been clear that the UV curve encodes basic data about them, such as the field content (in the case of Lagrangian theories), couplings, and their behavior under S-duality.  
For example, the UV curve $C$ associated to the 4d $\CN = 4$ supersymmetric Yang-Mills theory is a torus. Its complex structure modulus corresponds to the (complexified) coupling constant of the theory, and the S-duality group is identified with modular transformations of the torus.

The role of UV curves has -- in a sense -- \emph{evolved} over time;
It was soon understood that they can provide much more information about the 4d gauge theory if, instead of considering just their basic geometric features, one endows them with some extra structure.
A spectacular example of this idea is to consider Liouville theory on $C$, with operator insertions at punctures: 
in the so-called ``AGT correspondence''~\cite{Alday:2009aq}, this was argued to compute instanton partition functions of the 4d gauge theory~\cite{Nekrasov:2002qd,Nekrasov:2003rj,Pestun:2007rz}.
UV curves also play a central role in a range of studies of these theories: from the classification of line defects, of surface defects, to the study of BPS spectra, and the classification of class $\CS$ theories themselves, to name a few (see \cite{Drukker:2009tz, Alday:2009fs, Drukker:2009id, Gaiotto:2009fs, Gaiotto:2009hg, Gaiotto:2012rg, Chacaltana:2010ks, Chacaltana:2012zy} for a sample of early references).

In this paper the UV curve will once again play a central role. It will enable us derive novel connections between two rather different objects: Schur indices and BPS monodromies. The Schur index $\cI(q)$ was introduced in~\cite{Gadde:2011ik,Gadde:2011uv}, as a particular limit of the full 4d $\CN=2$ superconformal index~\cite{Kinney:2005ej,Romelsberger:2005eg}, and -- in analogy with the AGT correspondence -- it was shown that $\cI(q)$ arises as a correlator of a 2d TQFT on the UV curve $C$. 
The BPS monodromy $M(q)$ arose in the work of Kontsevich and Soibelman as a wall-crossing invariant of BPS spectra~\cite{Kontsevich:2008fj}. A series of developments for studying BPS spectra through geometric techniques on $C$~\cite{Gaiotto:2009hg, Gaiotto:2012rg}, led to a construction of $M(q)$ based on topological data of a certain ribbon graph~$\CG$, embedded in $C$~\cite{Longhi:2016wtv, Gabella:2017hpz}.

A relation between Schur indices and BPS monodromies was conjectured in~\cite{Cordova:2015nma}, following previous related work~\cite{Cecotti:1992rm, Cecotti:2010qn, Cecotti:2010fi, Iqbal:2012xm}. The conjecture, which we will henceforth refer to as ``the IR formula'', states that
\be\label{eqn:intro:IRformula}
	\cI (q) \ = \ \( q;q \)_{\infty}^{2r} \, \hTr \, M(q) \,.
\ee
The left-hand-side computes a refined count of Schur operators of the 4d $\CN=2$ theory (see Appendix~\ref{app:Schur} for a definition), while the right-hand side is obtained purely from data of the IR theory on the Coulomb branch, whose complex dimension is denoted by $r$ (see Section~\ref{sec:IRreview} for a review).\footnote{We remark that given the proposal in~\cite{Dumitrescu:20xx}, the left-hand-side is supposed to be computable via localization from a Lagrangian description of a not-necessarily conformal $\CN=2$ theory.} 
There is a wild zoo of different IR descriptions of a given UV theory, and in order for the above relation to make sense, the right-hand side ought to be invariant under the choice of IR Lagrangian. 
In fact, a key property of $M(q)$ is that it is invariant up to conjugation (which is taken into account by $\widehat\Tr$) across the whole Coulomb branch~\cite{Gaiotto:2008cd, Dimofte:2009tm}.

We take some steps towards proving (\ref{eqn:intro:IRformula}), by arguing that both quantities appearing in the IR formula share certain common properties.
In the following, we restrict to class $\CS$ theories of type $A_1$, and consider Riemann surfaces with at least one regular puncture (and possibly irregular ones too).
A basic property shared by $\CI(q)$ and $\widehat\Tr M(q)$, is the fact that they are both symmetric under the exchange of identical punctures.
The Schur index is a symmetric function of the flavor fugacities, thanks to crossing invariance of TQFT correlators. The fact that BPS monodromies are invariant under exchange of punctures is an immediate consequence of their construction from ``BPS graphs''~\cite{Longhi:2016wtv}.
In this paper, we show that $\CI(q)$ and $M(q)$ share another, far less trivial, property: under deformations of the theory induced by certain types of surface defects, both change in the same way, described by a universal set of $q$-difference operators.

For the Schur index of a 4d theory $\CT$, this conclusion can be reached by UV reasoning~\cite{Gaiotto:2012xa}. 
One starts by considering a ``larger'' UV theory $\CT'$, obtained by gauging an SU(2) flavor symmetry of $\CT$ corresponding to a puncture of $C$.
Then, one flows back to the original theory by ``Higgsing'' the new degrees of freedom through a position-dependent vacuum expectation value (VEV) for a (charged) baryonic operator.
At small energies (compared to the VEV), this gives back $\CT$ with the insertion of a UV vortex defect carrying vorticity $v\in \IZ$ with respect to the flavor symmetry that was (un)gauged.
By applying this operation to the Schur index $\CI'(q)$ of ${\CT'}$, one ends up with a new index $\CI_v(q)$ for the theory $\CT$ in the presence of a vortex defect~\cite{Gaiotto:2012xa}.\footnote{More precisely, in~\cite{Gaiotto:2012xa} this analysis was carried out for the full superconformal index, and its specialization to the Schur index follows straightforwardly~\cite{Alday:2013kda}.}
Furthermore, in the same reference, it was argued that this takes the form
\be
	\CI_v(q)  \ = \  \mathfrak{S}_v^{UV}  \circ \CI(q) \,,
\ee
where $\mathfrak{S}_v^{UV}$ is a certain $q$-difference operator acting on the Schur index (see Appendix~\ref{app:Schur} for more details). An important consequence of this observation is that Schur indices of arbitrary theories -- engineered by changing the choice of UV curve -- are automatically fixed, by a recursive ``bootstrap'' construction, in terms of these operators~\cite{Gaiotto:2012xa}. 

The analogous statement for the BPS monodromy is novel, and its derivation is our main result.
Instead of gauging and Higgsing, we can work directly with the theory $\CT$, and consider the insertion of suitable vortex surface defects.
In the presence of line defects, the BPS spectrum features new ``framed'' BPS states~\cite{Gaiotto:2010be, Gaiotto:2011tf}.
In the case of surface defects, the BPS monodromy gets replaced by a new wall-crossing invariant, the ``2d-4d BPS monodromy'', $M_v(q)$. This object and various properties of 2d-4d BPS spectra were introduced in~\cite{Gaiotto:2011tf} as a synthesis of Kontsevich-Soibelman wall-crossing in 4d and Cecotti-Vafa wall-crossing in 2d~\cite{Cecotti:1992rm}.
Extensions of the IR formula (\ref{eqn:intro:IRformula}) to include defects have been explored recently in~\cite{Cordova:2016uwk,Cordova:2017ohl,Cordova:2017mhb, Neitzke:2017cxz}.
 
\begin{figure}[h!]
\begin{center}
\includegraphics[width=0.3\textwidth]{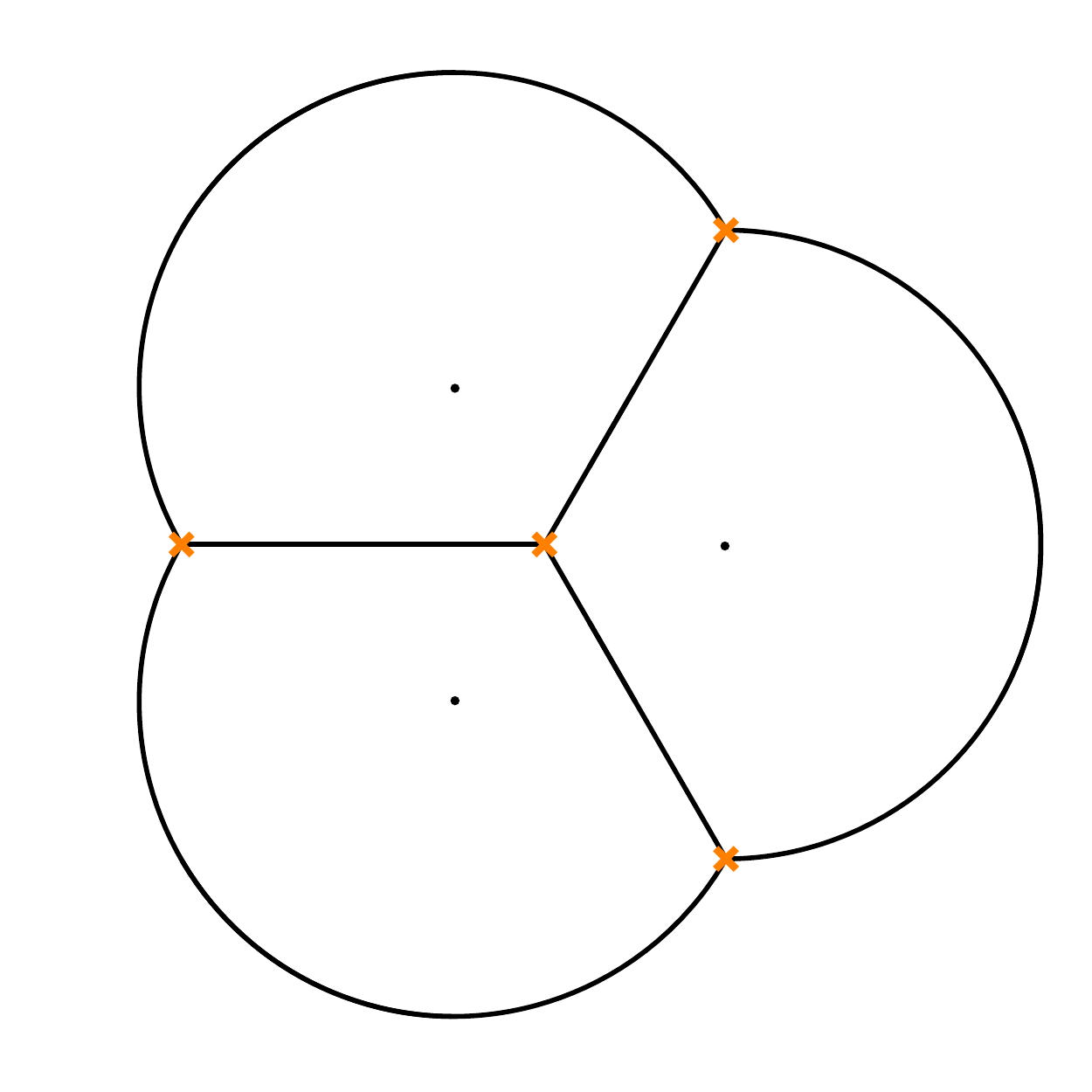}
\caption{The BPS graph of $\cN=2$ SQCD, \emph{i.e.} the $SU(2)$ gauge theory with $N_f=4$ fundamental hypermultiplets. The UV curve $C$ is given by a sphere with four punctures, one of which is placed at infinity. The BPS graph $\CG$ decomposes $C$ into four regions, each one is a punctured disk.}
\label{fig:BPSgraph-example}
\end{center}
\end{figure}

Every class $\CS$ theory comes equipped with a ``canonical'' surface defect, parameterized by the choice of a point $z\in C$~\cite{Gaiotto:2011tf}.
A construction of (classical) 2d-4d BPS monodromies in the presence of canonical defects, for all class $\CS$ theories of type $A_1$,  was derived in~\cite{Longhi:2012mj}. If we choose $z$ to be ``close enough'' to a regular puncture, the canonical defect corresponds to a vortex defect with unit vorticity $v=1$.
When the UV curve has multiple regular punctures, it becomes necessary to establish a criterion to determine which puncture the defect is ``close to''.
An elegant solution to this problem is offered by BPS graphs~\cite{Gabella:2017hpz}.
In $A_1$ theories, a BPS graph $\CG$ is a graph embedded in $C$, whose edges define different patches of $C$, each of which contains exactly one puncture.\footnote{
One should keep in mind, that BPS graphs arise from spectral networks at special loci in the Coulomb branch -- they arise from degenerate Strebel differentials~\cite{Longhi:2016wtv,Gabella:2017hpz} (Also see~\cite{Hollands:2013qza} for another relation between Strebel differentials and ``Fenchel-Neilsen networks''). However, this is not an issue for computing $M_v(q)$, since the latter is by definition invariant across the moduli space.
Moreover, BPS graphs arise from WKB spectral networks at a critical phase. It follows from basic principles of 2d-4d wall-crossing that their edges enclose regions with different 2d-4d BPS spectra~\cite{Gaiotto:2011tf,Longhi:2012mj}.
}
For example, the BPS graph of the SU(2), $N_f=4$ gauge theory is shown in Figure~\ref{fig:BPSgraph-example}.

Another important feature of BPS graphs is that they arise from ``spectral networks''. For this reason they carry information about 2d-4d BPS spectra and, by extension, about the 2d-4d BPS monodromy.
Thus, we are able to compute the \emph{full} spectrum of the 2d-4d system in the presence of a canonical surface defect placed in any region of $C\setminus \CG$ containing a regular puncture.
We actually compute the \emph{refined} spectrum, using technology developed in~\cite{Galakhov:2014xba}; this includes information about the spin of BPS states which is crucial for applications to the IR formula.
We further extend this result to higher-vorticity defects, with $v>1$. 
To engineer them, we adopt ideas of~\cite{Hori:2013ewa, Longhi:2016rjt, Longhi:2016bte}, and replace the Seiberg-Witten curve with Hitchin spectral curves in higher symmetric representations (of $A_1$), of dimension $N=v+1$.\footnote{Spectral networks in higher symmetric representations for $A_n$ theories have not been developed previously, therefore we give a self-contained treatment of these as well.}
By direct computation we are then able to prove that the 2d-4d BPS monodromy in the presence of a vortex defect of vorticity $v$ is related to the 4d BPS monodromy as follows
\be\label{eq:app:IR-defec-action}
	\widehat \Tr M_v(q)  \ = \  \mathfrak{S}_v^{IR}  \circ \widehat \Tr M(q)  \,.
\ee
Here $\mathfrak{S}_v^{IR}$ are certain $q$-difference operators, acting on the flavor fugacity associated to the puncture closest to the surface defect.
We obtain the explicit form of these difference operators, and find that 
\be
	\boxed{~\mathfrak{S}_v^{IR} \ \propto \  \mathfrak{S}_v^{UV}}
\ee
up to an overall inessential proportionality constant.

Given that our IR construction of vortex surface defects reproduces the action of UV vortex defects defined in terms of the Schur index, we can apply them to ``bootstrap'' the ``IR Schur index'', \emph{i.e.} $(q,q)_\infty^{2r}\, \widehat \Tr M(q)$.
We develop this procedure for the whole range of rank-one class $\CS$ theories engineered by UV curves $C$ with at least one regular puncture, and possibly including irregular punctures as well. By leveraging certain properties of the IR Schur index, derived exclusively from IR considerations, we can explicitly identify the UV Schur index derived from 2d TQFT methods for a large class of 4d class $\CS$ theories in~\cite{Gadde:2011ik,Gadde:2011uv}, with the one derived via the right-hand side of equation~\eqref{eqn:intro:IRformula} from purely IR, Coulomb branch considerations. 
The argument relies on the existence of surface defects which can be recast in terms of difference operators acting on the Schur index, which is proved by (\ref{eq:app:IR-defec-action}). 
These difference operators correspond to a degeneration limit of Macdonald $q$-difference operators, whose eigenfunctions are given by Schur polynomials.\footnote{More precisely, they are related via conjugation by vector multiplet contributions, see~\cite{Gaiotto:2012xa,Alday:2013kda}.} 
The IR defects act locally on, and irrespective of, the puncture. 
This implies that the index factorizes in terms of an infinite sum of eigenfunctions -- \emph{i.e.} Schur polynomials (upon conjugation). 
Given this factorization, we inductively derive the form of the Schur index, and find that it precisely agrees with the one computed from the 2d TQFT perspective in~\cite{Gadde:2011ik,Gadde:2011uv}.

\vspace{.5 cm}

The remainder of this paper is structured as follows. In Section~\ref{sec:IRreview}, we provide an in-depth review of the IR formalism, including the IR Schur formula and its version including surface defects. Subsequently, in Section~\ref{sec:BPS-spectrum}, we discuss general features of the 4d, 2d-4d and 2d BPS spectra of rank-one class $\CS$ theories. 
In particular, we discuss a novel way to describe vortex surface defects, including defects with arbitrary vortex numbers (\emph{i.e.} in higher dimensional representations of $SU(2)$), show that they act locally on the (regular) punctures UV curve, and discuss the corresponding 2d particles. Then, in Section~\ref{sec:2d-4d-macdonald}, we turn towards explicitly evaluating the trace of the BPS monodromies. We prove that 2d-4d BPS states cancel in the evaluation and explicitly obtain a general formula for vortex surface defects acting as difference operators on the Schur index. 
After these rather technical sections, we discuss some explicit examples in Section~\ref{sec:examples}. 
In Section~\ref{sec:one-puncture}, we treat the case with a single puncture separately, as it evades our general discussion in the previous sections. 
We start with the example of the $\CN=2^*$ theory, and then turn towards a general discussion of rank-one theories of class $\CS$ associated to UV curves with a single regular puncture. 
In Section~\ref{sec:irregular}, we extend our results to the case of theories with irregular punctures of type $I_{2,M}$. 
For instance, such irregular punctures appear in Argyres-Douglas theories of type $(A_1,D_M)$. Finally, in Section~\ref{sec:bootstrap} we use our results from earlier sections to ``bootstrap'' the Schur index for arbitrary rank-one class $\CS$ theories with regular punctures and possibly irregular punctures of type $I_{2,M}$. We explicitly show that the resulting IR Schur indices -- derived from the IR formalism -- precisely agree with the known results of the index. We conclude the main text with a brief discussion of our results and mention some future directions in Section~\ref{sec:disc}. In Appendix~\ref{app:Schur}, we provide some supplementary details on the Schur limit of the superconformal index, and in Appendix~\ref{sec:more-punctures}, we outline some more technical arguments to show that our surface defects indeed always act locally on the punctures of the UV curve.

\section{The IR Schur formula with surface defects}\label{sec:IRreview}

In this section, we review the IR Schur formula introduced in~\cite{Cordova:2015nma} for the index, and extended to include various types of defects in~\cite{Cordova:2016uwk,Cordova:2017ohl,Cordova:2017mhb,Neitzke:2017cxz}.\footnote{See also~\cite{Imamura:2017wdh}, for some results of an IR formula for the lens index.} We first start by recalling the IR Schur formula, and then turn towards the IR prescription for the Schur index in the presence of surface defects. In the later sections we will outline generalizations of the approach in~\cite{Cordova:2017ohl}, which will be important for computing the particular set of surface defects, which we then use to ``bootstrap'' the IR formula. For explicit examples, which may help to digest some technicalities, we refer to Section~\ref{sec:examples} and Section~\ref{sec:one-puncture}.

\subsection{The IR Schur formula}\label{sec:IRSchurRev}

We start by reviewing the proposal for the formula of the Schur index using IR Coulomb branch data. This formula was introduced in~\cite{Cordova:2015nma}, and we shall call it the ``IR formula'' or the ``IR Schur formula'' in the following. Notice, that there is an analogous formula relating the elliptic genera of two-dimensional $\cN=(2,2)$ theories with the BPS soliton degeneracies on the moduli space~\cite{Cecotti:1992rm,Gaiotto:2015aoa}.

The IR formula relates the Schur index to the trace of the BPS monodromy operator. The Schur index can be viewed as a particular limit of the superconformal index~\cite{Gadde:2011ik,Gadde:2011uv}. We refer to Appendix~\ref{app:Schur} for its definition and some results relevant for our purposes here. Then, the IR formula asserts that the Schur index is given as follows
\be\label{eqn:IRformula}
	\cI (q) \ = \ \( q;q \)_{\infty}^{2r} \, \hTr M(q) \,,
\ee
where 
\be
	\left(z;q\right)_\infty \ = \ \prod_{j=0}^{\infty} \( 1-z q^{i} \)
\ee
is the Pochhammer symbol, $r$ is the (complex) dimension of the Coulomb branch of the theory, and 
\be\label{eq:MSS}
	\bM(q) \ = \ \overline\bS(q) \bS(q)
\ee
is the quantum monodromy, with $\bS$ the ``quantum spectrum generator'' -- which was introduced in~\cite{Gaiotto:2009hg, Dimofte:2009tm, Gaiotto:2010be} for 4d systems and in~\cite{Gaiotto:2011tf} for 2d-4d systems as an extension of the Cecotti-Vafa~\cite{Cecotti:1992qh,Cecotti:1992vy,Cecotti:1992rm} and Kontsevich-Soibelman~\cite{Kontsevich:2008fj} formulae, and whose form, together with the notion of trace, we shall explain in the following.\footnote{The quantum spectrum generator also plays an important role in the description of BPS states in string theory~\cite{Dimofte:2009bv}.} To do so, we ought to move onto the Coulomb branch, and further introduce some IR technology.

Given a 4d $\cN=2$ theory of class $\CS$, we can move on the Coulomb branch by turning on a non-trivial choice of Coulomb vacuum, $u \in \CB$, in which the theory is rendered IR free. At a generic point $u \in \CB$, the resulting theory is then given by a particular abelian (interacting) $U(1)^{r}$ gauge theory, where $r$ is the complex dimension of the Coulomb branch $\CB$. Notice, that there are singular (complex codimension one) loci in $\CB$, where some BPS particles become massless and the gauge theory description breaks down. We shall henceforth focus on $\CB^{*}\subset \CB$ where the generic description remains valid. At a given generic $u\in \CB^*$, there exists a local system of lattices $\Gamma \to \CB^{*}$, on which we can define a Dirac-Schwinger-Zwanziger (DSZ) antisymmetric and integer-valued pairing, \emph{i.e.} $\langle \cdot , \cdot \rangle : \Gamma \times \Gamma \to \mathbb{Z}$. If there is a remnant flavor symmetry, the radical $\Gamma_{f}$ of the DSZ pairing is the sublattice $\Gamma_{f}\subset \Gamma$ of flavor charges. In fact, taking the quotient of $\Gamma_{g}=\Gamma/\Gamma_{f}$, the resulting lattice is of rank $2r$ containing electric and magnetic gauge charges. One can define the linear central charge function $Z \in {\rm Hom}(\Gamma, \mathbb{C})$ (holomorphic in $u$), such that for a local section $\gamma$ of $\Gamma$, the value $Z_{\gamma}(u)$ gives the central charge of the particle of charge $\gamma$. 

Geometrically, the IR data can be extracted from the Seiberg-Witten curve of the IR theory~\cite{Seiberg:1994rs,Seiberg:1994aj}; Namely, one can view the Seiberg-Witten curve $\Sigma$ as a branched cover over the UV curve, $C$, \emph{i.e.} $\Sigma \to C$.\footnote{As is well-known, the UV curve $C$ can naturally be viewed as the surface on which the 6d $\CN=(2,0)$ theory is compactified to give a 4d theory~\cite{Gaiotto:2009we,Gaiotto:2009hg}.} 
For instance, in the rank-one case, the Seiberg-Witten curve is a two-sheeted cover over $C$. Then, the charge lattice $\Gamma$ can be identified with a subquotient of $H_{1} (\Sigma, \mathbb{Z})$, where we identify two one-cycles $\gamma \sim \gamma'$ if and only if $Z_{\gamma} = Z_{\gamma'}$. 
Modulo this subquotient, the lattice of gauge charges $\Gamma_{g}$ is identified with $H_{1}(\bar \Sigma, \mathbb{Z})$, where $\bar \Sigma$ is the compact Riemann surface, where we fill in the punctures, and the lattice of flavor charges $\Gamma_{f}$ corresponds to small cycles around the punctures of $\Sigma$. Furthermore, the DSZ pairing corresponds to the intersection pairing of the corresponding homology classes on $\Sigma$.

A distinguishing feature of class $\CS$ theories is their relation to Hitchin systems~\cite{Hitchin}.
The Coulomb branch is identified with the base of the Hitchin fibration, and the Seiberg-Witten curve arises as the spectral curve of the integrable system, naturally embedded in $T^*C$.
The pullback of the Liouville one-form $\lambda$ is identified with the Seiberg-Witten differential on $\Sigma$, and its periods compute BPS central charges~\cite{Donagi:1995cf, Martinec:1995by, Gaiotto:2009hg}.

Now, to define the IR formula, we will need to compute the ``quantum'' or ``motivic'' 2d-4d spectrum generator~\cite{Gaiotto:2011tf, Cordova:2017ohl}. To do so, we introduce the (non-commutative) ``quantum torus algebra'', given by the elementary relations
\be\label{eq:quantum-torus-algebra-definition}
	X_\gamma  X_{\gamma'}  \ = \  q^{\frac{1}{2}\langle\gamma,\gamma'\rangle}  X_{\gamma+\gamma'}\,,
	\qquad  
	X_{-\gamma}  \ = \   X_{\gamma}^{-1}\,,
\ee
where $X_{\gamma}$ and $X_{\gamma'}$ are non-commutative variables associated to one-cycles $\gamma,\gamma'$ on $\Sigma$, or equivalently elements of the lattice, $\gamma,\gamma'\in\Gamma$, and $\langle\cdot,\cdot\rangle$ is the DSZ/intersection pairing. The algebra~(\ref{eq:quantum-torus-algebra-definition}) first appeared in connection to BPS spectra in~\cite{Kontsevich:2008fj}, and its physical origin was elucidated  in~\cite{Gaiotto:2010be}. In particular, we remark that the coordinates $X_{\gamma}$ of the quantum torus algebra can be interpreted as dyonic line defects of charge $\gamma$ in the IR abelian theory (see \emph{e.g.}~\cite{Gaiotto:2010be, Cordova:2013bza} for more details). Upon a conformal mapping to $S^{3}\times S^{1}$, they wrap the (temporal) $S^{1}$ factor and are aligned along a great circle of $S^{1}\subset S^{3}$, which explains their non-commutative nature (see \emph{e.g.}~\cite{Cordova:2016uwk}).

The mass $M$ of a particle of charge $\gamma \in \Gamma$ is constrained to be
\be
	M \ \geq \ \big| Z_{\gamma}(u) \big| \,, 
\ee
with equality if and only if it is BPS. Massive BPS states of a given $\cN=2$ theory belong to short multiples of the super-Poincar\'e symmetry. Thus, they lie in representations of  $SU(2)_{R}\times SU(2)_{J}$, where the latter factor is the little group. The one-particle Hilbert space $\cH$ at fixed $u\in \CB$ is graded by the charge $\gamma$ and decomposes as $\cH = \oplus_{\gamma \in \Gamma} \cH_{\gamma}$. Then, upon dividing out by the overall center-of-mass degrees of freedom, we get
\be
	\cH_{\gamma} \ = \ \left[ \( \mathbf{2}, \mathbf{1} \)\oplus \( \mathbf{1}, \mathbf{2} \)  \right]\otimes \fh_{\gamma} \,.
\ee
One can introduce an index~\cite{Gaiotto:2010be} -- the ``protected spin character'', which is the refined version of the ``second helicity supertrace'' -- counting the BPS states refined by $SU(2)_{R}\times SU(2)_{J}$, \emph{i.e.}
\be
	\Omega(\gamma,q) \ = \ \Tr_{\fh_{\gamma}} (-1)^{2R} \, q^{{J+R}}  \ = \ \sum_{n\in \IZ} \Omega_{n}(\gamma) q^{\frac{n}{2}} \,,
\ee
where $J$, $R$ are Cartan generators of $SU(2)_{J}$, $SU(2)_{R}$ respectively, and where the integers $\Omega_{n}(\gamma) \in \IZ$ -- the ``BPS degeneracy'' -- count BPS states with charge $\gamma \in \Gamma$ and spin $\frac{n}{2}$.\footnote{The ``no-exotics'' property asserts that BPS states have $R=0$ in 4d $\CN=2$ gauge theories~\cite{Gaiotto:2010be, Chuang:2013wt, DelZotto:2014bga}.} 
In $A_1$ theories of class $\CS$, the spectrum contains only BPS hypermultiplets and vectormultiplets~\cite{2013arXiv1302.7030B}.
For a massive $\cN=2$ hypermultiplet of charge $\gamma$, we have $\fh_\gamma = (\mathbf{1},\mathbf{1})$ and therefore $\Omega(\gamma,q)=+1$. 
For a vector multiplet, instead, we have $\fh_\gamma = (\mathbf{2},\mathbf{1})$ and therefore $\Omega(\gamma,q)=q^{-1/2}+q^{1/2}$. 

The BPS particle spectrum -- and thus $\Omega(\gamma,q)$ -- is locally constant in $u\in \CB^*$. However, it can jump when crossing (real) codimension-one interfaces in $\CB$ -- so-called ``walls of marginal stability'' -- along which the phases $\arg (Z_{\gamma})$ conspire and align for all BPS particles. Luckily, these jumps can be quantified by ``wall-crossing formulae''~\cite{Gaiotto:2008cd,Kontsevich:2008fj,Gaiotto:2009hg,Dimofte:2009tm}, which we shall now briefly review.\footnote{In practice, BPS spectra at arbitrary moduli can be rather involved, and by moving onto a ``simple'' sector of the theory, in which one can explicitly compute the protected spin character, the wall-crossing formulae allow to extract the BPS degeneracies in other sectors.}

Given the protected spin character, $\Omega(\gamma,q)$, one can repackage the coefficient $\Omega_n(\gamma)$ into the ``Kontsevich-Soibelman (KS) factor'', 
\be\label{eqn:KSfac}
	K\left( q; X_{\gamma}; \Omega_{i}(\gamma) \) \ \coloneqq \ \prod_{n\in \IZ} E_{q}\left( (-1)^{n} q^{n/2} X_{\gamma} \)^{(-1)^{n} \Omega_{n}(\gamma)}\,,
\ee
where $E_{q}(z)$ is the quantum dilogarithm, defined as follows
\be\label{eq:q-dilog-q-pochh}
	E_{q} (z) \ \coloneqq \ \big( -{q^{\frac{1}{2}}} z ;q \big)_{\infty}^{-1} \,.
\ee
Finally, from the KS factors we can define the quantum spectrum generator associated to an angular sector $\sphericalangle\subset \mathbb{R}/2\pi \mathbb{Z}$,
\be
	\bS_{\sphericalangle}(q) \ \coloneqq \ \prod_{\gamma \in \Gamma_{\sphericalangle}}^{\curvearrowleft} K\left( q; X_{\gamma}; \Omega_{i}(\gamma) \) \,,
\ee
where the product is restricted to charges $\gamma \in \Gamma_\sphericalangle \coloneqq \left\{ \gamma \in \Gamma \,:\,\arg\left( Z_\gamma \right) \in \sphericalangle \right\}$, and the ordering in the product is defined such that if $\arg Z_{\gamma_1}> \arg Z_{\gamma_2}$ KS factors associated to $\gamma_1$ are to the left of $\gamma_2$. The quantum spectrum generator and its conjugate are defined as follows:
\be\label{eqn:CSq}
	\bS(q) \ \coloneqq \ \bS_{[\pi, 0)}(q)\,, \qquad \overline{\bS} \left( q  \right) \ \coloneqq \ \bS_{[2\pi, \pi)}(q) \,.
\ee
Note the unconventional orientation of angular intervals, denoting the fact that lower values of the phase correspond to KS factors on the right.
More generally, if the angular sector is $\sphericalangle = [\vartheta+\pi , \vartheta)$, then $\bS_\sphericalangle(q)\sim \bS(q) $ is related to the quantum spectrum generator by conjugation induced by an overall rotation of $\vartheta$. The quantum BPS monodromy $\bM(q)$ is then the product of the quantum spectrum generator and its conjugate (\ref{eq:MSS}). Notice, that the splitting into $\bar \bS$ and $\bS$ naturally captures particles and their anti-particles. 
For later convenience, we also introduce the following definitions
\beaa
	\bS_{\vartheta \pm \pi/2} \left( q  \right) & \ \coloneqq \  \bS_{[\vartheta+\pi/2,\vartheta-\pi/2)}(q) \,, \qquad
	& \overline{\bS}_{\vartheta \pm \pi/2} \left( q  \right) & \ \coloneqq \ 
	{\bS}_{[\vartheta-\pi/2,\vartheta+\pi/2)}(q) \,.
\eeaa
It is obvious from the definitions, that $\overline{\bS}(q)  \bS(q)  \sim \overline{\bS}_{\vartheta \pm \pi/2}\bS_{\vartheta \pm \pi/2}$ are equivalent up to conjugation.

It is a crucial fact, that even though the KS factors may jump across walls of marginal stability in $\CB$, the quantum spectrum generators, $\bS_{\sphericalangle}$, is wall-crossing invariant as long as the phases of BPS states do not cross the boundary of the interval $\sphericalangle$~\cite{Gaiotto:2008cd,Kontsevich:2008fj,Gaiotto:2009hg,Dimofte:2009tm}. This allows us to define them irrespective of the reference moduli $u \in \CB^{*}$. This is of course a critical requirement for the matching with the Schur index from IR data. Furthermore, the formula~\eqref{eqn:CSq}, together with~\eqref{eqn:KSfac}, \emph{uniquely} determine the BPS degeneracies $\Omega_{n}(\gamma)$. Thus, given the quantum spectrum generators in any given  sector of the theory, one can determine the BPS states and their degeneracies in any other sector by relying on invariance of the spectrum generator across walls of marginal stability.

Finally, we should recall the definition of the trace appearing in~\eqref{eqn:IRformula}. Recall that the charges in the flavor lattice $\gamma_{f} \in \Gamma_{f}$ have trivial DSZ/intersection pairing with other elements of $\gamma \in \Gamma$. Thus, the corresponding element $X_{\gamma_{f}}$ is an element of the center of the quantum torus algebra. The trace then projects the quantum torus algebra to the central flavor elements, \emph{i.e.}
\be\label{eqn:hTr}
	\hTr \left( X_{\gamma} \right) \ = \ \left\{\begin{array}{ll} X_{\gamma}\,, \ & \gamma \in \Gamma_f \\ 0 \,, & \text{else} \,. \end{array} \right.
\ee
Thus, upon defining a basis $\{ \gamma_{f}^{i} \}_{i}$ of $\Gamma_{f}$, and taking the trace, the IR index in~\eqref{eqn:IRformula} reduces to a function of $q$ and the now commutative variables $X_{\gamma_{f}^{i}}$, which are related to the flavor fugacities of the Schur index.

The definition of the IR index is based on the BPS monodromy $M(q)$. As we have defined it, to compute $M(q)$ would require one to know the full BPS spectrum, at least in some sector of the Coulomb branch. In practice however, we will find it convenient to take an alternative geometric approach (see also Section~\ref{sec:examples} for explicit examples). 
We consider a ``spectral network'', defined as the web of trajectories $c:[0,1]\to C$ on the UV curve $C$ defined by the following differential equation~\cite{Gaiotto:2012rg}
\be\label{eq:S-wall}
	\partial_t \cdot \lambda_{ij} \ \in \ e^{\ii\vartheta} \IR_+\,,
\ee
where $\partial_t$ is the vector field along the curve $c$ and $\lambda_{ij}=\lambda_i-\lambda_j$ depends on the values taken by the Seiberg-Witten differential at a given point on $C$ (these are in 1:1 correspondence with the sheets of the cover $\Sigma\to C$). 
The topology of this network as a function of $\vartheta$ jumps at the phases of 4d BPS states~\cite{Gaiotto:2012rg}.
This therefore provides a way to compute $M(q)$. 
However, in practice, determining the BPS spectra at generic points in the moduli space is rather daunting. On the other hand, using the invariance under wall-crossing of the quantum spectrum generator allows us to evaluate it in ``simpler'' chambers. 
In this paper, we shall take advantage of the existence of a so-called ``Roman locus''~\cite{Gabella:2017hpz}, characterized by $u \in \CB$, such that the 4d BPS particles have central charges of common phase. 
One might worry that at this locus, the 4d BPS spectrum is ill-defined, since by definition 4d BPS states would be at marginal stability. Nevertheless, in~\cite{Longhi:2016wtv} it was argued that the 4d BPS spectrum generator is well-defined and can be characterized in terms of 2d-4d BPS states. 
Indeed, while 4d BPS states are at marginal stability, 2d-4d states are not, and their spectrum is well-defined and can be computed with spectral networks. At the Roman locus 4d central charges align at a critical ray with phase $\vartheta \equiv \vartheta_c$, and so do the 2d particles, while the 2d-4d states are off criticality.  
The spectral network, for $\vartheta=\vartheta_c$, becomes maximally degenerate, and defines a so-called ``BPS graph''~\cite{Gabella:2017hpz}. 
In turn this graph encodes the 4d spectrum generator~\cite{Longhi:2016wtv}.\footnote{In fact, in the case of $A_1$ theories, BPS graphs are dual to ideal triangulations of $C$~\cite{Gabella:2017hpz}. In~\cite{Gaiotto:2009hg} it was shown that the classical spectrum generator can be computed from a triangulation. The advantage of using BPS graphs is two-fold: it allows to compute the quantum spectrum generator, and more crucially it provides a way to compute 2d-4d BPS particles as well thanks to its underlying spectral network.}
Thus, throughout, we shall employ ``BPS graphs'' to explicitly compute the quantum spectrum generators $\bS$ and $\overline \bS$.\footnote{Alternatively, in~\cite{Cordova:2016uwk,Cordova:2017ohl,Cordova:2017mhb,Neitzke:2017cxz}, the authors leveraged the power of ``BPS quivers''~\cite{Cecotti:2010fi,Cecotti:2011rv,Alim:2011ae,Alim:2011kw} (which are related but not equivalent to BPS graphs), which describe the BPS spectrum in a strong-coupling chamber, where only a finite amount of hypermultiplets contribute; the associated charges can be described in terms of quiver data.}

\subsection{2d-4d wall-crossing and the IR Schur formula with surface defects}\label{sec:2d4dwcrev}

Supersymmetric quantum field theories can be decorated with general (global) defects, such as Wilson~\cite{Wilson:1974sk} and 't Hooft~\cite{tHooft:1977nqb,Kapustin:2005py} lines as well as their higher dimensional analogues. 
It has become clear that such extended defects are crucial in the study and classification of gauge theories; for example, there are known cases of disparate field theories that can only be distinguished by the insertion of defects (see \emph{e.g.}~\cite{Aharony:2013hda}). 
Half-BPS surface defects -- extended along a codimension-two manifold in the 4d spacetime -- in the 4d ${\cN = 4}$ supersymmetric Yang-Mills theory were introduced and studied in~\cite{Gukov:2006jk,Gukov:2008sn}, via two alternative descriptions; on the one hand, one may introduce a defect by specifying a codimension-two singularity in the field configuration, preserving half of the supersymmetries, on the other hand, they may be defined by coupling a given 4d theory to (purely) 2d degrees of freedom, supported on the surface defect. Upon integrating out the 2d fields in the path integral, it can sometimes be argued that the two constructions of surface defects are dual.\footnote{See \emph{e.g.}~\cite{Frenkel:2015rda,Ashok:2017odt,Gorsky:2017hro,Balasubramanian:2017gxc,Jeong:2018qpc} for some examples where different descriptions of surface defects in 4d $\CN=2$ theories are argued (and checked) to be equivalent.} Of course, it is possible to extend these considerations to the case of $\CN=2$ theories (see \emph{e.g.}~\cite{Gukov:2014gja} for a nice review).

We now turn to the IR index with the inclusion of surface defects. Its definition~\cite{Cordova:2017ohl} is an extension from the IR index without defects, now ``counting'' 2d, 2d-4d and 4d BPS states together, rather than only 4d BPS states. Here, we will introduce 2d wall-crossing as a warm-up and then turn towards 2d-4d wall-crossing. In this section, we shall mostly take an algebraic perspective, while in the later sections we are dealing with $A_{1}$-type class $\CS$ theories and thus employ a more geometric approach; in the process, we will recall important properties of the 2d-4d BPS states recast in terms of the UV and Seiberg-Witten curves. 

In Section~\ref{sec:near-punctures}, we provide a slightly alternative viewpoint on surface defects as the zero-length limit of a supersymmetric interface. 
This seems to provide a universal -- theory-independent -- way to treat surface defects near punctures from the IR perspective, and we explicitly calculate their action on the IR index in Section~\ref{sec:2d-4d-macdonald}.

\subsubsection{2d wall-crossing and the Cecotti-Vafa formula }

Let us now briefly pause and discuss $\cN=(2,2)$ theories and the ``Cecotti-Vafa wall-crossing formula''~\cite{Cecotti:1992qh,Cecotti:1992vy,Cecotti:1992rm}.\footnote{See~\cite{Gaiotto:2015aoa} for a categorification of the 2d wall-crossing formula.} 
Such theories allow for several kinds of relevant deformations.
If one requires the presence of a nontrivial central extension to the $\CN=(2,2)$ supersymmetry algebra, it turns out that it is not possible to preserve simultaneously $U(1)_A\times U(1)_V$, and one of the two symmetries must be sacrificed. In view of the main purpose of this work, which is to study 2d-4d systems, it turns out to be natural to sacrifice $U(1)_A$.
This implies that we can turn on a \emph{twisted} superpotential $\tCW$ for the theory, as well as \emph{twisted} masses $\tm$ for the 2d chiral multiplets~\cite{Witten:1993yc, Hanany:1997vm}.\footnote{In 2d-4d systems, the 4d vectormultiplet scalars couple to the 2d chirals in the guise of twisted masses. For this reason it is natural to stick to deformations that break $U(1)_A$.}  The masses $\tm$ couple to twisted current multiplets of charge $\gamma_{2d}$. 
Similar to 4d, such deformations trigger a flow out onto an $\cN=(2,2)$ IR theory, which we generically assume to be gapped with (finite) $N$ vacua labeled by an index $i$. 
Then, the spectrum of massive particles is made up from particles in a single vacuum $i$, as well as ``$ij$ solitions'' interpolating between two vacua $i$ and $j$, with $i \neq j$. Let us denote by $\tCW_{i}$ the value of the twisted superpotential at an isolated, non-degenerate, critical point $i$. 
The BPS bound for an $ij$ soliton then reads~\cite{Cecotti:1992qh}
\be
	M \ \geq \ \left|  Z_{2d} \right| \ = \ \left| \tCW_{j}- \tCW_{i} + \tm\cdot \gamma_{2d} \right| \,,
\ee
where $Z_{2d}$ is the 2d (twisted) central charge. The inequality is saturated if and only if the corresponding soliton is BPS. 
Notice that a particle in a fixed vacuum $i$ can be viewed as a soliton of type $ij$ with $i=j$, then the contribution from the twisted superpotential vanishes, and the central charge is solely proportional to the twisted mass. 

One can proceed to define an index counting the degeneracies of BPS states~\cite{Cecotti:1992qh}, 
\be
\begin{split}
	\mu_{ij, \gamma_{2d}} & \ = \  \Tr_{\cH_{ij},\gamma_{2d}} (-1)^{F_{2d}} F_{2d} \,,\\
	\omega_{i,\gamma_{2d}} & \ = \  \Tr_{\cH_{ii},\gamma_{2d}} ( -1 )^{F_{2d}} F_{2d}\,,
\end{split}
\ee
where the former counts $ij$ BPS solitons while the latter counts BPS states in a vacuum $i$, and $F_{2d}$ is the two-dimensional fermion number. Finally, the trace is taken over the Hilbert space of BPS states of a given sector of the theory labeled by the vacua $ij$ and the flavor charge $\gamma_{2d}$ which couples to the twisted masses.

The 2d BPS spectrum may jump when we cross walls of marginal stability, and the Cecotti-Vafa wall-crossing formula gives a way to relate different sectors in the parameter space of relevant deformations of the 2d $\CN=(2,2)$ theory. One first defines the analogues of the KS factors in 4d, 
\be\label{eq:2d-wall-crossing-operators}
\begin{split}
	S^{(2d)}_{ij;\gamma_{2d}} & \ \coloneqq \ \mathbbm{1} - \mu_{ij, \gamma_{2d}} u(\gamma) e_{ij} \,, \qquad i \neq j \,,\\
	K^{(2d)}_{ii;\gamma_{2d}} & \ \coloneqq \ \sum_{i=1}^{N} \(1-u(\gamma)\)^{-\omega_{i,\gamma_{2d}}} e_{ii} \,.
\end{split}
\ee
Here, $\mathbbm{1}$ is the $N\times N$ identity matrix, and $e_{ij}$ is the $N\times N$ matrix with a single non-zero entry (equal to $1$) in row $i$ and column $j$, and $u(\gamma_{2d})$ is the fugacity one may introduce for the flavor charge $\gamma_{2d}$. 
The notation is chosen with the upcoming geometric viewpoint in mind: $S^{(2d)}_{ij,\gamma_{2d}}$ resemble Stokes matrices, while the $K^{(2d)}_{ii;\gamma_{2d}}$ have a different role, associated with ``K-walls''  (see~\cite{Gaiotto:2009hg, Gaiotto:2012rg}).\footnote{Here, by Stokes matrices we refer to the ones arising in the WKB analysis of the Hitchin equations reformulated as flatness of a connection on a rank-$N$ bundle over $C$.
In the context of exact WKB analysis, the matrices $\CS$ arise in the ``connection formula'' as Stokes morphisms, while the $K$-factors appear in connection to a certain instance of the ``Delabaere-Dillinger-Pham'' formula~\cite{IwakiNakanishi1}.}
Then, the Cecotti-Vafa wall-crossing formula states that upon moving around in the parameter space of the 2d theory, while no BPS states exit the wedge $\sphericalangle=(\vartheta_1, \vartheta_2]$, the quantity
\be\label{eqn:2dgen}
	\bS^{(2d)}_{\sphericalangle} \ = \  \prod^{\curvearrowleft}_{\arg(Z_{2d})\in \sphericalangle} 
	S^{(2d)}_{ij;\gamma_{2d}} K^{(2d)}_{ii;\gamma_{2d}} 
\ee
is wall-crossing invariant. In~\eqref{eqn:2dgen}, the product is ordered by increasing phase $\arg (Z_{2d})$, as indicated by `$\curvearrowleft$'. 
Notice, that this gives various matrix identities upon crossing a wall of marginal stability. Finally, the Cecotti-Vafa formula asserts that upon taking the usual matrix-trace over the product $\bS^{(2d)}_{[\vartheta+2\pi, \vartheta+\pi)}\bS^{(2d)}_{[\vartheta+\pi)}$, one ends up with a specialization of the 2d $\cN=(2,2)$ elliptic genus.

\subsubsection{2d-4d systems and their BPS spectra}

A two-dimensional $\cN=(2,2)$ theory coupled to a four-dimensional $\cN=2$ theory may be viewed as a surface defect for the latter~\cite{Hanany:1997vm, Gukov:2006jk,Gukov:2008sn,Gaiotto:2009fs}. Together, they form a ``2d-4d system''. The BPS spectrum of this 2d-4d system is then a natural extension of the 2d and 4d theories by themselves, by including an additional sector of ``2d-4d BPS states''. More precisely, the spectrum of the 2d theory consists of 2d particles (that may be present in the isolated massive vacua of the 2d theory) as well as BPS solitons interpolating between two vacua. The coupling to the 4d theory has the effect of turning on twisted masses for both types of BPS states. We shall call 2d-4d BPS states those 2d BPS solitons with twisted masses having a 4d origin, while we will refer to 2d particles simply as 2d BPS states in the following. Akin to the 2d BPS soliton spectra and 4d BPS spectra, 2d-4d BPS states feature rich wall-crossing phenomena as moduli of the system are varied. This behavior was first described in~\cite{Gaiotto:2011tf}, where an invariant of 2d-4d wall-crossing was also introduced, known as the ``2d-4d spectrum generator”. In~\cite{Cordova:2017ohl}, a 2d-4d refined wall-crossing invariant was defined and it was proposed that (a certain notion of) the trace of the 2d-4d spectrum generator coincides with the Schur index in the presence of a certain surface defects in the 4d $\cN=2$ theory. The conjecture was shown to agree with defects engineered via ``ultraviolet methods'' in~\cite{Cordova:2017ohl, Cordova:2017mhb}  for some explicit examples.

We will focus on 2d-4d systems in the context of class $\CS$ theories, \emph{i.e.} arising from M5-M2 brane engineering~\cite{Witten:1997sc, Gaiotto:2009we, Gaiotto:2009hg, Hanany:1997vm, Gaiotto:2011tf}. 
These systems generally have branches of the moduli space of vacua where the 2d sector has isolated massive vacua. These vacua are determined by a point $u\in \CB$ in the Coulomb branch of the 4d theory, as well as an isolated, non-degenerate, critical point $i$ of the 2d twisted superpotential~\cite{Gaiotto:2011tf}. The BPS spectrum of a 2d-4d system consists of three sectors:
\begin{enumerate}
\item[(i)] {\bf 4d BPS states}, such as monopoles and dyons familiar from Seiberg-Witten theory. They are unaffected by the coupling to the 2d theory and are counted by $\Omega_{n}(\gamma)$ as discussed in detail in Subsection~\ref{sec:IRSchurRev}.
\item[(ii)] {\bf 2d BPS particles} in vacuum $i$ and ``spin'' $\frac{n}{2}$. Their number is counted by the BPS degeneracy $\omega_{i,\gamma, n}$, which is defined by a refinement, proposed by~\cite{Cordova:2016uwk}, of the CFIV index~\cite{Cecotti:1992qh}
\be\label{eqn:2dBPSind}
	\Tr_{\cH_{ii,\gamma}} \left[ (-1)^{F_{2d}} F_{2d}\, q^{{R-M_\perp}} \right] \ = \ \sum_{i\in\IZ} \omega_{i,\gamma, n} \,q^{n/2} \,.
\ee
Here, $F_{2d}$ is the fermion number in two dimensions, $R$ is an R-charge of the 4d theory, while $M_\perp$ is the generator of 4d rotations in the plane transverse to the defect. The combination $R-M_\perp$ generates a ${\mathfrak{u}}(1)$ global symmetry that commutes with the 2d $\CN=(2,2)$ subalgebra of the 4d $\CN=2$ super-Poincar\'e algebra, and therefore appears as a universal flavor symmetry of the 2d theory.
\item[(iii)] {\bf 2d-4d BPS solitons} interpolating between vacua $i$ and $j$ of the 2d theory, and carrying flavor charge $\gamma$. 
Their number is counted by 2d-4d BPS degeneracies $\mu_{ij,\gamma,n}$ defined as follows
\be
	\Tr_{\cH_{ij,\gamma}} \left[ (-1)^{F_{2d}} F_{2d}\, q^{{R-M_\perp}}  \right] \ = \ \sum_{n\in\IZ} \mu_{ij,\gamma, n} \,q^{n/2} \,.
\ee
As we will explain in Subsection~\ref{sec:spins}, it turns out that in class $\CS$ theories of type $A_1$ all 2d-4d solitons have zero spin and therefore only $\mu_{ij,\gamma,n=0}$ will contribute. Moreover, we will always find that this can be either $0$ or $1$ (this is not the case in higher-rank class $\CS$ theories). 
\end{enumerate}
Due to wall-crossing, each of the sectors of the BPS spectrum depends on a choice of moduli of the 2d-4d system, including the 4d Coulomb vacuum and couplings of the 2d twisted superpotential $\tCW$.

The 2d and 4d theories are coupled via gauging a 2d flavor symmetry by 4d vector multiplets, restricted to the surface defect~\cite{Gukov:2006jk,Gukov:2008sn}.\footnote{
See also~\cite{Gaiotto:2009fs} for the $\cN=2$ case,~\cite{Gukov:2014gja} for a nice review, and \emph{e.g.}~\cite{Gadde:2013ftv} for surface defects arising from explicitly coupling the 2d $\CN=(2,2)$ elliptic genus to the 4d $\CN=2$ superconformal index. The defects considered in the latter reference are precisely the ``vortex defects'' considered in~\cite{Gaiotto:2012xa} via the Higgsing procedure (see also Appendix~\ref{App:Higgsing}), and in the present paper via the IR formalism.} 
As a consequence of this coupling, the twisted masses of the 2d theory get identified with central charges of the 4d supersymmetry algebra. 
In fact, the central charges of the three types of BPS states can be schematically described by a general formula
\be\label{eq:central-charges}
	Z  \ = \  Z_{\gamma}+ \tCW_j - \tCW_i \,,
\ee
where $Z_\gamma$ is a 4d central charge, which corresponds to a 2d twisted mass, and $\tCW_i, \tCW_j$ are critical values of the 2d twisted superpotential. For 4d particles the last two terms do not contribute, leaving the standard 4d $\CN=2$ central charge.
For a 2d-4d soliton interpolating between vacua $i$ and $j$ and twisted flavor charge $\gamma$, one retains all three terms. 
A 2d particle in one of the 2d vacua (say $i$) has a pure flavor mass, and thus it can be viewed as an $ii$-soliton with central charge simply given by $Z_\gamma$ as for 4d BPS states. These 2d-4d BPS spectra feature extremely rich wall-crossing phenomena, intertwining 2d wall-crossing~\cite{Cecotti:1992rm} and 4d wall-crossing~\cite{Kontsevich:2008fj} in a highly nontrivial fashion. The full scope of these phenomena was analyzed in detail in~\cite{Gaiotto:2011tf}, which introduced an invariant of 2d-4d wall-crossing known as the ``2d-4d spectrum generator''. 

The 2d-4d IR formula proposed in~\cite{Cordova:2017ohl} relates the 2d-4d spectrum generator to the ``surface defect Schur index''. 
We will need a slight generalization of this setup, to consider several types of surface defects of ``vortex'' type coupled to the same 4d $\CN=2$ theory.
To this end, let us recall a few basic facts about the field theoretic description of 2d-4d systems.
As previously recalled, the coupling of the 2d theory to the 4d theory is realized by identifying a 2d flavor symmetry with a 4d (either gauge or flavor) symmetry.
A good prototype to keep in mind is that of a 2d GLSM with $N$ chiral multiplets transforming in the $N$-dimensional representation of an $SU(2)$ global 2d symmetry (aptly chosen to match the fact that we will couple to an $A_1$ theory of class $\CS$).
We then ``gauge'' this symmetry by coupling the 4d vectormultiplets (their restriction to the surface defect) to the 2d chiral fields, through the standard covariant derivative.
On the Coulomb branch of the 4d theory, the vacuum expectation values of the 4d adjoint scalars give a twisted mass to the 2d chirals.
These are therefore massive and can be integrated out to yield an effective description of the 2d theory, as a low-energy theory of the 2d vector-multiplet.
The gauge-invariant 2d field-strength is dual to a scalar $\sigma$, which is the top component of a twisted chiral multiplet (see e.g.~\cite{Witten:1993yc}).
The low energy theory is then described by a twisted superpotential for $\sigma$, which includes one-loop corrections from the massive 2d chirals.
Standard arguments lead to the conclusion that, generically, $\tCW$ has $N$ massive non-degenerate vacua.
Therefore, in the deep infrared the 2d theory is entirely gapped. The only massless degrees of freedom are Coulomb branch vector multiplets of the bulk theory. 
It can also be argued that,  given a suitable choice of how the 2d and 4d theories are coupled, the twisted vacuum manifold $\partial_\sigma\tCW=0$ can be identified with the Seiberg-Witten curve of the 4d theory~\cite{Gaiotto:2013sma}. This is especially natural in class $\CS$ theories where the 2d-4d system is engineered by M2 branes ending on a stack of M5 branes~\cite{Hanany:1997vm,Gaiotto:2011tf}. 
The key point is that each 2d vacuum gets identified with a certain point of the 4d Seiberg-Witten curve.

Despite the fact that the 2d theory is gapped in the infrared, its presence in a given massive vacuum nevertheless leaves a trace in the bulk (4d) dynamics. The defect sources a holonomy for the 4d vector fields, much like a supersymmetric solenoid whose internal magnetic flux is determined by the 2d vacuum~\cite{Gukov:2006jk}. At very low energies, this may be regarded as an extremely heavy vortex defect for the 4d theory~\cite{Gaiotto:2012xa}. 
In fact, both 2d-4d systems and vortex defects come in families parameterized by representations of $A_1$, whose dimension we will henceforth identify with $N$ (the number of 2d vacua)~\cite{Gaiotto:2012xa,Alday:2013kda,Bullimore:2014nla}.
In the case of 2d-4d systems we propose to identify $N$ with the representation defining the Hitchin spectral curve of the 4d theory, following~\cite{Hanany:1997vm,Hori:2013ewa,Gaiotto:2013sma, Longhi:2016bte}. 

We conclude this quick overview of BPS states in 2d-4d systems with some words on their geometric interpretation in the case of class $\CS$ theories.
Let us fix $N=2$ for the moment, as this is the case most discussed in the existing literature.
As already mentioned, the Seiberg-Witten curve is a double cover of the UV curve, since it arises as  the spectral curve of a Hitchin system.
There is a \emph{canonical} surface defect whose twisted superpotential $\tCW$ is parameterized by a point $z\in C$ and whose critical points are identified with the sheets of $\Sigma$ over $z$.\footnote{In fact, more precisely the Seiberg-Witten differential is identified with $\lambda \sim \sigma \, dz$ and sheets of $\Sigma$ are in 1:1 correspondence with the critical points $\sigma_i(z)$.}
2d solitons are then classified by relative homology classes of open paths connecting the two points/vacua on $\Sigma$. Conversely, it is well-known that charges of 4d BPS states correspond to closed homology cycles $\gamma\in H_1(\Sigma,\IZ)$. Let $a$ denote the path corresponding to a 2d soliton, there is a natural notion of concatenating $a$ and $\gamma$ (as homology classes). 
Therefore, 2d BPS solitons carry naturally 4d charges in 2d-4d systems of this type, and they have a geometric interpretation that is close to that of 4d particles.
 
Regarding the case of $N>2$, we will argue that a similar picture exists. The Seiberg-Witten curve gets replaced by a Hitchin spectral curve computed in the $N$-dimensional (irreducible) representation of $A_1$, and all the conceptual analogies between 2d-4d solitons and paths on this curve carry over. There are some small, but crucial, technical differences that will be discussed in detail below.

\subsubsection{The 2d-4d IR formula}

Thus far, we have reviewed the various types of BPS states that arise in 2d-4d systems. As already mentioned, they have a very rich wall-crossing behavior, which admits a nice description in terms of a 2d-4d wall-crossing formula~\cite{Gaiotto:2011tf}.
To state the formula, we introduce the analogues of 4d KS operators for each type of BPS state:
\begin{itemize}
\item	4d BPS particles are once again associated with Kontsevich-Soibelman operators as in \eqref{eqn:KSfac}, tensored with an $N\times N$ identity matrix:
\be
	K_\gamma^{(4d)} = K\left( q; X_{\gamma}; \Omega_{i}(\gamma) \)  \otimes \mathbbm{1} \,.
\ee
\item
2d-4d BPS solitons carrying 2d charge $ij$ and 4d charge $\gamma$ are associated with Stokes factors similar to those encountered in the context of 2d wall-crossing (recall (\ref{eq:2d-wall-crossing-operators})):
\be
\label{eqn:K2d4d}
	S^{(2d-4d)}_{ij;\gamma} \ = \ \mathbbm{1} - \sum_{k \in \mathbb{Z}} \mu_{ij,\gamma,k}  (-1)^{k}q^{\frac{k}{2}}X_{ij,\gamma} \otimes e_{ij} \,.
\ee
\item
2d BPS particles in a 2d vacuum $i$, and carrying 4d charge $\gamma$, are associated with `$K$' factors similar to those encountered in the context of 2d theories (recall (\ref{eq:2d-wall-crossing-operators})):
\be\label{eqn:K2d}
	K^{(2d)}_{\gamma}  
	\ = \ \sum_{i=1}^N \ K^{(2d)}_{ii;\gamma}
	\ = \ \sum_{i=1}^{N} \prod_{k\in \mathbb{Z}}\left( 1-(-1)^{k} q^{\frac{k}{2}} X_{\gamma} \right)^{-\omega_{i, \gamma,k} }  \otimes  e_{ii} \,.
\ee
\end{itemize}

The 2d-4d wall-crossing formula that governs the interactions among BPS states of 2d-4d systems can be succinctly stated as the invariance property of a ``quantum 2d-4d BPS spectrum generator''. 
The latter is defined as follows
\be\label{eqn:2d4dqgen}
	\bS^{2d-4d}_{\sphericalangle} \ = \  \prod_{ij;\gamma | \arg\left( Z \right) \in \sphericalangle}^{\curvearrowleft} S^{(2d-4d)}_{ij;\gamma} K^{(2d)}_{i,\gamma} K^{(4d)}_{\gamma} \,,
\ee
where factors are ordered by increasing phase of central charges towards the left.
As long central charges remain confined within the angular sector $\sphericalangle$, the wall-crossing formula asserts that $\bS^{2d-4d}_{\sphericalangle}$ is invariant.

Lastly, with the 2d-4d quantum spectrum generator in hand, we may now turn towards the corresponding 2d-4d IR formula. The conjecture in~\cite{Cordova:2017ohl} is that the Schur index in the presence of a surface defect, $\cI_{\mathbb{S}}$, is given by
\be\label{eqn:2d4dIRformula}
	\cI_{\mathbb{S}} (q) \ = \ \left( q;q \right)_{\infty}^{2r} \, \Tr \left[    \bS_{[\vartheta+2\pi, \vartheta+\pi)}^{2d-4d} {\bS}_{[\vartheta+\pi, \vartheta)}^{2d-4d} \right] \,.
\ee
The 2d-4d quantum spectrum generators are $N\times N$ matrices valued in the quantum torus algebra, and therefore the trace is a combination of the usual matrix-trace with the trace $\hTr$ of the quantum torus algebra, introduced in equation~\eqref{eqn:hTr}. In practice, we may perform the former followed by the latter. 

Let us point out that starting from the IR, we may couple various types of 2d theories, which then ought to correspond to different types of surface defects. The proposed mapping of IR and UV surface defects in~\eqref{eqn:2d4dIRformula} is rather striking, but it may not always be straightforward to identify the UV defect. As we shall see in the following, we modify the UV-IR prescription~\eqref{eqn:2d4dIRformula} by introducing surface defects that arise by taking the zero-length limit of a supersymmetric 1d interface in the 2d IR theory. By doing so, we can naturally match an infinite family of IR defects, which we show to be rather universal for class $\CS$ theories,\footnote{
In the sense that they can be expressed as universal operators acting on the Schur index of an $A_1$ class $\CS$ theory associated to a UV curve with at least one regular puncture.
} to the ``vortex surface defects'' of~\cite{Gaiotto:2012xa}, coming from the infinite tension limit of the vortex strings.

Finally, as in the purely 4d case, it is usually convenient to move to a particular point in the moduli space and evaluate the 2d-4d spectrum there. We shall mostly focus on the ``Roman locus'', where the 4d and 2d BPS states align with critical phase $\vartheta=\vartheta_c$, while the 2d-4d BPS solitons are at arbitrary phases (see Figure~\ref{fig:2d-4d-central-charges}). In particular, this removes the issue of ordering of the various contributions in~\eqref{eqn:2d4dqgen}. For instance, the full 2d-4d quantum generator can be written as
\be
	{\bS}_{\vartheta_c\pm\pi/2}^{2d-4d}  \ = \  \bS^{(2d-4d)}_{[\vartheta_c+\pi/2, \vartheta_c)}  \, \bS_{\vartheta_c}^{(2d),(4d)} \, \bS_{[\vartheta_c, \vartheta_c-\pi/2)}^{(2d-4d)} \,,
\ee
where the separate and commuting 2d and 4d contributions $\bS_{\vartheta_c}^{(2d),(4d)}$ are given by a separate product of the 2d Cecotti-Vafa factors,~\eqref{eqn:K2d}, and 4d Kontsevich-Soibelman factors,~\eqref{eqn:KSfac}, at critical phase $\vartheta_c$, while $\bS^{(2d-4d)}_{[\vartheta_c, \vartheta_c-\pi/2)}$ is the phase-ordered product over the 2d-4d factors,~\eqref{eqn:K2d4d}.

\section{2d-4d systems of $A_1$ class $\CS$ theories and vortex defects}\label{sec:BPS-spectrum}

In this section we explain how to use spectral networks to completely solve the BPS spectrum of the 2d-4d systems introduced in the previous section, and compute the quantum 2d-4d spectrum generator explicitly. This relies on a geometric interpretation of various quantities that were defined algebraically in the previous section. In fact, this is a much more convenient approach for certain types of computations.

\subsection{Spectral curves in higher symmetric representations}

Let us start by considering the spectral curve of an $A_1$ Hitchin system~\cite{Hitchin} in the $N$-dimensional representation 
\be\label{eq:A1-curve}
	\Sigma_N \ : \ {\det}_{N}\(\lambda - \varphi(z)\)  \ = \  0  \ \subset \ T^*C \,.
\ee
Here $\lambda$ is the Liouville one-form that plays the role of a coordinate on the fiber, and $\varphi$ is the Higgs field. An important property of this curve is that it factorizes into $\lfloor \frac{N+1}{2} \rfloor$ components. In fact, when $N=2$ the curve is simply $\lambda^2 + \phi_2 = 0$, and the Higgs field is determined by a quadratic differential $\phi_2$ (up to conjugation, and within contractible patches away from branch points)
\be
	\varphi(z)  \ = \  \(\begin{array}{cc} \sqrt{-\phi_2} & 0 \\ 0 & -\sqrt{-\phi_2} \end{array}\)\,.
\ee
Then, the sheets of the $N$-dimensional representation are given by
\be
	\rho_N\(\varphi(z)\)  \ = \  \(\begin{array}{ccccc} 
	{(N-1)}\sqrt{-\phi_2} & & & & \\
	& {(N-3)}\sqrt{-\phi_2} & & & \\
	& & \ddots & & \\
	& & & -{(N-3)}\sqrt{-\phi_2} &  \\ 
	& & & & -{(N-1)}\sqrt{-\phi_2}
	\end{array}\) \,.
\ee
The spectral curve factorizes into several pieces. Depending on the parity of $N$, these take the following form
\beaa\label{eq:curve-factorization}
	\Sigma_N \ : \ \ \prod_{k=1}^{N/2} \(\lambda^2 + (2k-1)^2 \phi_2\)  & \ = \  0 \ \sim \ \Sigma^{(1)} \times \Sigma^{(3)} \dots \Sigma^{(N-1)} \,, && \text{for} \quad N \in 		2\IN \,, \\
	\Sigma_N \ : \ \ \lambda \prod_{k=1}^{(N-1)/2} \(\lambda^2 + (2k)^2  \phi_2\)   & \ = \  0  \ \sim \ \Sigma^{(0)} \times \Sigma^{(2)} \dots \Sigma^{(N-1)}
	\,, && \text{for} \quad	N \in 2\IN+1 \,.
\eeaa
Each component $\Sigma^{(k)}$ is a two-sheeted cover of the UV curve $C$, except for $\Sigma^{(0)}$ which simply corresponds to the zero section of $T^*C$.

The branching locus for each of these covers is the set of points $z\in C$ where two sheets coincide. This requires $\phi_2(z)=0$, and therefore at a branch point \emph{all} sheets coincide at $\lambda = 0$. Despite the fact that all sheets coalesce together, their branching structure is very simple. It just consists of a Weyl reflection that takes $\lambda\to-\lambda$, due to the fact that each sheet is proportional  to $\pm\sqrt\phi_2$. Thus, we conclude that branching preserves the factorization of $\Sigma_N$, by acting as an involution on the $\Sigma^{(k)}$ components, while leaving $\Sigma^{(0)}$ fixed in the case when $N$ is odd.

\subsection{Spectral networks in higher symmetric representations}\label{sec:networks}

Spectral networks for higher symmetric representations of $A_1$ have not been investigated in previous literature, therefore we take a short detour and learn a few basic facts about them. We propose a natural definition, slightly generalizing~\cite{Gaiotto:2012rg}, which is in part based on ideas in~\cite{Longhi:2016rjt, Longhi:2016bte}. The main novelty is the appearance of ``collinear vacua'' for the 2d theory. To explain what this means, let $\Lambda_N$ be the weight system of the $N$ dimensional representation of $A_1$. We shall further order weights $\nu_i\in \Lambda_N$, such that
\be
	\nu_{i+1} - \nu_{i}  \ = \  \alpha \,,
\ee
where $\alpha$ is the positive root of $A_1$. 

The spectral curve $\Sigma_N$ arises as the manifold of vacua of a 2d-4d system. To describe this, one may choose a weak coupling regime for the class $\CS$ theory, and place the defect on one of the tubes in the pairs-of-pants decomposition of $C$. The 2d theory living on the surface defect consists of a $U(1)$ GLSM with a charged chiral multiplet, which is further coupled to 4d fields by  transforming in the $N$ dimensional representation of the $SU(2)$ gauge symmetry associated to the tube. 
We may likewise describe a surface defect placed near a puncture, as the decoupling limit of this setup where the tube becomes infinitely long~\cite{Gaiotto:2009we}. In that case, (part of) the 2d global symmetry is identified with the 4d flavor symmetry associated with the respective puncture.

At each $z\in C$ there are $N$ vacua corresponding to the sheets of $\Sigma_N$. These are the roots of~(\ref{eq:A1-curve}), and therefore correspond to points in the cotangent fiber $T^*_z C$ determined by
\be\label{eq:lambda-multiplicity}
	\lambda_i(z)  \ = \  \langle\nu_i ,\varphi(z)\rangle  \ = \  (N-2i+1)\sqrt{-\phi_2}\,.
\ee
Thus, it follows that the difference
\be
	\lambda_{i+1}(z) - \lambda_{i}(z)  \ = \  \langle \alpha,\varphi(z)\rangle \,, \qquad\forall i=1,\dots, N-1 \,,
\ee
is independent of $i$.

Recall that soliton charges in 2d systems are classified by pairs of vacua $ij$, whereas in 2d-4d systems the 2d charge lattice is extended by 4d gauge and flavor charges $\gamma\in H_1(\Sigma,\IZ)$. Since the splitting $(ij,\gamma)$ is not canonical and only locally well-defined on the moduli space $\CB\times C$ ($\CB$ is the Coulomb branch of the 4d theory), it is more appropriate to switch to a description of soliton charges in terms of open paths $a$ classified up to relative homology.  
The appropriate charge lattices for soliton charges are often denoted by 
\be
	\Gamma_{ij}(z) \ \coloneqq \ H_1^{\rel}(\Sigma, \IZ; \lambda_i(z),\lambda_j(z))\,.
\ee
We will mostly omit the explicit reference to a point on the curve $C$ unless necessary.

\begin{figure}[h!]
\begin{center}
\includegraphics[width=0.45\textwidth]{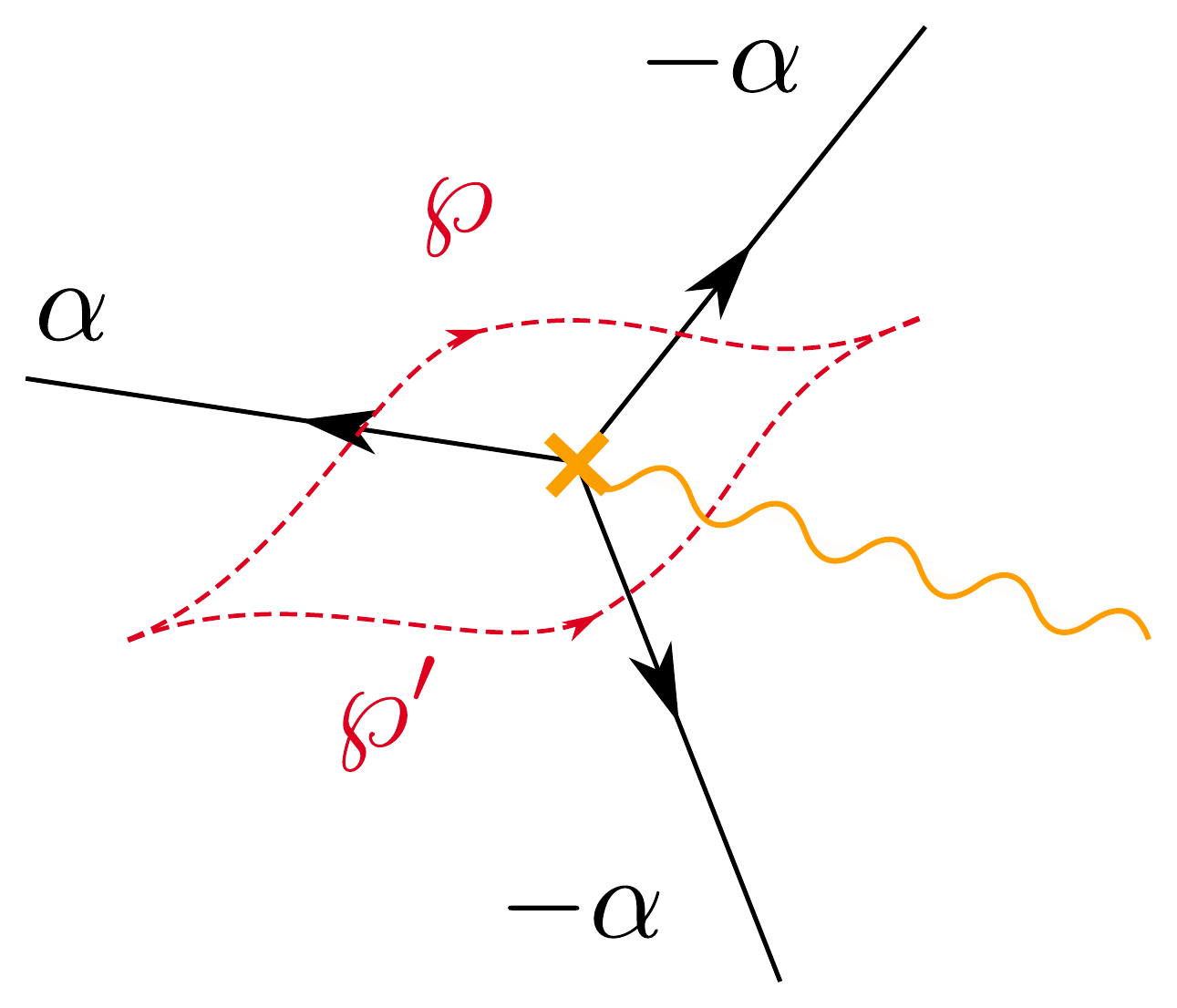}
\caption{A branch point with three emerging $\CS$-walls. Two homotopic paths $\wp,\wp'$ are also shown.}
\label{fig:branch-point-walls}
\end{center}
\end{figure}

Following~\cite{Gaiotto:2012rg, Longhi:2016rjt}, we define the spectral network as the web of trajectories on $C$ defined by the differential equation
\be\label{eq:S-wall}
	\(\partial_t , \langle\alpha,\varphi(z)\rangle\) \ \in \ e^{\ii\vartheta} \IR_+\,.
\ee
Here, $\partial_{t}$ is the vector field along the trajectory and by $(\cdot,\cdot)$ we denote the canonical inner product between vectors and one-forms in the fibers $T_{z}^{}C$ and $T_{z}^{*}C$. This differential equation is integrated starting from branch points of the covering $\Sigma_N\to C$, where $\langle\alpha,\varphi(z)\rangle=0$, to produce a web of trajectories $\CW$, dubbed ``spectral network''. Trajectories satisfying this condition are called $\CS$-walls. Their shape depends on a point $u\in \CB$ as well as on the phase $\vartheta$. Thus, we will denote the spectral network by $\CW(u,\vartheta)$. Since the location of branch points for $\Sigma_N$ is exactly the same as the branch points of $\Sigma_2$, and since the differential equation of $\CS$-walls is also independent of $N$, it follows  that the geometric shape of $\CW(\vartheta,u)$ is the \emph{same} for all $N\geq 2$. Near a branch point, where $\langle \varphi(z),\alpha\rangle = 0$, three walls of types $\pm\alpha$ emerge (see Figure~\ref{fig:branch-point-walls}). Note that in the case of $A_1$ networks (for arbitrary $N$) there are only two roots $\pm\alpha$, and this implies that two $\CS$-walls can never intersect transversely.

\begin{figure}[h!]
\begin{center}
\includegraphics[width=0.45\textwidth]{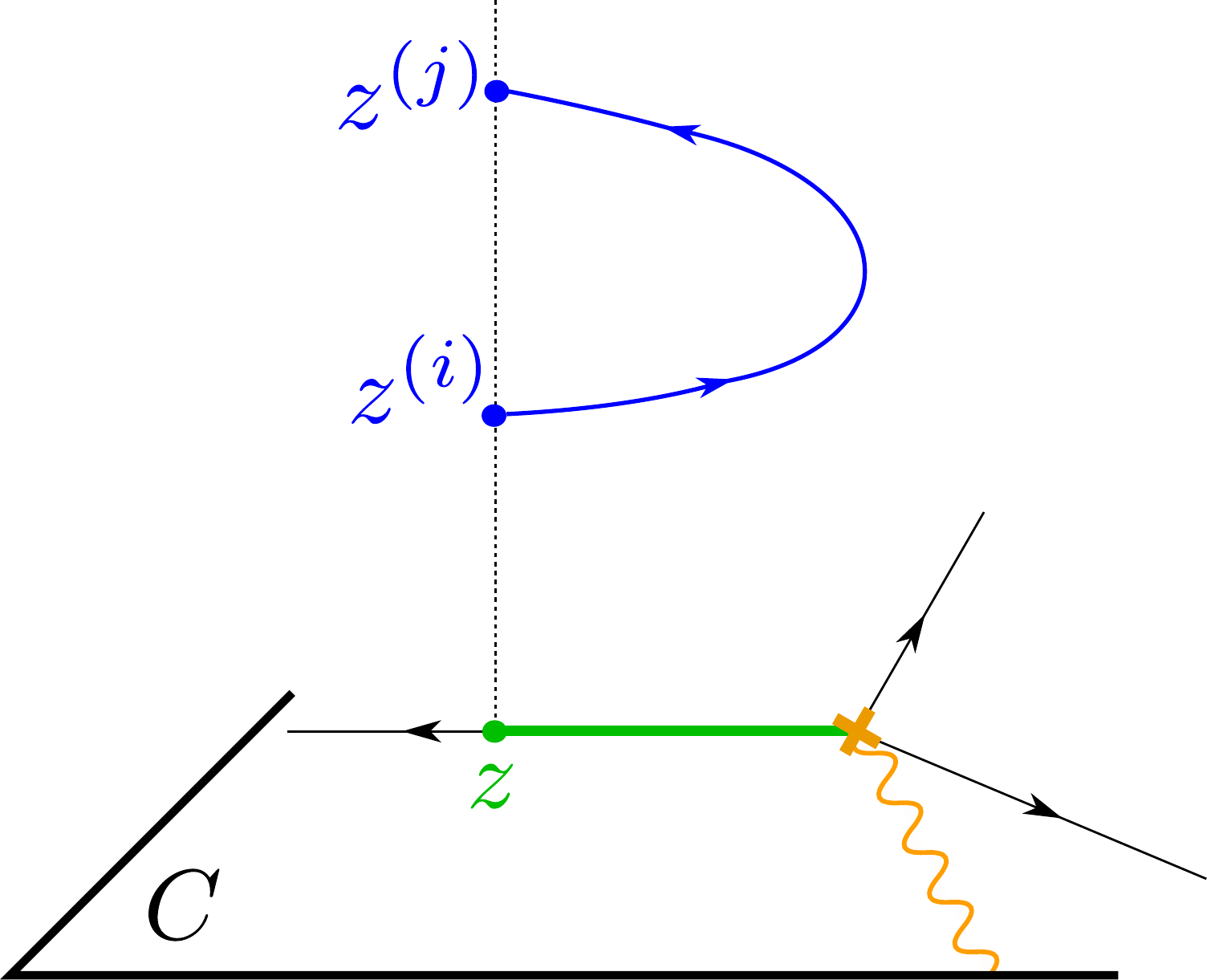}
\caption{The unique BPS soliton carried by an $\CS$-wall in the case of $N=2$.}
\label{fig:simpleton}
\end{center}
\end{figure}

There is another important piece of data that comes attached to a spectral network; namely the spectrum of 2d-4d BPS solitons of a surface defect placed at any point along an $\CS$-wall. The soliton data is encoded by a certain $N\times N$ Stokes matrix associated to each $\CS$-wall~\cite{Cecotti:1991me, Cecotti:1992rm, Dubrovin:1992yd, Gaiotto:2009hg, Gaiotto:2011tf, Gaiotto:2012rg}. For $\CS$-walls of $A_1$ theories with a surface defect in the fundamental representation ($N=2$), the soliton data consists of a single BPS state on each $\CS$-wall. Its charge is the relative homology class of a path that arises by lifting the wall to both sheets of $\Sigma$, and concatenating them at the branch point where the $\CS$-wall begins (see Figure~\ref{fig:simpleton}). As a consequence, for $N=2$, the Stokes matrices of $\CS$-walls are simply given by
\be\label{eq:detour-old}
	S_{\alpha}  \ = \  \(\begin{array}{cc} 1 & X_a \\ 0 & 1\end{array} \)\,,
	\qquad
	S_{-\alpha}  \ = \  \(\begin{array}{cc} 1 & 0 \\ X_b & 1\end{array} \)\,,
\ee
if $a$, $b$ is the soliton carried by an $\CS$-wall of type $\alpha$, $-\alpha$ respectively, \emph{i.e.}
\be
	a \in \Gamma_{12}(z) \,, \qquad \text{and} \qquad
	b\in \Gamma_{21}(z)\,.
\ee

Now, for $N>2$, the soliton data of each $\CS$-wall has not been determined, and it will be our task to take this step in the following. To this end, we start by recalling that the soliton data is uniquely fixed by a flatness constraint for a flat connection on $C$, which is conveniently characterized in terms of its parallel transport. 
Let $C\setminus\CW$ be the complement of the spectral network on $C$. Then, the parallel transport is diagonal within each connected component of the complement, and gets corrected by transition functions across $\CS$-walls. The transition functions are precisely the Stokes matrices. Demanding flatness of the transport enforces certain equations that determine these matrices, and consequently determine the soliton data encoded by them~\cite{Gaiotto:2012rg}. We shall take this as a principle, and use it to derive the soliton data for surface defects described by spectral curves for higher dimensional representations.

The only important constraint comes from homotopy invariance across a branch point. Consider two homotopic paths $\wp,\wp'$  as in Figure~\ref{fig:branch-point-walls}. Since the spectral curve factorizes into two-sheeted covers, it follows immediately that the parallel transport decomposes into a direct sum, with $2\times 2$ block-diagonal pieces corresponding to each component of the curve (for odd $N$, there is also a trivial diagonal entry corresponding to the ``zeroth sheet''). For this reason, the analysis on each component is identical to the $N=2$ case, which was studied in~\cite{Gaiotto:2012rg}. We will not repeat the analysis here but simply state the result in the following.

The Stokes matrices for $\CS$-walls of type $\alpha$ are 
\be\label{eq:S-wall-spectrum-1}
	S_\alpha 
	 \ = \  \mathbbm{1} + {\sum_{i=1}^{\lfloor N/2 \rfloor}} X_{a_{i\bar \imath}} e_{i\bar \imath}
	 \ = \   {\prod_{i=1}^{\lfloor N/2 \rfloor}} \( \mathbbm{1}  + X_{a_{i\bar \imath}} e_{i\bar \imath} \) \,,
\ee
where,  as before, $e_{ij}$ denotes the matrix with a single non-zero entry (equal to $1$) in row $i$ and column $j$, and $a_{ij}$ is soliton charge belonging to $\Gamma_{ij}(z)$.
Furthermore, we define 
\be\label{eqn:barimath}
	\bar \imath  \ \coloneqq \  N+1-i \,,
\ee
as the index of the weight related to $\lambda_i$ by a Weyl reflection 
\be
	\lambda_{\bar \imath}  \ = \  w_\alpha \circ \lambda_i \,, \qquad \bar \imath\geq i\,.
\ee 
If $N$ is odd, the zero-weight with labels $i=\bar \imath=(N+1)/2$ is (automatically) omitted from the sum.
Similarly, an $\CS$-wall of type $-\alpha$ has a Stokes matrix of the form
\be\label{eq:S-wall-spectrum-2}
	S_{-\alpha} 
	 \ = \  \mathbbm{1} + {\sum_{i=1}^{\lfloor N/2 \rfloor}} X_{a_{\bar \imath i}} e_{\bar \imath i} 
	 \ = \  {\prod_{i=1}^{\lfloor N/2 \rfloor}} \( \mathbbm{1} + X_{a_{\bar \imath i}} e_{\bar \imath i} \)\,.
\ee

In terms of 2d-4d BPS solitons, these Stokes matrices signal the presence of a soliton interpolating between vacua $i$ and $\bar \imath$ with charge $a_{i \bar \imath}$ or $a_{\bar \imath i}$.

\subsection{Universal features of 2d-4d BPS spectra from BPS graphs}\label{eq:universality}

\begin{figure}[h!]
\begin{center}
\includegraphics[width=0.9\textwidth]{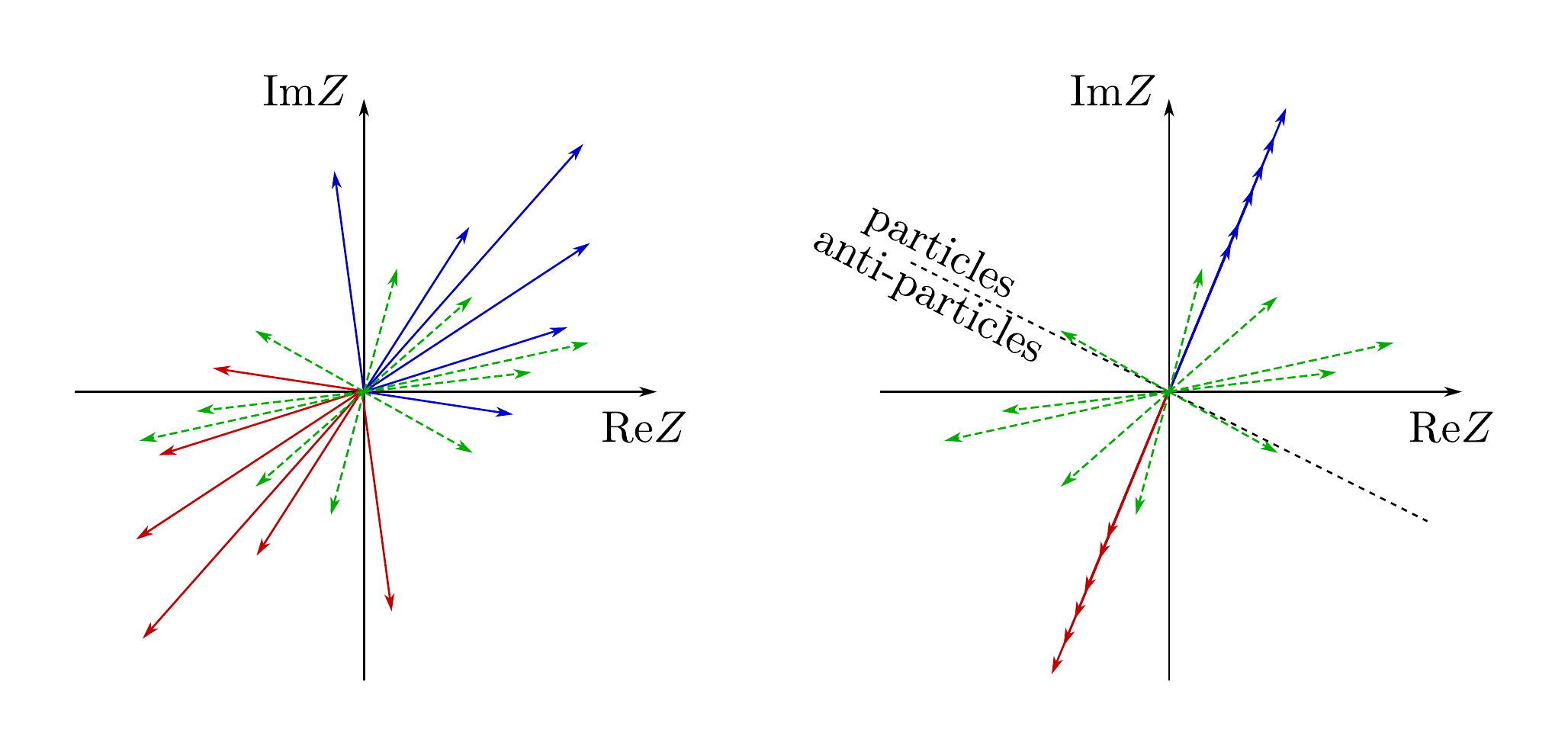}
\caption{We depict an example of the central charges of BPS states in a 2d-4d system, on the left, at a generic point in the Coulomb branch of the 4d theory, and on the right at the Roman locus. In the latter case the 4d central charges define a choice of half-plane (dashed black line), centered around the critical ray, which distinguishes 2d-4d particles from their CPT conjugates. We use solid arrows to denote central charges of the 4d theory, whereas dashed arrows are those of 2d-4d BPS solitons.}
\label{fig:2d-4d-central-charges}
\end{center}
\end{figure}

In order to make general statements about the BPS spectra of 2d-4d systems of rank-one class $\CS$ theories, we require full knowledge of the three types of BPS states, \emph{i.e.} 2d particles, 4d particles and 2d-4d solitons. Pursuing this question at generic points on the moduli space of vacua is rather challenging in general. However, as our ultimate goal is to compute the 2d-4d spectrum generators, we are free to make a convenient choice of vacuum in which to compute the spectrum. This is good enough, as the final answer is guaranteed to be wall-crossing invariant. We shall work at a special locus in the moduli space of vacua known as the ``Roman locus''~\cite{Gabella:2017hpz}, which is characterized by a choice of Coulomb branch moduli of the 4d theory such that the 4d BPS states have central charges with a \emph{common} phase. In fact, this condition implies that 4d BPS states are at marginal stability, and thus the (vanilla) 4d BPS spectrum is ill-defined.  Nevertheless, we are still free to tune the 2d moduli; a generic choice will ensure that central charges of 2d-4d states are phase-resolved, thanks to contributions from the 2d superpotential to~(\ref{eq:central-charges})~\cite{Longhi:2016wtv}. The configurations of central charges for 4d and 2d-4d states are depicted in Figure~\ref{fig:2d-4d-central-charges}. At the Roman locus all 4d central charges align along a critical ray whose phase we will denote as $e^{\ii\vartheta_c}$. Note that because of~(\ref{eq:central-charges}) (and in particular the discussion below it), the central charges of 2d BPS particles also fall along this ray. However, the central charges of 2d-4d states will instead (generically) lie off the critical ray.\footnote{As shown in Figure~\ref{fig:2d-4d-central-charges}, by CPT symmetry, each particle (or soliton) comes with a conjugate one whose central charge has the opposite sign.}

Thus, we conclude that at the Roman locus the 2d-4d spectrum is well-defined, while the 4d spectrum alone is ill-defined. This is not an issue since the 4d spectrum generator still makes sense at the Roman locus~\cite{Longhi:2016wtv}, and encodes all the information we need about 4d BPS states to compute the 2d-4d spectrum generator. In fact, at the Roman locus the spectral network becomes maximally degenerate, and defines a so-called ``BPS graph''~\cite{Gabella:2017hpz}. 
It was shown in~\cite{Longhi:2016wtv}, that this graph encodes the 4d spectrum generator and explicit formulae to compute this object were provided. Thus, by working at the Roman locus, we can regard the 4d contributions to the spectrum generator as given, and solely focus on the contributions of 2d particles and 2d-4d solitons.\footnote{A general construction of the 2d-4d spectrum generator for $A_1$ theories of class $\CS$ in the case  $N=2$ was already given in~\cite{Longhi:2012mj}. Here, we will explain how BPS graphs can further be employed to generalize this construction to larger $N>2$, and in fact to higher-rank theories of class $\CS$ as well.} For this purpose, we resort to the powerful geometric  framework of spectral networks.

The UV curve $C$ of a class $\CS$ theory can be interpreted as the moduli space of couplings for the surface defect theory. For example, in certain cases a local coordinate on $C$ may be identified with the Fayet-Ilioupolos coupling of a 2d $U(1)$ GLSM~\cite{Gaiotto:2009fs, Gaiotto:2011tf}. Given a surface defect with coupling $z\in C$, its 2d-4d soliton spectrum can be computed using spectral networks as follows.\footnote{Notice, that we will often say that the defect is ``placed'' at $z\in C$. This has its origin in the M5-brane setup, in which the surface defects  (treated in the present paper) are codimension-four, arising from M2-branes intersecting the M5-branes.} Recall from Section~\ref{sec:networks}, that the shape of spectral networks depends on a phase $\vartheta$. As we vary this phase between $0$ and $2\pi$, $\CS$-walls move around on the UV curve $C$, and some of them will swipe across $z$. The spectrum of 2d-4d BPS solitons on the surface defects is the union of ``soliton data''  carried by each of these walls~\cite{Gaiotto:2011tf, Gaiotto:2012rg, Longhi:2016bte}.

In general, the shape of the network at a given phase can be extremely complicated, and its evolution with varying $\vartheta$ can be rather hard to study. Moreover, this information depends on the global geometry of $\Sigma_N$, which changes from theory to theory. Therefore, it would seem that using spectral networks to make general statements about 2d-4d solitons is rather daunting. Instead, we will argue in the following that the problem becomes fully tractable when working at the Roman locus, thanks to some very special properties that spectral networks acquire there. 

\begin{figure}
    \centering
    \begin{subfigure}[b]{0.24\textwidth}
        \includegraphics[width=\textwidth]{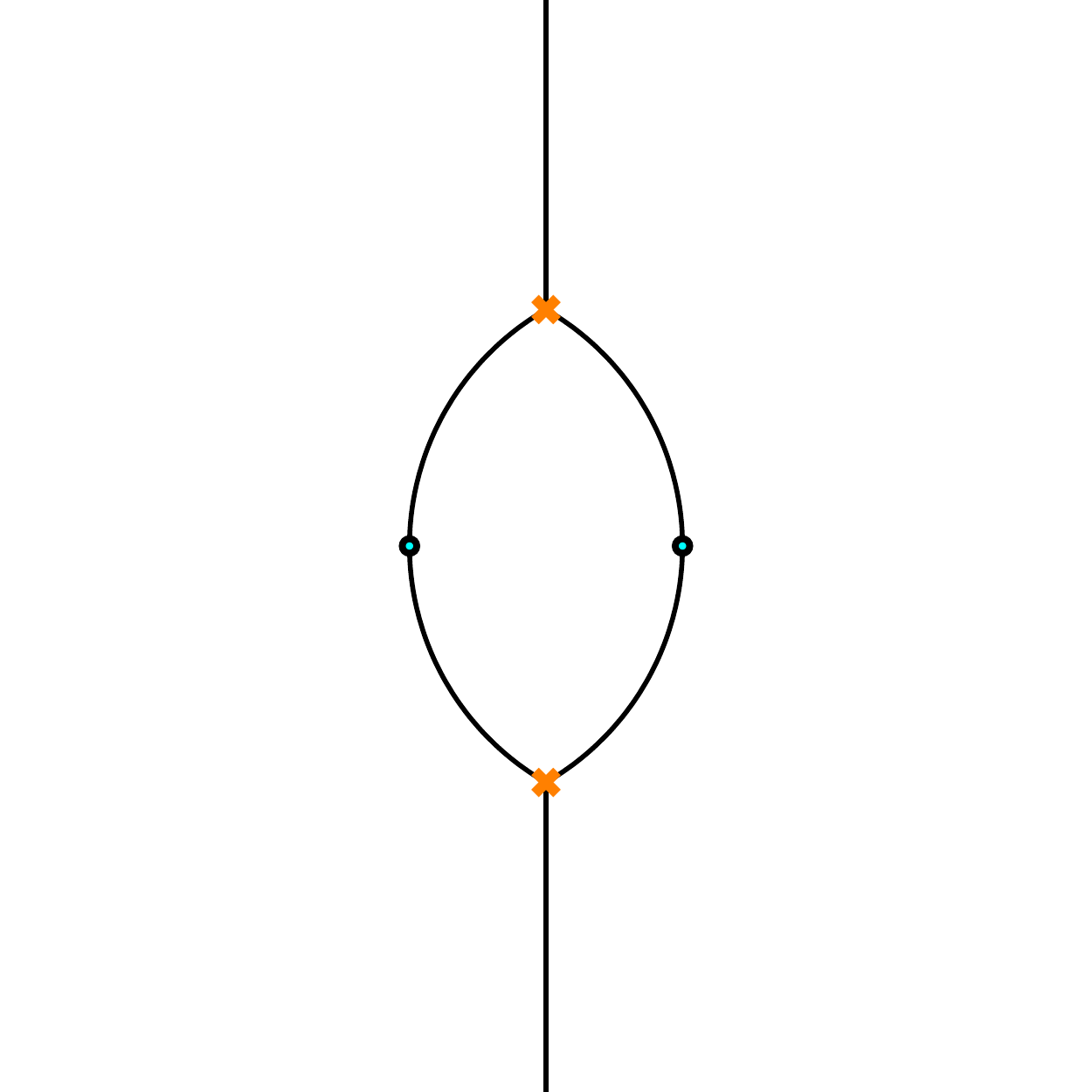}
        \caption{$\vartheta=\vartheta_c-\pi/2$}
    \end{subfigure}
    \begin{subfigure}[b]{0.24\textwidth}
        \includegraphics[width=\textwidth]{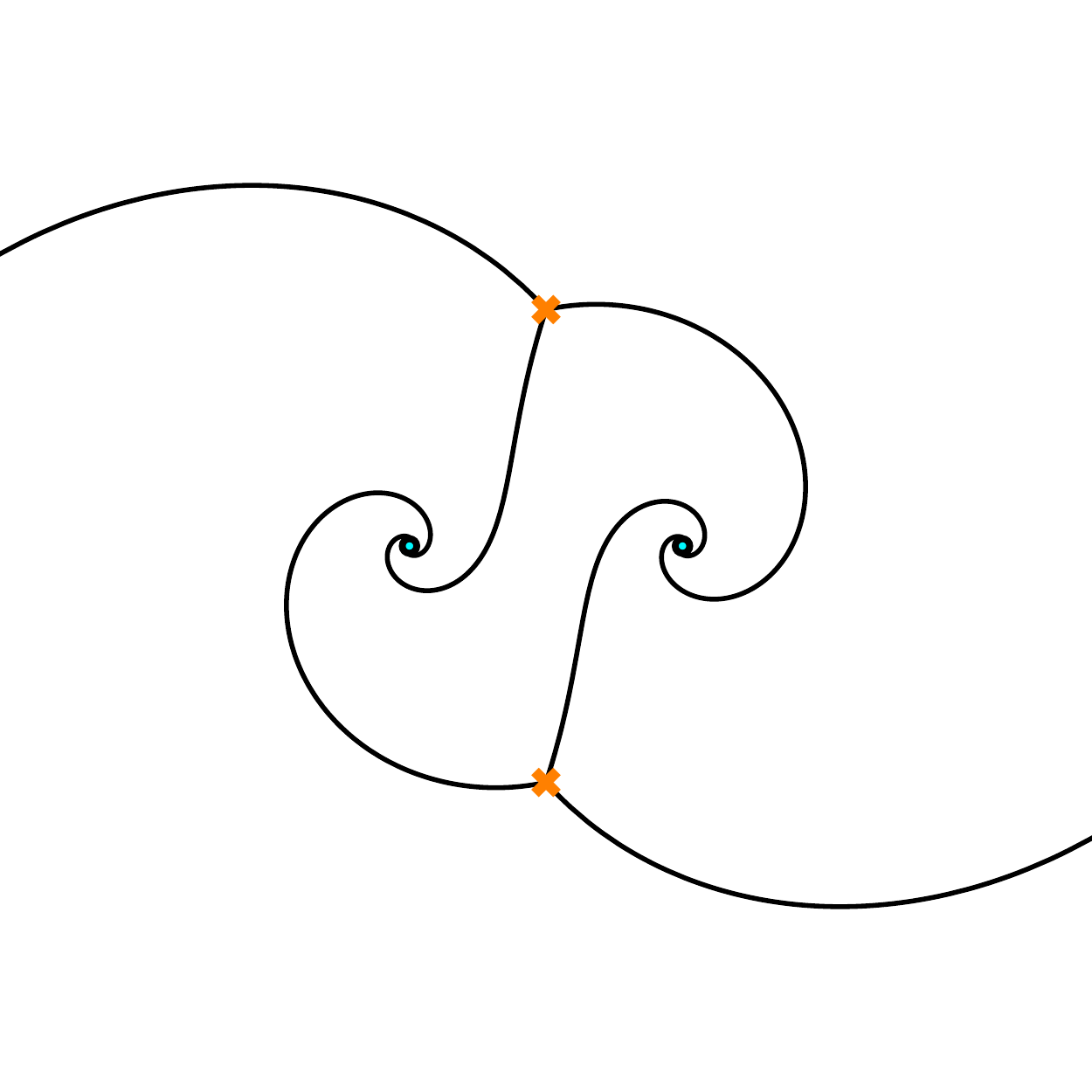}
        \caption{$\vartheta<\vartheta_c$}
    \end{subfigure}
    \begin{subfigure}[b]{0.24\textwidth}
        \includegraphics[width=\textwidth]{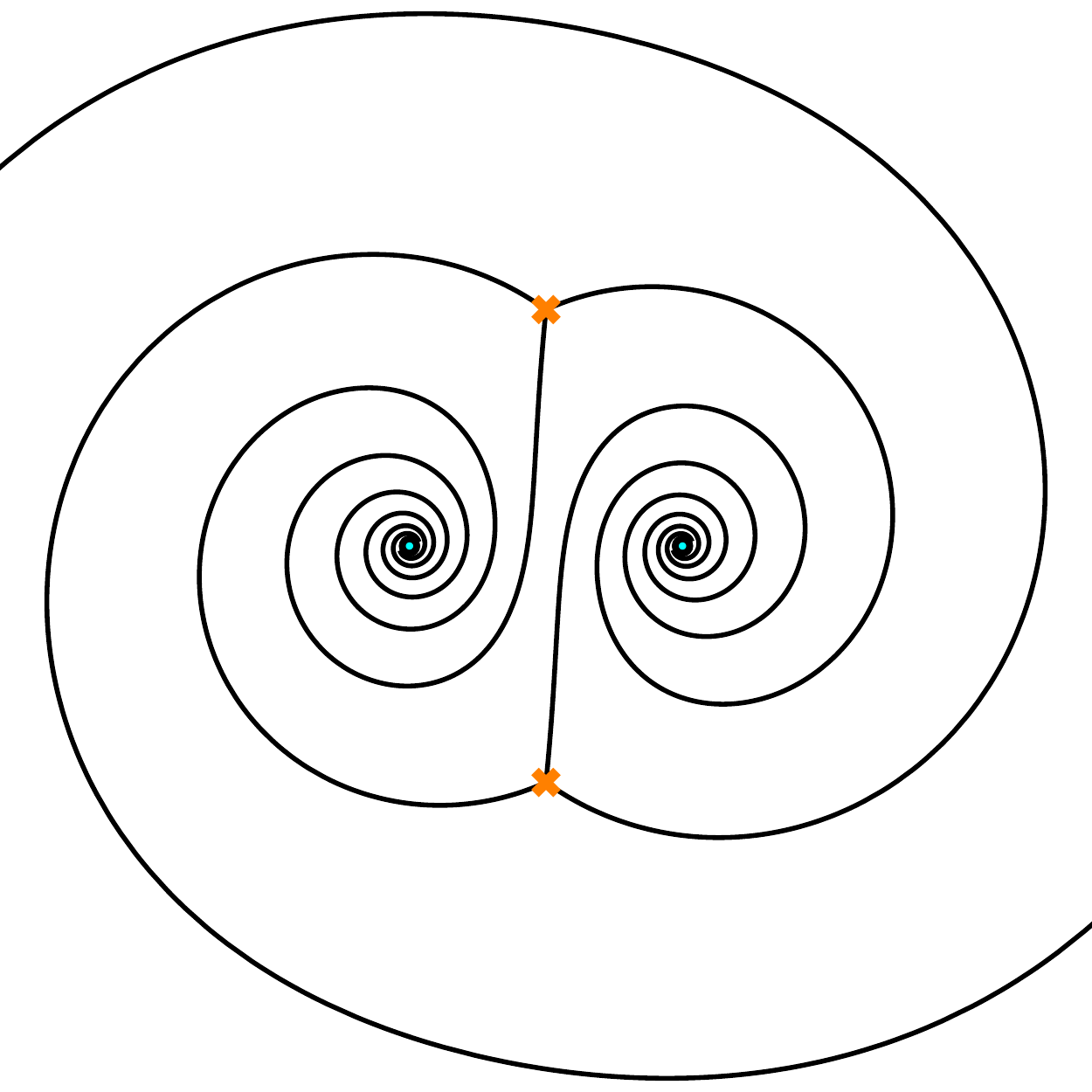}
        \caption{$\vartheta\lesssim\vartheta_c$}
    \end{subfigure}
    \begin{subfigure}[b]{0.24\textwidth}
        \includegraphics[width=\textwidth]{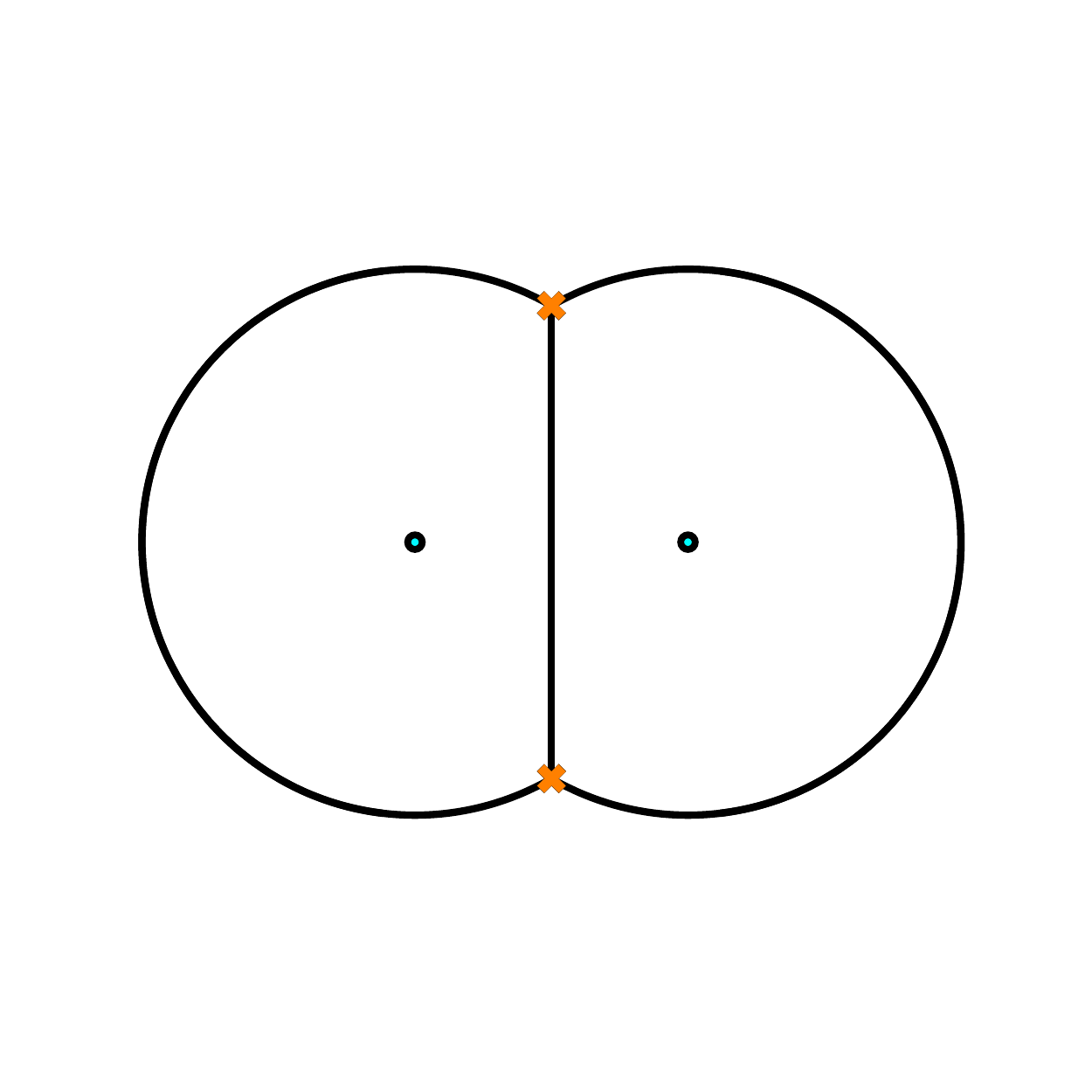}
        \caption{$\vartheta=\vartheta_c$}
    \end{subfigure}
    \caption{$A_1$ spectral networks on the three-punctured sphere. As the phase reaches $\vartheta_c$, all $\CS$-walls collapse into three double-walls. This is the BPS graph of the $T_2$ (trinion) theory.}
    \label{fig:T2-graph}
\end{figure}

Recall from equation~(\ref{eq:A1-curve}), that the Seiberg-Witten differential of an $A_1$ theory of class $\CS$ is determined by a single quadratic differential $\phi_2$ on $C$, for any choice of $N$. For class $\CS$ theories of type $A_1$, the moduli space of quadratic differentials on a punctured Riemann surface is identified with the Coulomb branch moduli space augmented by the moduli space of UV masses, which correspond to the residues of $\sqrt\phi_2$ at punctures. We will restrict to quadratic differentials with simple zeroes and double poles at punctures, corresponding to ``regular'' singularities. Strebel proved the existence of a locus -- the so-called ``Strebel locus'' -- on the moduli space of these differentials, where all critical leaves of the foliation of $e^{-2 \ii \vartheta_c}\phi_2(u)$ (the $\CS$-walls of $\CW(u,\vartheta_c)$) are compact~\cite{Strebel}. Compactness of the leaves means that all $\CS$-walls end on branch points, and thus the spectral network is highly degenerate at the Strebel locus. All walls are ``double-walls'' made of anti-parallel trajectories,  and the topology jumps abruptly from  $\vartheta_c-\epsilon$ to $\vartheta_c+\epsilon$ (see Figure~\ref{fig:T2-graph} for an example).\footnote{An interesting application of spectral networks at the Strebel locus was found in~\cite{Hollands:2013qza}, where it was showed that they encode Fenchel-Nielsen coordinates for moduli spaces of flat connections on $C$.} Moreover, it follows from the work of Liu~\cite{Liu}, that it is always possible to choose moduli such that $\vartheta_c$ is the \emph{only} critical phase of the foliation, if the Riemann surface has at least one puncture. 
That is, there is a sub-locus of the Strebel locus where all central charges of the 4d theory have the same phase, up to a sign. This is precisely the definition of  the Roman locus, whose existence is therefore guaranteed for $A_1$ theories of class $\CS$ (with the above-mentioned restriction to ``full'' punctures).\footnote{The same conclusion can be reached on physical grounds; class $\CS$ theories of type $A_1$ are \emph{complete theories} in the sense of~\cite{Cecotti:2011rv}, therefore there must exists a choice of moduli that realizes alignment of phases of central charges. This implies that the Roman locus is not empty.}

The results of Strebel and Liu are in fact much stronger than just predicting the existence of BPS graphs. They also imply that there exists a choice of moduli for $\phi_2$ such that the BPS graph takes certain shapes. We will take advantage of this prediction as follows. Let us fix a choice of puncture on $C$ and assume that the BPS graph encloses this puncture by two double-walls connecting two branch points. For example in Figure~\ref{fig:T2-graph}, this is true for all three punctures (recall that one puncture is at infinity). This is guaranteed to be possible if $C$ has at least two punctures, as will be argued in detail in Appendix~\ref{sec:more-punctures}. We will also deal with the case of surfaces with a single puncture later in Section~\ref{sec:one-puncture}.
Moreover, while these theorems concern the spectral network at the critical phase $\vartheta_c$, where the BPS graph appears, they also say something about the behavior of the network at other phases.
In particular, we are interested in the \emph{local} behavior of $\CS$-walls near the puncture of choice, for all values of $\vartheta$. We claim that this behavior is universal, and an argument for this will be presented in Appendix~\ref{sec:more-punctures}. The point is that a universal behavior of $\CS$-walls in some region of $C$ allows us to predict the 2d-4d soliton data for a surface defect located in that region. We will now turn to a detailed analysis of this statement.

\subsection{Surface defects near punctures}\label{sec:near-punctures}

\begin{figure}
\begin{center}
\includegraphics[width=0.22\textwidth]{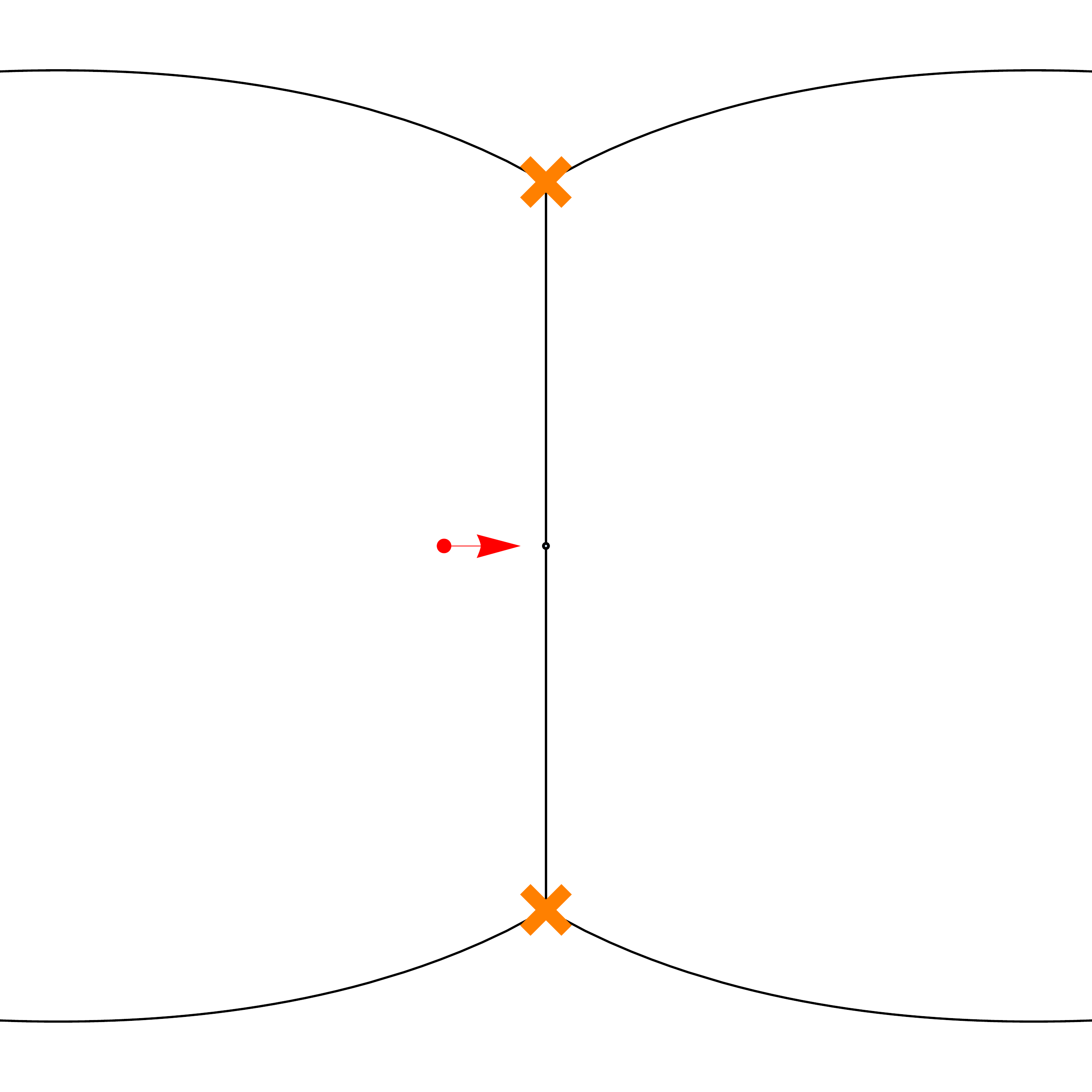}
\includegraphics[width=0.22\textwidth]{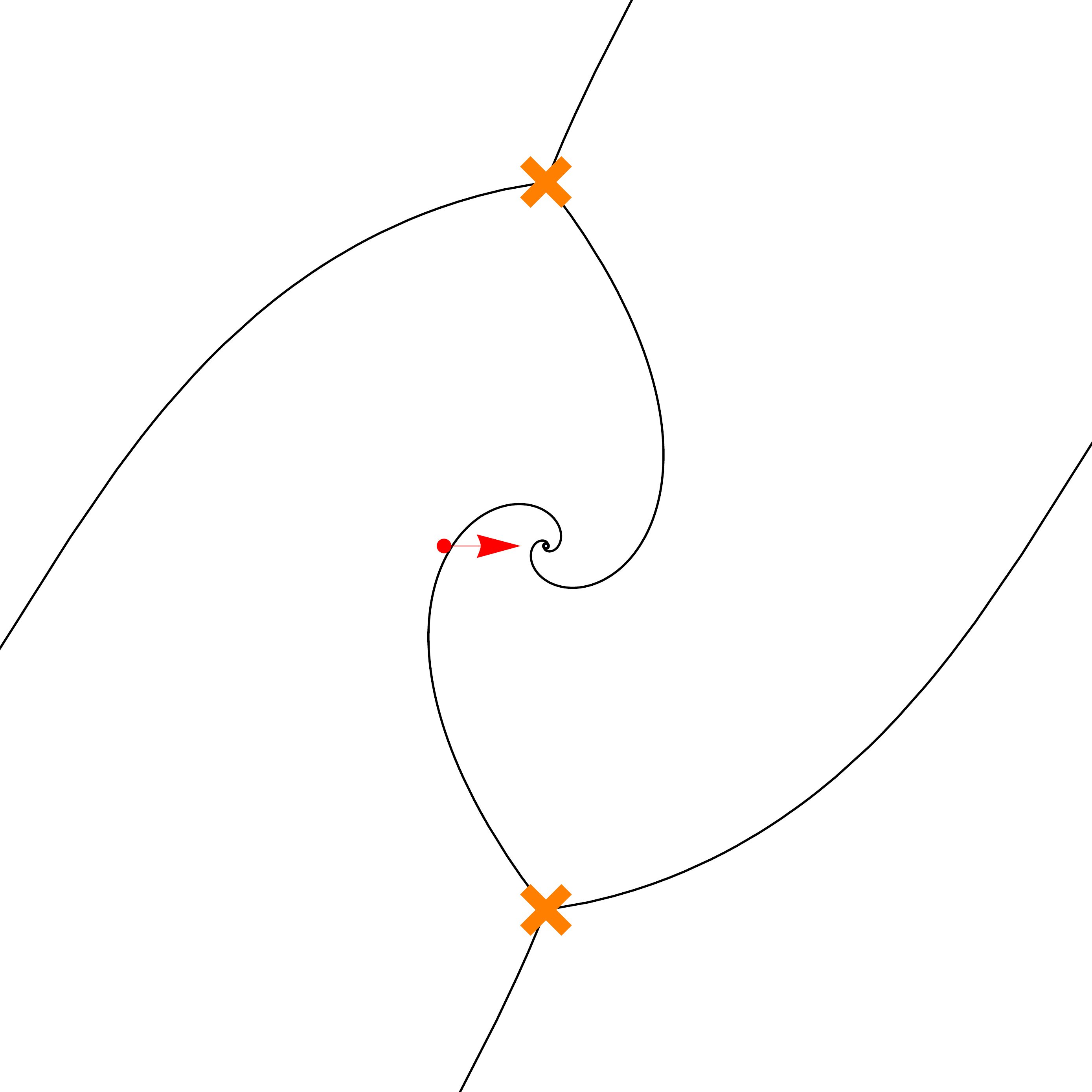}
\includegraphics[width=0.22\textwidth]{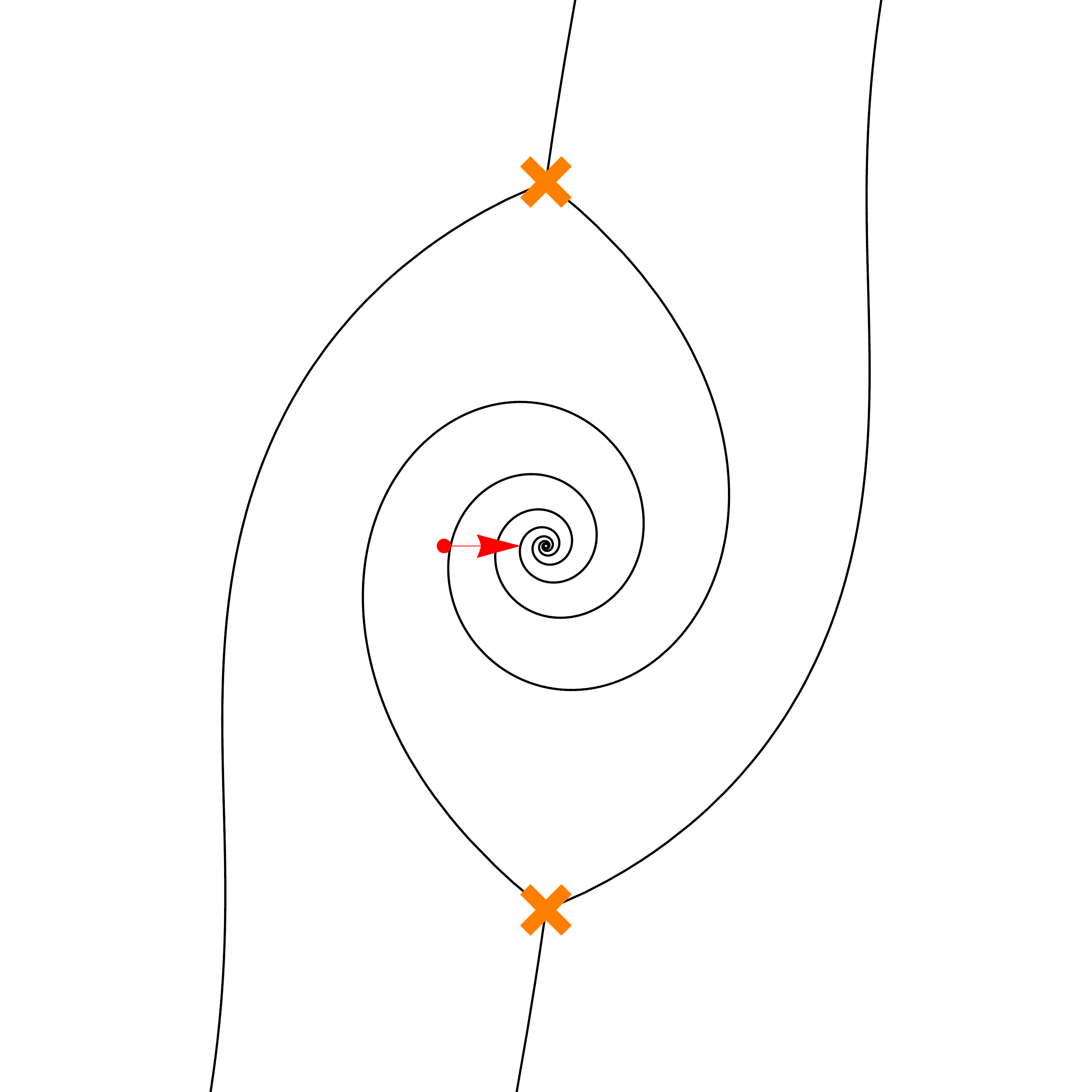}
\includegraphics[width=0.06\textwidth]{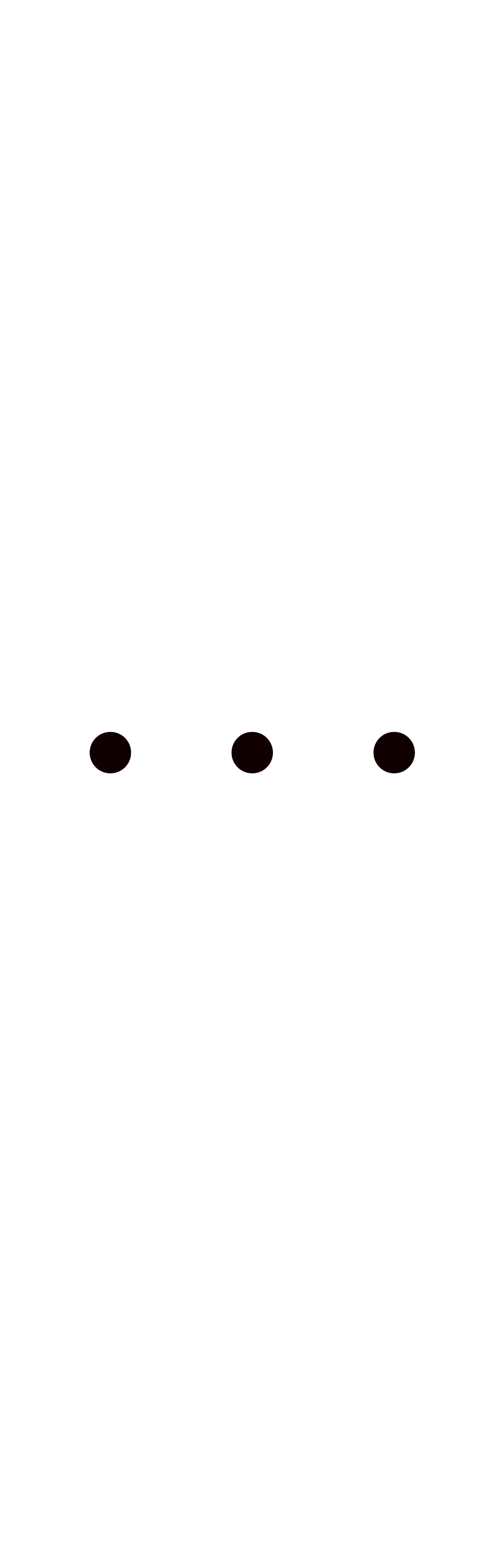}
\includegraphics[width=0.22\textwidth]{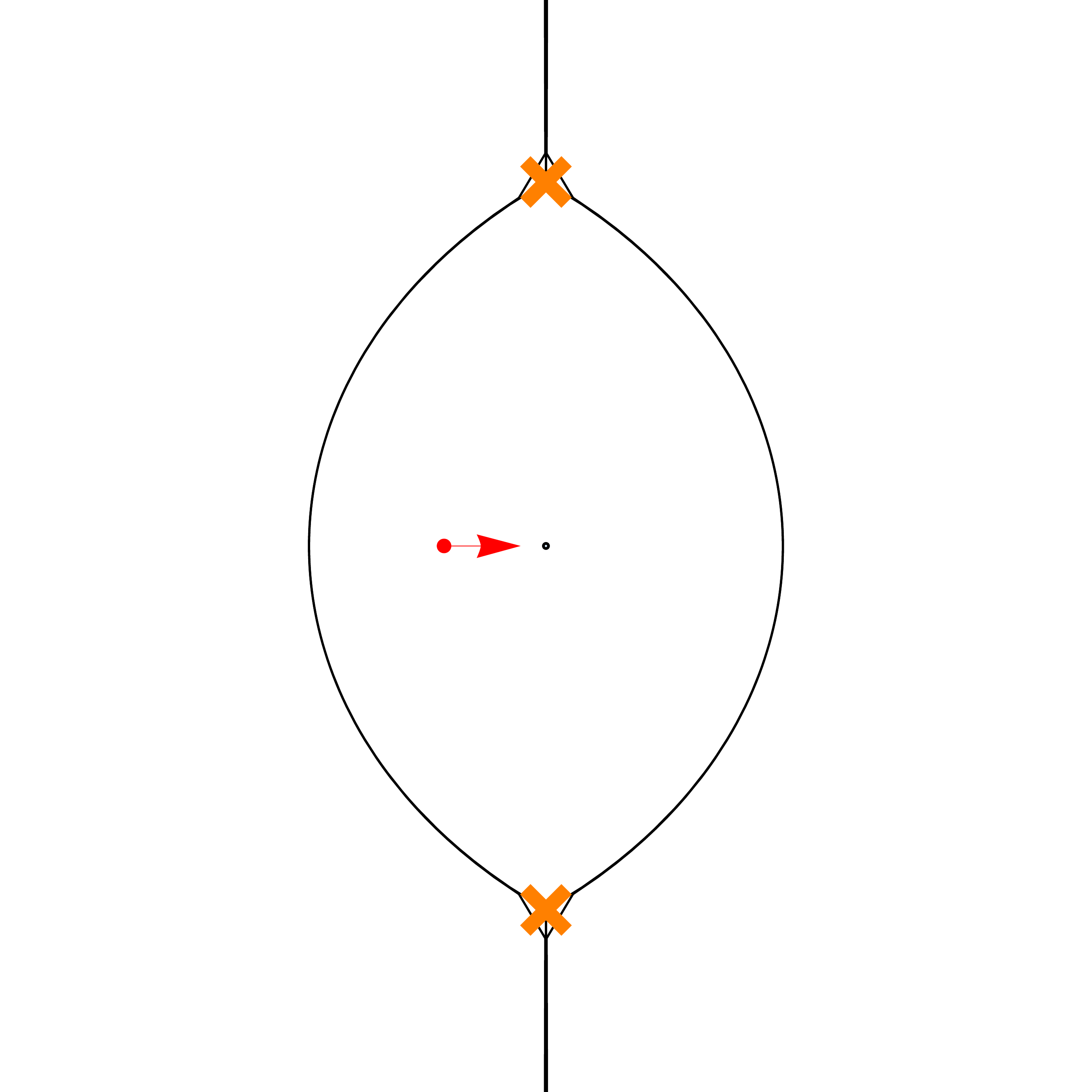}\\
\includegraphics[width=0.22\textwidth]{figures/puncture_network_31.pdf}
\includegraphics[width=0.06\textwidth]{figures/dots.pdf}
\includegraphics[width=0.22\textwidth]{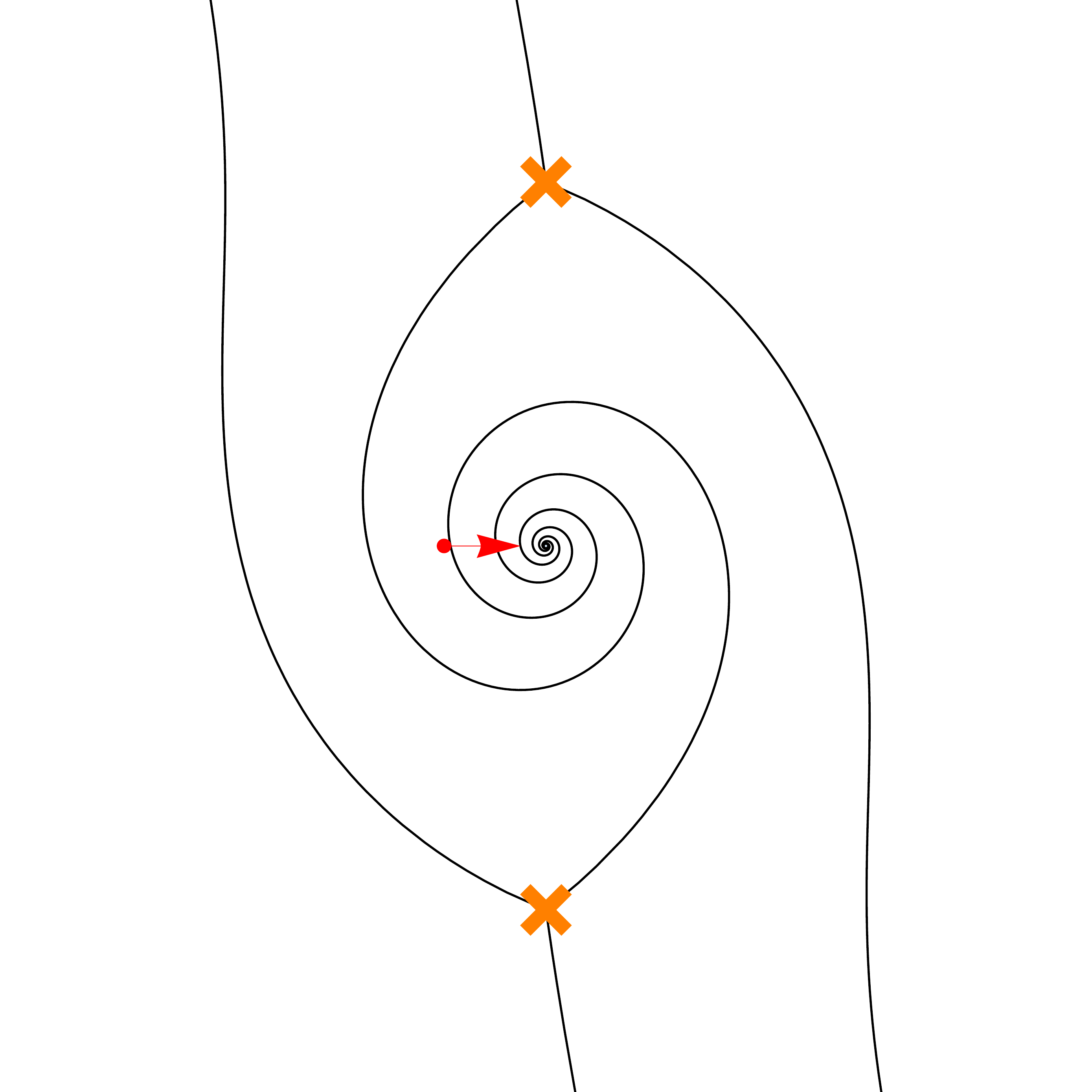}
\includegraphics[width=0.22\textwidth]{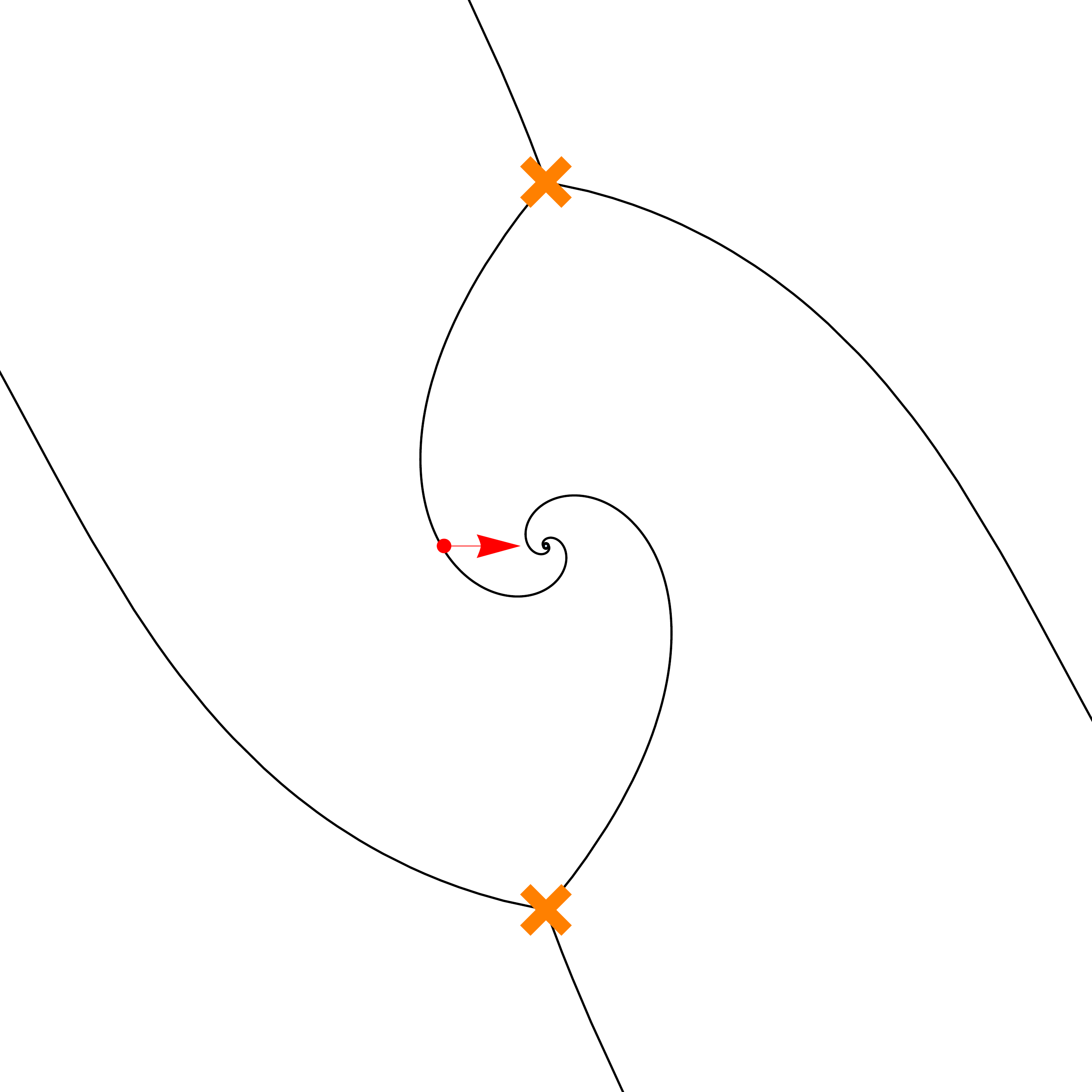}
\includegraphics[width=0.22\textwidth]{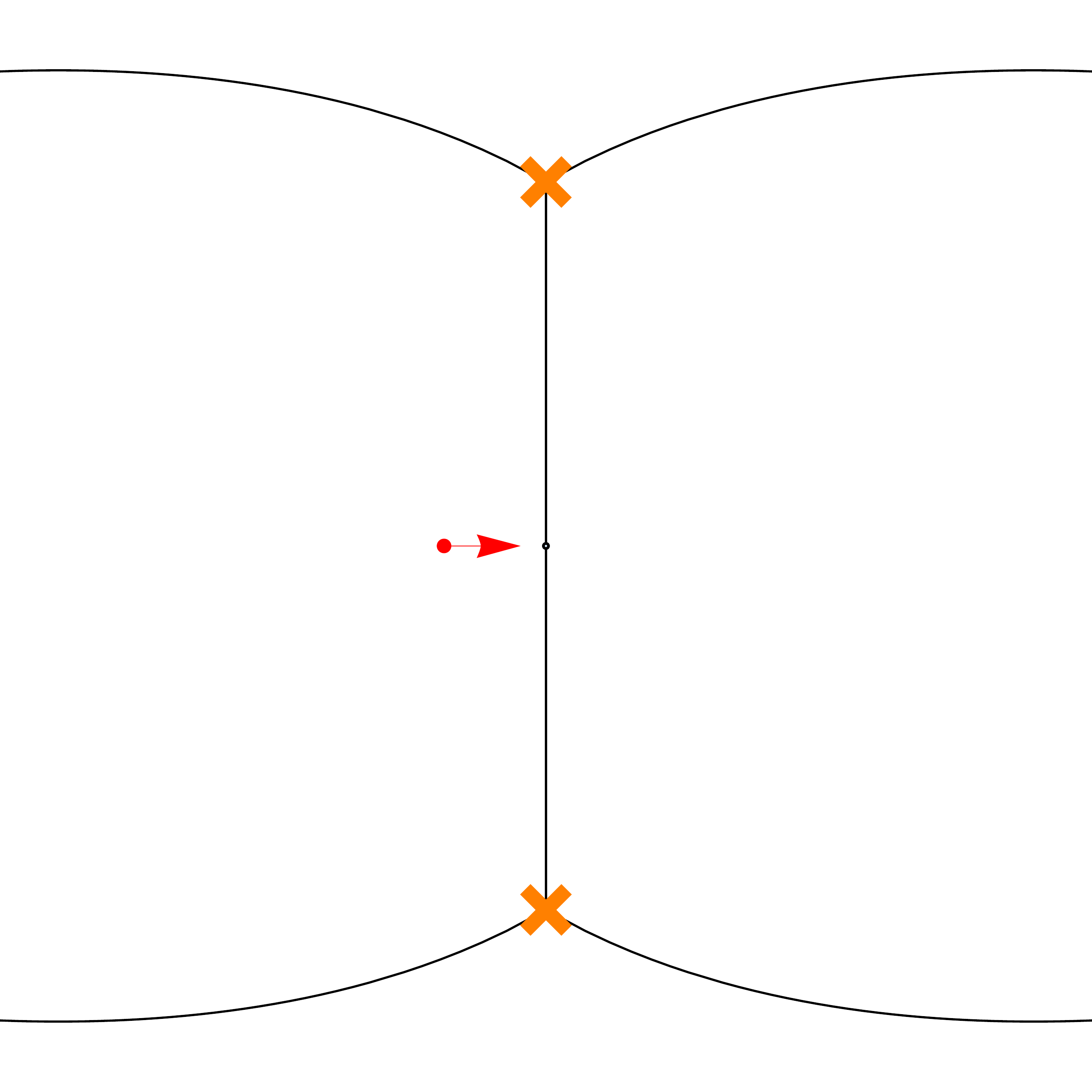}
\end{center}
\caption{
Sequence of spectral networks near a regular puncture. The red dot marks $z\in C$, the parameter of the surface defect, while the red arrow denotes a choice of ``resolution'' of the surface defect into an infinitesimal interface. The top line illustrates phases from sector $I$ as described in~(\ref{eq:2d4d-seq-1}), for phases ranging from $\vartheta_c-\pi/2$ (top-left) to $\vartheta_c$ (top-right). In the second frame, an $\CS$-wall passes through $z$, the charge of the soliton carried by the wall is  $a_{\bar \imath i}$, as defined in~(\ref{eq:a-i}). In the third frame, an $\CS$-wall passes through $z$, it carries solitons $a_{\bar \imath i} + \gamma_R^{(i)}$, as defined in~(\ref{eq:a-i-gamma-R}). The bottom line illustrates phases from sector $II$ as described in~(\ref{eq:2d4d-seq-2}), for phases ranging from $\vartheta_c$ (bottom-left) to $\vartheta_c+\pi/2$ (bottom-right). The $\CS$-wall passing through $z$ in the second-to-last frame carries a soliton with charge $-a_{\bar \imath i} + \gamma_L^{(i)}$, whereas the one in the third-to-last frame carries $-a_{\bar \imath i} + \gamma_L^{(i)}+ \gamma_R^{(i)}$, according to~(\ref{eq:2d4d-seq-2}).
}
\label{fig:A1-puncture-behavior}
\end{figure}

From the discussion of the previous subsection it follows that, under certain assumptions, it is possible to tune the moduli of the 4d theory, such as UV masses and Coulomb branch moduli, in such a way that the local behavior of spectral networks near a puncture of choice takes the universal form as shown in Figure~\ref{fig:A1-puncture-behavior}.\footnote{In the language of WKB triangulations (defined in~\cite{Gaiotto:2009hg}) this can be phrased as the statement that there is always a WKB triangulation such that a puncture is shared by \emph{exactly two} triangles and there are \emph{exactly two} edges ending on the puncture. 
This leaves out theories engineered by a UV curve $C$ with exactly one (regular) puncture (\emph{e.g.} the $\CN=2^{*}$ theory); in our arguments we excluded this case, and we will devote a separate analysis to it in Section~\ref{sec:one-puncture}.
}
More precisely, this is guaranteed for theories engineered by a UV curve with at least two punctures, as discussed in Appendix~\ref{sec:more-punctures}.
In the remainder of this section we will thus focus on theories with this property. 
The generalization to Riemann surfaces with a single puncture turns out to be qualitatively very similar, we will analyze it below in Section~\ref{sec:one-puncture}.

\begin{figure}[h!]
\begin{center}
\includegraphics[width=0.40\textwidth]{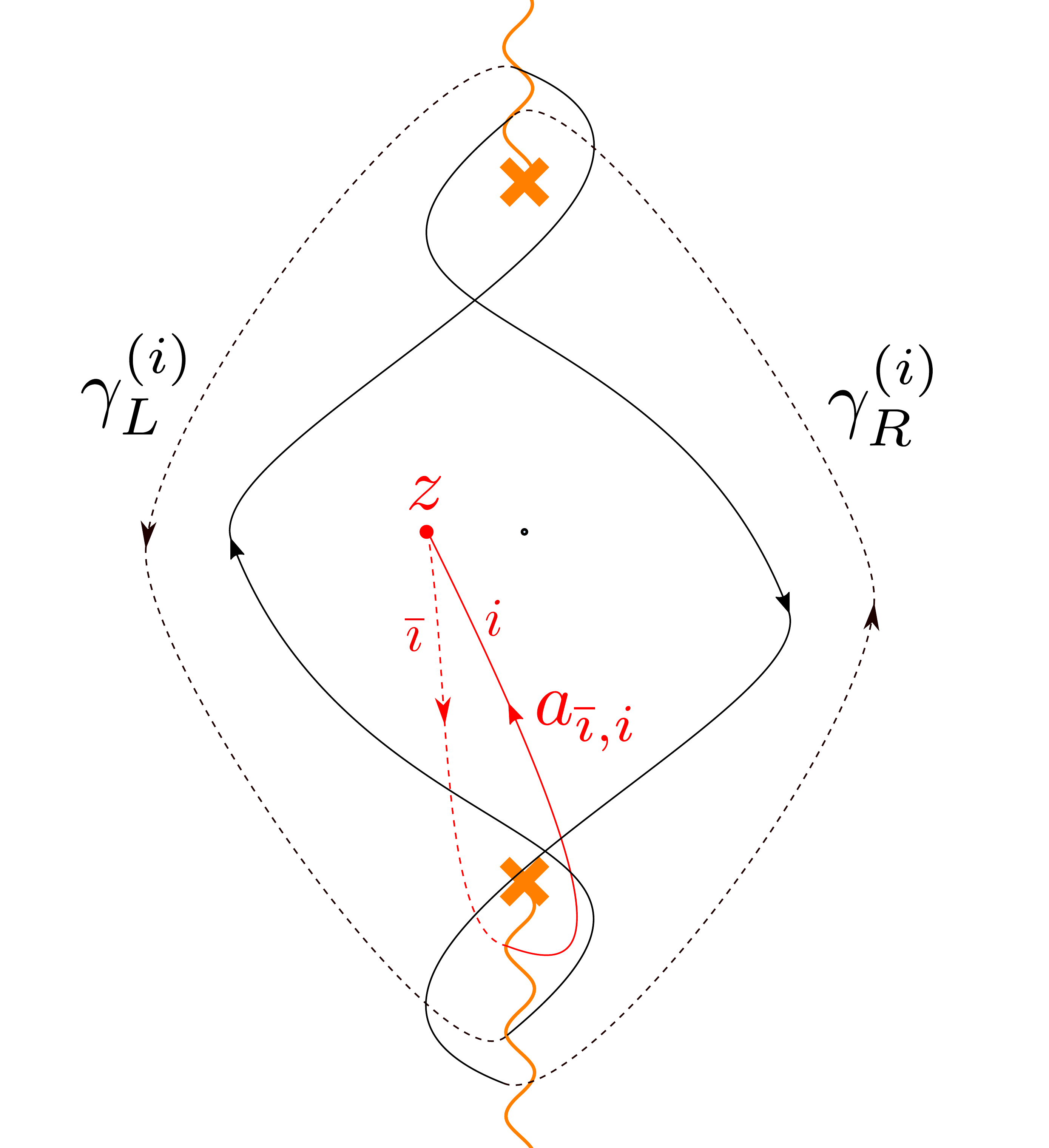}
\caption{Cycles of BPS states for a surface defect near a puncture. Each closed cycle represents a 4d BPS particle, the open path is a 2d-4d BPS soliton for the surface defect at $z \in C$.
Solid lines run on sheets labeled by $i$ while dashed lines run on sheets labeled by $\bar \imath$.}
\label{fig:puncture-cycles}
\end{center}
\end{figure}

At the critical phase $\vartheta_c$ all walls become degenerate, \emph{i.e.} they collapse into segments made of (anti-)parallel $\CS$-walls, of types $\bS_{\pm\alpha}$, running between two branch points. These are also known as ``double-walls''. As Figure~\ref{fig:A1-puncture-behavior} shows, there are two distinguished double-walls surrounding the puncture; let $p_{L,R}$ be the paths of the left, right double-wall connecting the two branch points, respectively, oriented from the bottom to the top.
Denoting by $p_{L,R}^{(i)}$ the lift of $p_{L,R}$ to the $i$-th sheet of $\Sigma_N$, we define 
\be\label{eq:gamma-i}
	\gamma_{L}^{(i)}  \ = \  \left[p_L^{(i)} - p_L^{(\bar \imath)}\right] \,,\ \
	\gamma_{R}^{(i)}  \ = \  \left[p_R^{(\bar \imath)} - p_R^{(i)}\right]
	\quad
	\in \ {H_1({\Sigma_N,\IZ})} \,,
\ee
for each $i=1,\dots, \lfloor \frac{N}{2} \rfloor$, automatically excluding the zeroth sheet when $N$ is odd. These cycles are illustrated in Figure~\ref{fig:puncture-cycles}. Note that equation~(\ref{eq:lambda-multiplicity}) implies
\be
	Z_{\gamma_L^{(i)}}  \ = \  \oint_{\gamma_{L}^{(i)}} \lambda  \ = \  \int_{p_L}\lambda_i - \lambda_{\bar \imath}  \ = \  - (N-2i+1) \int_{p_L}\langle\alpha,\varphi\rangle \,,
\ee
which depends on $i$ in a controlled way. For $N=2$ the index $i$ is restricted to the value $1$, and can for simplicity be omitted, therefore periods of these cycles for different values of $N$ are all multiples of the  ones for the $N=2$ case, \emph{i.e.}
\be\label{eq:mass-factor}
	Z_{\gamma_L^{(i)}}   \ = \  (N-2i+1) \cdot Z^{(N=2)}_{\gamma_L}  \,.
\ee
A similar relation holds among periods of cycles arising from $p_R$. The existence of these relations reflects the fact that periods of $\lambda$ lie in a sub-variety of ${\rm Jac}(\Sigma)$~\cite{Donagi:1993, Martinec:1995by, Longhi:2016rjt}, which is a consequence of the fact that the geometry of $\Sigma_N$ is entirely parameterized by a quadratic differential.

We are now ready to describe the 2d-4d soliton spectrum of a surface defect located at $z\in C$ as shown in Figure~\ref{fig:A1-puncture-behavior}.
Starting at the phase $\vartheta=\vartheta_c - \pi/2$ and increasing $\vartheta$, the first wall to cross $z$ is the $\CS$-wall starting from the lower branch-point and heading directly to $z$. We shall take this $\CS$-wall to be of type $-\alpha$, this fixes the root $\pm\alpha$ assigned to all other walls.\footnote{This choice can be made without loss of generality; it amounts to fixing the Weyl freedom in matching weights of $\Lambda_N$ with sheets of $\Sigma_N$. For a more complete discussion see~\cite[Appendix B]{Longhi:2016rjt}.}  According to equation~(\ref{eq:S-wall-spectrum-2}), the first $\CS$-wall carries 2d-4d solitons with charges
\be\label{eq:a-i}
	a_{\bar \imath i}~\in~\Gamma_{\bar \imath i}(z) \,, \qquad i \ = \ 1,\dots ,\left\lfloor{N/2}\right\rfloor\,.
\ee
It will be important to note that these charges have canonical representatives, namely actual paths on $\Sigma_N$ determined by the geometry of the $\CS$-wall.
Let $p$ be the oriented path of the $\CS$-wall running from the lower branch point to $z$, taken at phase $\vartheta$ when the wall passes through $z$. 
Then soliton paths are concatenations of lifts of $p$ with opposite orientations
\be\label{eq:lifting}
	a_{\bar \imath i} \ = \  p^{(i)} - p^{(\bar \imath)}
\ee
where $p^{(i)}$ denotes the lift of $p$ to the $i$-th sheet, and concatenation takes place at the ramification point as schematically depicted in  Figure~\ref{fig:simpleton}.
A representative path for $a_{\bar \imath i}$ is shown in Figure~\ref{fig:puncture-cycles}.

As we increase $\vartheta$ further, the next wall crossing $z$ is the one originating from the upper branch point, which runs half-way around the puncture before arriving at the position of the surface defect. This wall is also of type $-\alpha$ since it flows \emph{into} the puncture akin to the previous one (walls of type $\alpha$ would have the opposite orientation), therefore it also carries solitons of types $\Gamma_{\bar \imath i}$.  It is straightforward to see that its soliton content features the following charges
\be\label{eq:a-i-gamma-R}
	a_{\bar \imath i} + \gamma_R^{(i)} \ \in \  \Gamma_{\bar \imath i}(z)\,, \qquad i \ = \ 1,\dots, \lfloor N/2 \rfloor \,.
\ee
Proceeding by increasing $\vartheta$, the $\CS$-walls begin to spiral tighter and tighter into the puncture, leading to an infinite tower of times that each wall crosses $z$. Because of the spiraling behavior, the soliton charges get a contribution of $\gamma_L^{(i)}+\gamma_R^{(i)}$ at each turn. The overall spectrum of 2d-4d BPS solitons with central charges (whose phase is) in the interval $\vartheta_c-\pi/2<\vartheta<\vartheta_c$ is
\be\label{eq:2d4d-seq-1}
\begin{array}{l|l}
I:\,(\vartheta_c-\pi/2<\vartheta<\vartheta_c) \qquad\qquad &
\begin{split}
	& \qquad\qquad\vdots  \\
	~~\vartheta\qquad & a_{\bar \imath i} + k \( \gamma_L^{(i)} + \gamma_R^{(i)} \) \\
	~~\downarrow\qquad & a_{\bar \imath i} + k \( \gamma_L^{(i)} + \gamma_R^{(i)} \) + \gamma_R^{(i)} \\
	& \qquad\qquad\vdots
\end{split}
\end{array}
\ee
for $k\geq 0$ increasing upon increasing the phase $\vartheta$.

A similar analysis yields the spectrum of 2d-4d states for phases beyond $\vartheta_c$, \emph{i.e.} it is given by another infinite tower of states 
\be\label{eq:2d4d-seq-1b}
\begin{split}
	& \qquad\qquad\vdots \\
	\vartheta\qquad& b_{i \bar \imath} + \ell \( \gamma_L^{(i)} + \gamma_R^{(i)} \) + \gamma_R^{(i)} \\
	\downarrow\qquad& b_{i \bar \imath} + \ell \( \gamma_L^{(i)} + \gamma_R^{(i)} \) \\
	& \qquad\qquad\vdots 
\end{split}
\ee
with $\ell\geq 0$ now decreasing with increasing phase $\vartheta$. The tower ends with the set of solitons $b_{i \bar \imath}$ supported on the last wall sweeping through $z$ before $\vartheta$ reaches $\vartheta_c+\pi$ (the second-to-last frame in Figure~\ref{fig:A1-puncture-behavior}).  As is evident from the pictures, these paths are clearly related to $a_{\bar  \imath i}$ by
\be
	a_{\bar \imath i} + b_{i \bar \imath}  \ \simeq \ \gamma_L^{(i)} \,,
\ee
where we used additive notation $a+b$ to denote concatenation of paths. Therefore, we may re-express the second tower of solitons in~(\ref{eq:2d4d-seq-1b}) as follows
\be\label{eq:2d4d-seq-2}
\begin{array}{l|l}
II:\,(\vartheta_c<\vartheta<\vartheta_c+\pi/2) \qquad\qquad &
\begin{split}
	&\qquad\qquad\qquad \vdots \\
	~~\vartheta \qquad& -a_{\bar \imath i} + (\ell+1) \( \gamma_L^{(i)} + \gamma_R^{(i)} \)  \\
	~~\downarrow\qquad & -a_{\bar \imath i} + \ell \( \gamma_L^{(i)} + \gamma_R^{(i)} \) + \gamma_L^{(i)} \\
	&\qquad\qquad\qquad \vdots 
\end{split}
\end{array}
\ee

By symmetry, we can easily derive the 2d-4d solitons for the remaining phases. It is clear that $\vartheta\to\vartheta+\pi$ changes $\alpha\to-\alpha$ in~(\ref{eq:S-wall}), and thus we expect the following towers
\be\label{eq:2d4d-seq-3-4}
\begin{array}{l|l}
III:\,(\vartheta_c+\pi/2<\vartheta<\vartheta_c+\pi) \qquad &
\begin{split}
	& \qquad\qquad\quad\vdots \\
	~~\vartheta\qquad& -a_{\bar \imath i} - k \( \gamma_L^{(i)} + \gamma_R^{(i)} \) \\
	~~\downarrow\qquad& -a_{\bar \imath i} - k \( \gamma_L^{(i)} + \gamma_R^{(i)} \) - \gamma_R^{(i)} \\
	& \qquad\qquad\quad\vdots \\
\end{split}
	\\
	\hdashline[2pt/2pt]
IV:\,(\vartheta_c+\pi<\vartheta<\vartheta_c+3\pi/2) \qquad &
\begin{split}
	& \qquad\qquad\quad\vdots \\
	~~\vartheta \qquad& a_{\bar \imath i} - (\ell+1) \( \gamma_L^{(i)} + \gamma_R^{(i)} \)  \\
	~~\downarrow\qquad & a_{\bar \imath i} - \ell \( \gamma_L^{(i)} + \gamma_R^{(i)} \) - \gamma_L^{(i)} \\
	& \qquad\qquad\quad\vdots 
\end{split}
\end{array}\hspace{5pt}
\ee
with $k\geq 0$ increasing, and $\ell\geq 0$ decreasing, as $\vartheta$ increases. This completes the study of 2d-4d BPS solitons that are supported on the surface defect at $z$. 

\subsection{Quantum torus algebra and spin}\label{sec:spins}

For the purpose of studying the IR formula in the context of 2d-4d systems, we will need to compute the ``quantum'' or ``motivic'' 2d-4d spectrum generator. For this purpose we consider a non-commutative deformation of the algebra of the formal variables $X_a,X_\gamma$ used to describe the soliton data of spectral networks.

Formal variables associated with closed homologies $\gamma,\gamma'\in H_1(\Sigma_N,\IZ)$ obey the following relations
\be\label{eq:q-torus-alg}
	X_\gamma  X_{\gamma'}  \ = \  q^{\frac{1}{2}\langle\langle\gamma + \gamma'\rangle\rangle}   X_{\gamma+\gamma'}\,,
	\qquad  
	X_{-\gamma}  \ = \   X_{\gamma}^{-1}\,,
\ee
where 
\be
	\langle \langle \,\cdot\,,\,\cdot\,\rangle\rangle  \ = \  k_\gamma  \langle \,\cdot\,,\,\cdot\,\rangle
\ee
is a positive integer multiple $k_\gamma \in \IN$ of the intersection pairing. Recall that the Dirac-Schwinger-Zwanziger electromagnetic pairing is identified with the intersection of homology cycles. We will return to this shortly.

The algebra~(\ref{eq:q-torus-alg}) appeared in~\cite{Kontsevich:2008fj}, and its physical origin was elucidated in~\cite{Gaiotto:2010be}. We will need an extension of  it to include open paths, whose definition was proposed in~\cite{Galakhov:2014xba}. It takes the following form: given any two open paths $a$ and $b$ on $\Sigma_N$
\be\label{eq:q-torus-alg-oo}
X_{a}X_{b}  \ = \  \left\{
	\begin{array}{ll}
	q^{\frac{1}{2}\langle\langle a,b\rangle\rangle} X_{ab} \,, \qquad & \text{end}(a) = \text{beg}(b)\,,\\
	0\,, & \text{otherwise} \,.
	\end{array}
\right.
\ee
Moreover, the two sets of variables can mix with one another according to the following rule
\be\label{eq:q-torus-alg-oc}
	X_a X_\gamma  \ = \  q^{\frac{1}{2}\langle\langle a,\gamma\rangle\rangle} X_{a+\gamma}\,.
\ee

Let us now explain the origin of the slightly unusual pairing $\langle\langle\, \cdot\, , \,\cdot\, \rangle\rangle$. Recall from~(\ref{eq:curve-factorization}), that the spectral curve always factorizes into $\left\lfloor (N+1)/2 \right\rfloor$ components. It means that the homology lattice is naturally graded
\be\label{eq:lattice-decomposition}
	H_1(\Sigma_N,\IZ)  \ = \  
	\left\{
	\begin{split}
		& H_1(\Sigma^{(1)},\IZ)  \oplus H_1(\Sigma^{(3)},\IZ) \oplus \cdots \oplus H_1(\Sigma^{(N-1)},\IZ) \,, \qquad N\in 2\IN \,, \\
		& H_1(\Sigma^{(0)},\IZ)  \oplus H_1(\Sigma^{(2)},\IZ) \oplus \cdots \oplus H_1(\Sigma^{(N-1)},\IZ) \,, \qquad N\in 2\IN+1 \,. 
	\end{split}
	\right.
\ee
This implies two things: Firstly, it implies that there is a discrete label $k_\gamma$ assigned to each homology cycle $\gamma$ that keeps track of the grading. Secondly, we have that 
\be
	\langle \gamma,\gamma'\rangle  \ = \ 0\,, \quad \text{if} \quad  k_{\gamma} \neq k_{\gamma'} \,,
\ee
since the intersection pairing is block-diagonal in the decomposition~(\ref{eq:lattice-decomposition}). Hence, we may unambiguously define 
\be
	\langle\langle \gamma,\gamma'\rangle\rangle \ \coloneqq \ k_\gamma \langle \gamma,\gamma'\rangle \ \equiv \  k_{\gamma'} \langle \gamma,\gamma'\rangle \,.
\ee
Analogous considerations hold for open paths; their (relative) homology lattice is graded by components of $\Sigma_N$ just like closed cycles, therefore they come with a canonical label $k_a$ attached to them.

We shall discuss $k_\gamma$ and $k_a$ momentarily. However, first, let us fix their values. For a path (either closed or open) on the component $\Sigma^{(i)}$ of $\Sigma_N$, we impose 
\be\label{eq:intersection-correction}
	k_\gamma \ \ \text{or} \ \ k_a \ \equiv \  (N-2 i +1)^2 \,.
\ee

In the literature $k_\gamma$ (and $k_a$) is always assumed to be $1$, and this is consistent with the fact that the vast majority of works focus on the fundamental representation (where $N=2$ and $i=1$), or for higher rank cases on minuscule representations (whose $\Sigma$ has one connected component). The modification of the quantum torus algebra proposed here is novel, and will play a crucial role in what follows. Thus, let us stop and explain its motivation. It is not a coincidence, that the factor~(\ref{eq:intersection-correction}) is exactly the square of the one appearing in~(\ref{eq:lambda-multiplicity}). In the classical limit (\emph{i.e.} $q\to 1$), the formal variables $X_\gamma$ (and $X_a$) are avatars of local coordinates $\CX_\gamma$ on a moduli space of flat abelian connections, $\nabla^{\rm ab}$, on $\Sigma_N$.  Concretely, this means that they are holonomies
\be
	\CX_\gamma \ \sim \ \exp \int_\gamma \nabla^{\rm ab} \,.
\ee

However, the abelian connection on $\Sigma$ is related to a \emph{non}-abelian connection of $C$ by the ``non-abelianization'' map defined via spectral networks~\cite{Gaiotto:2012rg}. Locally, the relation between the two is roughly
\be
	\nabla^{\rm n.a.}_{z} \ \sim \  \bigoplus_{i=1}^{N}   \nabla^{\rm ab}_{z^{(i)}}  \,,
\ee
where $z\in C$ is a generic point (chosen away from $\CS$-walls, branch points or punctures), and $z^{(i)}$ is its lift to the $i$-th sheet of $\Sigma_N$. However, the non-abelian connection (roughly) corresponds to the flat connection constructed from a solution to Hitchin's equations\footnote{For notation and background see \emph{e.g.}~\cite{Gaiotto:2009hg}.}
\be
	\nabla^{\rm n.a.} \ \sim \ \frac{\varphi}{\zeta} + A + \zeta\overline\varphi \,. 
\ee

Akin to~(\ref{eq:lambda-multiplicity}), the non-abelian connection in the $N$-dimensional representation is related to the one in the fundamental representation by an overall rescaling
\be
	\nabla^{\rm ab}_{z^{(i)}}  \ = \   (N-2 i + 1) \cdot \nabla^{{\rm ab}\ (N=2)}_{z^{(1)}} \,.
\ee

Let us now return to the (classical) coordinates $\CX_\gamma$. For $N>2$ each cycle $\gamma$ comes as a family $\gamma^{(i)}$ with $i=1,\dots, \left\lfloor N/2\right\rfloor$ as in~(\ref{eq:gamma-i}). Different members of the family simply live on different components of the cover, essentially ``above'' or ``below'' each other according to the projection $\pi:\Sigma_N\to C$. Consequently, we have
\be\label{eq:holonomy-powers}
	\CX_{\gamma^{(i)}}  \ = \  \(\CX_{\gamma^{(N=2)}}\)^{N-2i+1} \,.
\ee
In order to move to the non-commutative version of these variables, it was proposed in~\cite{Galakhov:2014xba}, to apply deformation quantization to the Weil-Petersson symplectic form on the moduli space of flat abelian connections on $\Sigma_N$. In particular, for the $N=2$ case this means that Add: upon quantization, we get
\be
	\{ \log  \CX^{(N=2)}_{\gamma} , \log \CX^{(N=2)}_{{\gamma'} }  \}  \ = \  \langle \gamma  , {\gamma'} \rangle 
	\quad
	\to
	\quad
	[ \log \hat  \CX^{(N=2)}_{\gamma} , \log \hat \CX^{(N=2)}_{{\gamma'} }  ]  \ = \  \hbar\langle \gamma  , {\gamma'} \rangle  \,,
\ee
where $\hbar = \log q$.\footnote{We remark, that in~\cite{Galakhov:2014xba} the quantum variable is $y_{there} = q_{here}^{1/2}$.}
This is equivalent to 
\be
	\hat  \CX^{(N=2)}_{\gamma} \cdot   \hat \CX^{(N=2)}_{{\gamma'} }    \ = \  q^{\frac{1}{2} \langle \gamma  , {\gamma'} \rangle } \hat  \CX^{(N=2)}_{\gamma + \gamma'} \,. 
\ee
In the case $N>2$, the symplectic form takes a block-diagonal form (viewed as a bi-linear on vector fields) due to the decompositions~(\ref{eq:curve-factorization}) and~(\ref{eq:lattice-decomposition}).
In fact combining this observation with the relation between holonomies~(\ref{eq:holonomy-powers}) leads to the following Poisson bracket
\be
\begin{split}
	\{ \log  \CX_{\gamma^{(i)}} , \log \CX_{{\gamma'}^{(j)} }  \} 
	&  \ = \  \delta_{ij}  (N-2i +1)^2 \, \{ \log  \CX^{(N=2)}_{\gamma^{(i)}} , \log \CX^{(N=2)}_{{\gamma'}^{(j)} }  \} \\
	&  \ = \  \delta_{ij}  (N-2i +1)^2 \,  \langle \gamma  , {\gamma'} \rangle  \\
	&  \ = \  (N-2i +1)^2 \,  \langle \gamma^{(i)}  , {\gamma'}^{(j)} \rangle  \,.
\end{split}
\ee
By deformation quantization, this leads directly to
\be
\begin{split}
	\hat  \CX_{\gamma^{(i)}} \cdot   \hat \CX_{{\gamma'}^{(j)} }  & \ = \ q^{\frac{1}{2}   (N-2i +1)^2 \, \langle \gamma^{(i)}  , {\gamma'}^{(j)} \rangle } \hat  					\CX_{\gamma^{(i)} + {\gamma'}^{(j)}}  \\
	& \ = \  q^{\frac{1}{2}    \langle\langle \gamma^{(i)}  , {\gamma'}^{(j)} \rangle\rangle } \hat  \CX_{\gamma^{(i)} + {\gamma'}^{(j)}}  \,,
\end{split}
\ee
thus providing our derivation for~(\ref{eq:q-torus-alg}),~(\ref{eq:q-torus-alg-oo}) and~(\ref{eq:q-torus-alg-oc}).\footnote{\label{foot:actual-paths}
A word of caution is due at this point. Several expressions involve intersections of open paths, both with other open paths and with closed paths. Such intersections are ill-defined on homology classes. In~\cite{Galakhov:2014xba}, a refinement of the classification of soliton charges by \emph{(relative) regular homotopy} (defining $\langle a,b\rangle$ by using the notion of \emph{writhe}) was introduced. 
Another option is to work with \emph{actual paths}, \emph{i.e.} with actual geometric representatives of soliton charges, instead of abstract relative homology classes. Having the exact shape of a path implies that we can make sense of its intersection with other paths.
We will stick to the second option in this paper. This uses more information than necessary, but allows us to avoid introducing somewhat ``byzantine'' definitions. It is nevertheless possible to repeat our story in the language of~\cite{Galakhov:2014xba}.
As should be clear to the attentive reader, the actual path of a soliton is defined by lifting the actual path of an $\CS$-wall (which lives on $C$) to $\Sigma_N$, whose shape is described by~(\ref{eq:S-wall}), and whose lift was already introduced  in~(\ref{eq:lifting}). This definition is natural since~(\ref{eq:S-wall}) is nothing but the geometric reformulation of the BPS soliton equations of 2d $\CN=(2,2)$ massive theories~\cite{Cecotti:1992rm, Klemm:1996bj,Gaiotto:2009hg}.
}

Finally, let us collect a few useful intersection numbers that will show up in our computations. We are interested in open and closed paths that arise near a puncture, similar to the ones depicted in Figure~\ref{fig:puncture-cycles}. By direct inspection, we find
\be\label{eq:puncture-cycles-intersections}
\begin{split}
	\langle \gamma_L^{(i)} , \gamma_R^{(j)} \rangle &  \ = \  0 \,,\qquad
	\langle a_{\bar \imath i} , \gamma_L^{(j)} \rangle  \ = \  \delta_{ij} \,,\qquad
	\langle a_{\bar \imath i} , \gamma_R^{(j)} \rangle  \ = \  \delta_{ij} \,,
\end{split}
\ee
and consequently we obtain
\be\label{eq:kappa-i}
	\langle\langle a_{\bar \imath i} , \gamma_L^{(j)} \rangle\rangle  \ = \  \kappa_i \delta_{ij}  \ = \  \langle\langle a_{\bar \imath i} , \gamma_R^{(j)} \rangle\rangle \,,
\ee
where 
\be
	\kappa_i \ = \ (N-2i+1)^2
\ee
is the factor introduced in~(\ref{eq:intersection-correction}).

The non-commutative deformation of the algebra of formal $X$-variables is related to the spin of BPS states~\cite{Gaiotto:2010be}. 
Indeed, the interpretation of these variables as holonomies led to the proposal that the spin of BPS states is captured by self-intersections of paths on $\Sigma$~\cite{Galakhov:2014xba}. More precisely, to determine self-intersections one should refine the classification of paths from homology classes to regular homotopy classes. This refinement is possible since soliton paths have canonical representatives built from lifts of $\CS$-walls. Therefore, given a soliton path $a$ carried by an $\CS$-wall, we consider a representative for $a$ made of lifts of the wall and compute the self-intersections of such path. The self-intersections are counted with signs, by an invariant known as \emph{writhe}, see Figure~\ref{fig:particle-spin}. It is easy to see that 2d-4d solitons always have zero self-intersections in $A_1$ spectral networks, since they are simply lifts of $\CS$-walls, which never self-intersect (see \emph{e.g.} Figure~\ref{fig:simpleton}). On the other hand, closed paths, corresponding to charges of 2d BPS particles and 4d BPS particles, in general have non-vanishing writhe, as shown in Figure~\ref{fig:particle-spin}. We then say that 2d-4d solitons have zero spin, whereas the spin of 2d particles will be determined by the writhe of their charge paths, as will become evident below. The ``spin'' is actually a flavor symmetry from the viewpoint of the 2d theory, as it corresponds to the representation of 2d states under space-like rotations of the 4d theory in the plane transverse to the surface defect (see discussion in Subsection~\ref{sec:2d4dwcrev}).

\subsection{2d BPS particles}\label{sec:2d-particle}

Having determined the full spectrum of 2d-4d solitons for a surface defect near a puncture, we now turn to the more subtle issue of computing the 2d BPS particle spectrum. This is more subtle because spectral networks were not specifically designed for this purpose. Nevertheless, in the following, we will argue that they capture this information.

As explained at the beginning of this section, 2d BPS particles are part of the spectrum in a 2d vacuum labeled by $i$ (or $\bar \imath$), and carry ``pure flavor'' charges from the 2d viewpoint, which coincide with 4d charges $\gamma\in H_1(\Sigma_N,\IZ)$. For each 2d particle, $\sigma$, in a vacuum $i$, we wish to determine its charge $\gamma_\sigma$, its spin $ j_\sigma$, and its degeneracy $\omega_{i,\gamma_\sigma, j_\sigma}$ (recall equation~\eqref{eqn:2dBPSind} and the discussion there). To this end, we take advantage of the factorization property of the spectral curve~(\ref{eq:curve-factorization}), and of the consequent factorization property of Stokes matrices~(\ref{eq:S-wall-spectrum-1}). This essentially implies that the whole spectral network, including its soliton data, consists of $\left\lfloor N/2\right\rfloor$ non-interacting copies of a standard $A_1$ spectral network for the fundamental representation (\emph{i.e.} $N=2$), up to scaling factors such as in~(\ref{eq:mass-factor}) that distinguish among the different copies.

As explained in subsections~\ref{eq:universality} and~\ref{sec:near-punctures}, the surface defects we wish to study are placed near a puncture. In the case of the fundamental representation (\emph{i.e.} $N=2$), and if the defect is close enough to the puncture, the 2d theory is well-approximated by an $\cN=(2,2)$ supersymmetric $\ICP^1$ sigma model~\cite{Gaiotto:2011tf, Longhi:2016bte}. In this case, the spectrum of 2d BPS particles as well as their spin has already been computed, both via spectral networks in~\cite{Gaiotto:2011tf}, and via field theoretic techniques in~\cite{Dorey:1998yh}. Since the two computations agree, we will review the one using spectral networks, as it naturally lays the foundation for the generalization to $N>2$. 

Recall from~(\ref{eq:central-charges}), that the central charges of 2d BPS particles have the same phase as those of 4d BPS particles. Thus, we expect 2d BPS particles to manifest themselves within the spectral network at the critical phase $\vartheta_c$. This is precisely when the network becomes highly degenerate, and turns into the BPS graph. The analysis of soliton data for BPS graphs has been carried our in detail in~\cite{Longhi:2016wtv}, and we can directly borrow results from this reference. In particular, from~\cite[eq. (4.61)]{Longhi:2016wtv}, we immediately read off the result for one type of 2d particle relevant to our choice of defect.\footnote{\label{foot:2d-particles}
In~\cite{Longhi:2016wtv}, one should look at factors appearing in the denominators of the functions $Q^{(\pm)}$. These functions are the generating series of framed 2d-4d BPS states, and contain information both about 4d and 2d BPS particles. 2d BPS particles most often appear in the denominators of these expressions. Then, focusing on denominators, the expressions from~\cite{Longhi:2016wtv} feature three types of charges. Only one of these however corresponds to a cycle surrounding the puncture where the defect is placed, we therefore focus on that one.} 
Its charge is given by
\be
	\gamma_\sigma  \ = \  \gamma_L+\gamma_R \,,
\ee
where the notation $\gamma_L,\gamma_R$ refers to the notation in Figure~\ref{fig:puncture-cycles}. Following an argument outlined in~\cite[p.126]{Gaiotto:2011tf}, we take $\omega_{a_{12},\gamma_\sigma, j_{\sigma}} = \langle\gamma_\sigma,a_{12}\rangle$ to compute
\be
	\omega_{a_{12},\gamma_L+\gamma_R, j_{12}}  \ = \  \omega_{-a_{21},\gamma_L+\gamma_R, j_{12}}  \ = \  2 \,.
\ee
Moreover, by the same argument as given in~\cite{Gaiotto:2011tf}, we expect by symmetry (and by consistency with~\cite[eq. (4.61)]{Longhi:2016wtv}), that\footnote{\label{ft:omegas}Note that our convention is slightly different from~\cite{Gaiotto:2011tf}, justifying the signs in this formula. For us, a soliton of type $ij$ is a path from sheet $i$ to sheet $j$, whose central charge is $Z_{ij} = Z_j-Z_i$, where $Z_i\sim \tCW_i$ is the $i$-th vacuum value of the superpotential of the 2d theory. Thus, in our conventions $\gamma_{ij} = \gamma_j - \gamma_i$. The signs in~(\ref{eq:omega-vacua}) follow from splitting $a_{12}$ this way.}
\be\label{eq:omega-vacua}
	\omega_{1,\gamma_{\sigma_1},j_{\sigma_1}} =-1\,,\qquad 
	\omega_{2,\gamma_{\sigma_2},j_{\sigma_2}}=1\,.
\ee
Therefore, we find two BPS particles,  one for each vacuum, both with charge $\gamma_{\sigma_1}=\gamma_{\sigma_2}=\gamma_L+\gamma_R$, and with opposite degeneracies.

\begin{figure}[h!]
\begin{center}
\includegraphics[width=0.95\textwidth]{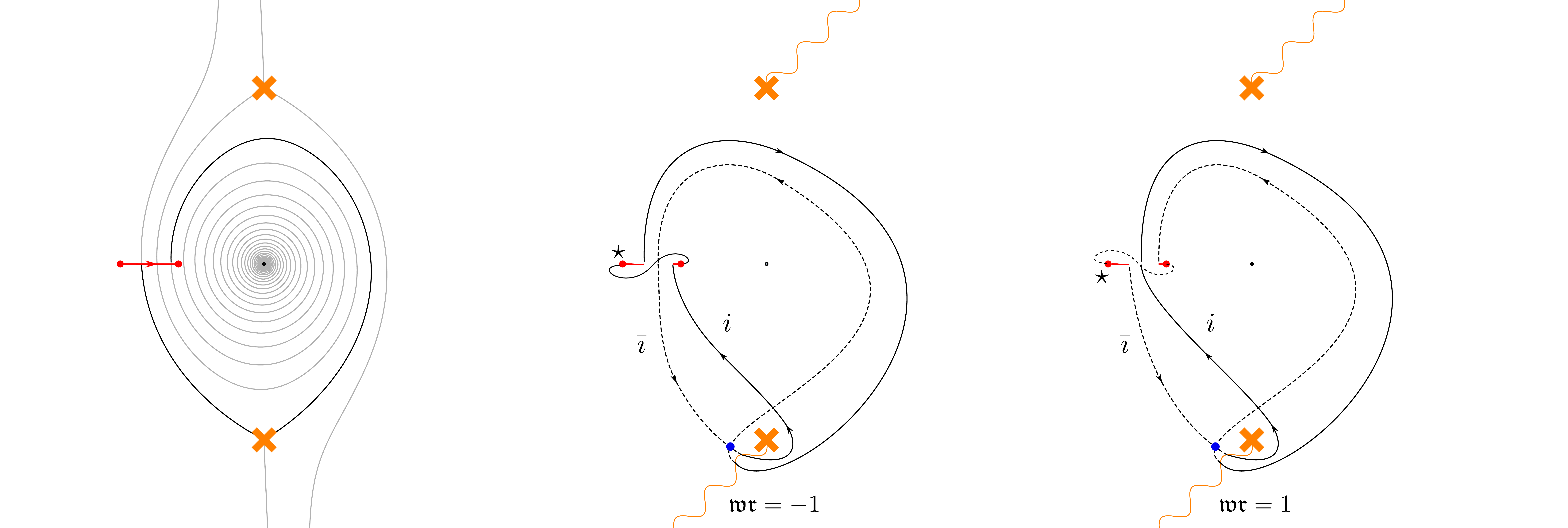}
\caption{2d BPS particles arising as concatenations of $\bar \imath i$ and $i\bar \imath$ 2d-4d solitons. The self-intersection of the path comes from a single point (the blue dot), and its sign depends on whether the concatenation begins on sheet $i$ or $\bar \imath$, corresponding to a particle in one vacuum or in the other.}
\label{fig:particle-spin}
\end{center}
\end{figure}

We can also fix the spins of these BPS particles, by recalling that this is captured by self-intersections of soliton paths~\cite{Galakhov:2014xba}. For the sake of illustration, let us consider a supersymmetric interface between two surface defects, as shown in Figure~\ref{fig:particle-spin}. We can recover a surface defect by sending the length of that interface to zero. However, it is easier to understand certain steps if we keep its length finite.\footnote{2d BPS particles appear as families of closed trajectories around the puncture when $\vartheta=\pi/2$, and in that limit one can replace this longer defect with an infinitesimal one sitting anywhere within the region near the puncture. All considerations made here then apply automatically to the infinitesimal interface.}
We will focus on the concatenation of detours from the $\CS$-walls pictured in black, and neglect the rest. The 2d-4d solitons supported on these walls are $a_{\bar \imath i}$ from sequence~(\ref{eq:2d4d-seq-1}) and $-a_{\bar \imath i} + \gamma_L^{(i)}+\gamma_R^{(i)}$ from sequence~(\ref{eq:2d4d-seq-2}). Therefore, their concatenation is $\gamma_L+\gamma_R$, which is the charge of the 2d BPS particles we wish to study.

There are two ways that these paths can be concatenated;\footnote{In spectral networks, these concatenations occur in the so-called ``American'' and ``British'' resolutions (see~\cite{Gaiotto:2012rg} for this terminology). But at the critical phase $\vartheta_c$ they both appear simultaneously for a surface defect modeled by an infinitesimal interface within the region near the puncture.} starting on sheet $1$ (denoted $i$ in Figure~\ref{fig:particle-spin}), or starting on sheet $2$ (denoted $\bar \imath$).
In either case the homology class of the concatenation is the same, however, the self-intersection of the path \emph{changes}. This is known as the writhe $\mathfrak{wr}$ and it is well-defined upon fixing a starting point, which is marked by a star, $\star$, in the picture, according to the orientation of the line interface.
The path that begins in vacuum $1$ has $\mathfrak{wr}=-1$, while the one that begins on sheet $2$ has $\mathfrak{wr}=+1$. 
This means that in vacuum $1$ there is a 2d particle with charge $\gamma_{\sigma_1}=\gamma_L+\gamma_R$ and spin $j_{\sigma_1}=-1$, whereas the one in vacuum $2$ has the same charge but opposite spin, \emph{i.e.}
\be\label{eq:particle-spin-N-eq-2}
	j_{\sigma_1} \ = \  -1\,, \qquad
	j_{\sigma_2} \ = \  1\,.
\ee

Thanks to the factorization property of spectral networks for higher symmetric representations, it is now straightforward to extend the above analysis to $N>2$,. 
There is simply one 2d BPS particle for each vacuum, with charges 
\be
	\gamma_{\sigma_{i}}  \ = \  \gamma_{\sigma_{\bar \imath}}  \ = \  \gamma^{(i)}_L+\gamma^{(i)}_R\,.
\ee
Note that these are pure-flavor charges from the viewpoint of the 4d theory, in the sense that their intersection with any closed cycle vanishes.\footnote{\label{foot:flavor}
To see this, it suffices to take representatives of these cycles, such as those shown in Figure~\ref{fig:puncture-cycles}, and to check that their sum is homologous to small circles surrounding the puncture.}
The BPS degeneracies of 2d particle states are determined by a trivial uplift of the analysis of~\cite{Longhi:2016wtv}, generalizing~(\ref{eq:omega-vacua}) to
\be\label{eq:2d-omega-N}
	\omega_{i,\gamma_{\sigma_{i}},j_{\sigma_{i}}}  \ = \  -1\,,\qquad 
	\omega_{ \bar \imath, \gamma_{\sigma_{\bar \imath}}, j_{\sigma_{\bar \imath}}}  \ = \ 1\,.
\ee

The geometric picture of 2d BPS particles also suggests a natural value of their spin. Recall that in the case $N=2$ we found in~(\ref{eq:particle-spin-N-eq-2}) that 
$j^{(N=2)}_{\sigma_1} = -1$, by adopting the dictionary proposed in~\cite{Galakhov:2014xba}, which identifies spin with self-intersections of soliton paths. 
We also argued in Section~(\ref{sec:spins}) that the $q$-power counting (self-)intersections gets corrected by a quadratic coefficient $(N-2k+1)^2$ on sheet $k$.
Together, we are thus led to the following ansatz for generic $N$
\be\label{eq:spin-proposal}
	j^{(N)}_{\sigma_k}  \ = \  (N-2k+1)^2 j^{(N=2)}_{\sigma_1}  \ = \  -(N-2k+1)^2 \,,
\ee
where $k$ is restricted to run over the first ``half'' of vacua $1\leq k\leq \left\lfloor (N+1)/2 \right\rfloor$. For the second half of vacua we expect $j^{(N)}_{\sigma_{\bar k}} = -j^{(N)}_{\sigma_{k}}$, since for $N=2$ the two spins have opposite sign~(\ref{eq:particle-spin-N-eq-2}).

This concludes our analysis of 2d BPS states for general surface defects in $N$-dimensional representations near punctures. In the next section, we turn towards explicit calculations of such surface defects, which in the following allows us to ``bootstrap'' the IR formalism.

\section{The 2d-4d IR formula and Macdonald operators}\label{sec:2d-4d-macdonald}

A remarkable relation between 2d-4d BPS spectra and defect Schur indices was proposed in~\cite{Cordova:2017ohl} (see also Section~\ref{sec:IRreview}). 
The ``2d-4d IR formula'' asserts that the trace of the 2d-4d spectrum generator coincides with the Schur index of the 4d theory with the insertion of a surface defect.

In the previous section, we considered a family of 2d-4d systems involving generic class $\CS$ theories of $A_1$ type, whose 2d sectors flow in the infrared to vortex defects. 
Using spectral networks we completely solved for the 2d-4d BPS spectra and computed the quantum 2d-4d spectrum generator explicitly.
We will now turn towards the evaluation of its trace, and show that it reproduces exactly the (degeneration limit of) Macdonald difference operators acting on the Schur index of the 4d theory, thereby deriving the striking relation conjectured in~\cite{Cordova:2017ohl} for this infinite and theory-independent class of defects.

\subsection{A deformation of the 2d-4d spectrum generator}

The 2d-4d spectrum generator is a product of Stokes factors corresponding to 2d particles, 2d-4d solitons, and 4d particles. 
The product is ordered according to phases of central charges, which depend on a choice of moduli of the 2d-4d system.
At the Roman locus there is a natural choice of half-plane in the $Z$-plane, which divides the whole spectrum into  particles and anti-particles.
We choose particles to be those BPS states whose central charges fall in the half-plane centered around the critical ray, and anti-particles to be the rest (see Figure~\ref{fig:2d-4d-central-charges}). If a particle lies on the boundary, it is sufficient to perturb the 2d moduli $z$ slightly to resolve the ambiguity.

We recall from Section~\ref{sec:IRreview}, that the 2d-4d BPS spectrum generator takes the form
\be\label{eq:2d-4d-spectrum-gen}
	\overline{\bS}_{\vartheta_c\pm\pi/2}^{2d-4d} {\bS}_{\vartheta_c\pm\pi/2}^{2d-4d} \,,
\ee
where ${\bS}_{\vartheta_c\pm\pi/2}^{2d-4d} $ contains the contributions of particles, whereas $\overline{\bS}_{\vartheta_c\pm\pi/2}^{2d-4d}$ contains the contributions of anti-particles. Given the choice of moduli discussed above, we can furthermore take each of these contributions to have the following general form
\be\label{eq:half-monodromy-decomposition}
	{\bS}_{\vartheta_c\pm\pi/2}^{2d-4d}  \ = \  {\bS}_{[\vartheta_c+\pi/2,\vartheta_c)}^{(2d-4d)} \, \bS_{\vartheta_c}^{(2d),(4d)} \, \bS^{(2d-4d)}_{[\vartheta_c,\vartheta_c-\pi/2)}\,,
\ee
where $\bS^{(2d-4d)}_{[\vartheta_2,\vartheta_1)}$ is the contribution of 2d-4d solitons with phases in the  interval $[\vartheta_1,\vartheta_2)$ (with $\vartheta_1<\vartheta_2$), and $ \bS_{\vartheta_c}^{(2d),(4d)} $ is the contribution of 2d particles and  4d  particles whose central charges all have the same critical phase $\vartheta_c$.

The spectrum generator~(\ref{eq:2d-4d-spectrum-gen}) contains information about BPS spectra, but alone it does \emph{not} compute the expectation value of a surface operator in the theory. To include the insertion of a surface operator we adopt the following trick: we consider a supersymmetric interface between two surface operators placed at $z_1, z_2$, which is modeled by a path $\wp$ on $C$ oriented from $z_1$ to $z_2$.
The advantage is that expectation values of supersymmetric interfaces can be computed using spectral networks data.
Eventually we take a limit where $z_1,z_2\to z$ and $\wp$ shrinks to a point, corresponding to the trivial interface on a unique surface defect placed at $z$, as illustrated in Figure~\ref{fig:A1-puncture-behavior}. 
We shall define the expectation value of the surface defect at $z$ as the shrinking limit of the interface expectation value.

A supersymmetric interface is parameterized both by a choice of path $\wp$ and by a phase $e^{\ii \vartheta}$, the latter determining which linear combination of supercharges is preserved by the interface.  The expectation value of the interface operator is computed by the parallel transport construction of spectral networks, denoted by $F(\wp,\vartheta)$. Without loss of generality we choose $\vartheta = {\vartheta_c-3\pi/2}$, and consider the insertion of this operator in the 2d-4d spectrum generator, \emph{i.e.}
\be\label{eq:2d-4d-spectrum-gen-with-defect}
	F(\wp,\vartheta_c-3\pi/2) \, \overline{\bS}_{\vartheta_c\pm\pi/2}^{2d-4d} {\bS}_{\vartheta_c\pm\pi/2}^{2d-4d} \,.
\ee
For ease of notation, we will sometimes omit the phase. In fact the position of $F(\wp)$ captures this piece of information due to the phase-ordering encoded in the definition.

\subsection{2d-4d cancellation mechanism}\label{sec:stokes-factors}

To evaluate~(\ref{eq:2d-4d-spectrum-gen-with-defect}), we shall start by studying the phase-ordered products of Stokes factors associated with 2d-4d solitons.
According to the analysis of the 2d, 4d, and 2d-4d BPS spectra carried out in Section~\ref{sec:BPS-spectrum}, the quantum spectrum generator~(\ref{eq:half-monodromy-decomposition}) and its conjugate take the following form
\be\label{eq:S-bar-S}
\begin{split}
	{\bS}_{\vartheta_c\pm\pi/2}^{2d-4d}  & \ = \  \bS_{II}^{(2d-4d)} \, \bS_{\vartheta_c}^{(2d),(4d)} \, \bS^{(2d-4d)}_{I}
	\,,\\
	\overline{\bS}_{\vartheta_c\pm\pi/2}^{2d-4d}  & \ = \  \overline{\CS}_{IV}^{(2d-4d)} \, \overline{\bS}_{\vartheta_c}^{(2d),(4d)} \, \overline{\CS}^{(2d-4d)}_{III}
	\,,
\end{split}
\ee
where the 2d-4d Stokes factors correspond to the four sectors $I-IV$ of 2d-4d solitons of surface defects near punctures that were analyzed in detail in Subsection~\ref{sec:near-punctures}.
Using our knowledge of the spectrum, these factors can be computed in closed form.
For example, the Stokes factor of sector $I$ contains two types of terms
\be\label{eq:stokes-overall-I}
	\bS_I^{(2d-4d)}  \ = \  \cdots \Upsilon_k \Delta_k \cdots \Upsilon_1 \Delta_1 \Upsilon_0 \Delta_0 \,,
\ee
where $\Upsilon,\Delta$ correspond to Stokes factors of $\CS$-walls sourced at the upper and lower branch points respectively. 
Their explicit form is 
\be
\begin{split}
	\Upsilon_{k} &  \ = \  \mathbbm{1} +  \sum_i X_{a_{\bar \imath i}  + \gamma_R^{(i)} + k (\gamma_L^{(i)}  + \gamma_R^{(i)})}  \,,    \\
	\Delta_{k} &  \ = \  \mathbbm{1} +  \sum_i X_{a_{\bar \imath i}  + k (\gamma_L^{(i)}  + \gamma_R^{(i)})}    \,,
\end{split}
\ee
where, for simplicity, we omitted the matrices $e_{\bar \imath i}$ from~(\ref{eq:S-wall-spectrum-1}).
These are now superfluous since their role is embodied by the algebra of formal variables associated with open paths (\ref{eq:q-torus-alg-oo}).
The whole product is therefore
\be\label{eq:stokes-sector-I}
\begin{split}
	\bS_I^{(2d-4d)} 
	&  \ = \ 
	\mathbbm{1} + \sum_{k\geq 0} \sum_{i=1}^{\lfloor N/2 \rfloor} X_{a_{\bar \imath i}  +  \gamma_R^{(i)} + k (\gamma_L^{(i)}  + \gamma_R^{(i)})} 
	+ X_{a_{\bar \imath i}  + k (\gamma_L^{(i)}  + \gamma_R^{(i)})}   
	\\
	&  \ = \  
	\mathbbm{1} 
	+ \sum_{k\geq 0} \sum_{i=1}^{\lfloor N/2 \rfloor} 
	\(
	1 + 
	q^{\frac{1}{2} \kappa_i }  X_{\gamma_R^{(i)}}  
	\) 
	q^{k\cdot \kappa_i}X_{ k (\gamma_L^{(i)}  + \gamma_R^{(i)})}   
	X_{a_{\bar \imath i}}   
	\\
	&  \ = \  
	\mathbbm{1} 
	+ \sum_{i=1}^{\lfloor N/2 \rfloor} 
	\frac{
	1 + q^{\frac{1}{2} \kappa_i }  X_{\gamma_R^{(i)}}  
	}{
	1 - q^{\kappa_i}X_{  \gamma_L^{(i)}  + \gamma_R^{(i)}}   
	}
	X_{a_{\bar \imath i} }   \,,
	\\
\end{split}
\ee
where we used the quantum torus algebra rules derived in Subsection~\ref{sec:spins}, and $\kappa_i$ was defined in~(\ref{eq:kappa-i}).
A similar computation yields
\be
\begin{split}
	\bS_{IV}^{(2d-4d)} 
	&  \ = \  
	\mathbbm{1} 
	+ \sum_{i=1}^{\lfloor N/2 \rfloor} 
	\frac{
	1 + q^{-\frac{1}{2} \kappa_i }  X_{-\gamma_R^{(i)}}  
	}{
	1 - q^{- \kappa_i}X_{  -\gamma_L^{(i)}  - \gamma_R^{(i)}}   
	}
	q^{-\frac{1}{2}  \kappa_i}X_{  -\gamma_L^{(i)}}\, 
	X_{a_{\bar \imath i} }   
	\\
	&  \ = \ 
	\mathbbm{1} 
	- \sum_{i=1}^{\lfloor N/2 \rfloor} 
	\frac{
	1 + q^{\frac{1}{2} \kappa_i }  X_{\gamma_R^{(i)}}  
	}{
	1 - q^{ \kappa_i}X_{  \gamma_L^{(i)}  + \gamma_R^{(i)}}   
	}
	X_{a_{\bar \imath i}}   
	\,.
\end{split}
\ee
If follows immediately that
\be\label{eq:canc-I-IV}
	\bS_{I}^{(2d-4d)} \overline\bS_{IV}^{(2d-4d)}   \ = \  1\,.
\ee
An analogous computation verifies the cancellation of the two other Stokes factors of 2d-4d states, \emph{i.e.}
\be\label{eq:canc-II-III}
	\overline\bS_{III}^{(2d-4d)} \bS_{II}^{(2d-4d)}   \ = \  1\,.
\ee

We shall argue in a moment that 
\be\label{eq:F-S-commutativity}
	F(\wp) \overline\bS_{IV}^{(2d-4d)}  \ = \  \overline\bS_{IV}^{(2d-4d)} F(\wp)\,.
\ee
Our final goal is to evaluate the trace of~(\ref{eq:2d-4d-spectrum-gen-with-defect}). Thus, combining~(\ref{eq:F-S-commutativity}) with the cyclicity-property of the trace, we can take advantage of the cancellations in equations~(\ref{eq:canc-I-IV}) and~(\ref{eq:canc-II-III}), thus leaving us with the remaining computation of (the trace of)
\be\label{eq:2d-4dspec-gen-bare}
	F(\wp)  \, \overline{\bS}_{\vartheta_c}^{(2d),(4d)} \, \bS_{\vartheta_c}^{(2d),(4d)} \,.
\ee

To derive~(\ref{eq:F-S-commutativity}), we note that $\overline\bS_{IV}^{(2d-4d)}$ contains contributions from 2d-4d solitons whose paths have a well-defined geometry. 
As previously discussed, they are lifts -- as defined in~(\ref{eq:lifting}) -- of the $\CS$-walls that end on the surface defect at $z$ at various phases in sector $IV$. 
However, recall that we take $\wp$ to be a very short path whose endpoints are infinitesimally close to $z$.
When an $\CS$-wall ends at $z$ (for a certain value of the phase $\vartheta$), it does not intersect $\wp$ due to the fact that the latter is infinitesimally short (see Figure~\ref{fig:A1-puncture-behavior}). 
This implies that none of the lifts $\wp^{(i)}$ will intersect \emph{any} of the open paths comprising the soliton data of the $\CS$-wall, at least for the given phase. 
Since all intersections are trivial, all $X$-variables in $\overline\bS_{IV}^{(2d-4d)} $ commute with those in $F(\wp)$, according to the algebra rules reviewed in Subsection~\ref{sec:spins}.\footnote{
For this argument to work it is of course crucial that we use \emph{actual paths} (as opposed to homology classes) to compute intersection pairings (see also footnote~\ref{foot:actual-paths}).}

\subsection{Macdonald operators from 2d particle spectra}

The remarkable cancellation of 2d-4d soliton Stokes factors, discussed in the previous subsection, leads to the much simpler problem of computing the trace of~(\ref{eq:2d-4dspec-gen-bare}). Furthermore, we have already explicitly computed the 2d particle spectrum in Subsection~\ref{sec:2d-particle}, and found that all 2d particles carry 4d charges that correspond to pure 4d flavor charges of the form $\gamma_L^{(i)}+\gamma_R^{(i)}$. In particular, this implies that all Stokes factors of 2d particles commute with those of 4d particles, leading to the following additional factorization
\be
\begin{split}
	\bS_{\vartheta_c}^{(2d),(4d)} &  \ = \  \bS_{\vartheta_c}^{(2d)} \,  \bS_{\vartheta_c}^{(4d)} \,,\\
	\overline{\bS}_{\vartheta_c}^{(2d),(4d)} &  \ = \  \overline{\bS}_{\vartheta_c}^{(2d)} \,  \overline{\bS}_{\vartheta_c}^{(4d)} \,.
\end{split}
\ee
Thus, the trace of~(\ref{eq:2d-4dspec-gen-bare}) takes the general form
\be\label{eq:spec-gen-particles-repeat}
\begin{split}
	& \Tr \(F(\wp)  \, \overline{\bS}_{\vartheta_c}^{(2d)}\,\bS_{\vartheta_c}^{(2d)} \,  \overline{\bS}_{\vartheta_c}^{(4d)} \,  \bS_{\vartheta_c}^{(4d)} \)\,.
\end{split}
\ee

Notice that 2d BPS particles come in CPT conjugate pairs, and the Stokes factors for each pair are given by
\be
	\(1- q^{\frac{j_\sigma}{2}} X_{\gamma_{\sigma}}\)^{\omega_{ i,\gamma_{\sigma},j_{\sigma}}}
	\(1- q^{- \frac{j_\sigma}{2}} X_{-\gamma_{\sigma}}\)^{\omega_{i,-\gamma_{\sigma}, -j_\sigma}} \,.
\ee
Using the general property $\omega_{i,-\gamma_\sigma,-j_\sigma} = -\omega_{i,\gamma_\sigma,-j_\sigma}$ as well as the inversion property in equation~(\ref{eq:q-torus-alg}), each pair of Stokes factors simplifies to a monomial
\be\label{eq:2d-part-antipart}
	\(- q^{\frac{j_\sigma}{2}} X_{\gamma_\sigma}\)^{\omega_{i,\gamma_\sigma,j_\sigma}}\,.
\ee

The trace in~(\ref{eq:spec-gen-particles-repeat}) involves both a sum over 2d vacua and a trace in the quantum torus algebra. Carrying out the former leads to
\be\label{eq:spec-gen-particles-repeat-simplified}
\begin{split}
	 \sum_{i=1}^{N}
	\hTr \left[
	X_{\wp^{(i)}}\,  
	\(- q^{\frac{j^{(N)}_{\sigma_i}}{2}} X_{\gamma_{\sigma_i}}\)^{\omega_{i,\gamma_{\sigma}, j_\sigma}}
	\overline{\bS}_{\vartheta_c}^{(4d)} \,  \bS_{\vartheta_c}^{(4d)} 
	\right]
	\,.
\end{split}
\ee
The sum runs over 2d vacua, in each of which the BPS spectrum contains (in addition to the 4d BPS spectrum) a single 2d particle (and its anti-particle), as discussed in Subsection~\ref{sec:spins}. Indeed, the spins $j^{(N)}_{\sigma_i}$ are the ones given in~(\ref{eq:spin-proposal}), and $\omega_{i,\gamma_{\sigma}, j_\sigma}$ is given in~(\ref{eq:2d-omega-N}).
The variable $X_{\wp^{(i)}}$ is the (matrix) element of $F(\wp)$ acting on the vacuum $i$, and $\hTr$ denotes the trace in the quantum torus algebra.
The 4d BPS spectrum is encoded in the operators $\overline{\bS}_{\vartheta_c}^{(4d)} \,  \bS_{\vartheta_c}^{(4d)}$, whose details depend on the 4d theory, and we will not make any specific assumptions about it, except that its trace evaluates to the Schur index of the 4d theory.\footnote{It is crucial to note that 2d Stokes factors are \emph{pure-flavor} and therefore do not alter the trace of the 4d Stokes factors, by virtue of how the trace is defined on the quantum torus algebra.}

Let us write down~(\ref{eq:spec-gen-particles-repeat-simplified}) a bit more explicitly; it takes the following form
\beaa\label{eq:even-intermediate}
	&
	 - \hTr \Bigg[
	\bigg\{
		X_{\wp^{(1)}}\,  \Big(q^{\frac{j^{(N)}_{\sigma_1}}{2}} X_{\gamma_{L}^{(1)}+ \gamma_{R}^{(1)}}\Big)^{-1} 
		+ \cdots
		+X_{\wp^{(N/2)}}\,  \Big(q^{\frac{j^{(N)}_{\sigma_{N/2}}}{2}} X_{\gamma_{L}^{(N/2)}+ \gamma_{R}^{(N/2)}} \Big)^{-1}
		\\
	& \qquad\quad  
		+X_{\wp^{(N/2+1)}}\,  q^{\frac{j^{(N)}_{\sigma_{N/2+1}}}{2}} X_{\gamma_{L}^{(N/2)}+ \gamma_{R}^{(N/2)}} 
		+ \cdots
		+X_{\wp^{(N)}}\,  q^{\frac{j^{(N)}_{\sigma_N}}{2}} X_{\gamma_{L}^{(1)}+ \gamma_{R}^{(1)}} 
	\bigg\}  
	\overline{\bS}_{\vartheta_c}^{(4d)} \,  \bS_{\vartheta_c}^{(4d)} 
	\Bigg]
	\,,
\eeaa
if $N$ is even, while it takes the following form
\beaa\label{eq:odd-intermediate}
	&
	 - \hTr \Bigg[
	\bigg\{
		X_{\wp^{(1)}}\, \Big( q^{\frac{j^{(N)}_{\sigma_1}}{2}} X_{\gamma_{L}^{(1)}+ \gamma_{R}^{(1)}} \Big)^{-1}
		+ \cdots
		+X_{\wp^{((N-1)/2)}}\,  \Big(q^{\frac{j^{(N)}_{\sigma_{(N-1)/2}}}{2}} X_{\gamma_{L}^{((N-1)/2)}+ \gamma_{R}^{((N-1)/2)}} \Big)^{-1}
		\\
	& \quad
		+X_{\wp^{((N-1)/2+1)}}\\
	&\quad
		+X_{\wp^{((N-1)/2+2)}}\,  q^{\frac{j^{(N)}_{\sigma_{(N-1)/2+2}}}{2}} X_{\gamma_{L}^{((N-1)/2)}+ \gamma_{R}^{((N-1)/2)}}
		+ \cdots
		+X_{\wp^{(N)}}\,  q^{\frac{j^{(N)}_{\sigma_N}}{2}} X_{\gamma_{L}^{(1)}+ \gamma_{R}^{(1)}}  
	\bigg\}
	\overline{\bS}_{\vartheta_c}^{(4d)} \,  \bS_{\vartheta_c}^{(4d)} 
	\Bigg]
	\,,
\eeaa
if $N$ is odd. 

Now, let us remark upon an important detail: how are we supposed to deal with the operator $X_{\wp^{(i)}}$?  According to the rules of the quantum torus algebra (suitably extended to open paths as discussed in Subsection~\ref{sec:spins}), the effect of inserting $X_{\wp^{(i)}}$ is to replace each variable 
\be\label{eq:q-shifts-wp}
	X_{\gamma} \ \to \ q^{\frac{1}{2}\langle\langle\wp^{(i)},\gamma\rangle\rangle} X_{\wp^{(i)}+\gamma} \,.
\ee
To compute the expectation value of a surface defect we eventually take the limit of shrinking $\wp$ to a point $z$. 
In this limit, the above expression eventually approaches $ q^{\frac{1}{2}\langle\langle\wp^{(i)},\gamma\rangle\rangle} X_{\gamma}$.

Recall that the trace $\widehat\Tr$ on the quantum torus algebra singles out gauge-neutral operators. As a matter of fact, it replaces them with the fugacities associated to flavor symmetries, as dictated by the flavor charge of each operator (see~\cite{Cecotti:2010fi, Cordova:2016uwk}).
Since $\gamma_L^{(i)}+\gamma_R^{(i)}$ are pure-flavor cycles, they can be directly replaced with suitable fugacities. Actually, applying the relation~(\ref{eq:holonomy-powers}), leads to the following identification
\be\label{eq:X-gamma-a-rel}
	X_{\gamma_{L}^{(i)}+ \gamma_{R}^{(i)}}   \ = \  \(X^{(N=2)}_{\gamma_{L}+ \gamma_{R}}\)^{N-2i+1}  \ = \  a^{2(N-2i+1)} \,,
\ee
where the last equality defines the fugacity $a$, associated to the puncture (near the surface defect) we are studying. Namely, $a=e^{m/2}$, if the Higgs field of the $A_1$ Hitchin system behaves as follows near the puncture
\be\label{eq:simple-pole}
	\varphi \ \sim \ \frac{dz}{z} \(\begin{array}{cc}-m/2 & \\ & m/2\end{array}\) \,.
\ee
Note that~(\ref{eq:X-gamma-a-rel}) is compatible with~(\ref{eq:mass-factor}). From Figure~\ref{fig:puncture-cycles} one readily observes that
\be\label{eq:flavor-cycles}
	\gamma^{(i)}_L+\gamma^{(i)}_R  \ = \  C^{(\bar \imath)} - C^{(i)} \,,
\ee
where $C^{(i)}, C^{(\bar \imath)}$ are counter-clockwise cycles around the lift of the puncture to sheets $i,\bar \imath$, respectively.
Since $\bar \imath = N + 1 - i$, we obtain 
\be\label{eq:sheets-near-puncture}
	\lambda_i \ \sim \ -\frac{1}{z} (N-2i+1) \frac{m}{2} \,,
	\qquad
	\lambda_{\bar \imath} \ \sim \ \frac{1}{z} (N-2i+1) \frac{m}{2}\,,
\ee
and it follows that 
\be
	X_{ C^{(\bar \imath)}}  \ = \  X_{ - C^{(i)}}  \ = \  e^{\frac{m}{2}(N-2i+1)}  \ = \  a^{N-2i+1}\,,
\ee
and therefore $X_{\gamma^{(i)}_L+\gamma^{(i)}_R} = X_{C^{(\bar \imath)}} X_{ - C^{(i)}}$ agrees with~(\ref{eq:X-gamma-a-rel}).

To compute the $q$-shifts arising from equation~(\ref{eq:q-shifts-wp}), we use the geometry of the cycles $\gamma^{(i)}_L+\gamma^{(i)}_R$, at which we arrive by lifting the flow of the quadratic differential. Now, by direct inspection of Figure~\ref{fig:puncture-cycles}, we propose 
\be
	\langle\wp^{(i)} , C^{(j)} \rangle  \ = \  -\delta_{ij} \,,
\ee
where $1\leq i,j\leq N$ run over all sheets. 

In computing these intersections, it is essential to regard $C^{(i)}$ as \emph{families} of circular paths around the puncture, that foliate the punctured disk bounded by double-walls $p_L,p_R$. This is consistent with the interpretation of the foliation as a solution to the 2d BPS equations~\cite{Gaiotto:2011tf}, and leads us to infer that $\wp^{(i)}$ intersects at least \emph{some} leaves of the foliation.\footnote{Again, we remind the reader of comments in footnote~\ref{foot:actual-paths}.} In turn, this means that the formal parallel transport along $\wp$ picks up a correction due to the corresponding detour.

Thus, using this information about the intersections, we can finally compute the $q$-shifts. They read
\be
\begin{split}
	X_{\wp^{(i)}} X_{C^{(j)}} 
	&  \ = \  q^{\frac{1}{2} \langle\langle \wp^{(i)} , C^{(j)} \rangle\rangle } 
	X_{\wp^{(i)} +C^{(j)} } 
	 \ = \  q^{-\frac{1}{2} \delta_{ij} \(N-2i+1\)^2 } 
	X_{\wp^{(i)} + C^{(j)} } \,.
\end{split}	
\ee
Upon taking the limit in which the supersymmetric interface approaches a surface defect (\emph{i.e.} shrinking  $\wp$ to a point $z$), this means that these $q$-shifts result in the explicit substitution
\be
a \ \to \ q^{\frac{1}{2}  \(N-2i+1\) } \cdot a \,.
\ee
of the flavor fugacity at the puncture.
 
Finally, we arrive at the conclusion that if $N$ is even, then~(\ref{eq:even-intermediate}) evaluates to\footnote{The $q$-shifts of $a$-powers in the prefactor are doubled because we bring $X_{\wp^{i}}$ across these variables of type $X_{\gamma_L^{(i)}+\gamma_R^{(i)}}$.} 
\be\label{eq:even-final}
\begin{split}
	&
		q^{-\frac{j^{(N)}_{\sigma_1}}{2} - (N-1)^2} \, a^{-2(N-1)} \cdot
		\CI(a\to a q^{\frac{1}{2}(N-1)})
		+ \cdots
		+ q^{-\frac{j^{(N)}_{\sigma_{N/2}}}{2} - 1} \,a^{-2} \cdot  
		\CI(a\to a q^{\frac{1}{2}})
		\\
	& \qquad  
		+ q^{\frac{j^{(N)}_{\sigma_{N/2+1}}}{2} - 1} \,a^{2} \cdot  
		\CI(a\to a q^{-\frac{1}{2}})
		+ \cdots +
		q^{\frac{j^{(N)}_{\sigma_N}}{2} - (N-1)^2} \, a^{2(N-1)} \cdot
		\CI(a\to a q^{-\frac{1}{2}(N-1)})
	\,,
\end{split}
\ee
up to an overall sign. Here, $\cI$ is the Schur index for an arbitrary rank-one class $\CS$ theory (with at least two regular punctures),\footnote{\label{foot:othercases} 
Extension to the cases of class $\CS$ theories associated once-punctured UV curves and (particular) including irregular punctures are provided in Sections~\ref{sec:one-puncture} and~\ref{sec:irregular}.}  
\emph{i.e.}
\be
	\cI \ = \ (q,q)_\infty^{2r}  \,
	\hTr \left[
	\overline{\bS}_{\vartheta_c}^{(4d)} \,  \bS_{\vartheta_c}^{(4d)} 
	\right]
	\,.
\ee
Similarly, if $N$ is odd,~(\ref{eq:odd-intermediate})  evaluates to$^{\ref{foot:othercases}}$
\be\label{eq:odd-final}
\begin{split}
	& q^{-\frac{j^{(N)}_{\sigma_1}}{2} - (N-1)^2} \,  a^{-2(N-1)} \cdot
		\CI(a\to a\, q^{\frac{1}{2}(N-1)})
		+ \cdots
		+ q^{-\frac{j^{(N)}_{\sigma_{(N-1)/2}}}{2} - 2^2} a^{-4} \cdot
		\CI(a\to a\, q)
		\\
	&  \qquad + \CI(a)
		\\
	&  \qquad + q^{\frac{j^{(N)}_{\sigma_{(N-1)/2+2}}}{2} - 2^2} a^{-4} \cdot
		\CI(a\to a\, q^{-1})
		+ \cdots
		+q^{\frac{j^{(N)}_{\sigma_N}}{2} - (N-1)^2} \,  a^{2(N-1)} \cdot
		\CI(a\to a\, q^{-\frac{1}{2}(N-1)})
	\,.
\end{split}
\ee
Using the explicit form of $j^{(N)}_{\sigma_i}, j^{(N)}_{\sigma_{\bar \imath}} $ in equation~(\ref{eq:spin-proposal}),
we can now write down a general formula, valid for any positive integer $N$
\be\label{eq:general-operator-formula}
	\boxed{
	\mathfrak{S}^{IR}_{v} \circ \CI \  : = \ 
	\sum_{k=1}^{N} 
	q^{-\frac{1}{2}(N-2k+1)^2}  a^{-2(N-2k+1)} \CI\(a q^{\frac{1}{2} \(N-2k+1\)}\)\,.
	}
\ee
This expression coincides with the action of difference operators, as expected from the insertion of vortex surface defects of vorticity $v=N-1$ in the evaluation of the Schur index~\cite{Gaiotto:2012xa,Gadde:2013ftv,Alday:2013kda} (modulo an inessential overall factor  $q^{\frac{1}{2} \(N^2-1\right)}$, see (\ref{HiggsSDs})).

\section{Examples}\label{sec:examples}

We shall now provide some explicit (detailed) examples of vortex surface defects for a given set of theories. The computation of the Schur indices for these examples will provide the ``base case'' for the induction arguments in Section~\ref{sec:bootstrap}. Another purpose of this section is to illustrate the somewhat technical discussion developed in earlier sections with some simple examples.

\subsection{The $T_2$ theory}\label{sec:T2}

The so-called $T_2$ theory is engineered as a class $\CS$-theory by a sphere with three punctures. It admits a Lagrangian description as the free theory of a hypermultiplet in the bifundamental representation of $SU(2)$ with flavor symmetry given by $SU(2) \times SU(2) \times SU(2)$. The Coulomb branch is trivial, and there is a unique 4d BPS spectrum, consisting of four particles and four anti-particles, which, taken all together, furnish the trifundamental representation of the flavor symmetry. The quadratic differential 
\be  
	\phi_2 
	 \ = \ 
	\frac{\mu_2^a z_{ab}z_{ac}(z-z_b)(z-z_c)  + ( \text{cycl}) } {(z-z_a)^2(z-z_b)^2(z-z_c)^2  } \, \diff z^2 \,,
\ee
is parameterized by (squares of) UV masses $\{ \mu_{2}^{a}, \mu_{2}^b, \mu_{2}^c \}$, and  positions of punctures $\{ z_a, z_b, z_c \}$. We also used the notation $z_{ab}= z_a-z_b$, \emph{et cetera}.

\begin{figure}[h!]
\begin{center}
\includegraphics[width=0.35\textwidth]{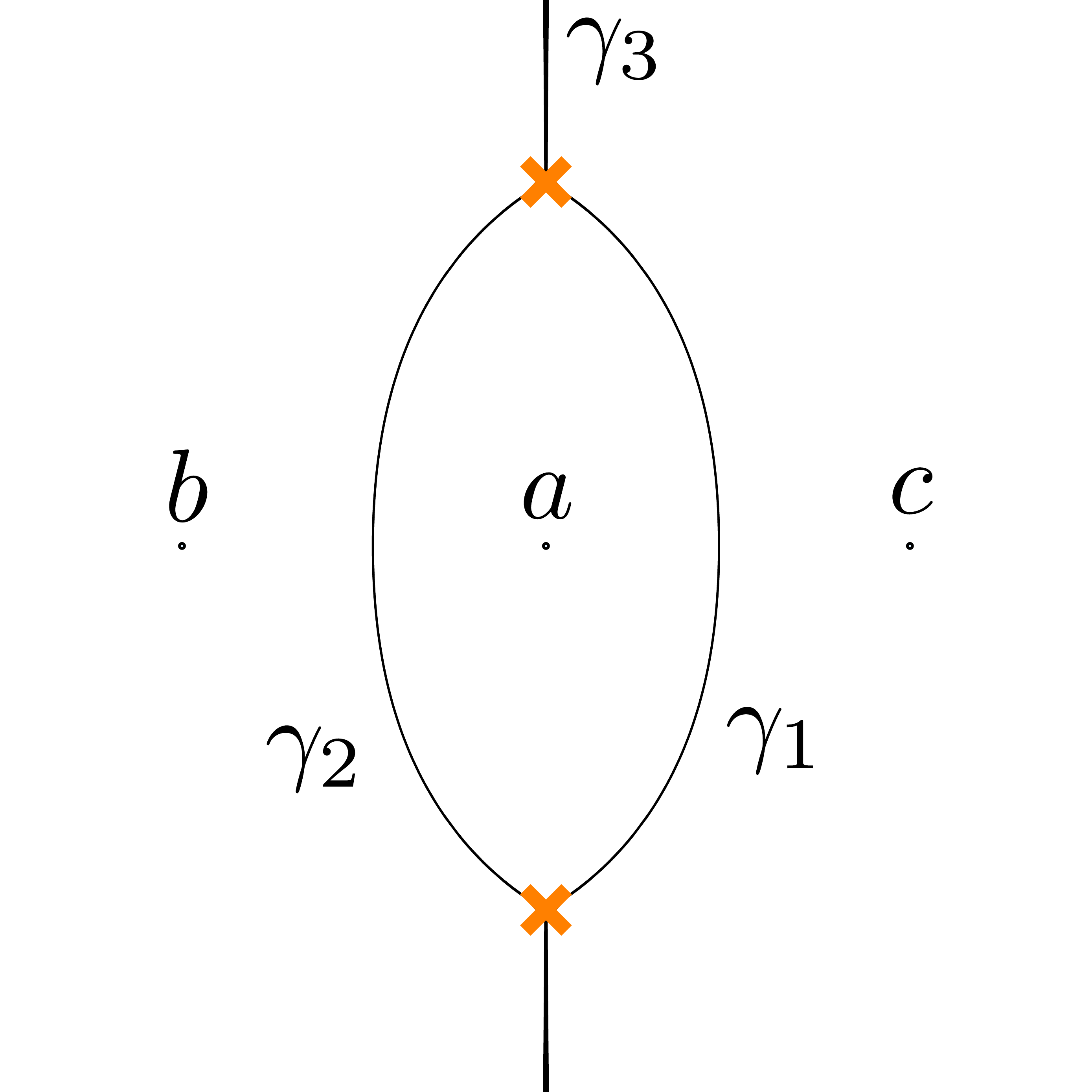}
\hspace{0.15\textwidth}
\includegraphics[width=0.35\textwidth]{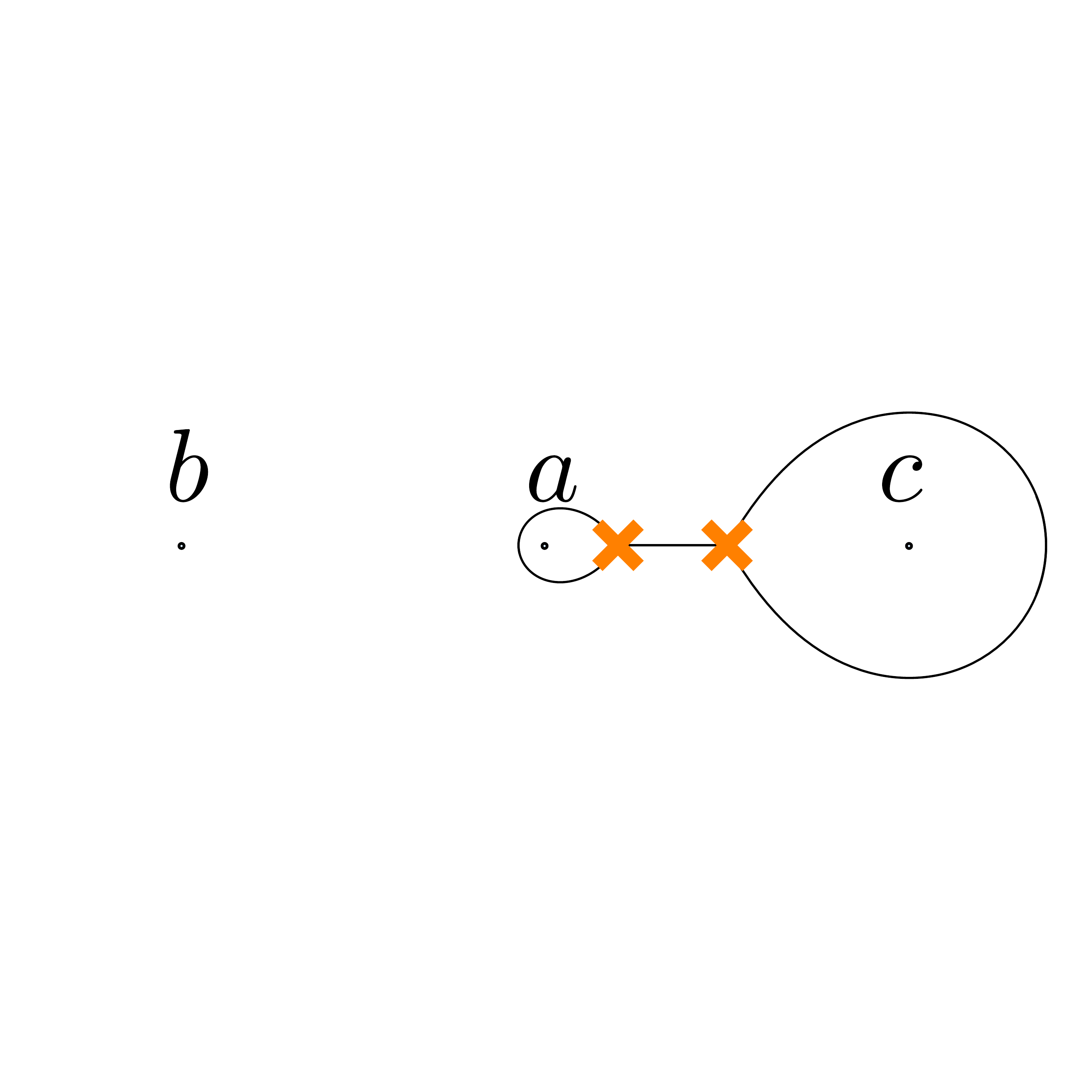}
\caption{Two BPS graphs of the $T_2$ theory. In the left graph, the edge labeled by $\gamma_3$ runs around $\ICP^1$ and connects the two branch points.}
\label{fig:T2-BPSg}
\end{center}
\end{figure}

The Roman locus is defined by requiring all masses to have identical phase, $\arg(\mu_2^{a})=\arg(\mu_2^{b})=\arg(\mu_2^{c})$. For instance, taking $\mu_2^a\in\IR_{>0}$, all central charges are real. Thus, plotting the spectral network at $\vartheta_c = 0$, we find one of the BPS graphs shown in Figure~\ref{fig:T2-BPSg}. Setting $\mu_2^a = \mu_2^b=\mu_2^c=1$ gives the BPS graph on the left, while setting $\mu_2^a =\mu_2^c=1,\ \mu_2^b = 9/2$ gives the one on the right.

\begin{figure}
\begin{center}
\includegraphics[width=0.22\textwidth]{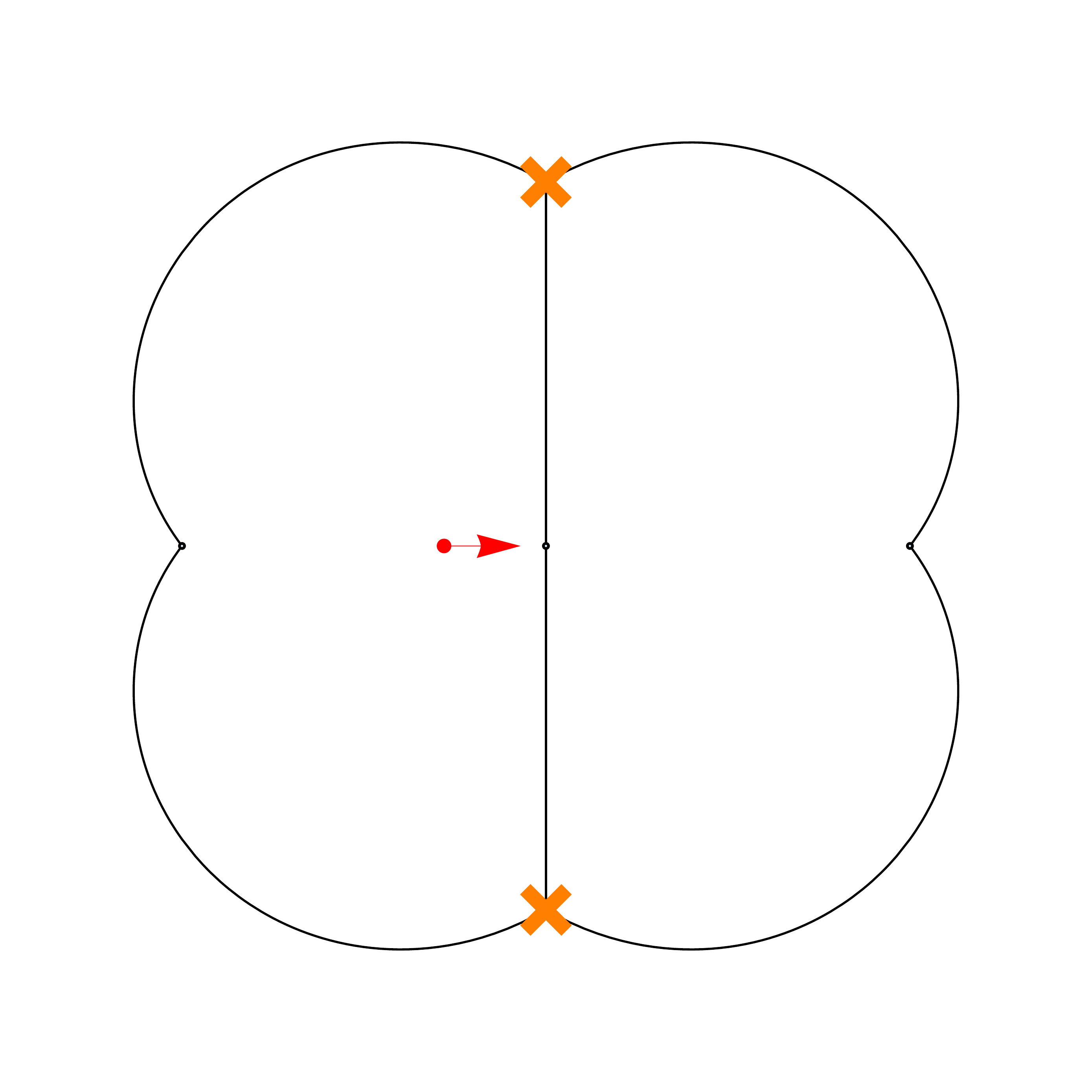}\hfill
\includegraphics[width=0.22\textwidth]{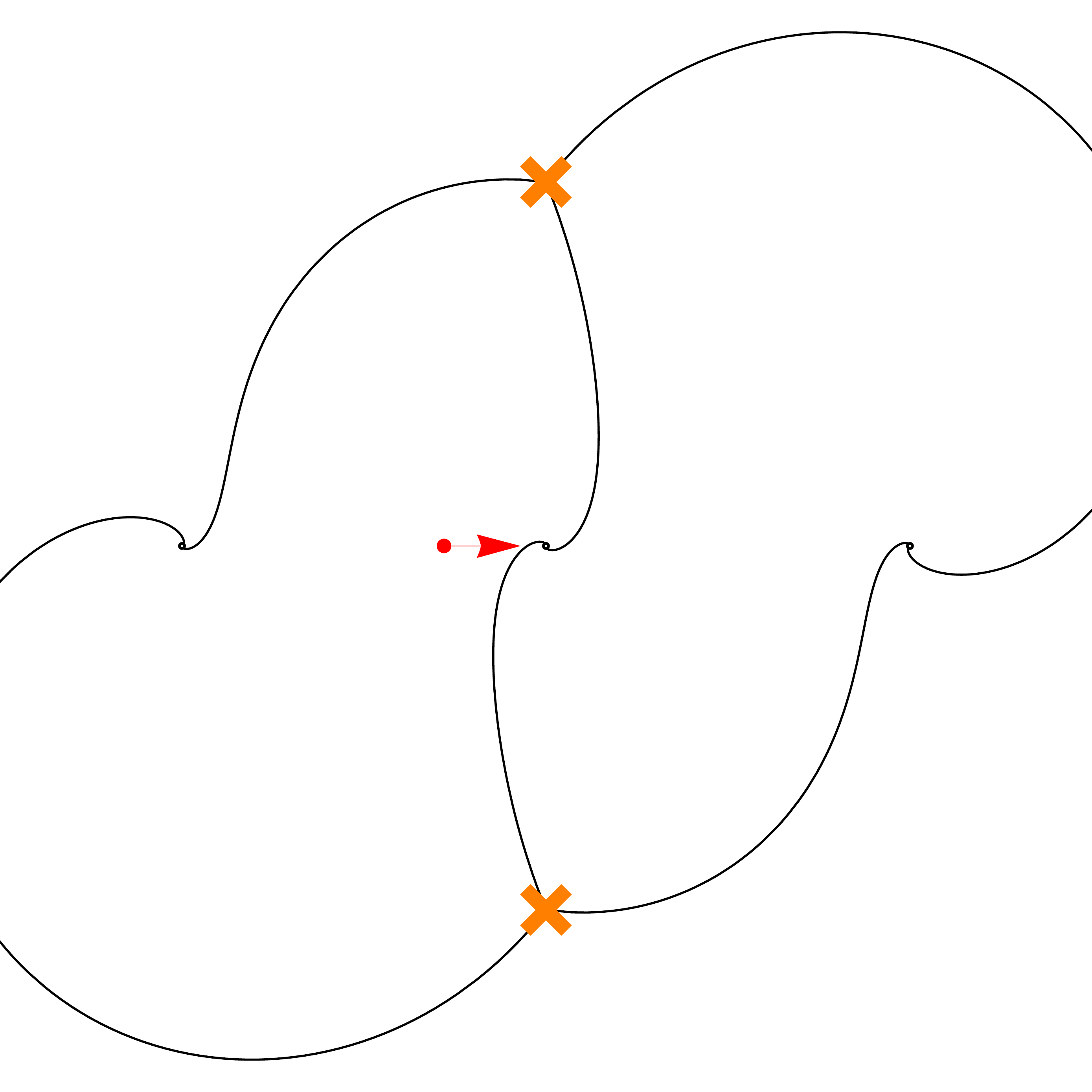}\hfill
\includegraphics[width=0.22\textwidth]{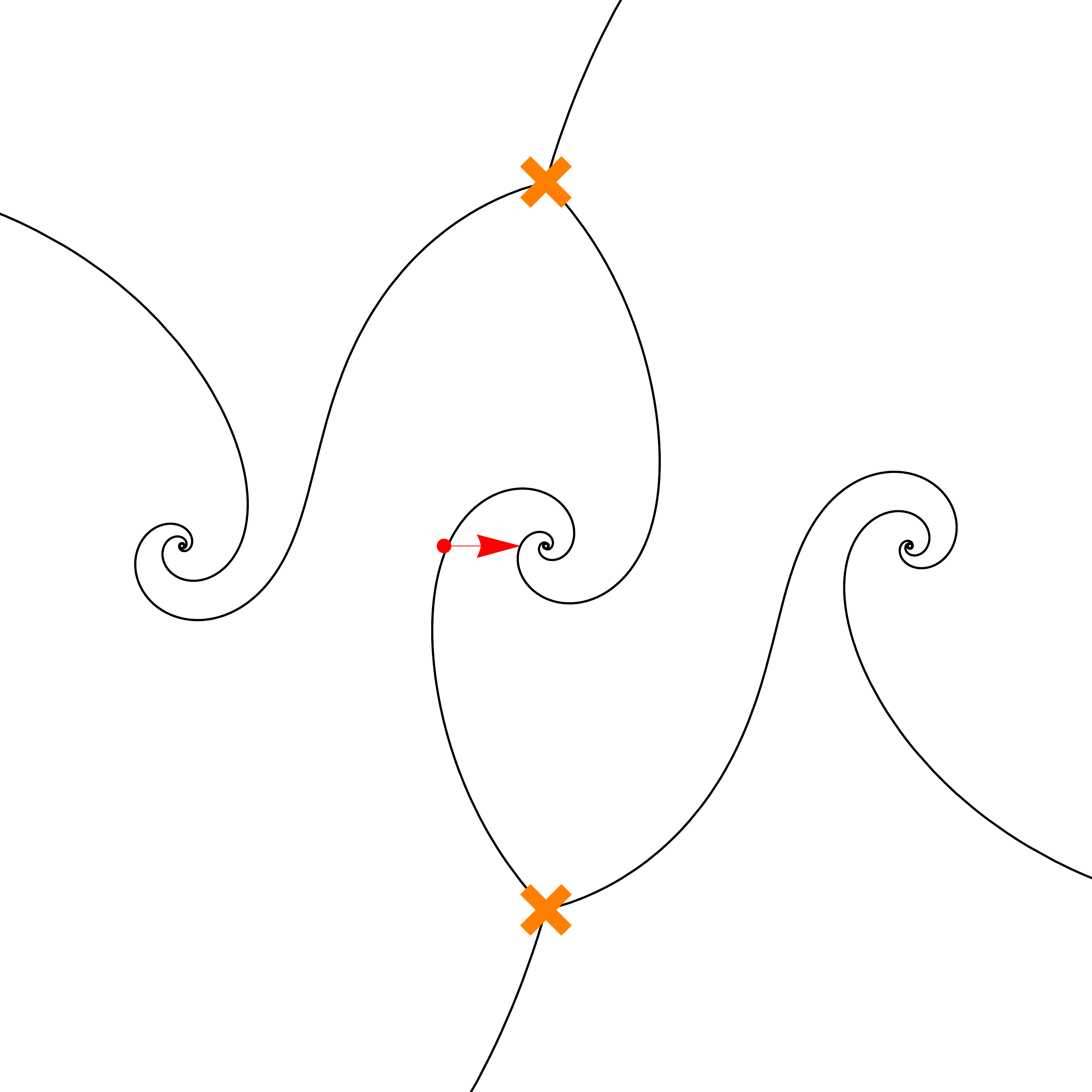}\hfill
\includegraphics[width=0.22\textwidth]{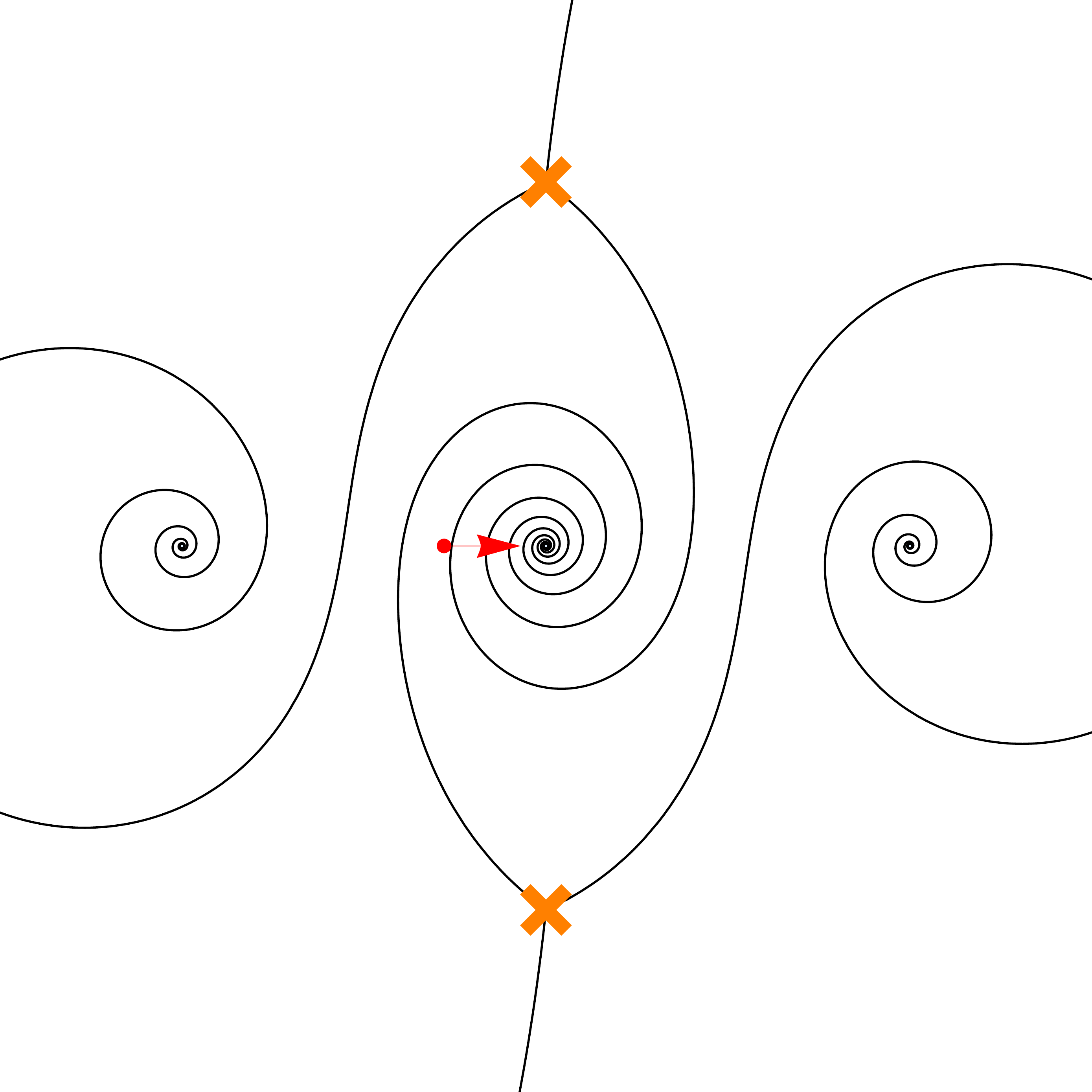}\\[5pt]
\includegraphics[width=0.22\textwidth]{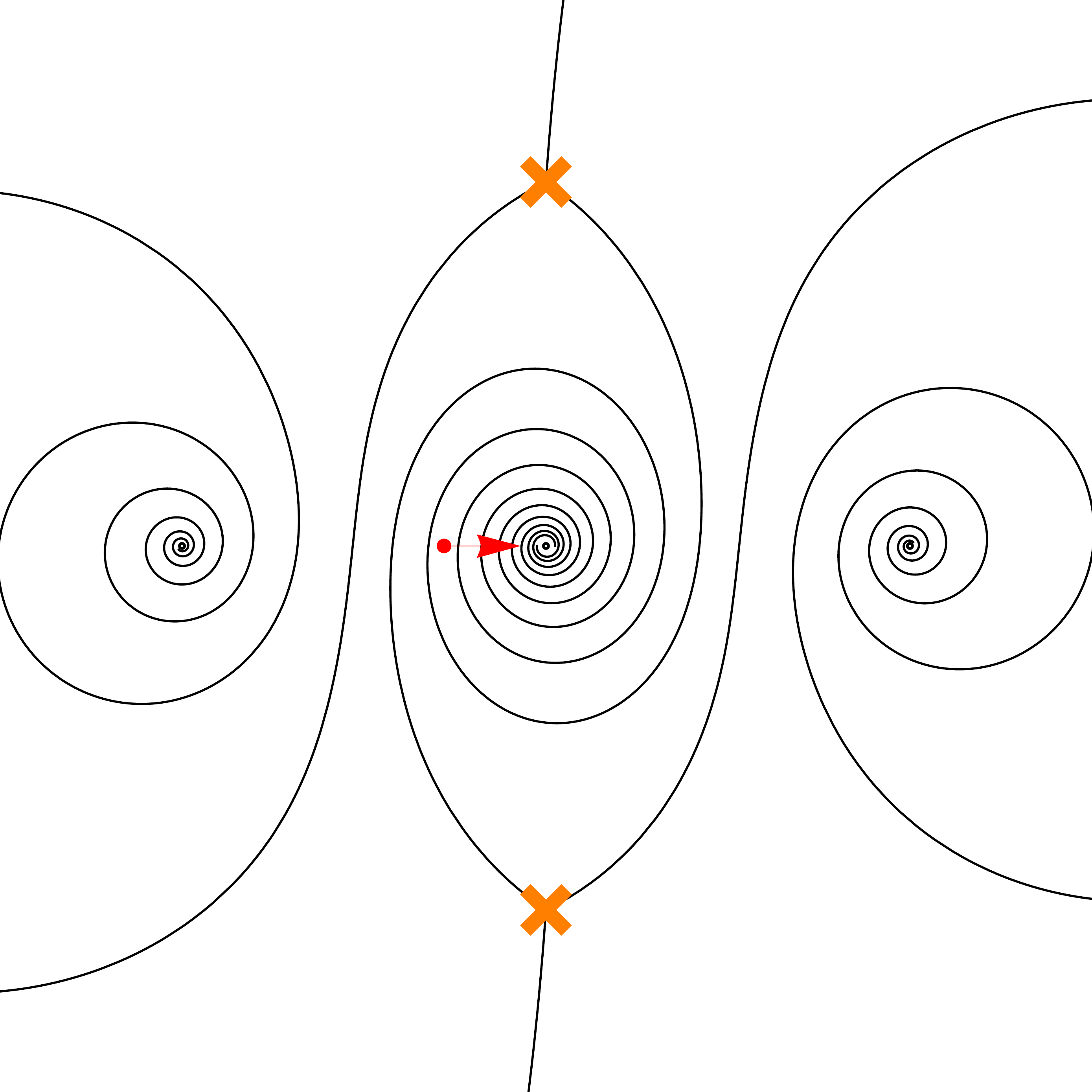}
\hfill\includegraphics[width=0.06\textwidth]{figures/dots.pdf}\hfill
\includegraphics[width=0.22\textwidth]{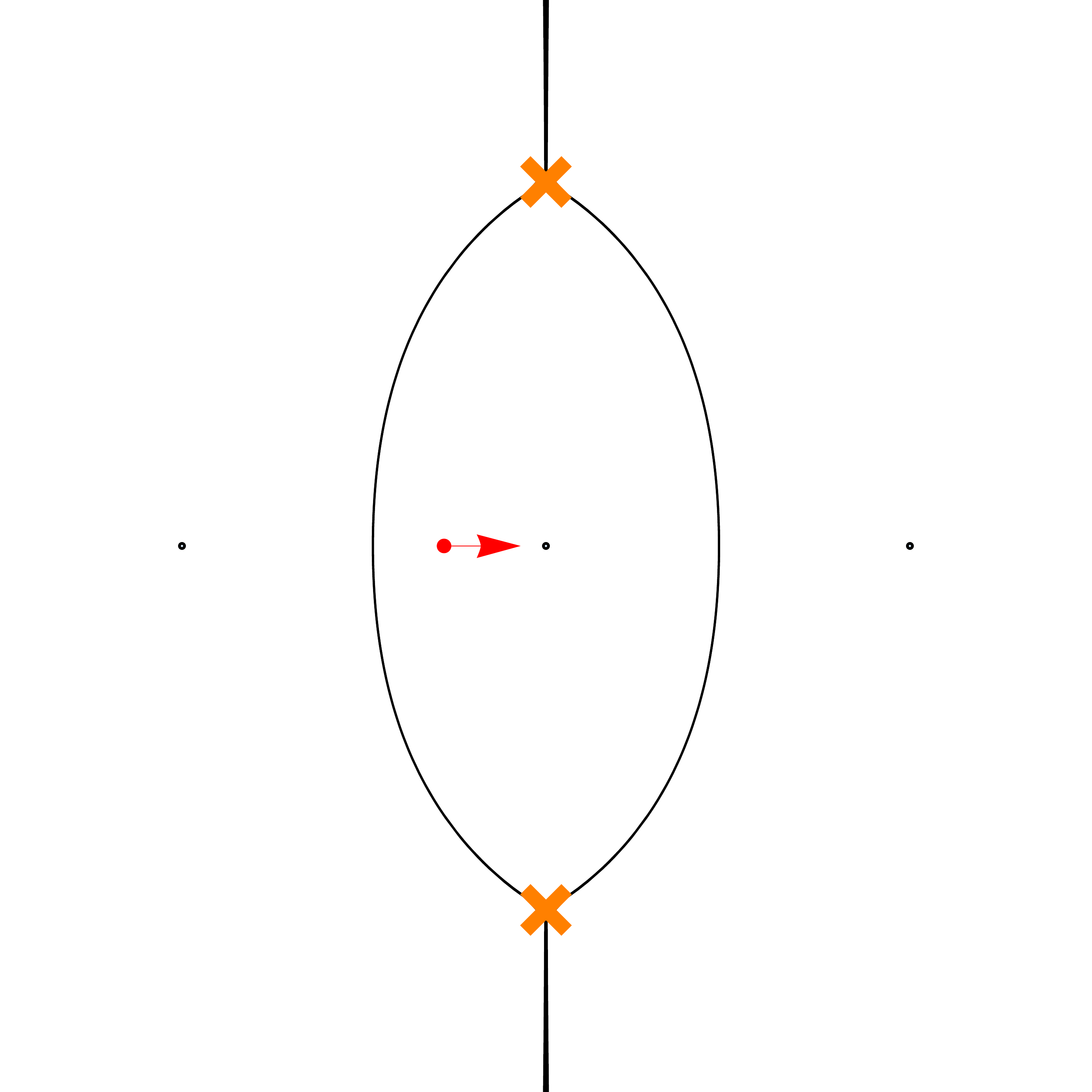}
\hfill\includegraphics[width=0.06\textwidth]{figures/dots.pdf}\hfill
\includegraphics[width=0.22\textwidth]{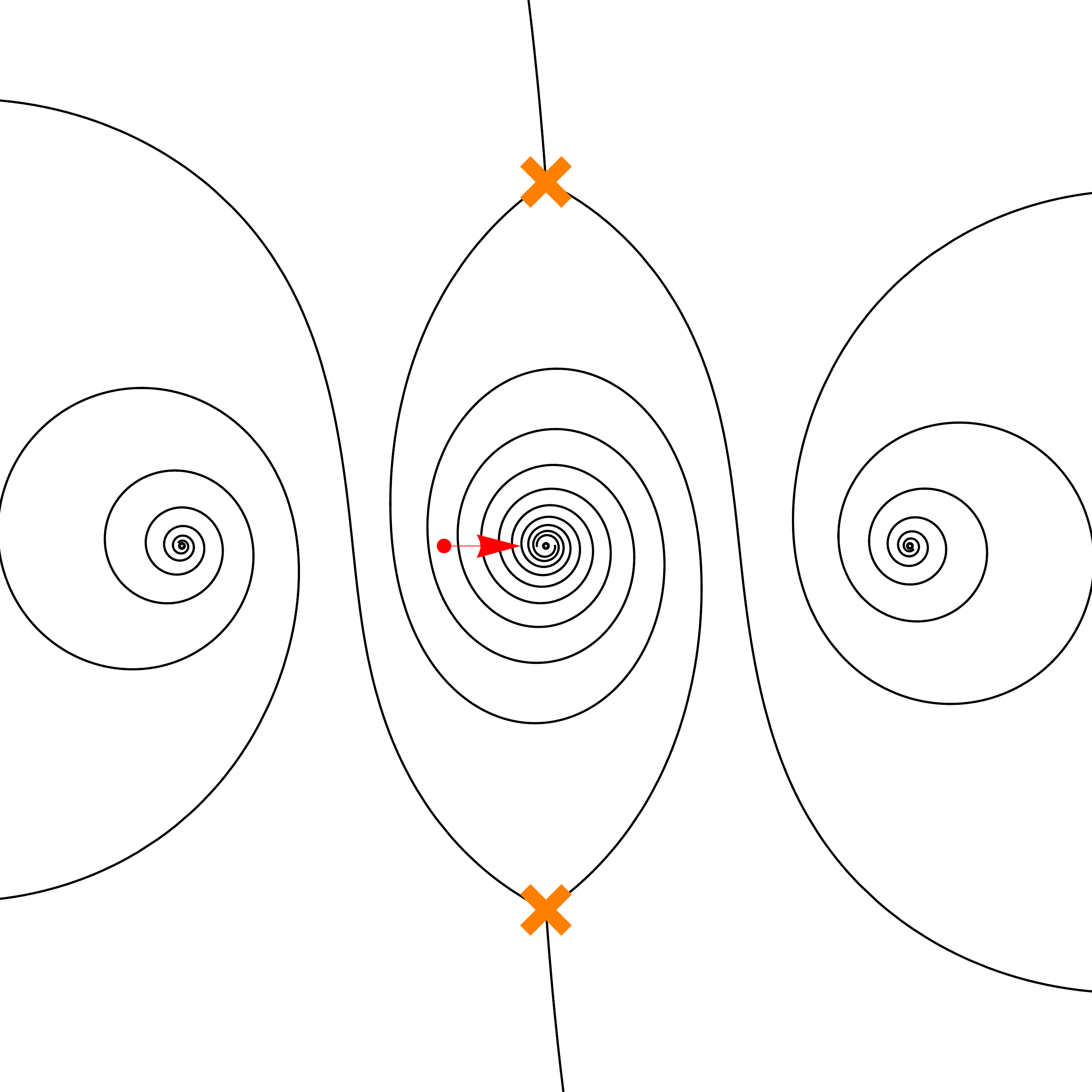}\\[5pt]
\includegraphics[width=0.22\textwidth]{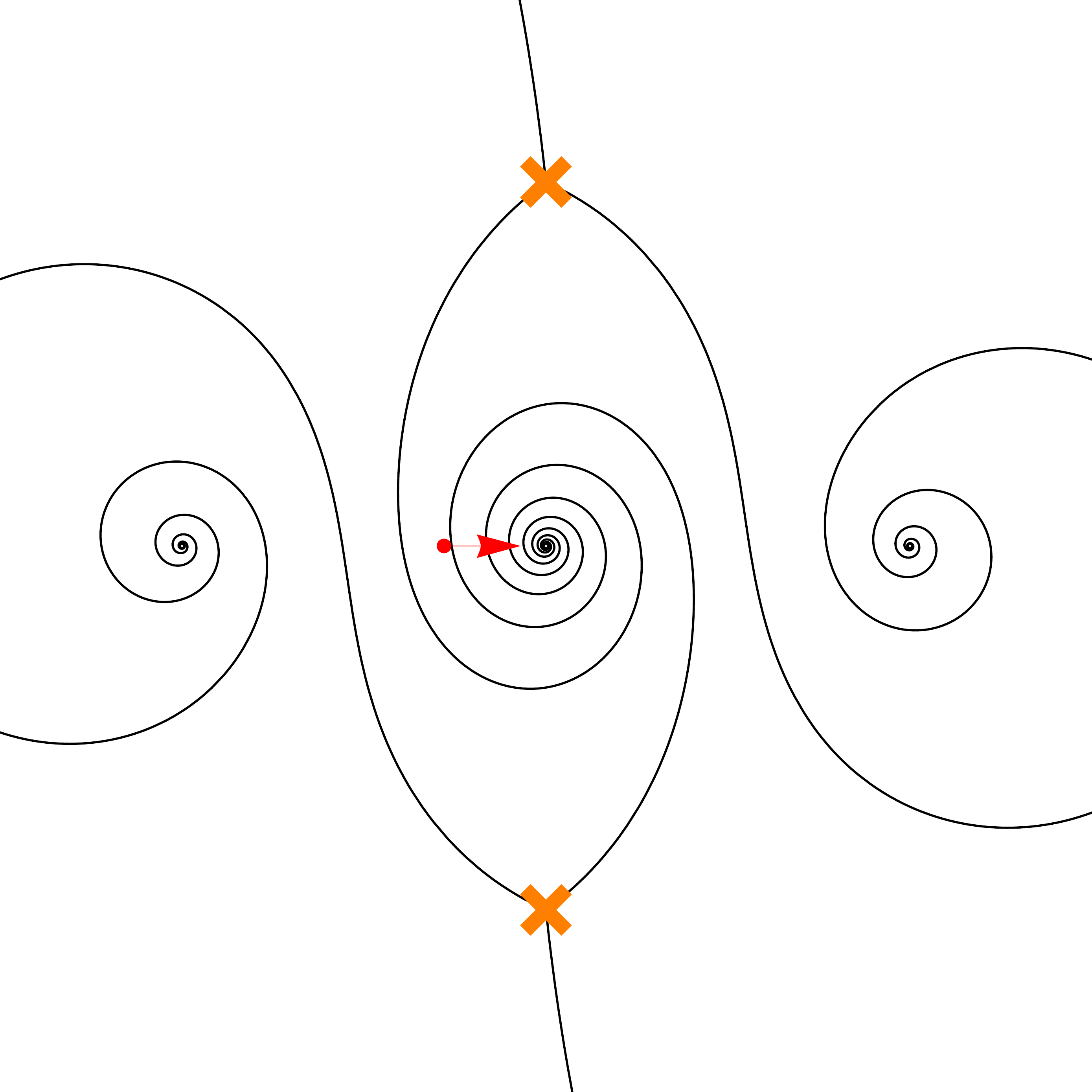}\hfill
\includegraphics[width=0.22\textwidth]{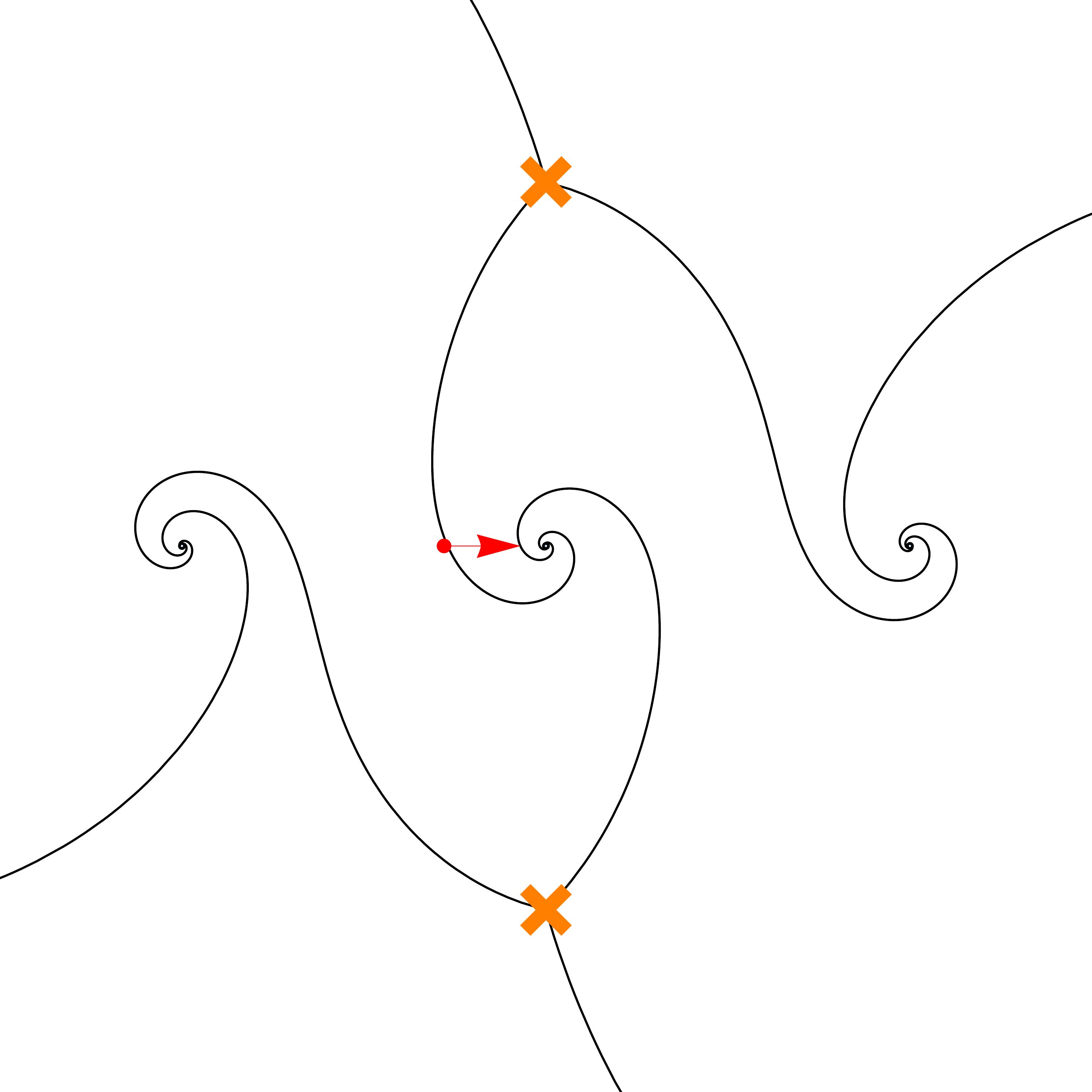}\hfill
\includegraphics[width=0.22\textwidth]{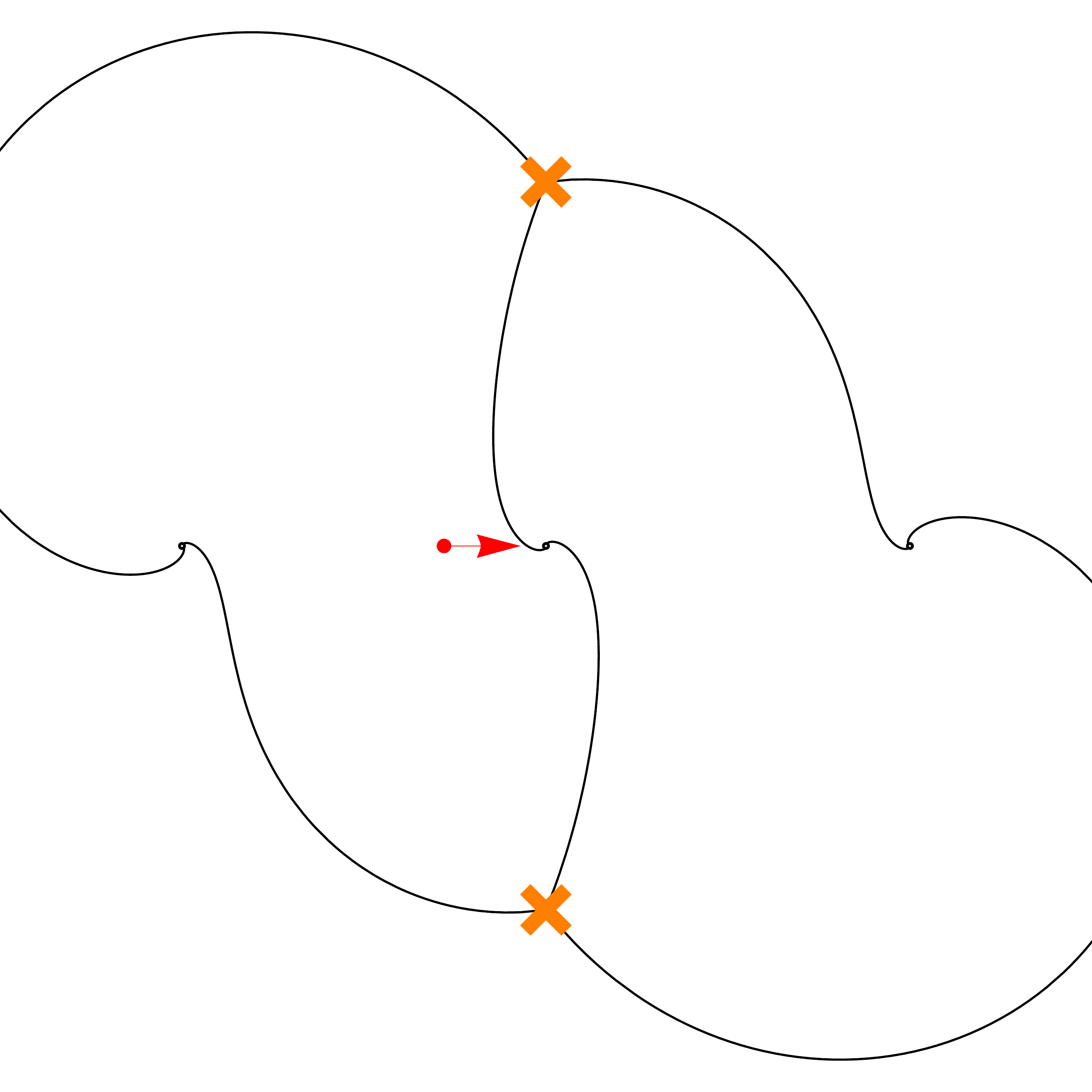}\hfill
\includegraphics[width=0.22\textwidth]{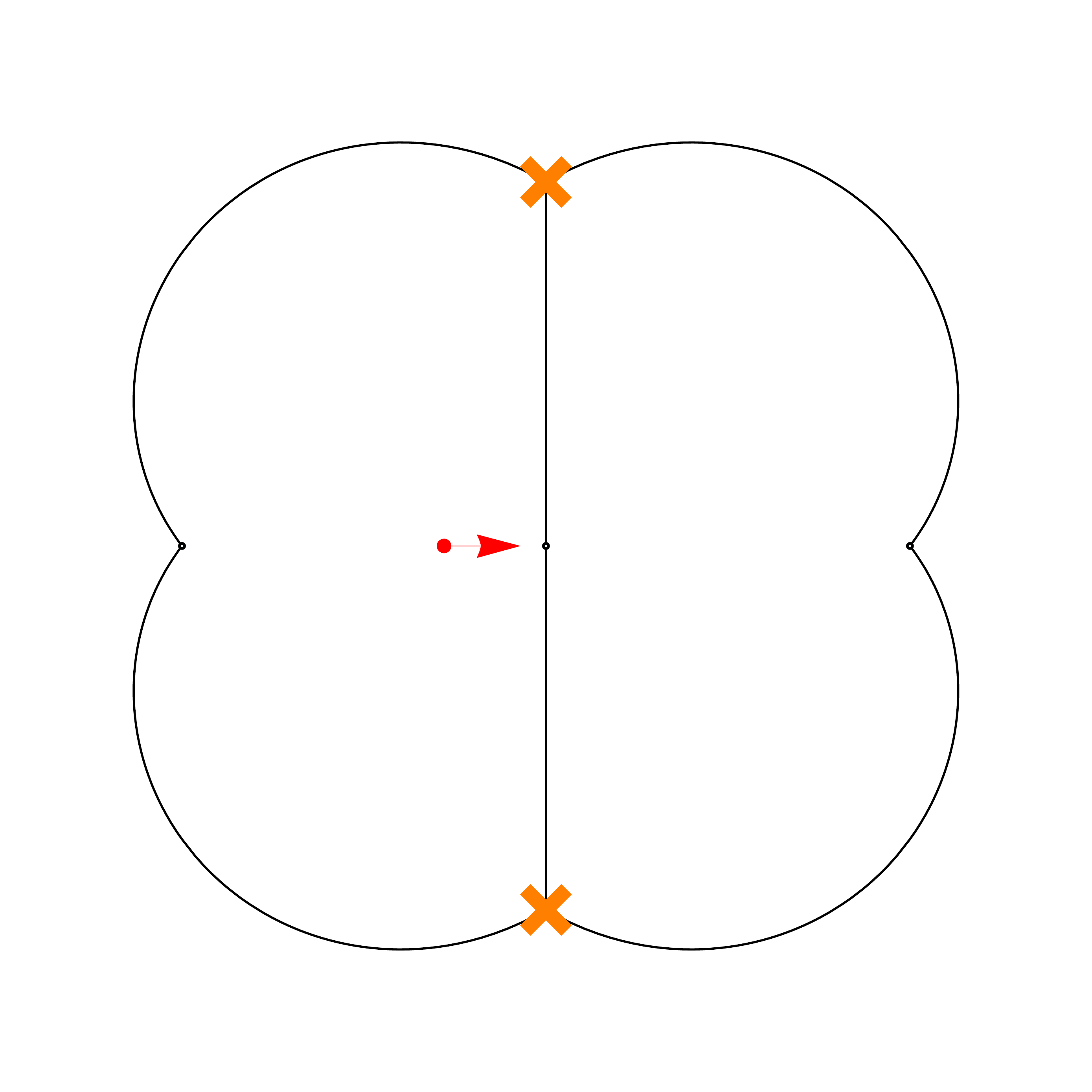}
\end{center}
\caption{Sequence of spectral networks of the $T_2$ theory.
The phases begin with sector $I$ as described in~(\ref{eq:2d4d-seq-1}), for phases ranging from $\vartheta_c-\pi/2$ (top-left) up to $\vartheta_c$ (center), and proceeds with phases from sector $II$, as described in~(\ref{eq:2d4d-seq-2}), ranging from $\vartheta_c$ to $\vartheta_c+\pi/2$ (bottom-right).
}
\label{fig:T2-networks}
\end{figure}

We will choose to work with the quadratic differential giving rise to the BPS graph on the left of Figure~\ref{fig:T2-BPSg}, while the role of the other BPS graph will be discussed below. For generic values of the phase, the network changes as shown in Figure~\ref{fig:T2-networks}. It is evident, that the sequence of $\CS$-walls crossing the surface defect is exactly identical to the one considered in the generic situation of Section~\ref{sec:BPS-spectrum}.  In particular, both the 2d-4d and the 2d BPS spectra must coincide with the ones obtained in that section, and the whole analysis goes through.

At each puncture the Higgs field has a pole as given in~(\ref{eq:simple-pole}). By a small abuse of notation, let us denote $e^{m/2}$ by $a,b,c$ respectively for the three punctures labeled by $a,b,c$ in Figure~\ref{fig:T2-BPSg}. Furthermore, we shall study the fundamental representation in detail and hence fix $N=2$.
The charge lattice is generated by three charges $\gamma_1,\gamma_2,\gamma_3$ associated to the three edges of the BPS graph.
By direct inspection, all cycles have vanishing mutual pairing $\langle\gamma_i,\gamma_j\rangle=0$, and the corresponding formal variables are related to flavor fugacities via
\be\label{eq:T2-var-change}
	X_{\gamma_1}  \ = \  \frac{ac}{b}\,,\qquad X_{\gamma_2}  \ = \  \frac{ab}{c}\,, \qquad X_{\gamma_3}  \ = \  \frac{bc}{a}\,,
\ee
such that $X_{\gamma_1+\gamma_2} = a^2$, which is reflecting the fact that $\gamma_1+\gamma_2$ is equivalent to a pair of cycles around the puncture.

The 4d spectrum generator is given by
\be
	\bS^{(4d)} \ = \ 
	E_q(X_{\gamma_1}) 
	E_q(X_{\gamma_2}) 
	E_q(X_{\gamma_3}) 
	E_q(X_{\gamma_1+\gamma_2+\gamma_3})\,, 
\ee
where we introduced the quantum dilogarithm, defined as follows
\be\label{eq:quantum-dilog}
	E_q(x)  \ \coloneqq \ \( -{q^\frac{1}{2}} x ; q \)^{-1}_{\infty} \,.
\ee

The trace in the quantum torus algebra acts trivially, since there are no gauge charges in this theory. Therefore, the Schur index is obtained by simply applying the change of variables given in~(\ref{eq:T2-var-change}), and we obtain
\be\label{eq:T2-4d-index}
	\CI (a,b,c)  \ = \  
	\prod_{\pm}\prod_{m\geq 0}(1+q^{1/2+m} a^{\pm} b^{\pm} c^{\pm})^{-1} \,.
\ee

\subsubsection{Surface defects of the $T_2$ theory}

Although, our general description of vortex defects from the previous sections applies to this example, we nevertheless deem it worth to provide it here, illustrating the various general technical discussions more ``hands-on''.

The 2d spectrum of this theory was described in Section~\ref{sec:BPS-spectrum}, and consists of one particle in each vacuum with opposite degeneracies and spins.
Using this, the 2d-4d spectrum generator has the following explicit form
\be\label{eq:spec-gen-T2-first-part}
\begin{split}
	\bS_{\vartheta_c\pm\pi/2}^{2d-4d} 
	\ = \  & \bS^{(2d-4d)}_{II} \bS^{(2d)} \bS^{(4d)}  \bS^{(2d-4d)}_I
	\\
	\ = \ & 
	\left( \begin{array}{cc}
	1 & X_{-a_{21}+  \gamma_2}\\ 
	0 & 1
	\end{array}\right) 
	\left( \begin{array}{cc}
	1 & X_{-a_{21}+ \gamma_1+\gamma_2}\\ 
	0 & 1
	\end{array}\right)
	\cdots
	\left( \begin{array}{cc}
	(1-q^{-1/2}X_{\gamma_1+\gamma_2})^{-1}  & 0\\ 
	0 & (1-q^{1/2} X_{\gamma_1+\gamma_2})^{}  
	\end{array}\right)
	\\
	& 
	\left( \begin{array}{cc}
	E_q(X_{\gamma_1}) & 0\\ 
	0 & E_q(X_{\gamma_1})
	\end{array}\right)
	\left( \begin{array}{cc}
	E_q(X_{\gamma_2}) & 0\\ 
	0 & E_q(X_{\gamma_2})
	\end{array}\right)\\
	&
	\left( \begin{array}{cc}
	E_q(X_{\gamma_3}) & 0\\ 
	0 & E_q(X_{\gamma_3})
	\end{array}\right)
	\left( \begin{array}{cc}
	E_q(X_{\gamma_1+\gamma_2+\gamma_3}) & 0\\ 
	0 & E_q(X_{\gamma_1+\gamma_2+\gamma_3})
	\end{array}\right)
	\\
	& \cdots 
	\left( \begin{array}{cc}
	1 & 0\\ 
	X_{a_{21}+\gamma_1+\gamma_2} & 1
	\end{array}\right)
	\left( \begin{array}{cc}
	1 & 0\\ 
	X_{a_{21} + \gamma_1} & 1
	\end{array}\right)
	\left( \begin{array}{cc}
	1 & 0\\ 
	X_{a_{21}} & 1
	\end{array}\right) 
	\\
	 \ = \ & 
	\left( \begin{array}{cc}
	1 & X_{-a_{21}} \, q^{1/2}X_{\gamma_2} \frac{1 + q^{1/2}X_{\gamma_1}}{1-q X_{\gamma_1+\gamma_2}}\\ 
	0 & 1
	\end{array}\right) 
	\\
	& 
	\left( \begin{array}{cc}
	{(1-q^{-1/2}X_{\gamma_1+\gamma_2})}^{-1} & 0\\ 
	0 & (1-q^{1/2}X_{\gamma_1+\gamma_2})
	\end{array}\right) \left[ \prod_{i=1}^{3} E_q(X_{\gamma_i}) \right] E_q(X_{\gamma_1+\gamma_2+\gamma_3})
	\\
	& 
	\left( \begin{array}{cc}
	1 & 0\\ 
	 \frac{1 + q^{1/2}X_{\gamma_1}}{1-q^{} X_{\gamma_1+\gamma_2}}\, X_{a_{21}} & 1
	\end{array}\right)\,,
\end{split}
\ee
where we used the quantum torus algebra to introduce $q$-shifts. 
The upper triangular matrices in the second line of~\eqref{eq:spec-gen-T2-first-part} enter via the 2d-4d generator in sector $II$, $\bS^{(2d-4d)}_{II}$, while the diagonal (last) matrix is the contribution from $\bS^{(2d)}$. The matrices in the third and forth line are due to $\bS^{(4d)}$, and the product of elements in the sixth line is due to $\bS^{(2d-4d)}_{I}$.

Analogously, the conjugate spectrum generator for anti-particles reads
\be\label{eq:spec-gen-T2-second-part}
\begin{split}
	\overline{\bS}_{\vartheta_c\pm\pi/2}^{2d-4d}  \ = \ &
	\left( \begin{array}{cc}
	1 & 0\\ 
	q^{-1/2}X_{-\gamma_2} \left[ \frac{1 + q^{-1/2}X_{-\gamma_1}}{1-q^{-1} X_{-\gamma_1-\gamma_2}} \right] \, X_{-a_{21}}& 1
	\end{array}\right) 
	\\
	& 
	\left( \begin{array}{cc}
	(1-q^{1/2} X_{-\gamma_1-\gamma_2})^{}& 0\\ 
	0 & (1-q^{-1/2} X_{-\gamma_1-\gamma_2})^{-1}
	\end{array}\right)\left[  \prod_{i=1}^{3} E_q(X_{-\gamma_i}) \right] E_q(X_{-\gamma_1-\gamma_2-\gamma_3})
	\\
	&  
	\left( \begin{array}{cc}
	1 & X_{-a_{21}} \, \frac{1 + q^{-1/2}X_{-\gamma_1}}{1-q^{-1} X_{-\gamma_1-\gamma_2}} \\ 
	0 & 1
	\end{array}\right)\,.
\end{split}
\ee

Let us now consider the expectation value of an infinitesimal interface as described in Section~\ref{sec:2d-4d-macdonald}; We introduce an infinitesimally small path $\wp$ localized near $z$, whose role it is to make sense of the index formula providing the necessary regularization for computing different sorts of $q$-shifts.
Our choice of interface is illustrated by the red dot with an arrow in Figure~\ref{fig:T2-networks}. The arrow should be understood as taking a short segment between two infinitesimally close points, oriented accordingly. Then, the corresponding trace reads
\be\label{eq:T2-trace-2d4d}
\begin{split}
	& \Tr\left[  F(\wp)\, \overline{\bS}_{\vartheta_c\pm\pi/2}^{2d-4d}   \bS_{\vartheta_c\pm\pi/2}^{2d-4d} \right]
	\\
	& \ = \  	\Tr  
	\(\begin{array}{cc}  X_{\wp^{(1)}} & 0 \\ 0 & X_{\wp^{(2)}}  \end{array}\)
	\, 
	\(\begin{array}{cc}  1 & 0 \\ \bar F_{21} & 1  \end{array}\)
	\(\begin{array}{cc}  \bar D_1 &  \\  & \bar D_2  \end{array}\)
	\(\begin{array}{cc}  1 & \bar F_{12} \\ 0 & 1  \end{array}\)
	\(\begin{array}{cc}  1 & F_{12} \\ 0 & 1  \end{array}\)
	\(\begin{array}{cc}  D_1 &  \\  & D_2  \end{array}\)
	\(\begin{array}{cc}  1 & 0 \\ F_{21} & 1  \end{array}\)	
	\\
	&  \ = \  
	\Tr \(\begin{array}{cc}  X_{\wp^{(1)}} (\bar D_1 D_1 +  \bar D_1 \bar F_{12} D_2  F_{21} +  \bar D_1 F_{12} D_2  F_{21}) & \star \\ \star &  X_{\wp^{(2)}} (\bar D_2 D_2 +   \bar F_{21} \bar D_1 \bar F_{12} \bar D_2  +   \bar F_{21} \bar D_1 F_{12} \bar D_2)   \end{array}\) 
	\\
	&  \ = \  \hTr\( X_{\wp^{(1)}} \bar D_1 D_1 +  X_{\wp^{(2)}} \bar D_2 D_2 \) \,,
\end{split}
\ee 
where in the first line we defined the following pieces
\be
\begin{split}	
	D_1 &  \ = \  (1-q^{-1/2} X_{\gamma_1+\gamma_2})^{-1}   E_q( X_{\gamma_1})E_q(X_{\gamma_2})E_q(X_{\gamma_3}) E_q(X_{\gamma_1+\gamma_2+\gamma_3})\,,
	\\ 
	D_2 & \ = \  (1- q^{1/2}  X_{\gamma_1+\gamma_2})^{} E_q(X_{\gamma_1}) E_q( X_{\gamma_2})E_q(X_{\gamma_3}) E_q(X_{\gamma_1+\gamma_2+\gamma_3})\,,
	\\
	\bar D_1 & \ = \  (1- q^{1/2} X_{-\gamma_1-\gamma_2})^{}  E_q(X_{-\gamma_1})E_q(X_{-\gamma_2})E_q(X_{-\gamma_3}) E_q(X_{-\gamma_1-\gamma_2-\gamma_3})\,,
	\\
	\bar D_2 &  \ = \  (1-q^{-1/2}  X_{-\gamma_1-\gamma_2})^{-1}   E_q( X_{-\gamma_1})E_q( X_{-\gamma_2})E_q( X_{-\gamma_3}) E_q(   X_{-\gamma_1-\gamma_2-\gamma_3})\,,
	\\
	F_{12} &  \ = \  X_{-a_{21}} \, q^{1/2}X_{\gamma_2} \frac{1 + q^{1/2}X_{\gamma_1}}{1-q X_{\gamma_1+\gamma_2}}\,,
	\\
	F_{21} &  \ = \  \frac{1 + q^{1/2}X_{\gamma_1}}{1-q^{} X_{\gamma_1+\gamma_2}}\, X_{a_{21}} \,,
	\\
	\bar F_{12} &  \ = \  X_{-a_{21}} \, \frac{1 + q^{-1/2}X_{-\gamma_1}}{1-q^{-1} X_{-\gamma_1-\gamma_2}} \,,
	\\
	\bar F_{21} &  \ = \  q^{-1/2}X_{-\gamma_2} \frac{1 + q^{-1/2}X_{-\gamma_1}}{1-q^{-1} X_{-\gamma_1-\gamma_2}} \, X_{a_{21}} \,.
\end{split}
\ee 
The second line in~\eqref{eq:T2-trace-2d4d} then follows an explicit computation, with the off-diagonal terms being irrelevant. Finally, in the third line we take into account the following nontrivial cancellations
\be
	F_{12} + \bar F_{12} \ \equiv \  0 \,, \quad \text{and} \quad F_{21} + \bar F_{21} \ \equiv \ 0 \,,
\ee
which can be seen by direct inspection.

To evaluate the remaining trace $\hTr$  in the quantum torus algebra, we need to specify what to do with $X_{\wp^{(i)}}$. Since $\wp$ is infinitesimal, these variables are essentially one, except that they pick up nontrivial intersections with the cycle associated to the puncture $C_a^{(2)}-C_a^{(1)}=\gamma_1+\gamma_2$ (see~(\ref{eq:flavor-cycles})), \emph{i.e.}
\be
	\langle \wp^{(1)} , C_a^{(1)} \rangle  \ = \  -1  \ = \  -\langle  \wp^{(2)} , C_a^{(2)} \rangle \,.
\ee
We then use the quantum torus algebra to absorb $X_{\wp^{(i)}}$ by shifting $X_{\gamma}\to q^{\frac{1}{2} \langle\wp^{(i)},\gamma\rangle}X_{\gamma}$, and arrive at the final answer
\be
\begin{split}
	\Tr\( X_{\wp^{(1)}} \bar D_1 D_1 +  X_{\wp^{(2)}} \bar D_2 D_2 \)
	 \ = \ & (1-q \, q^{-1/2} X_{\gamma_1+\gamma_2})^{-1} (1-q^{-1} q^{1/2} X_{-\gamma_1-\gamma_2})^{} \CI(a\to a q^{1/2}) \\ 
	& + (1-q^{-1} q^{1/2} X_{\gamma_1+\gamma_2})^{} (1-q^{} \, q^{-1/2} X_{-\gamma_1-\gamma_2})^{-1} \CI(a\to a q^{-1/2}) \\
	 \ = \ &-q^{1/2}\(a^{-2} \, \CI(a\to a q^{1/2})  + a^2 \,  \CI(a\to a q^{-1/2})\) \,,
\end{split}
\ee 
in agreement with the 2d-4d Schur index obtained by acting with a vortex surface defect operator on the 4d index~(\ref{eq:T2-4d-index}).

\subsection{Superconformal QCD}\label{sec:su2nf4}

\begin{figure}[h!]
\begin{center}
\hfill
\includegraphics[width=0.45\textwidth]{figures/tetrahedron.pdf}
\hfill
\includegraphics[width=0.45\textwidth]{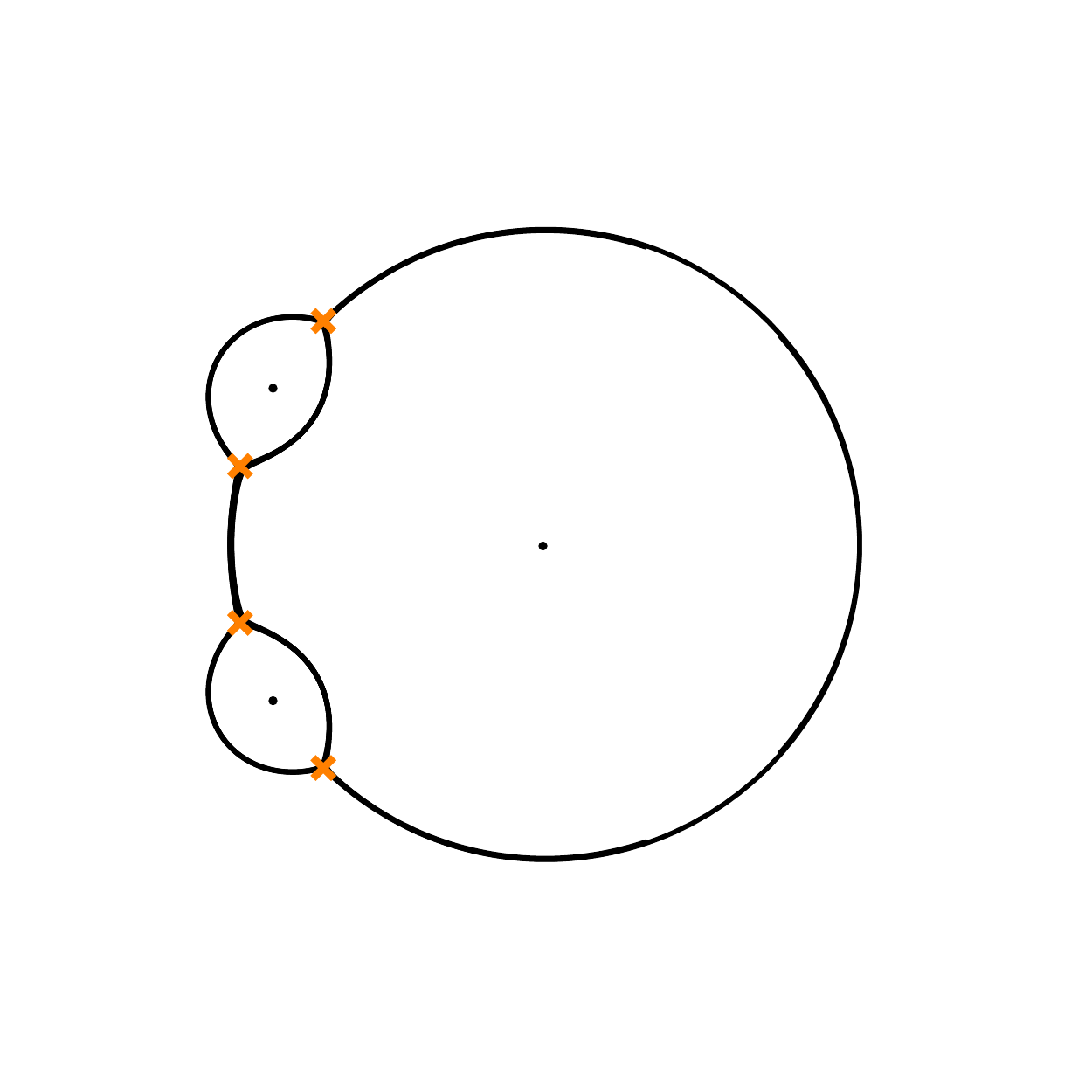}
\hfill
\caption{Two BPS graphs for the $SU(2)$, $N_f=4$ theory.}
\label{fig:nf4}
\end{center}
\end{figure}

Next, let us consider the massive deformation of superconformal QCD, \emph{i.e.} the $SU(2)$ gauge theory with $N_{f}=4$ fundamental hypermultiplets.  As a class $\CS$ theory, it is engineered by a sphere with four regular punctures. We will fix the positions of the punctures at the three roots of unity and at infinity,  leading to the following quadratic differential
\be
\begin{split}
	\phi_2  \ = \  &
	M_{a} z^4+ \(-3 e^{\frac{\pi  \ii}{3}} M_{c}+3 e^{\frac{2 \pi  \ii}{3}} M_{d}+3 M_{b}-2 u\) z^3 \\
	& + \(3 e^{\frac{\pi  \ii}{3}} M_{c}-3 e^{\frac{2 \pi  \ii}{3}} M_{d}+3 M_{b}-3 (M_{c}+M_{d})\) z^2\\
	&
	+ (3 (M_{b}+M_{c}+M_{d})-M_{a}) z + u \,,
\end{split}
\ee
where $M_i$, with $i \in \{a,b,c,d\}$ are the squares of UV masses. On the one hand, setting
\be
	M_a \ = \ M_b \ = \ M_c \ = \ M_d \ = \ 3 \,, \quad \text{and}\quad  u \ = \ 0\,,
\ee
we find the BPS graph in the left frame of Figure~\ref{fig:nf4}. On the other hand, setting
\be
	M_a \ = \ M_b \ = \ 3\,,\qquad M_c \ = \ M_d  \ = \  0.2 \,, \quad \text{and}\quad  u \ = \ 3.028\,,
\ee
the BPS graph changes and turns into the one in the right frame of Figure~\ref{fig:nf4}.

\begin{figure}
\begin{center}
\includegraphics[width=0.22\textwidth]{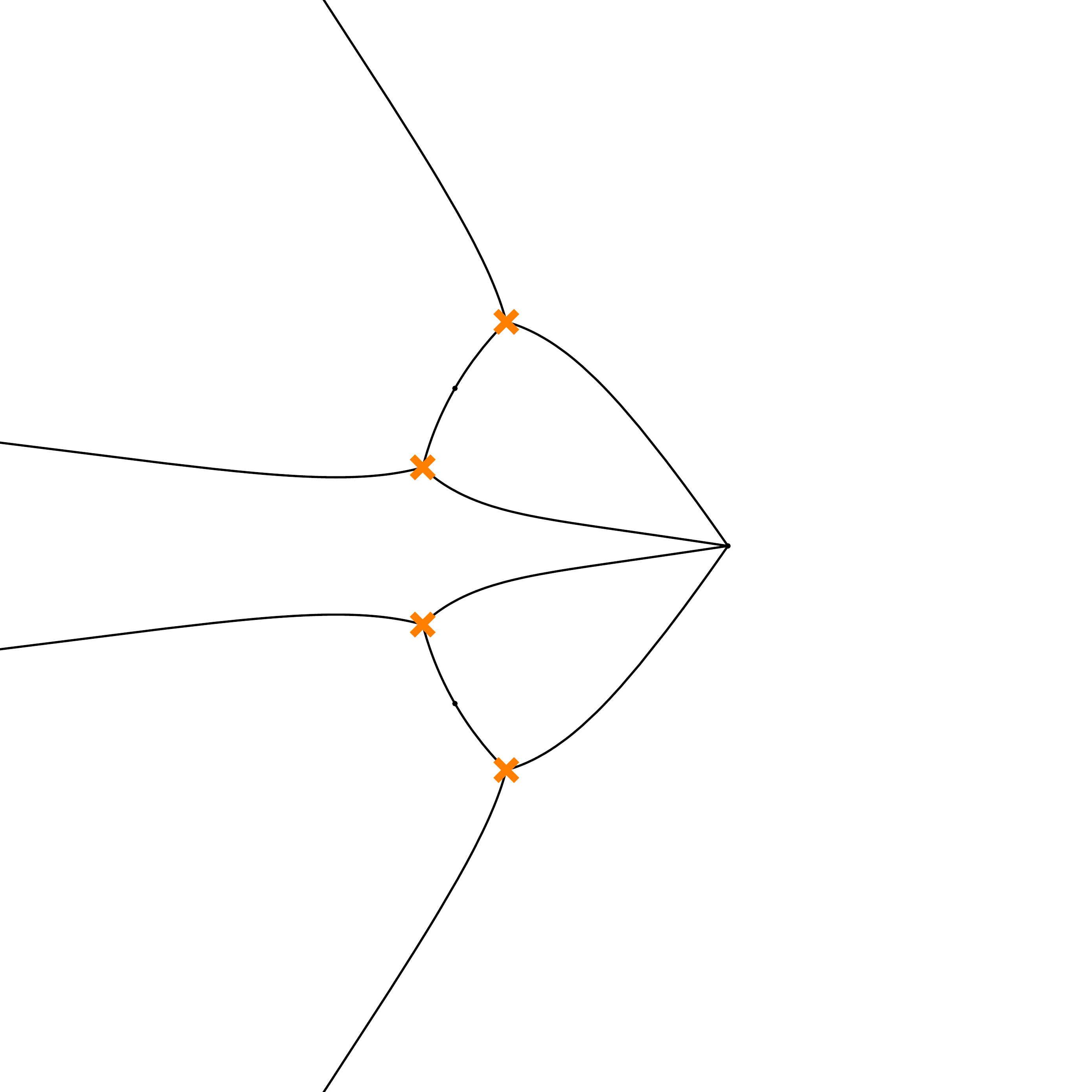}\hfill
\includegraphics[width=0.22\textwidth]{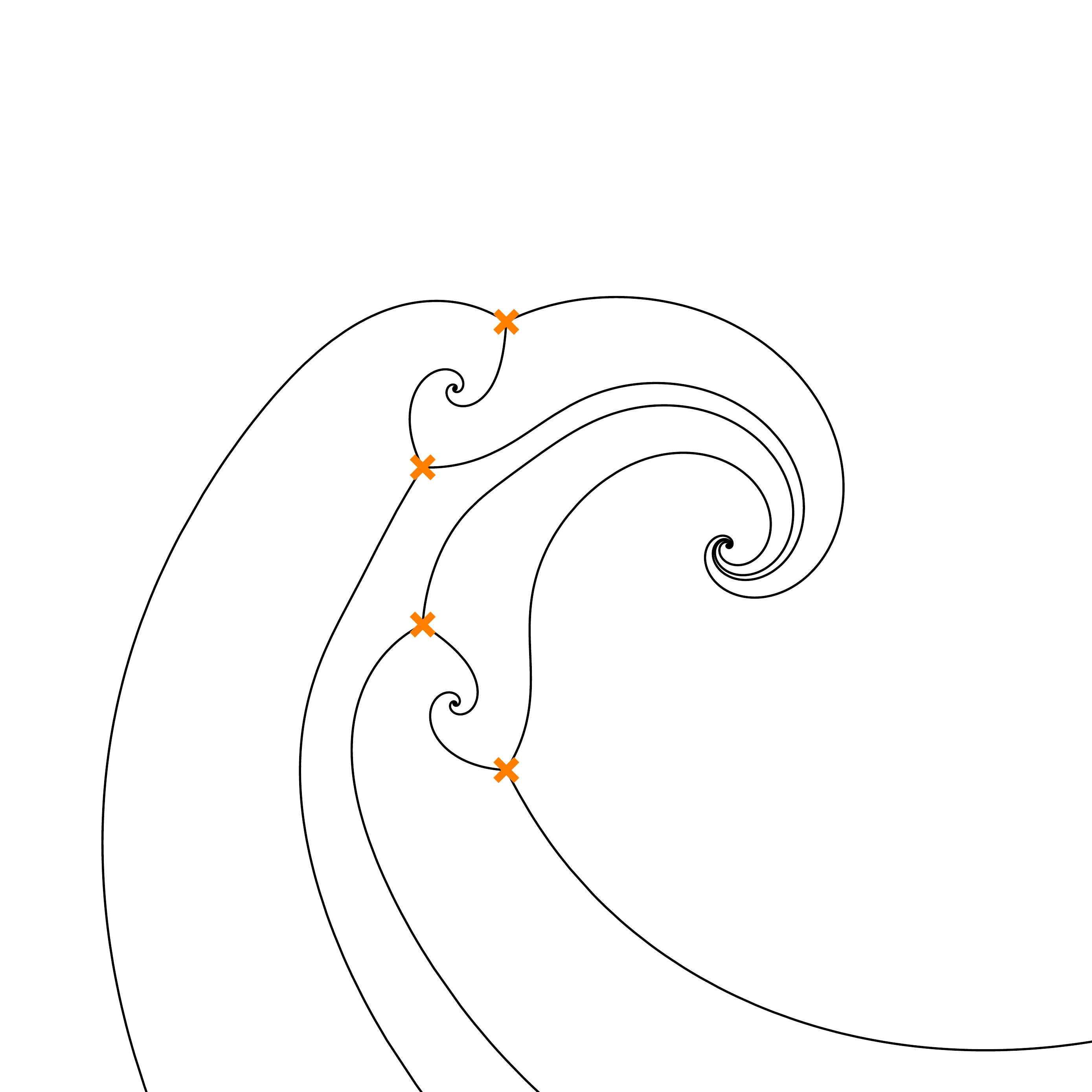}\hfill
\includegraphics[width=0.22\textwidth]{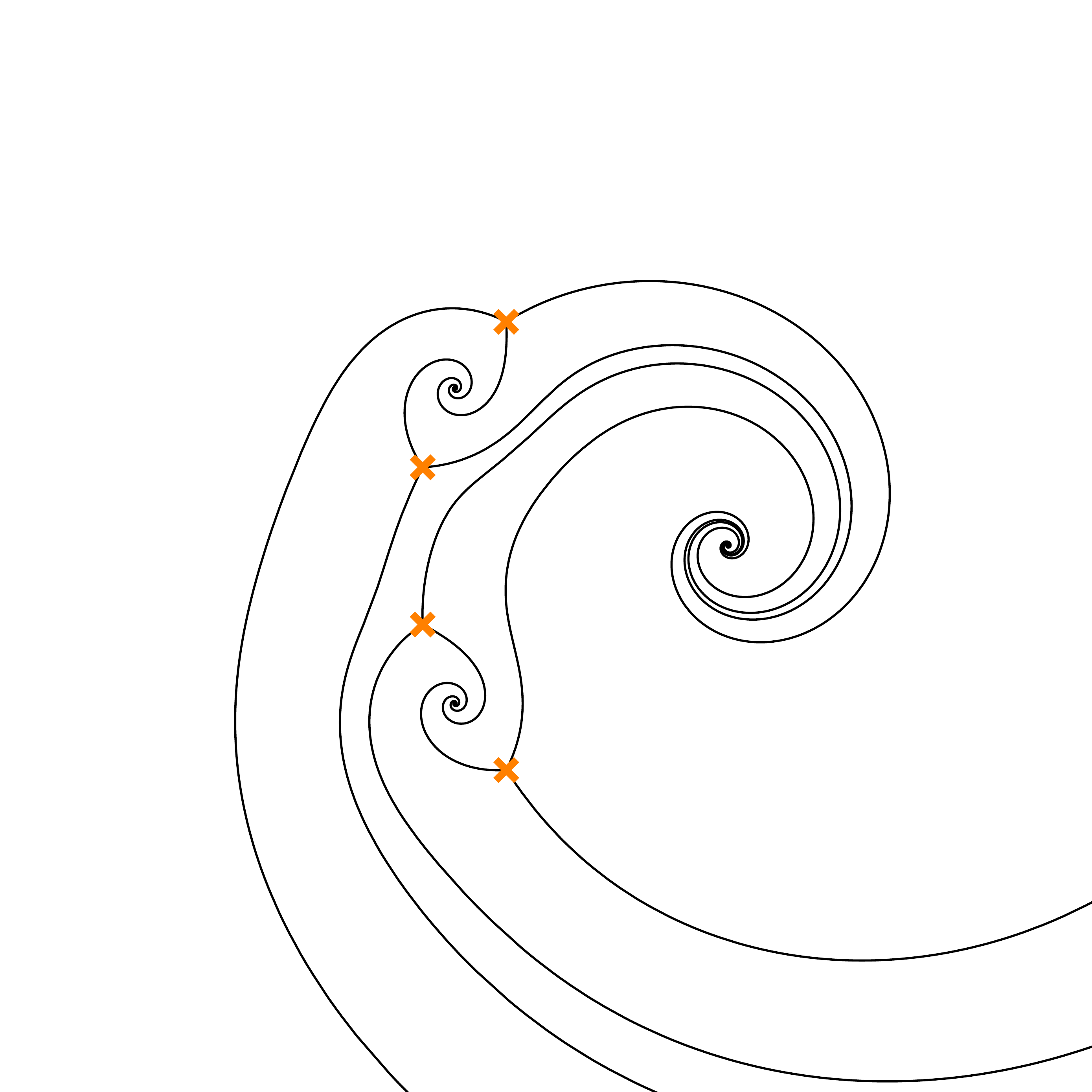}\hfill
\includegraphics[width=0.22\textwidth]{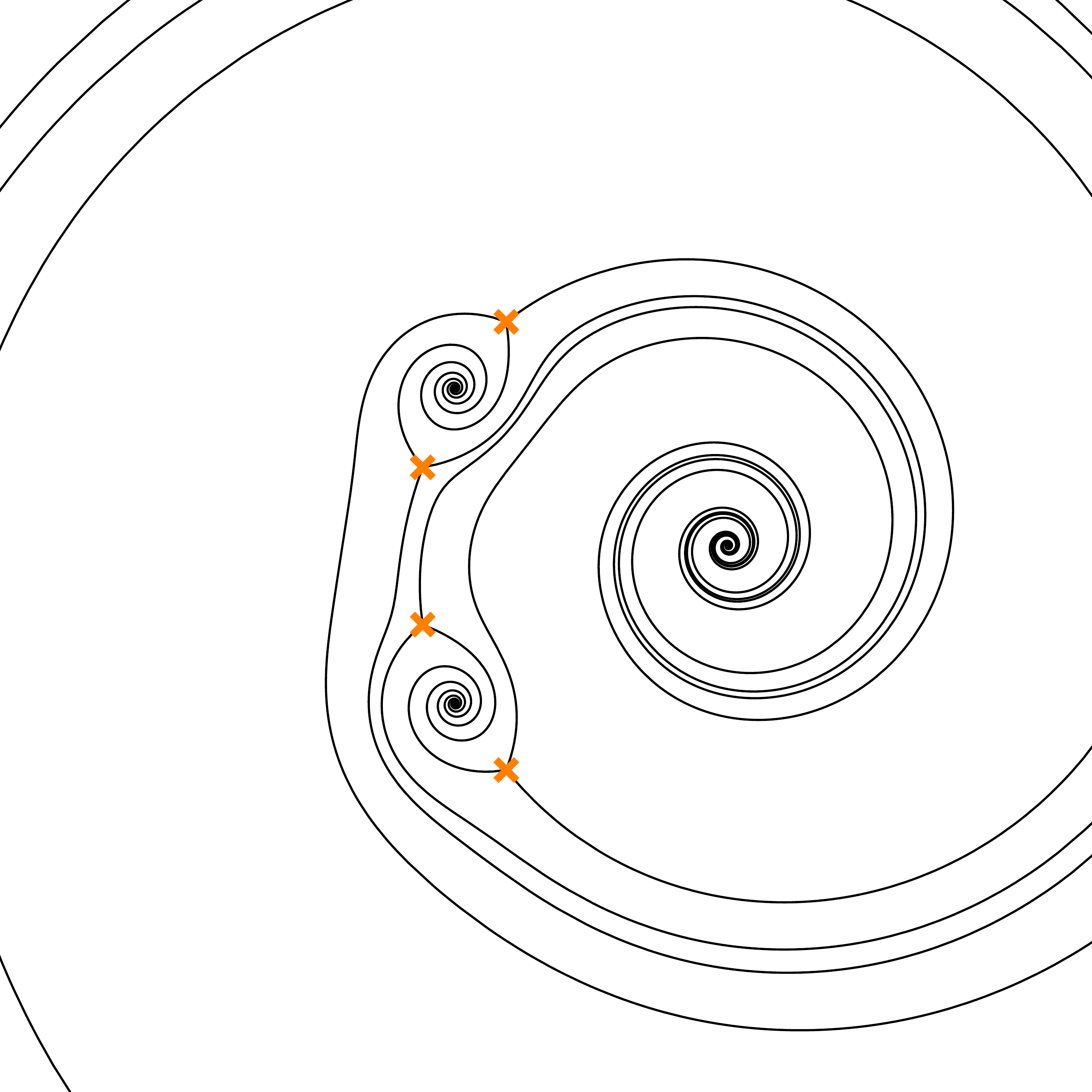}\\[5pt]
\includegraphics[width=0.22\textwidth]{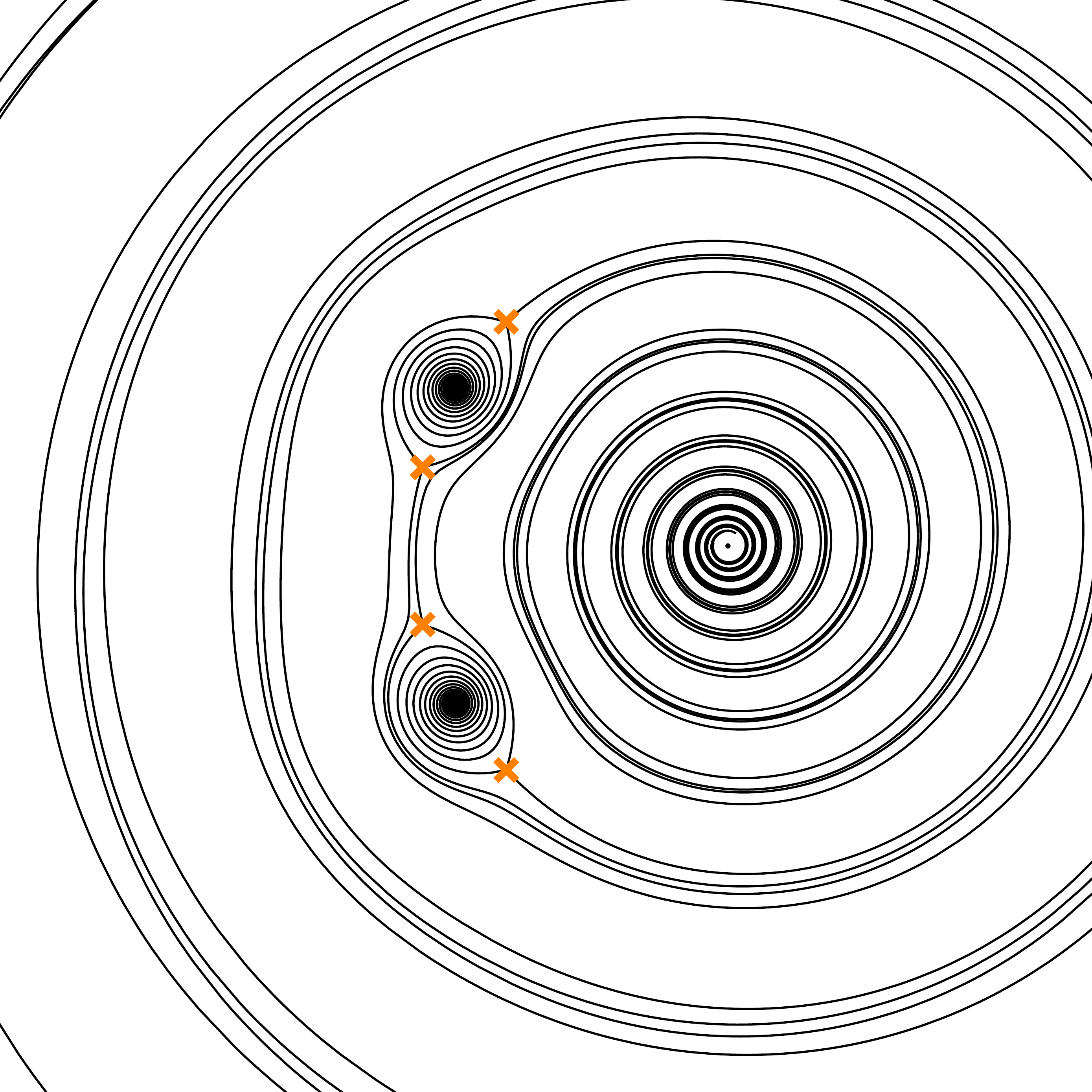}
\hfill\includegraphics[width=0.06\textwidth]{figures/dots.pdf}\hfill
\includegraphics[width=0.22\textwidth]{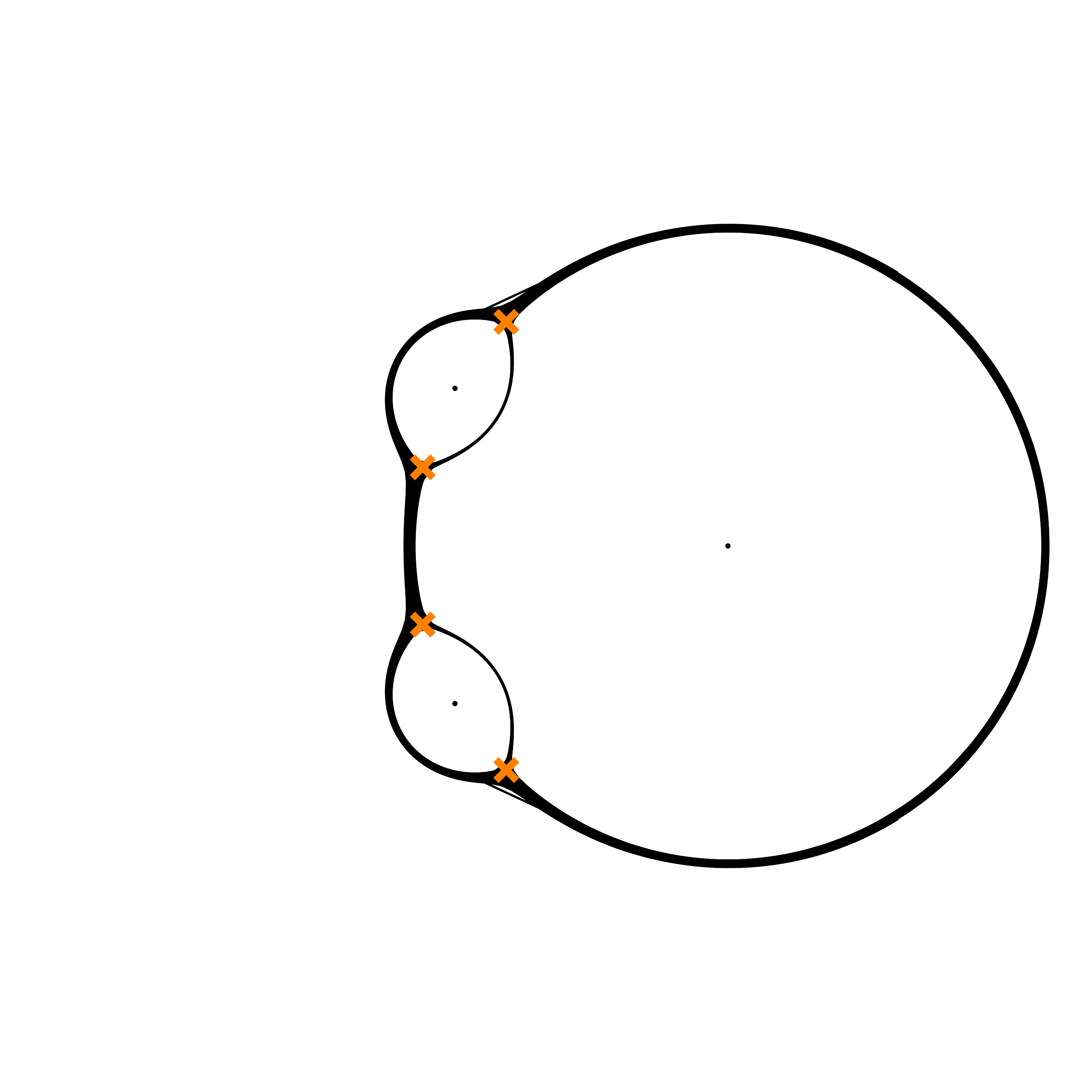}
\hfill\includegraphics[width=0.06\textwidth]{figures/dots.pdf}\hfill
\includegraphics[width=0.22\textwidth]{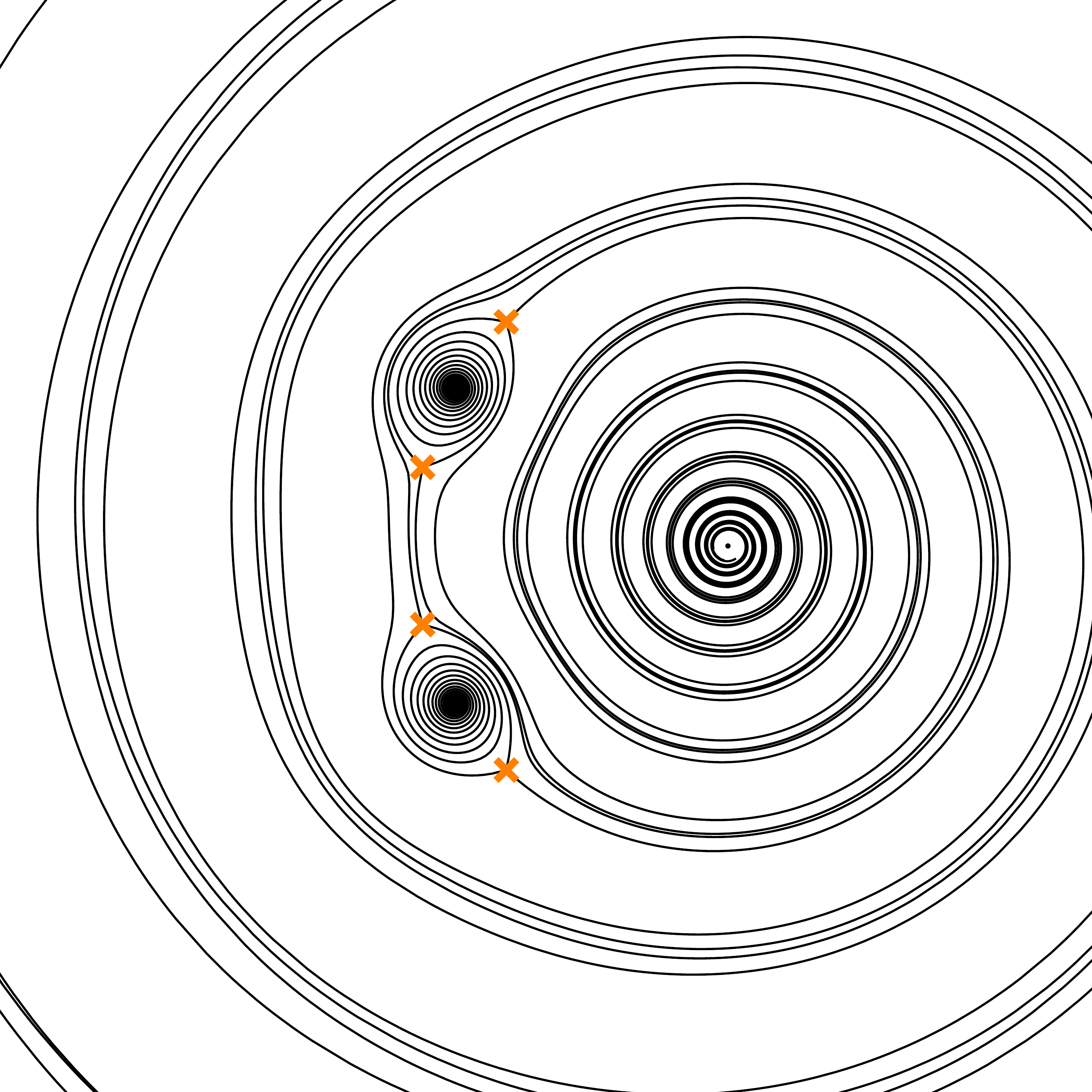}\\[5pt]
\includegraphics[width=0.22\textwidth]{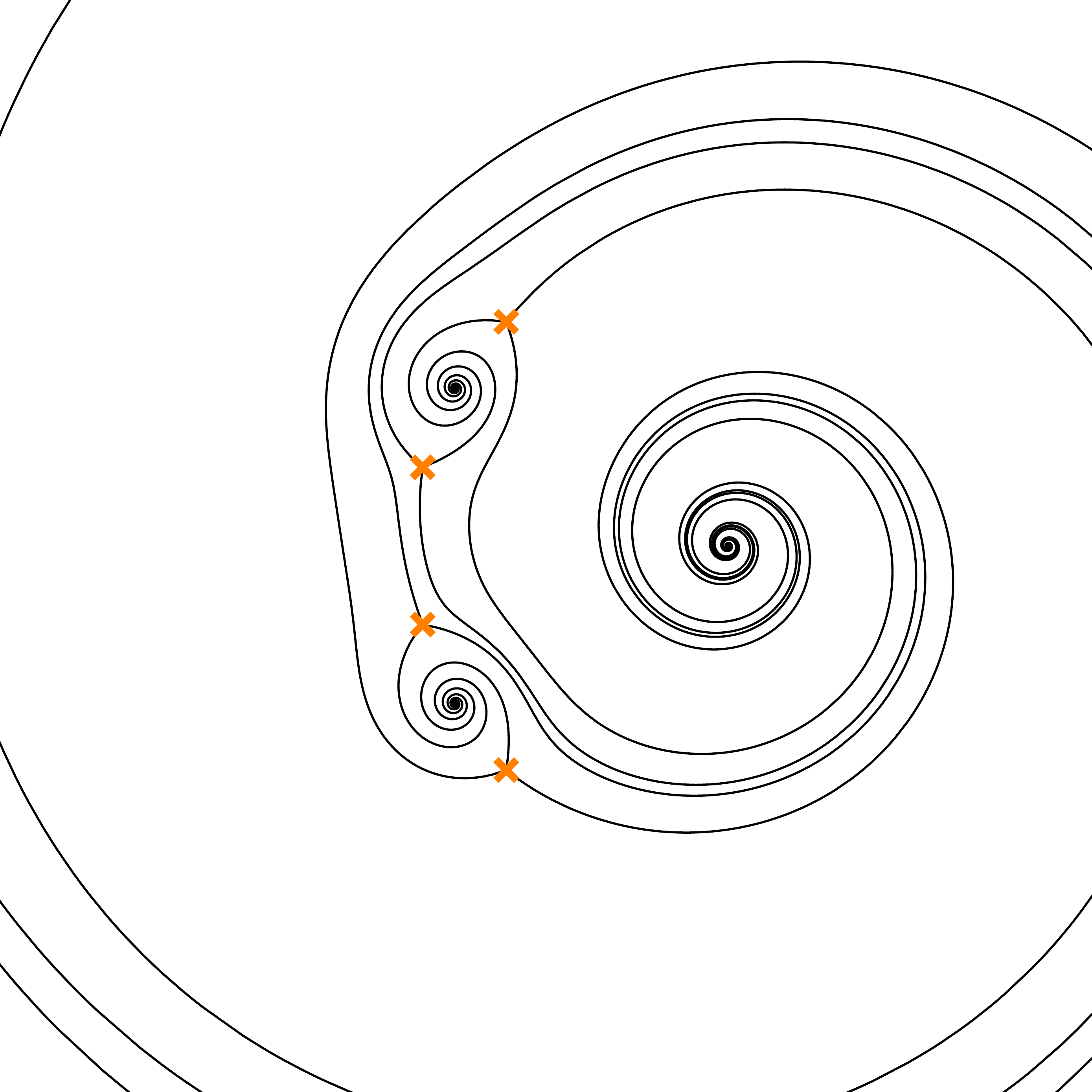}\hfill
\includegraphics[width=0.22\textwidth]{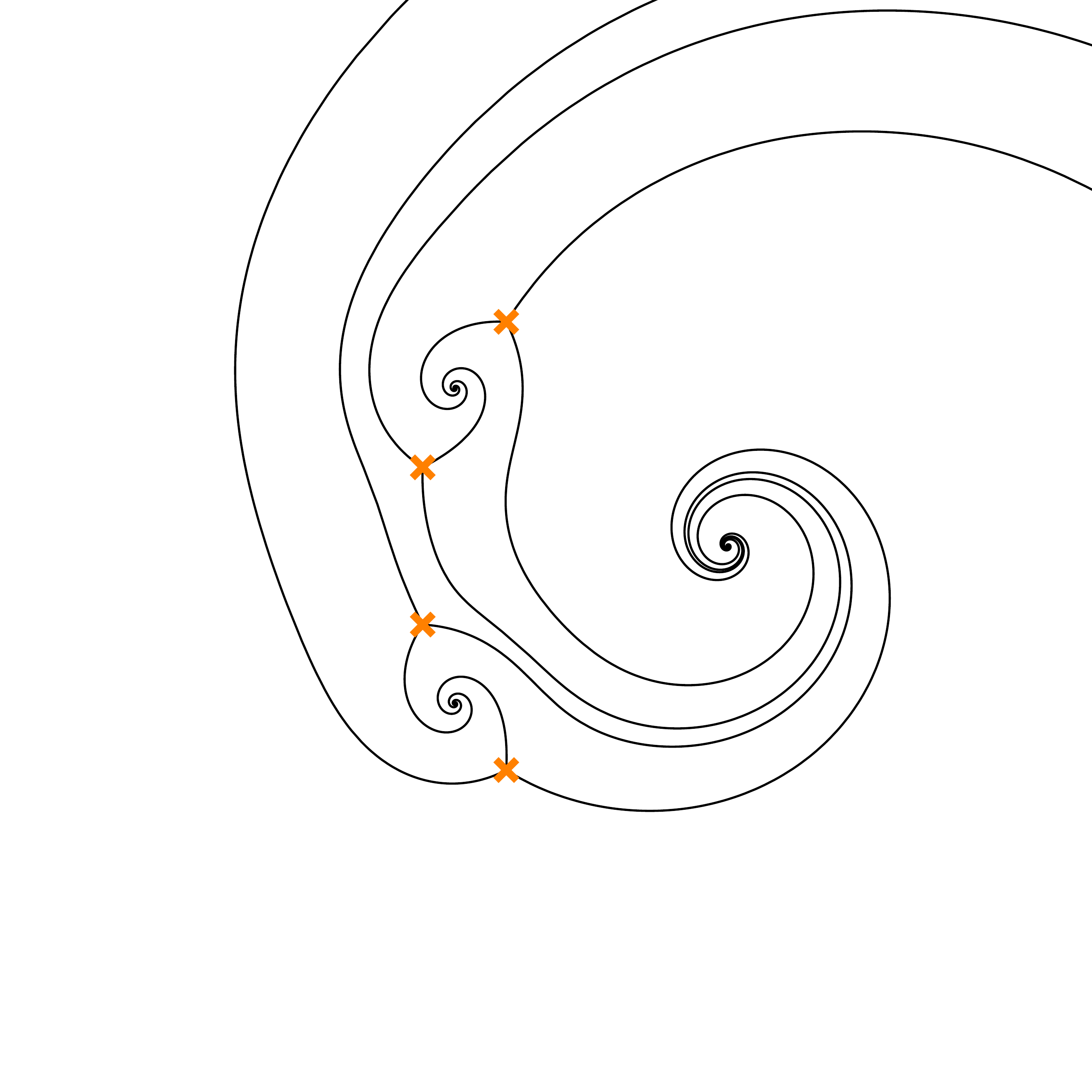}\hfill
\includegraphics[width=0.22\textwidth]{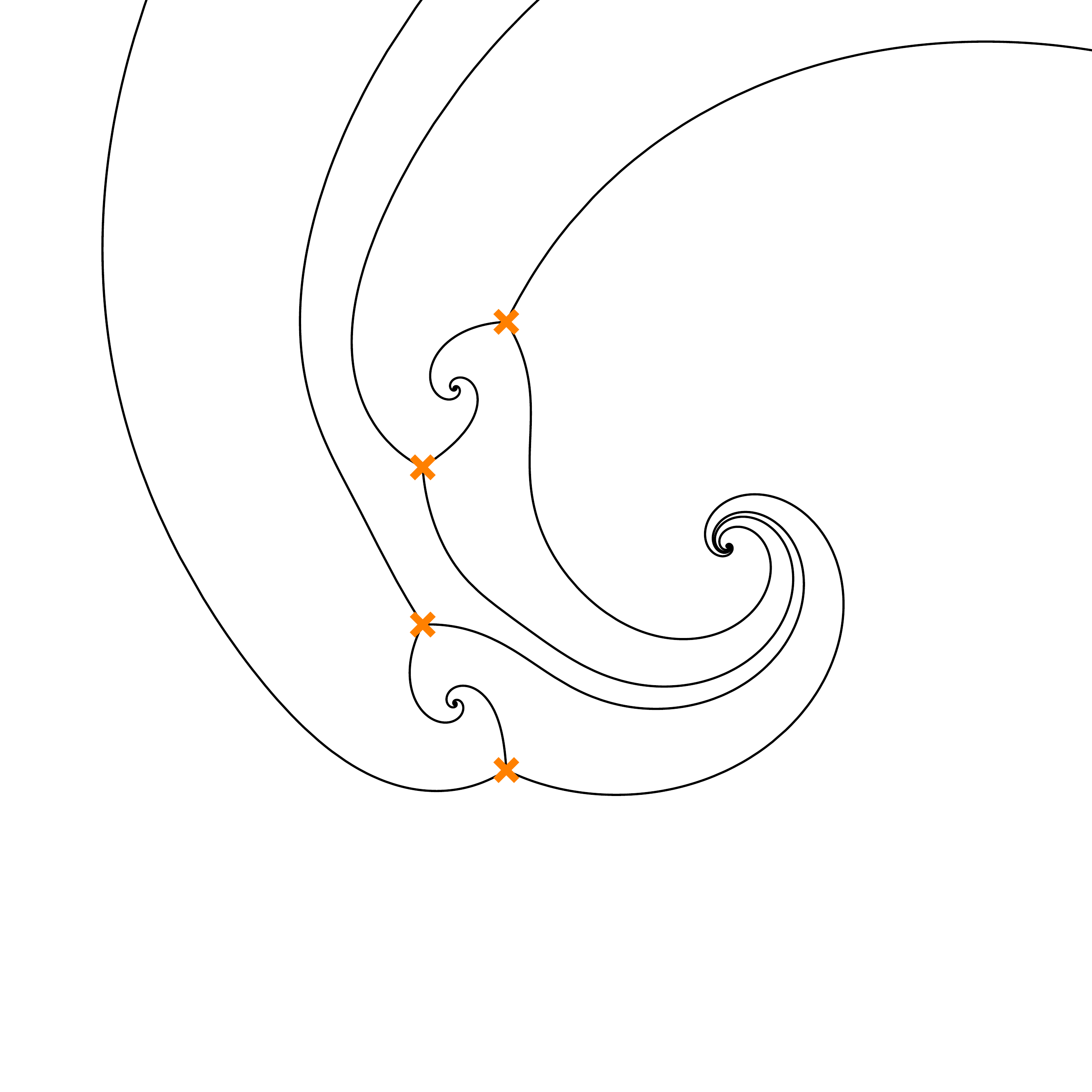}\hfill
\includegraphics[width=0.22\textwidth]{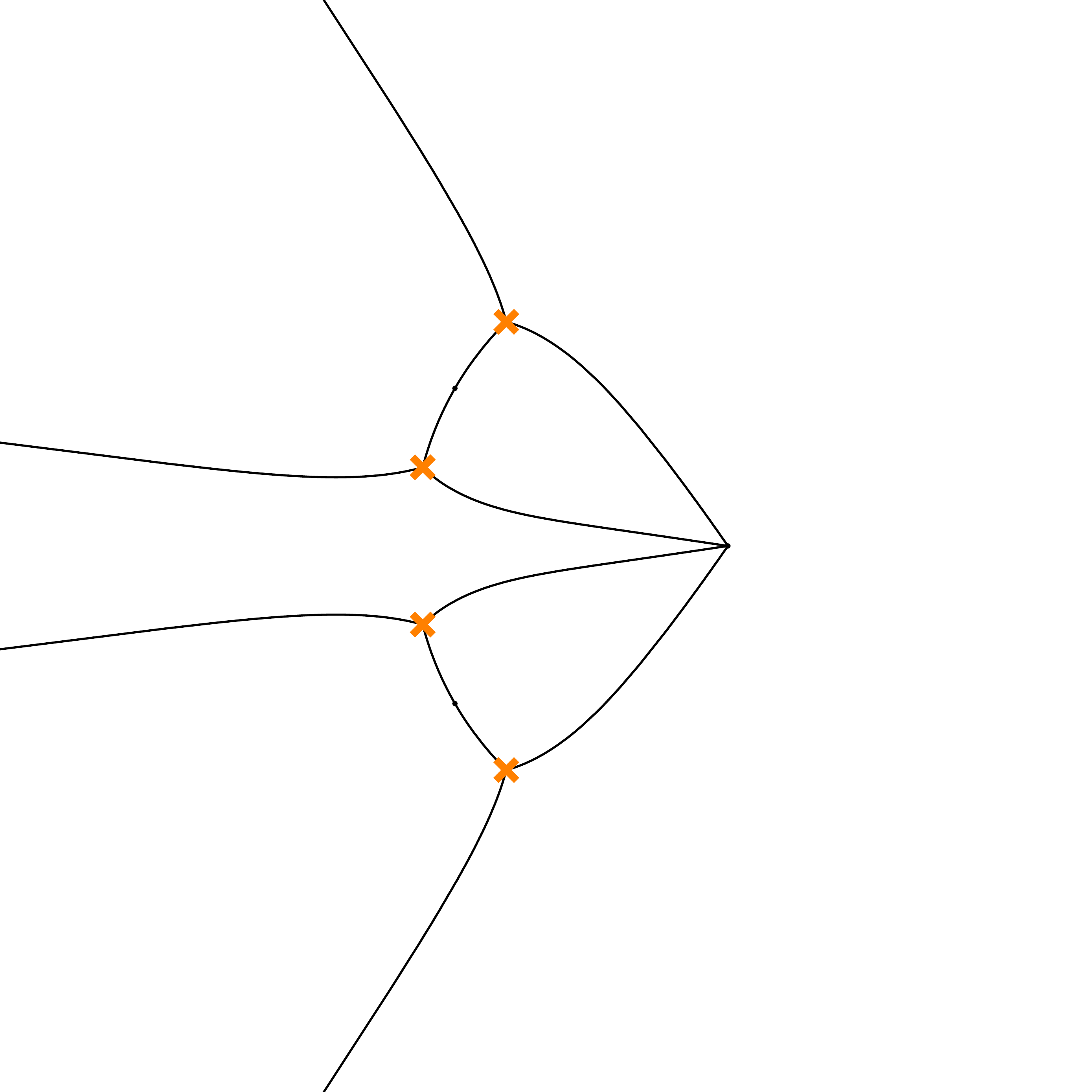}\\
\end{center}
\caption{Sequence of spectral networks of the $SU(2)$, $N_f=4$ theory. The phases begin with sector $I$ as described in~(\ref{eq:2d4d-seq-1}), for phases ranging from $\vartheta_c-\pi/2$ (top-left) up to $\vartheta_c$ (center), and proceeds with phases from sector $II$ as described in~(\ref{eq:2d4d-seq-2}) ranging from $\vartheta_c$ to $\vartheta_c+\pi/2$ (bottom-right).
}
\label{fig:nf4-networks}
\end{figure}

If we were to study the BPS spectrum for a surface defect using the first BPS graph, this would require an entirely new analysis. However, for the second BPS graph we can directly apply the general results from Section~\ref{sec:near-punctures}, for the behavior of networks near a puncture surrounded by a bigon. This is analogous to the situation encountered in the $T_2$ theory (see Figure~\ref{fig:T2-BPSg}). Indeed, direct inspection confirms that $\CS$-walls near the puncture have the universal behavior, as shown in Figure ~\ref{fig:nf4-networks}.
The key point is that near each of the punctures surrounded by bigons, there is a spiral made of two $\CS$-walls sourced by two nearest branch points. No other $\CS$-walls from other regions ever come near. We shall comment more on the role of the other BPS graph of this theory in Appendix~\ref{sec:more-punctures}.

Having shown that the BPS graph of the $SU(2)$, $N_f=4$ theory has the universal behavior studied above near a puncture, the whole derivation of the 2d-4d spectrum generator and its trace follow through unchanged.

\section{Once-punctured surfaces}\label{sec:one-puncture}

In this section, we cover the (remnant) case of a theory of class $\CS$ associated to once-punctured UV curves. This particular class of theories was not covered in the previous general discussion. The arguments we present in the following section allows us to extend our discussion of IR vortex surface defects to \emph{generic} $A_1$ theories of class $\CS$ with $n>0$ regular punctures. We start with the example of the  4d $\cN=2^{*}$ theory, and then generalize to arbitrary once-punctured curves.

\subsection{The $\mathcal{N}=2^*$ theory}\label{sec:N2star}

The rank-one $\CN=2^*$ theory is engineered as a class $\CS$-theory associated to a torus with one  regular puncture. Therefore, this is the first example where the assumptions about the local behavior of $\CS$-walls near a puncture, for which we argued in Section~\ref{sec:near-punctures}, does not hold. For this reason, this theory will require a novel in-depth treatment. However, we will see that many of the key features from Section~\ref{sec:BPS-spectrum} still hold, although they enter via a slightly different guise. 

The classical spectrum generator of the $\CN=2^*$ theory was obtained in~\cite{Gaiotto:2009hg}, and the quantum spectrum generator was obtained in~\cite{Longhi:2016wtv} from the BPS graph. In fact, the 4d BPS spectrum of this theory has been explicitly worked out at special loci on the Coulomb branch in~\cite{Longhi:thesis}.
The quadratic differential for this theory is given as follows
\be
\phi_2  \ = \  u + m^2 \wp(z|\tau)\,,
\ee
where $u$ is a Coulomb branch modulus, $m$ is a UV mass for the adjoint hypermultiplet, and $\tau$ is the gauge coupling corresponding to the torus modular parameter. There is a double pole at $z=0$.

\begin{figure}[h!]
\begin{center}
\includegraphics[width=0.35\textwidth]{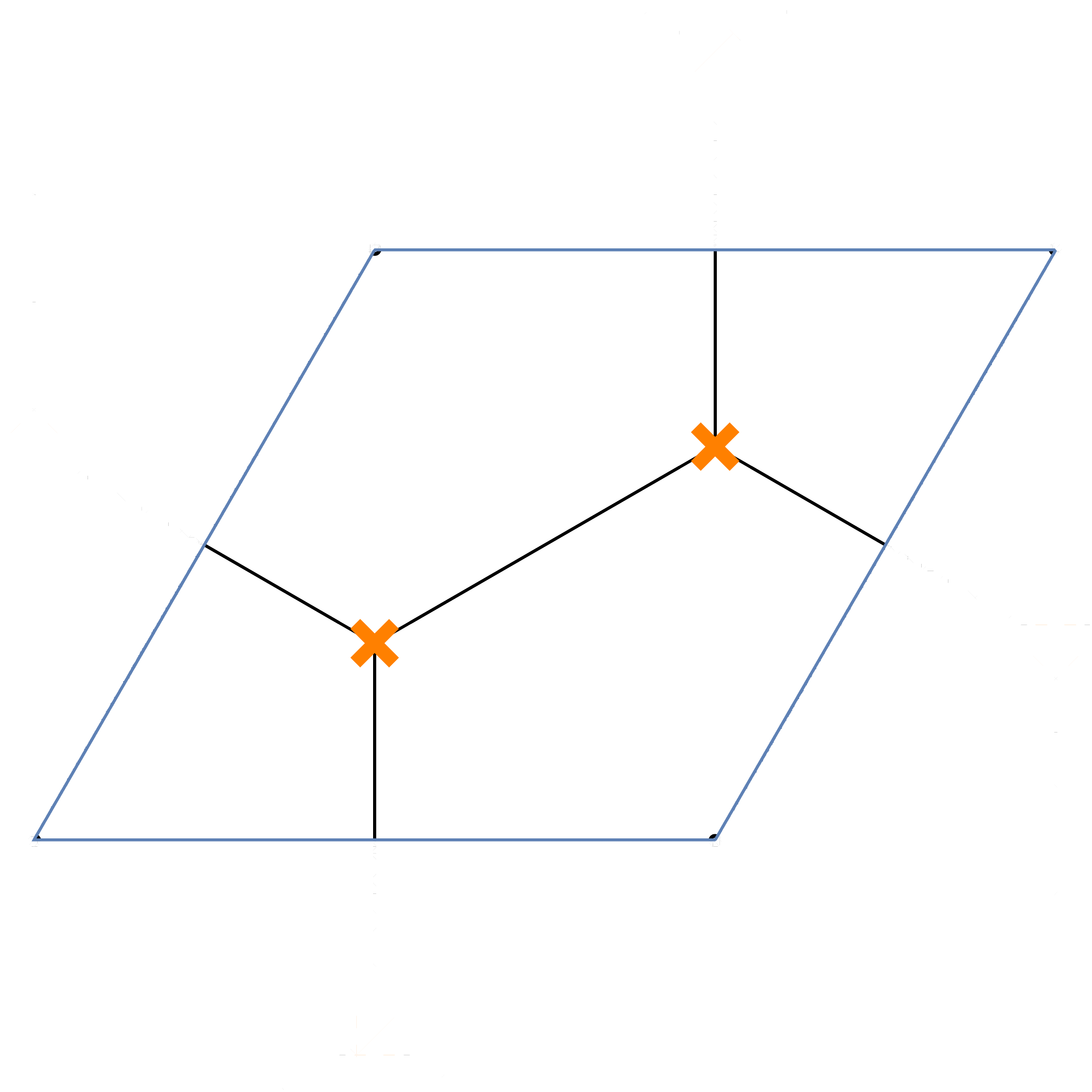}
\caption{The BPS graph of the $SU(2)$, $\CN=2^*$ theory.}
\label{fig:N2star-BPSg}
\end{center}
\end{figure}

\begin{figure}
\begin{center}
\includegraphics[width=0.22\textwidth]{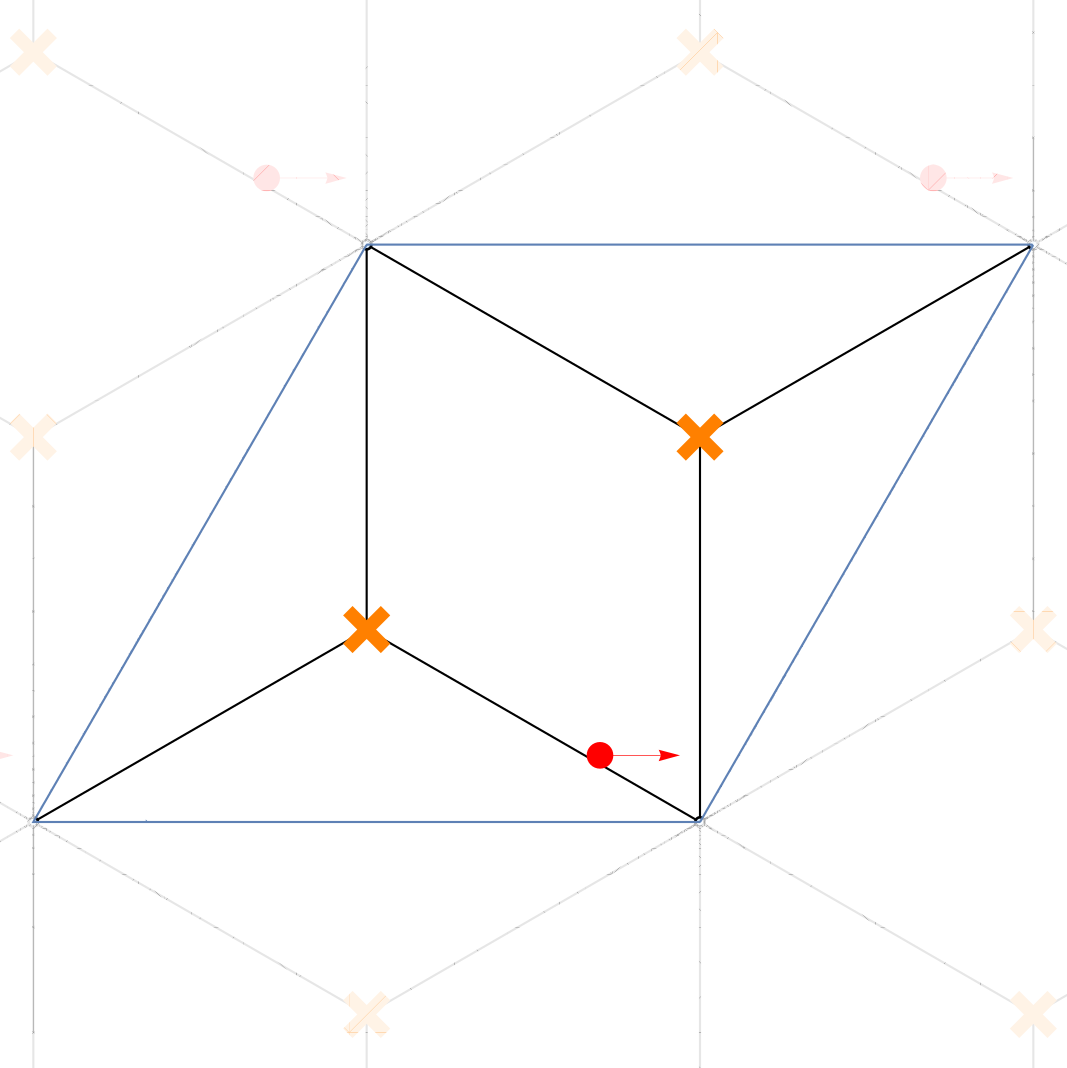}\hfill
\includegraphics[width=0.22\textwidth]{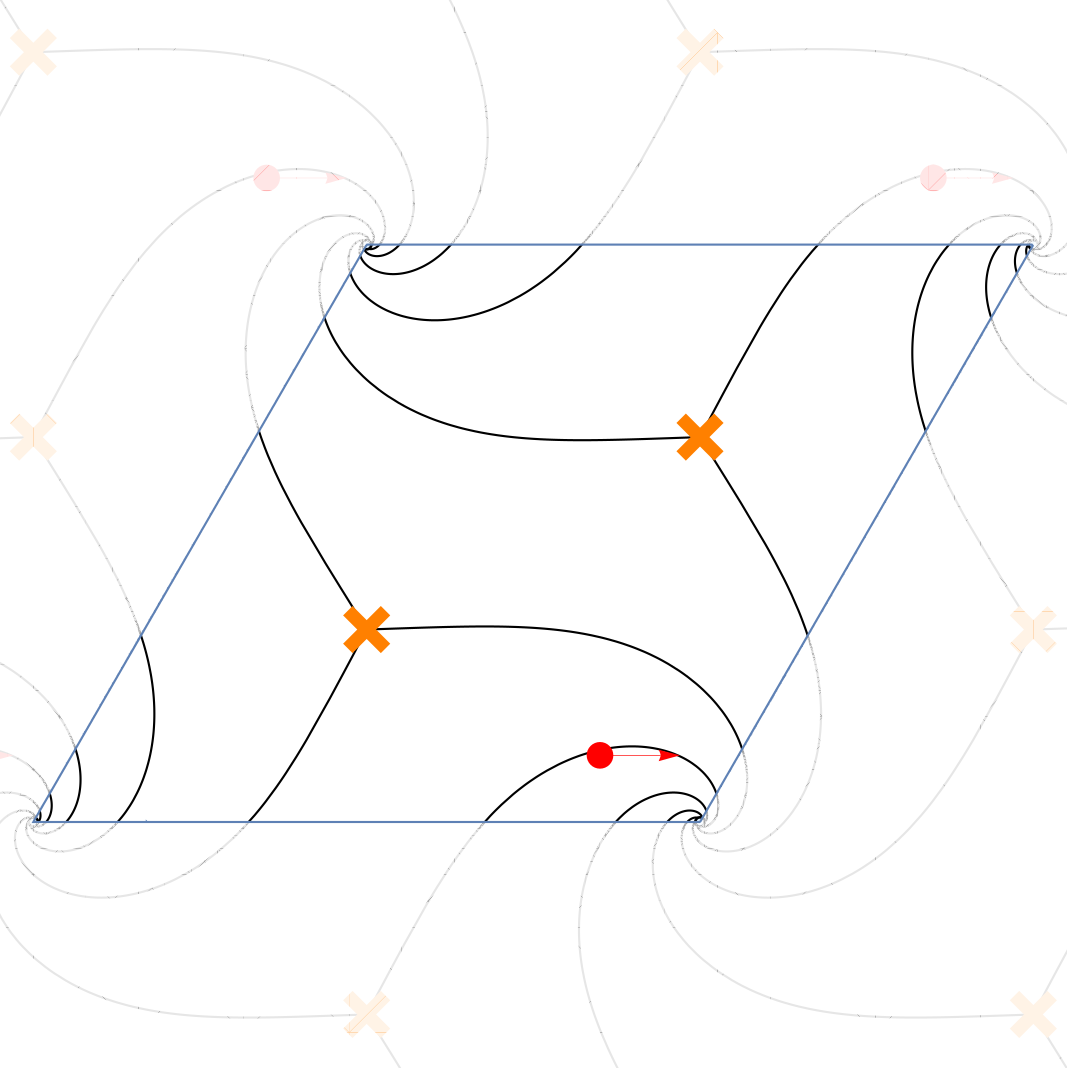}\hfill
\includegraphics[width=0.22\textwidth]{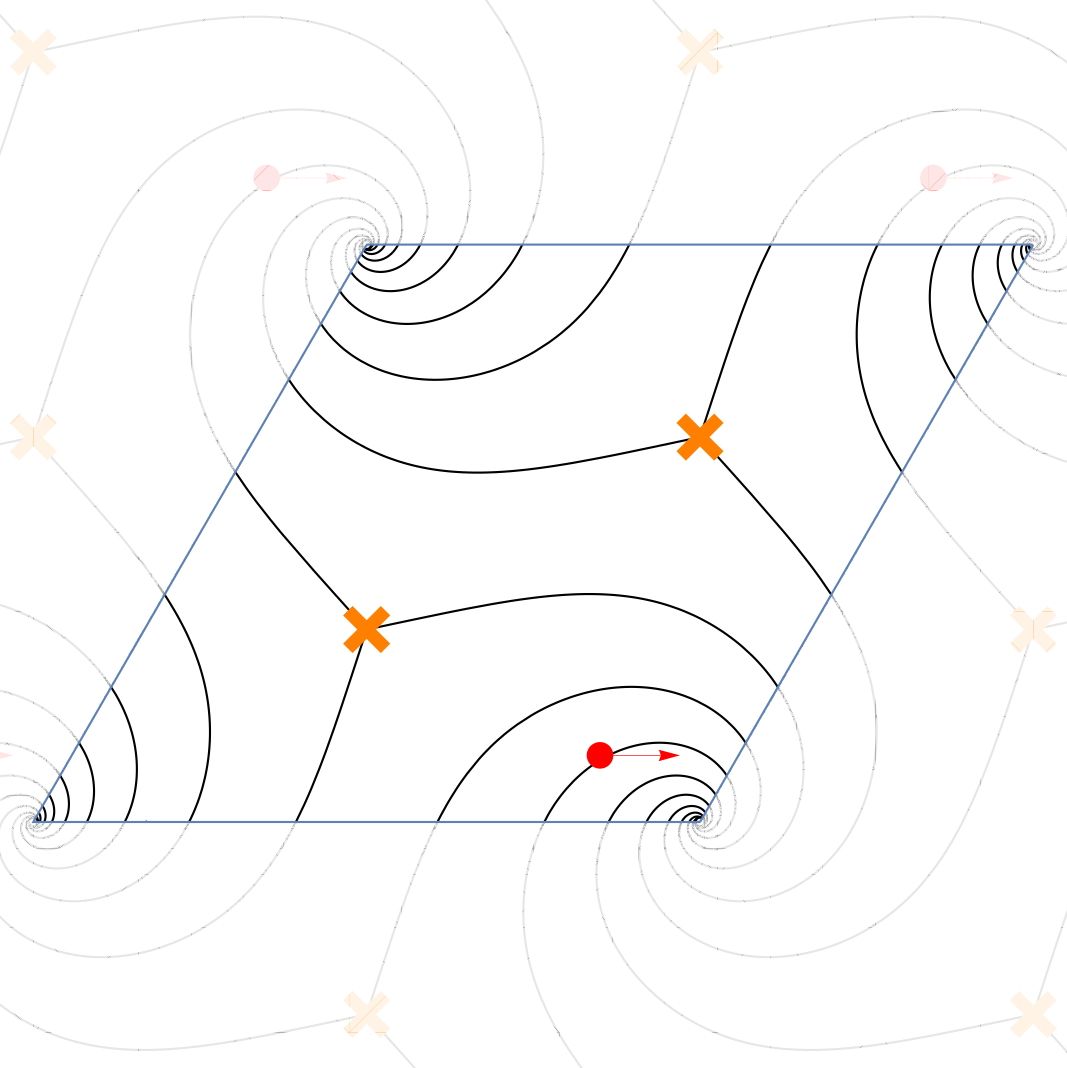}\hfill
\includegraphics[width=0.22\textwidth]{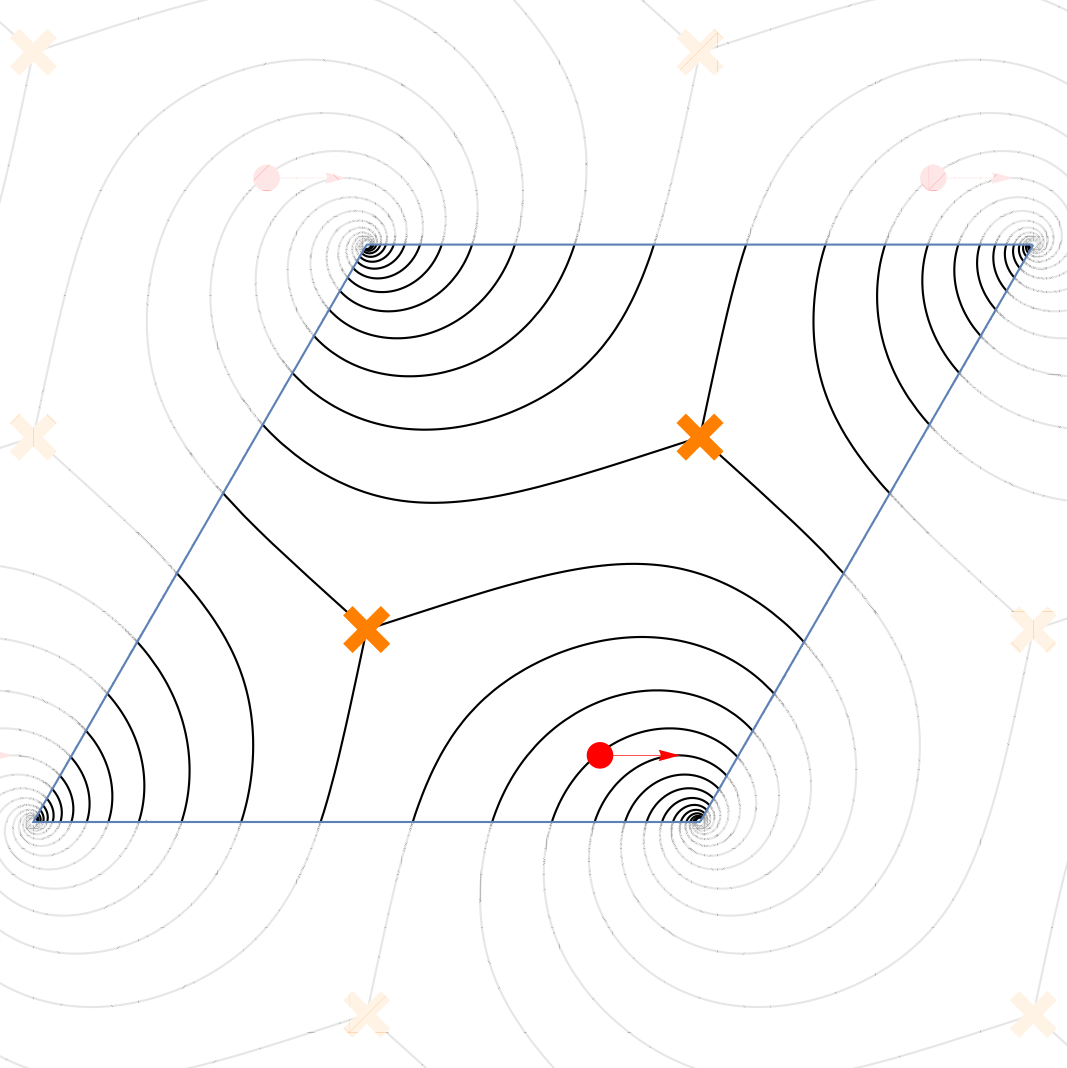}\\
\includegraphics[width=0.22\textwidth]{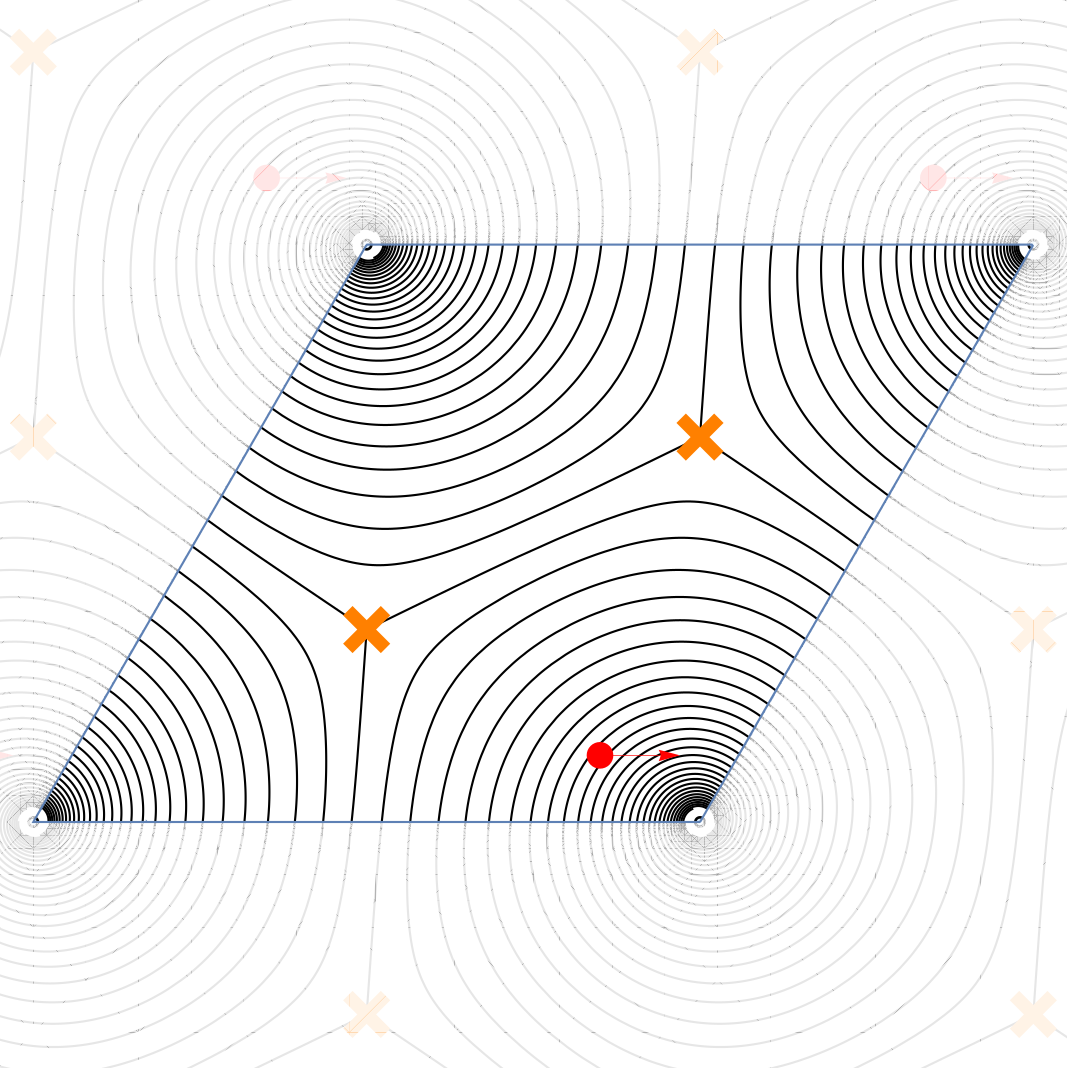}
\hfill\includegraphics[width=0.06\textwidth]{figures/dots.pdf}\hfill
\includegraphics[width=0.22\textwidth]{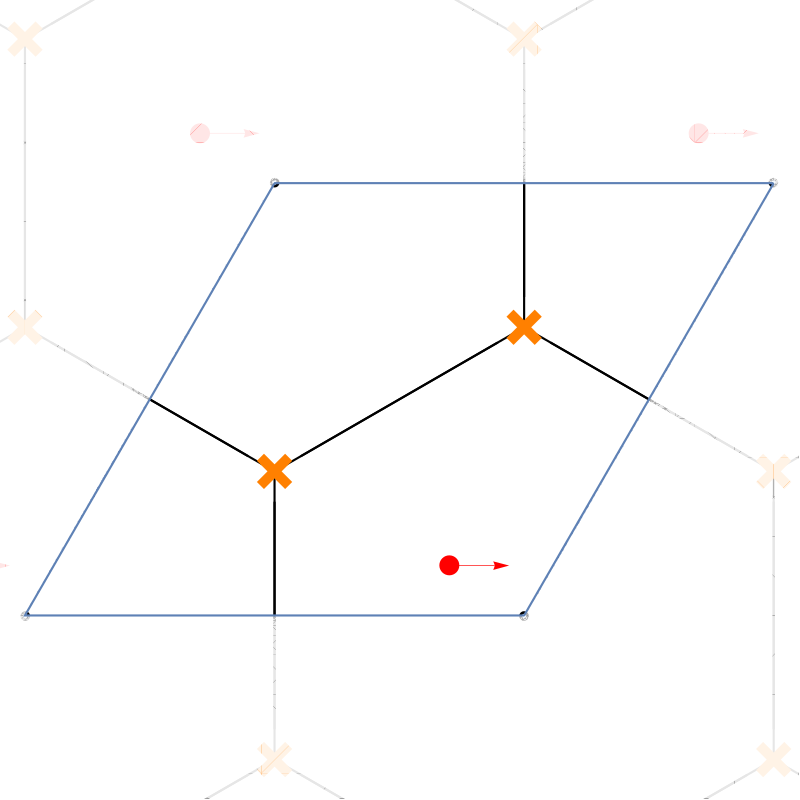}
\hfill\includegraphics[width=0.06\textwidth]{figures/dots.pdf}\hfill
\includegraphics[width=0.22\textwidth]{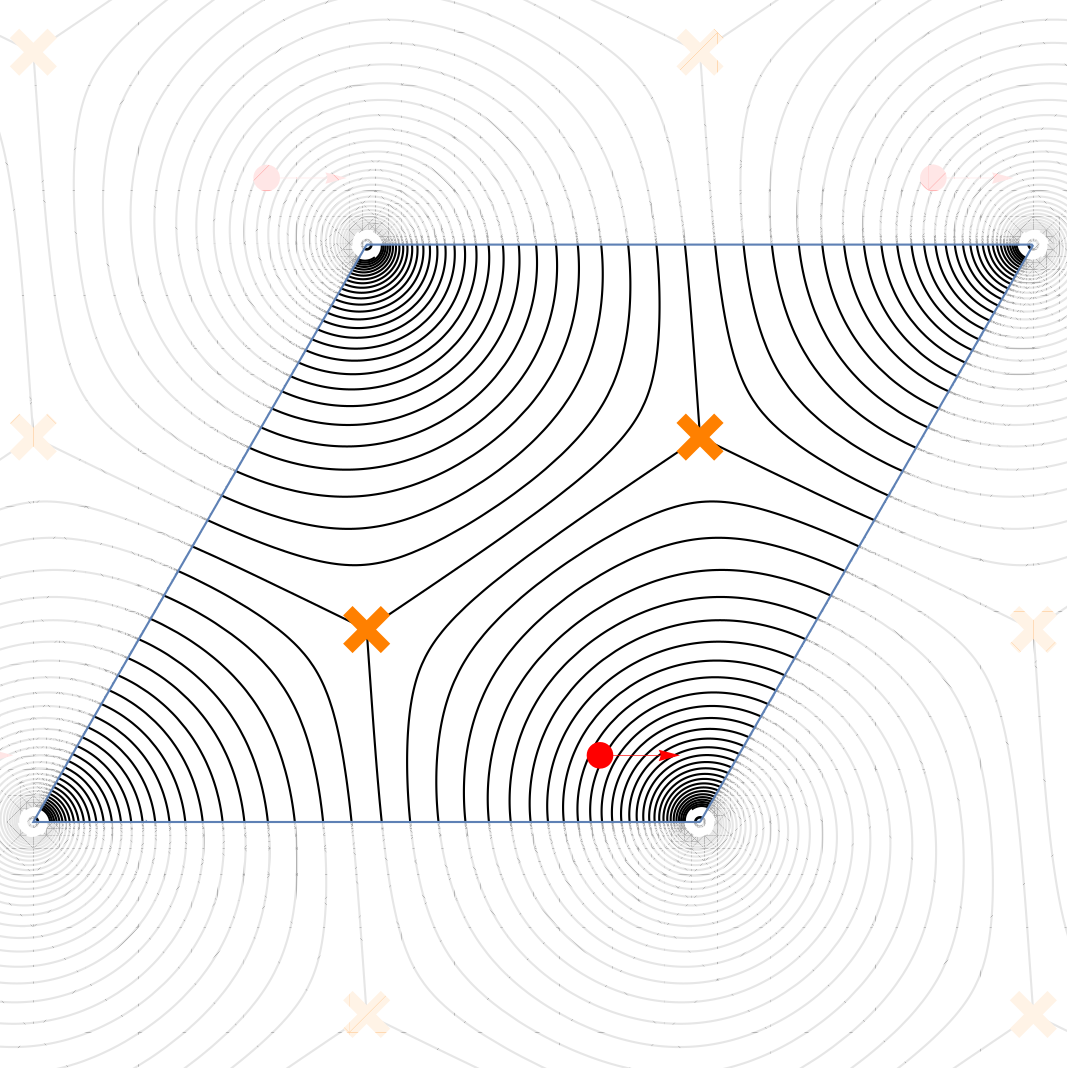}\\
\includegraphics[width=0.22\textwidth]{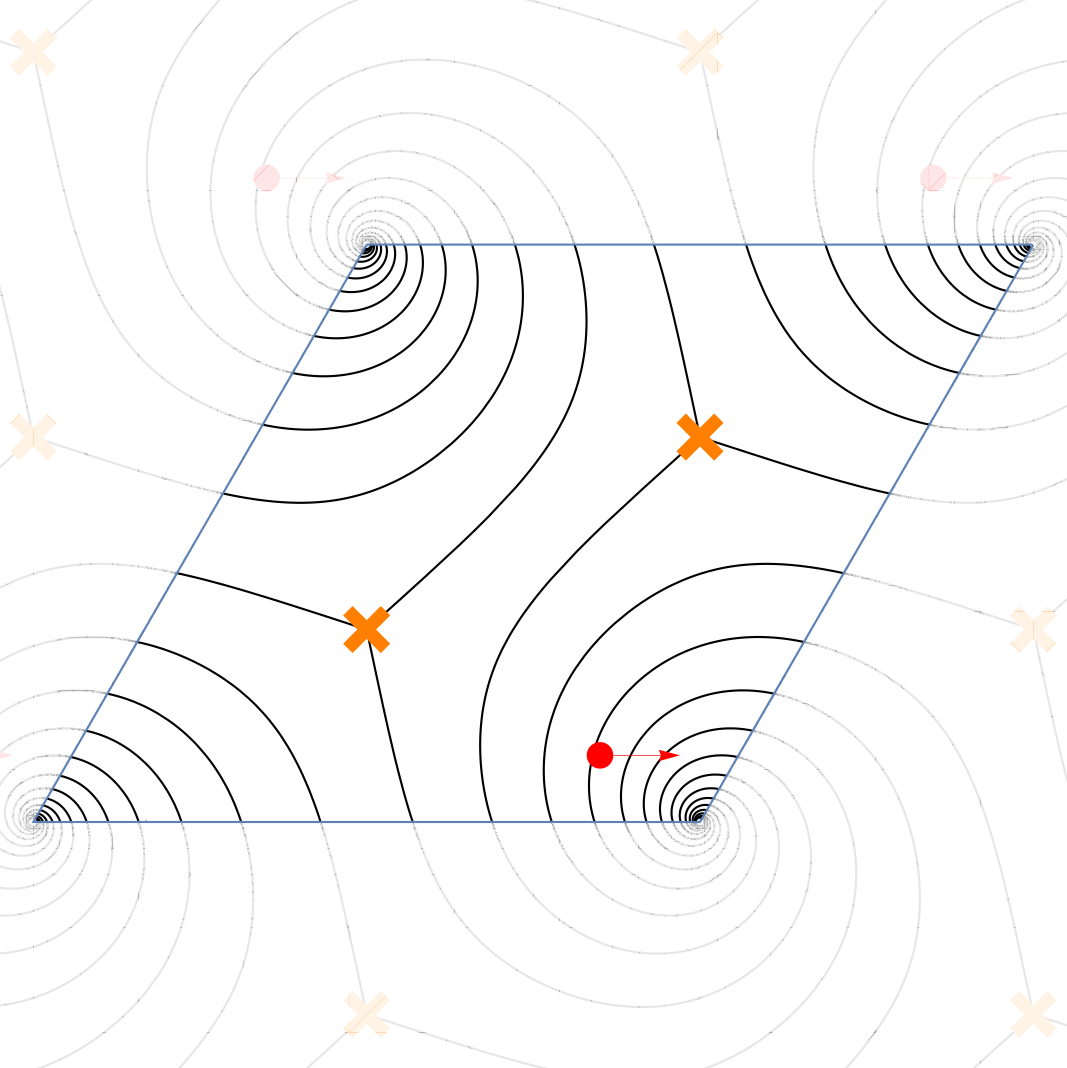}\hfill
\includegraphics[width=0.22\textwidth]{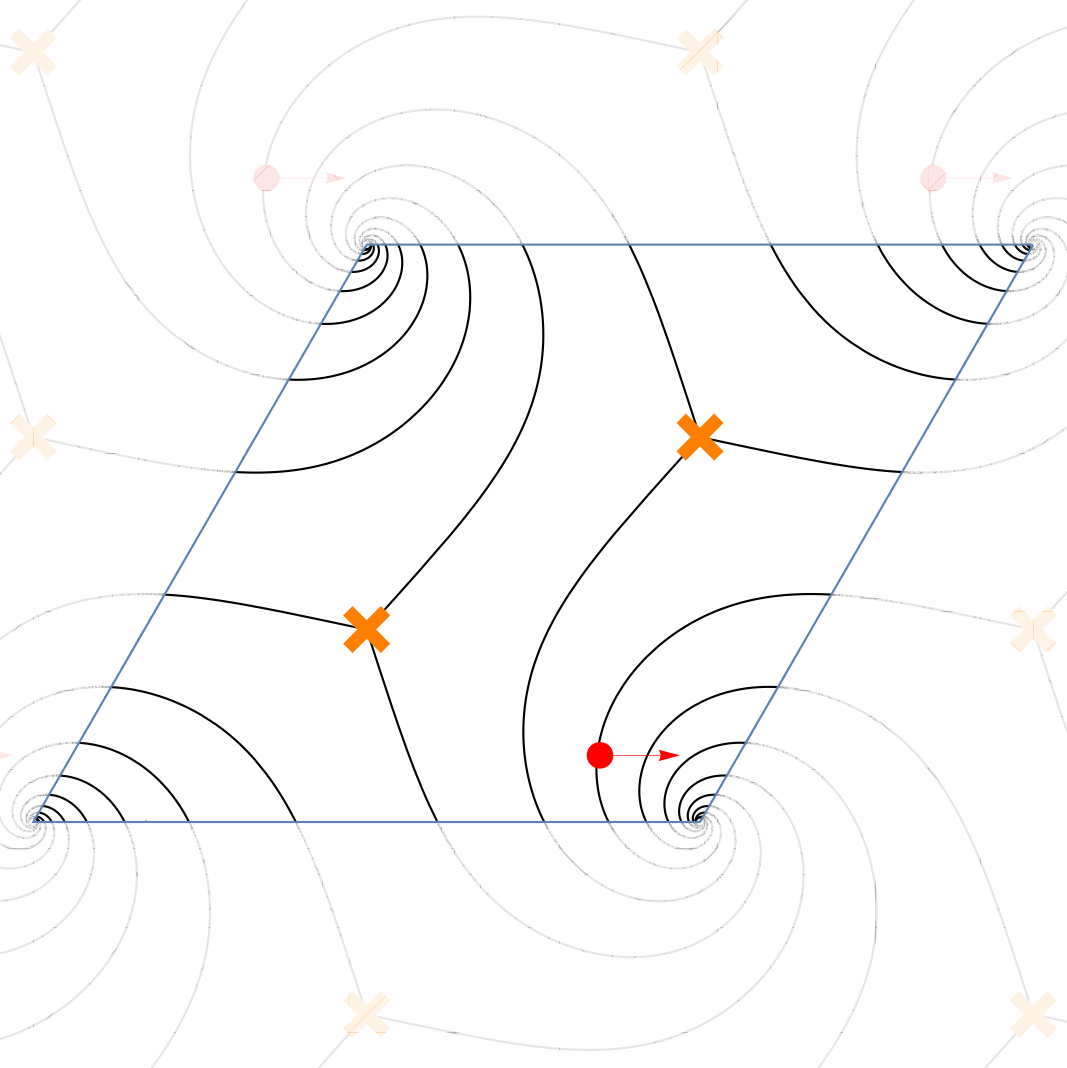}\hfill
\includegraphics[width=0.22\textwidth]{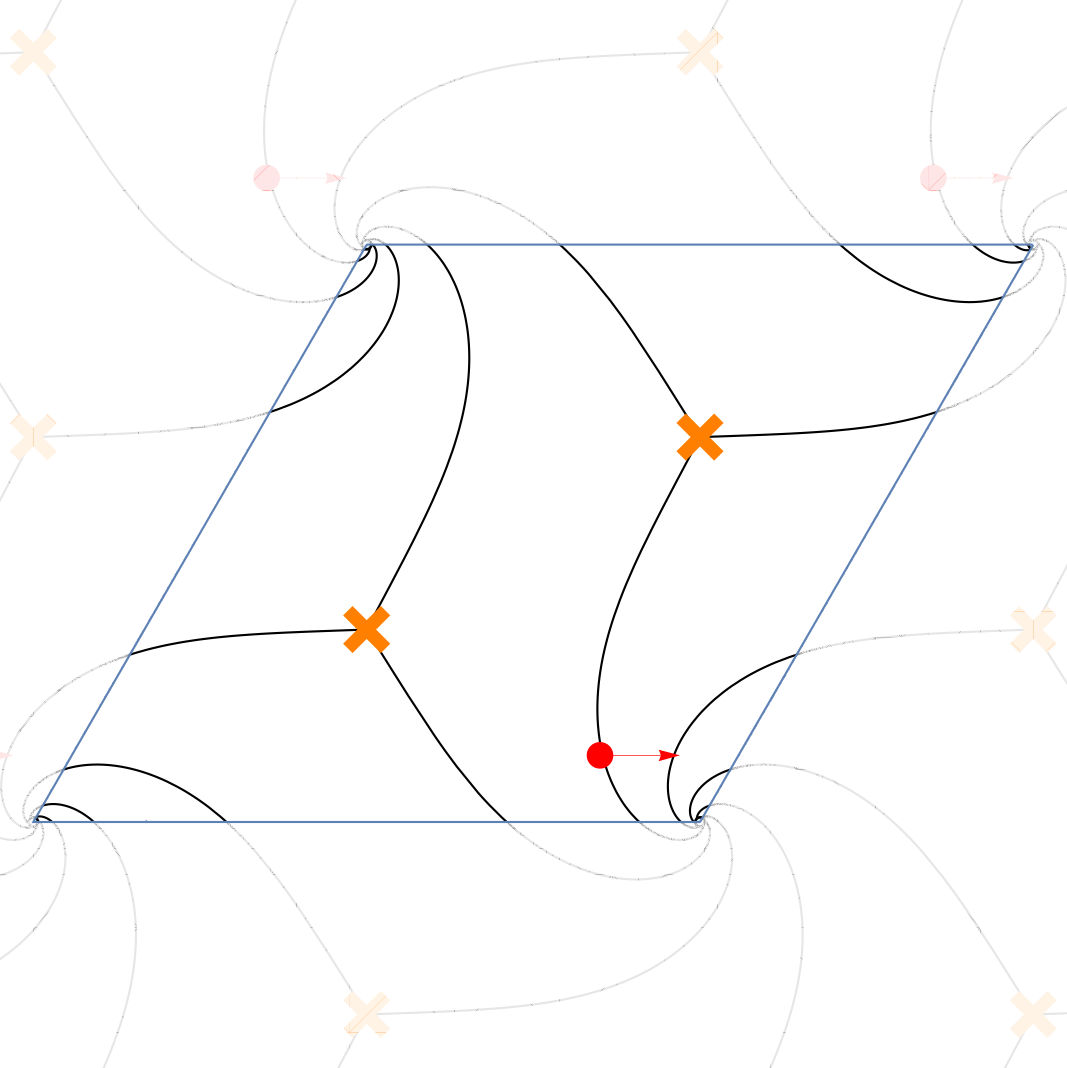}\hfill
\includegraphics[width=0.22\textwidth]{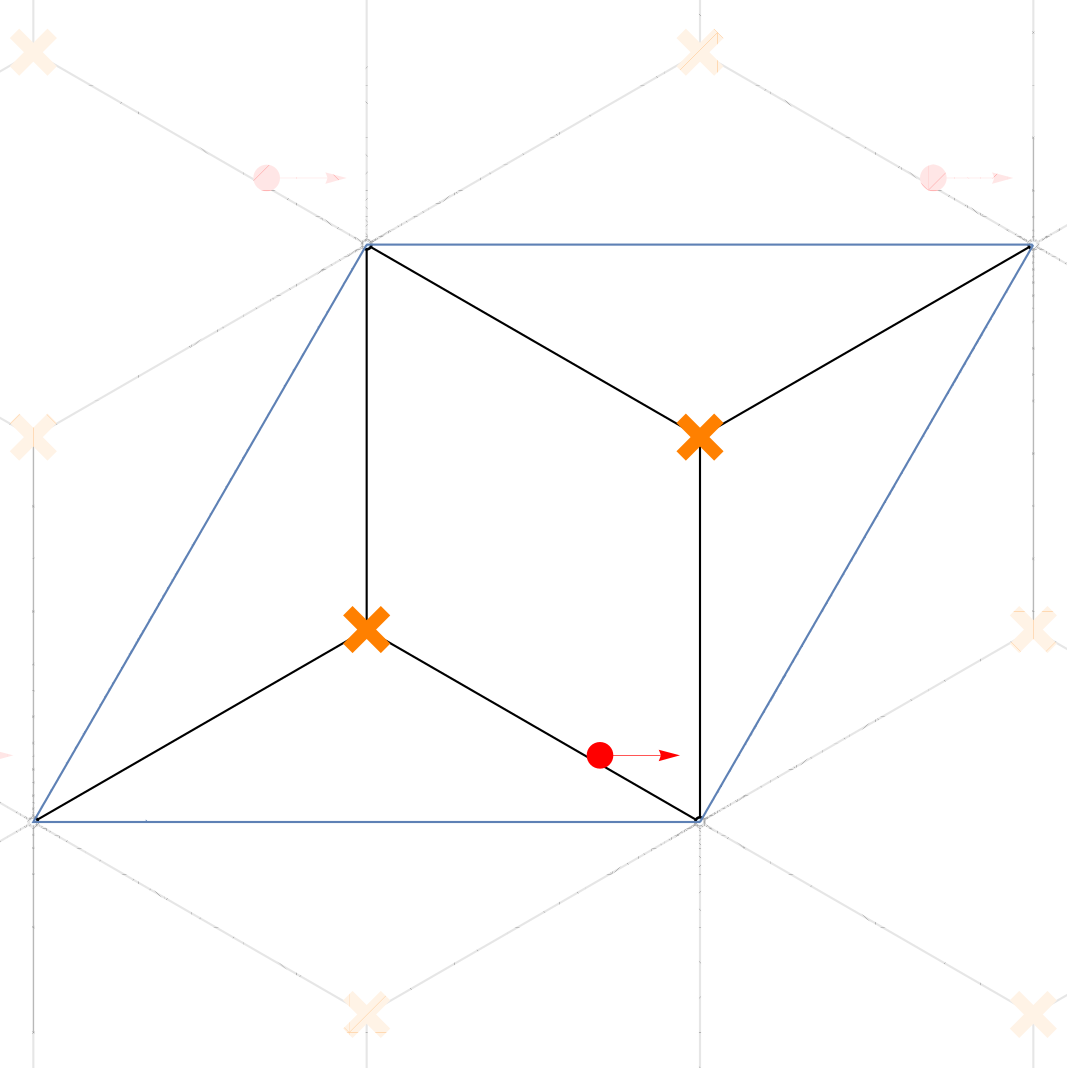}\\
\end{center}
\caption{Sequence of spectral networks of the $SU(2)$, $\CN=2^*$ theory. The phases begin with sector $I$ as described in~(\ref{eq:2d4d-seq-1}), for phases ranging from $\vartheta_c-\pi/2$ (top-left) up to $\vartheta_c$ (center), and proceeds with phases from sector $II$ as described in~(\ref{eq:2d4d-seq-2}) ranging from $\vartheta_c$ to $\vartheta_c+\pi/2$ (bottom-right).
}
\label{fig:N2-star-networks}
\end{figure}

If we fix $m=1$ and $\tau = e^{\pi \ii / 3}$, the Roman locus turns out to be at $u=0$. With this choice of moduli all central charges are real. Indeed, plotting the spectral network at $\vartheta_c = 0$ we find the BPS graph shown in Figure~\ref{fig:N2star-BPSg}.
For generic values of the phase, the network evolves as shown in Figure~\ref{fig:N2-star-networks}.

\begin{figure}
\begin{center}
\includegraphics[width=0.22\textwidth]{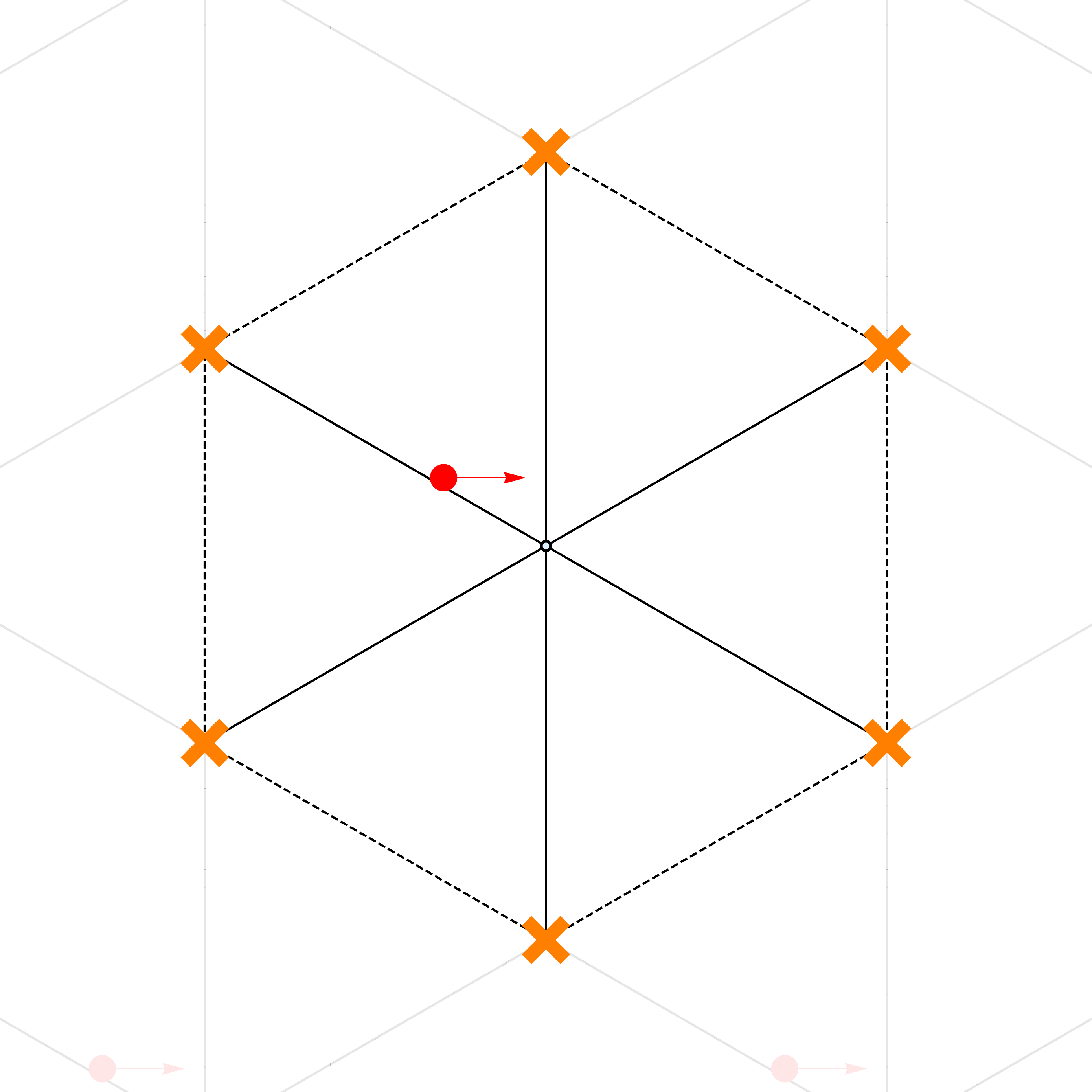}\hfill
\includegraphics[width=0.22\textwidth]{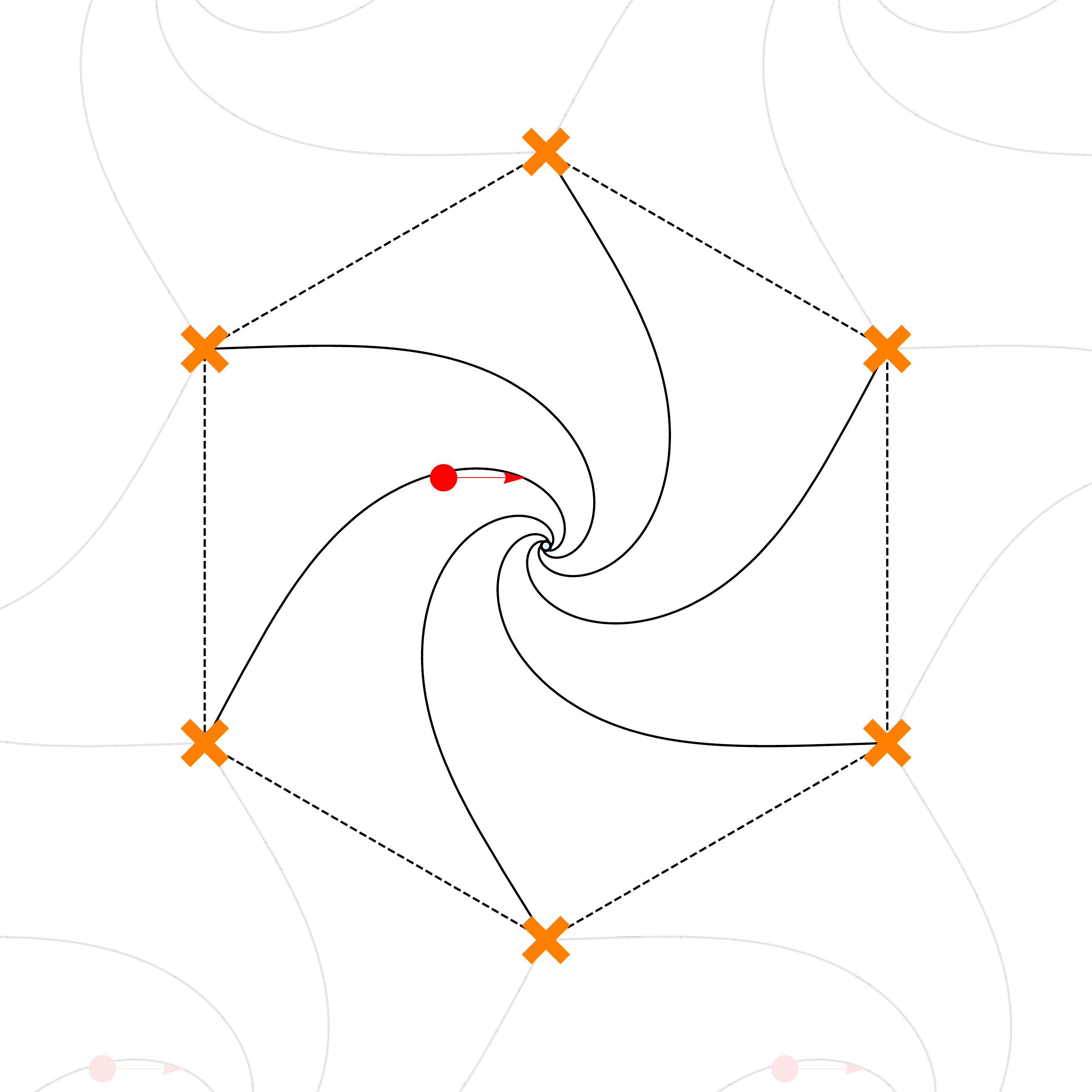}\hfill
\includegraphics[width=0.22\textwidth]{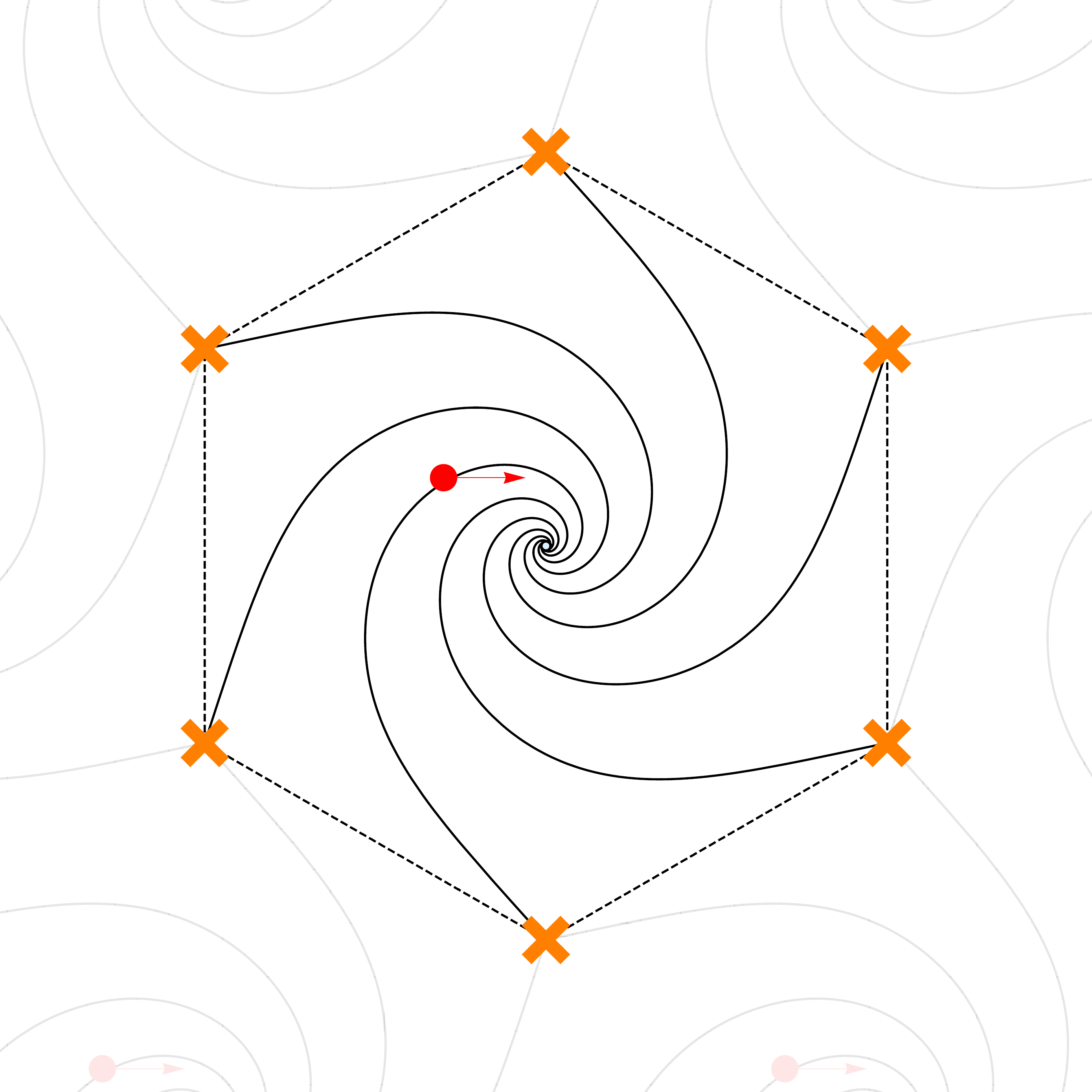}\hfill
\includegraphics[width=0.22\textwidth]{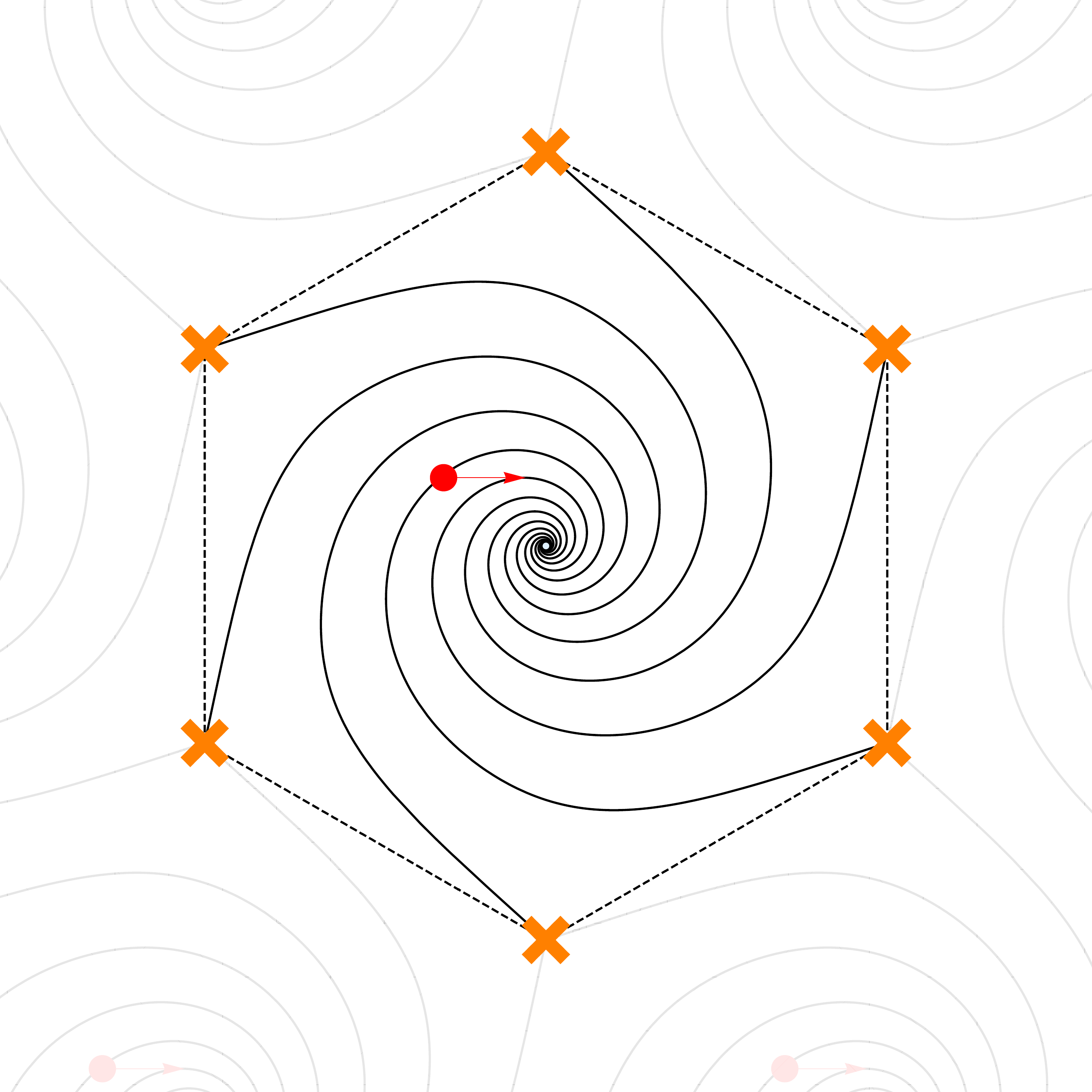}\\
\includegraphics[width=0.22\textwidth]{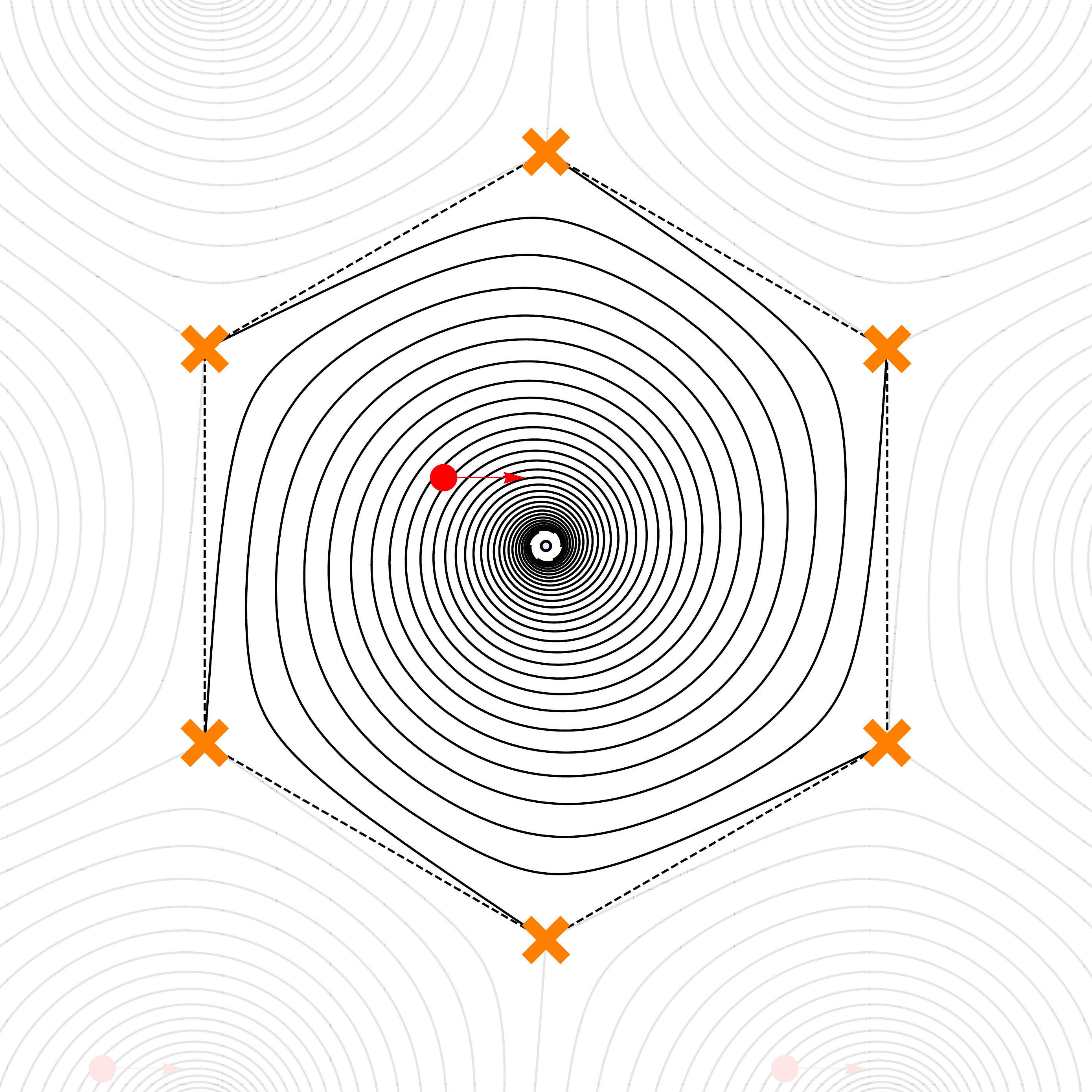}
\hfill\includegraphics[width=0.06\textwidth]{figures/dots.pdf}\hfill
\includegraphics[width=0.22\textwidth]{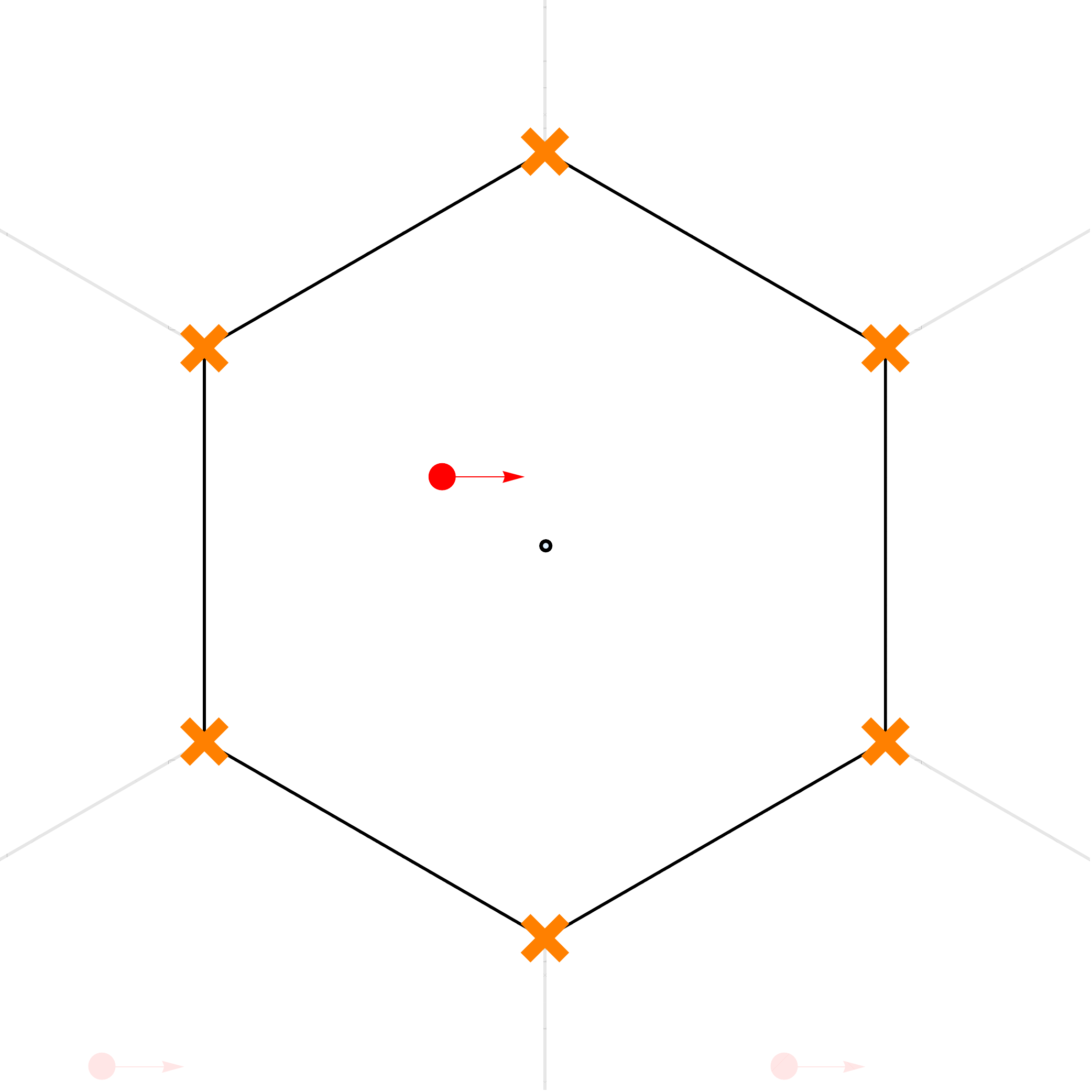}
\hfill\includegraphics[width=0.06\textwidth]{figures/dots.pdf}\hfill
\includegraphics[width=0.22\textwidth]{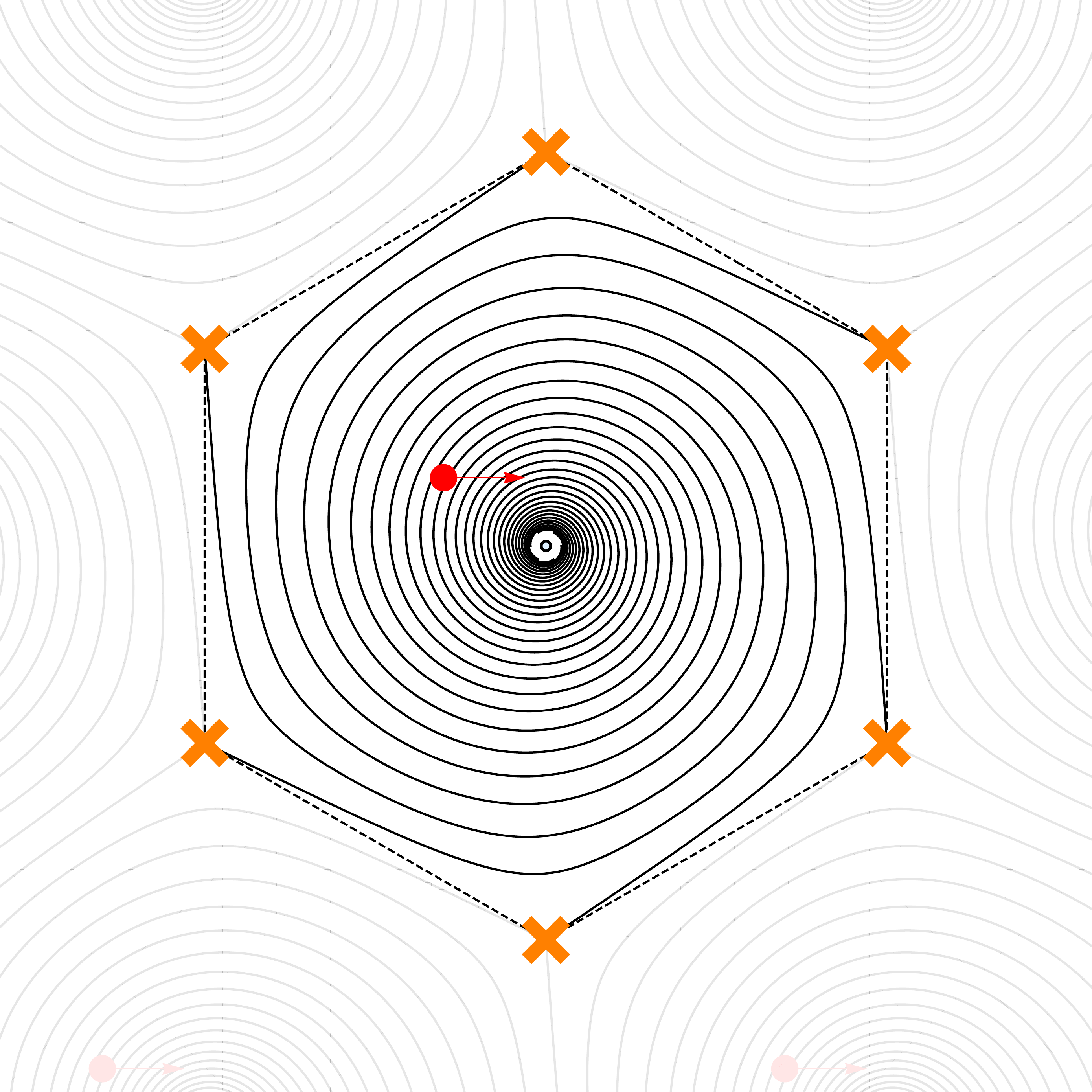}\\
\includegraphics[width=0.22\textwidth]{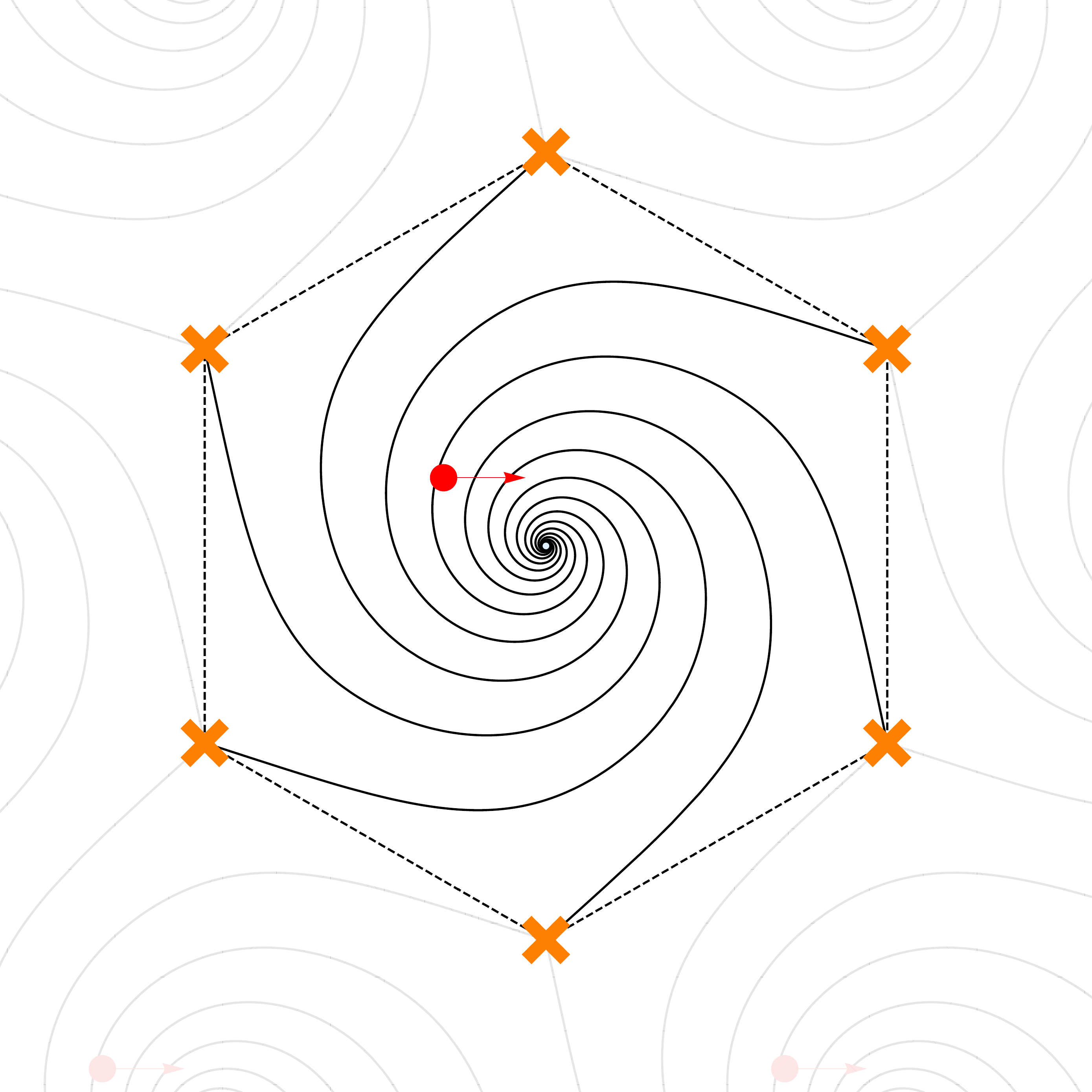}\hfill
\includegraphics[width=0.22\textwidth]{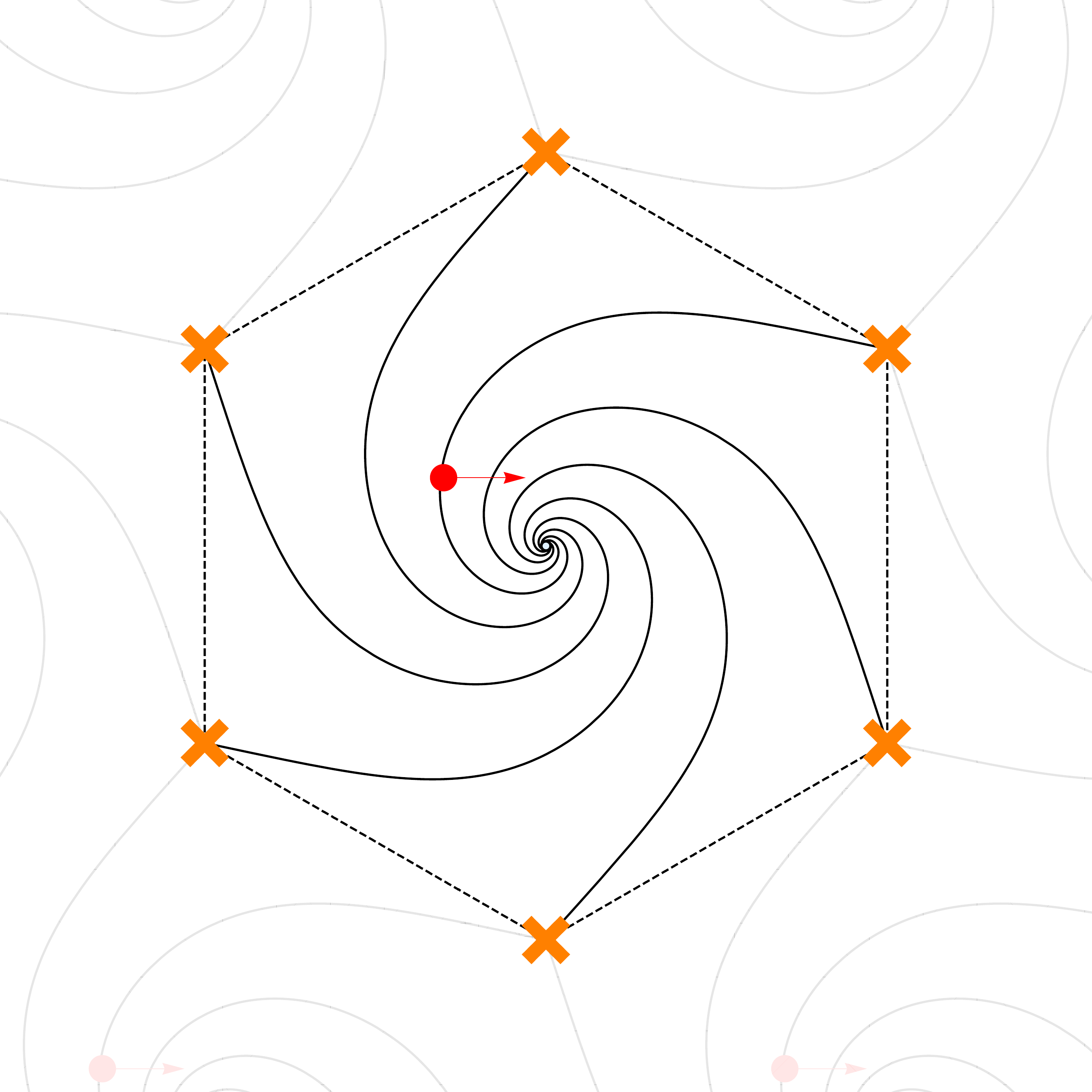}\hfill
\includegraphics[width=0.22\textwidth]{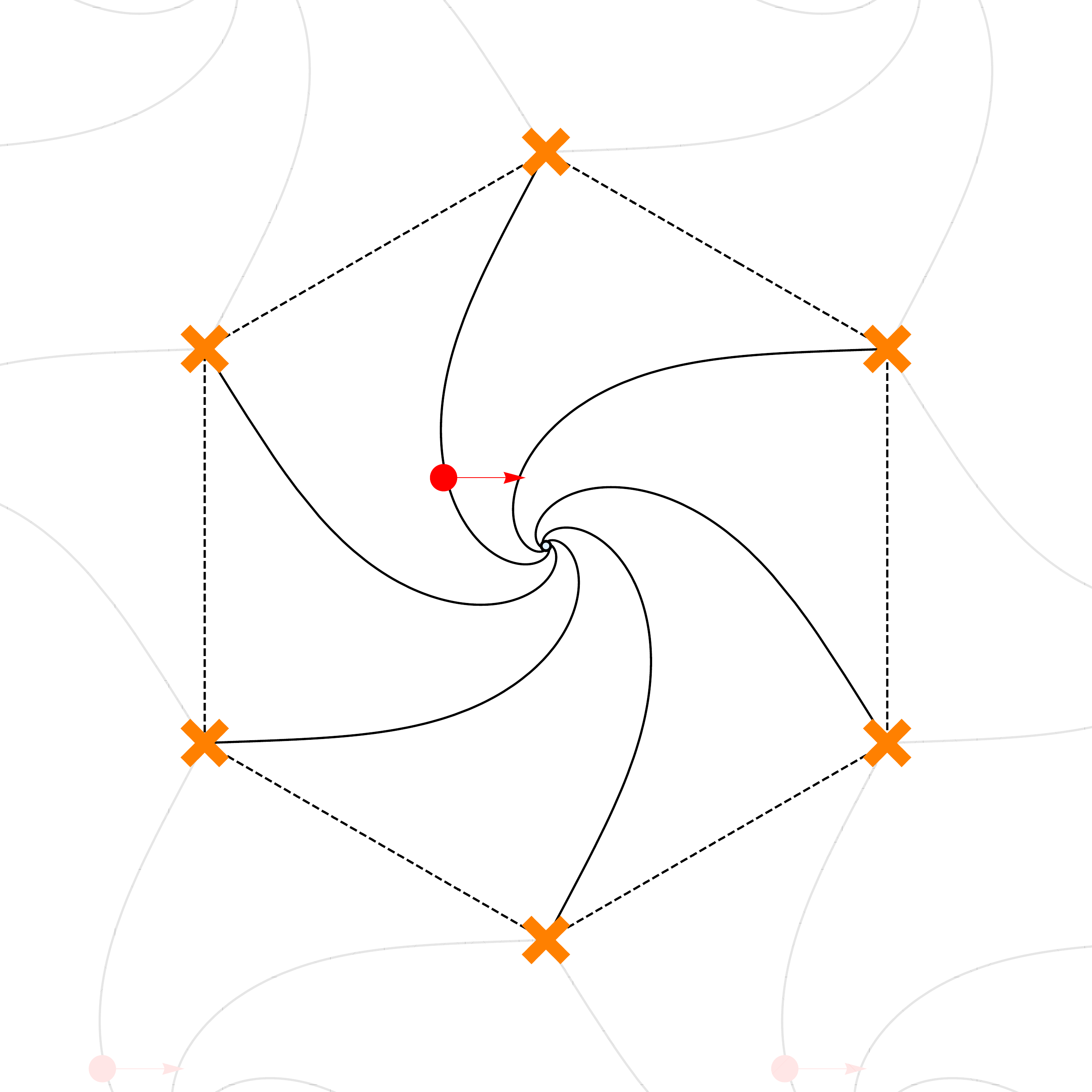}\hfill
\includegraphics[width=0.22\textwidth]{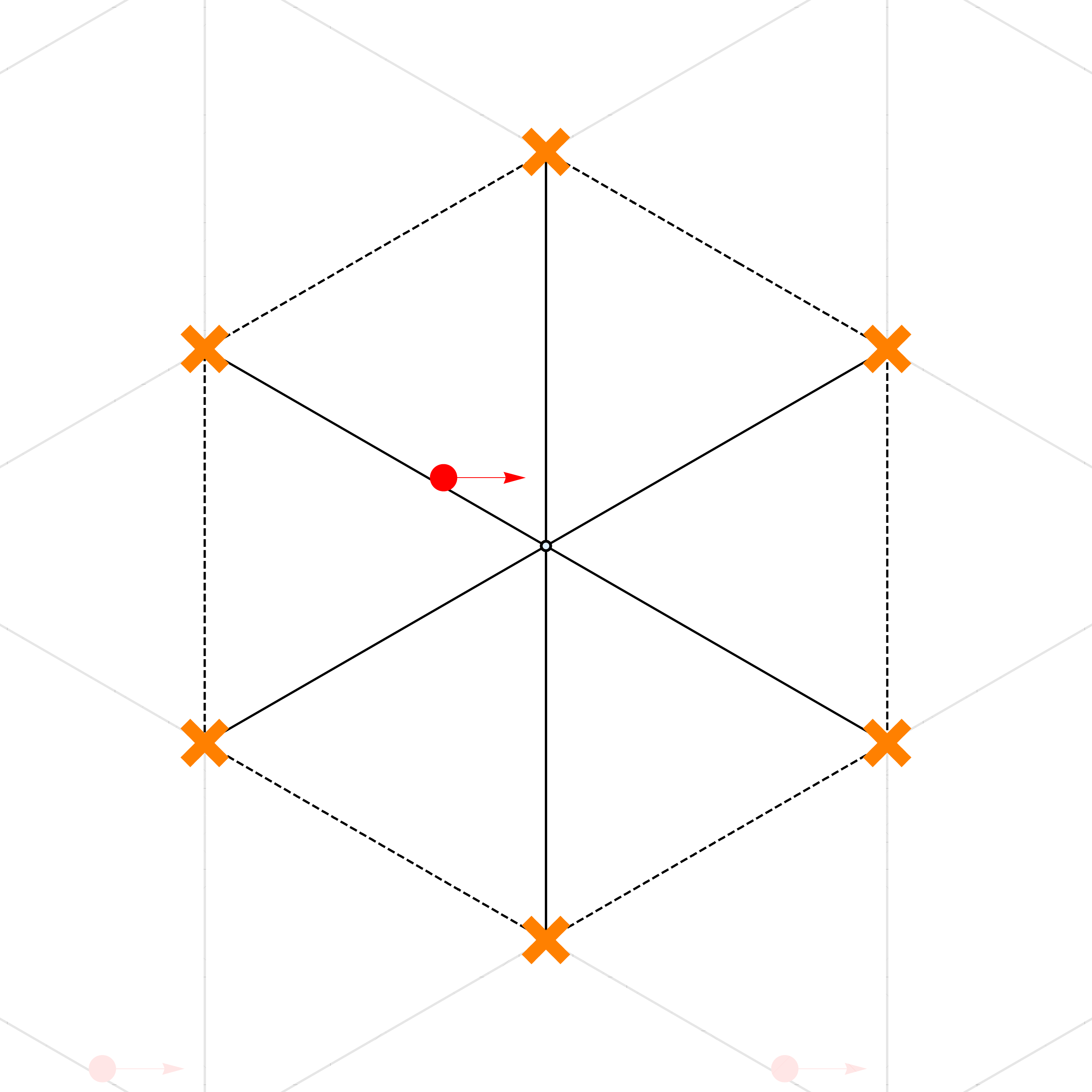}\\
\end{center}
\caption{Sequence of spectral networks of the $SU(2)$, $\CN=2^*$ theory, presented as a glued hexagon.
The phases begin with sector $I$ as described in~(\ref{eq:2d4d-seq-1}), for phases ranging from $\vartheta_c-\pi/2$ (top-left) up to $\vartheta_c$ (center), and proceeds with 
phases from sector $II$ as described in~(\ref{eq:2d4d-seq-2}) ranging from $\vartheta_c$ to $\vartheta_c+\pi/2$ (bottom-right).
}
\label{fig:N2-star-hexagon-networks}
\end{figure}

Let us fix a surface defect at $z$, marked by the red dot in our figures. For the purpose of discussing its 2d-4d BPS spectrum, it is actually more convenient to pass to another presentation of the network, which is shown in Figure~\ref{fig:N2-star-hexagon-networks}.


\begin{figure}[h!]
\begin{center}
\includegraphics[width=0.45\textwidth]{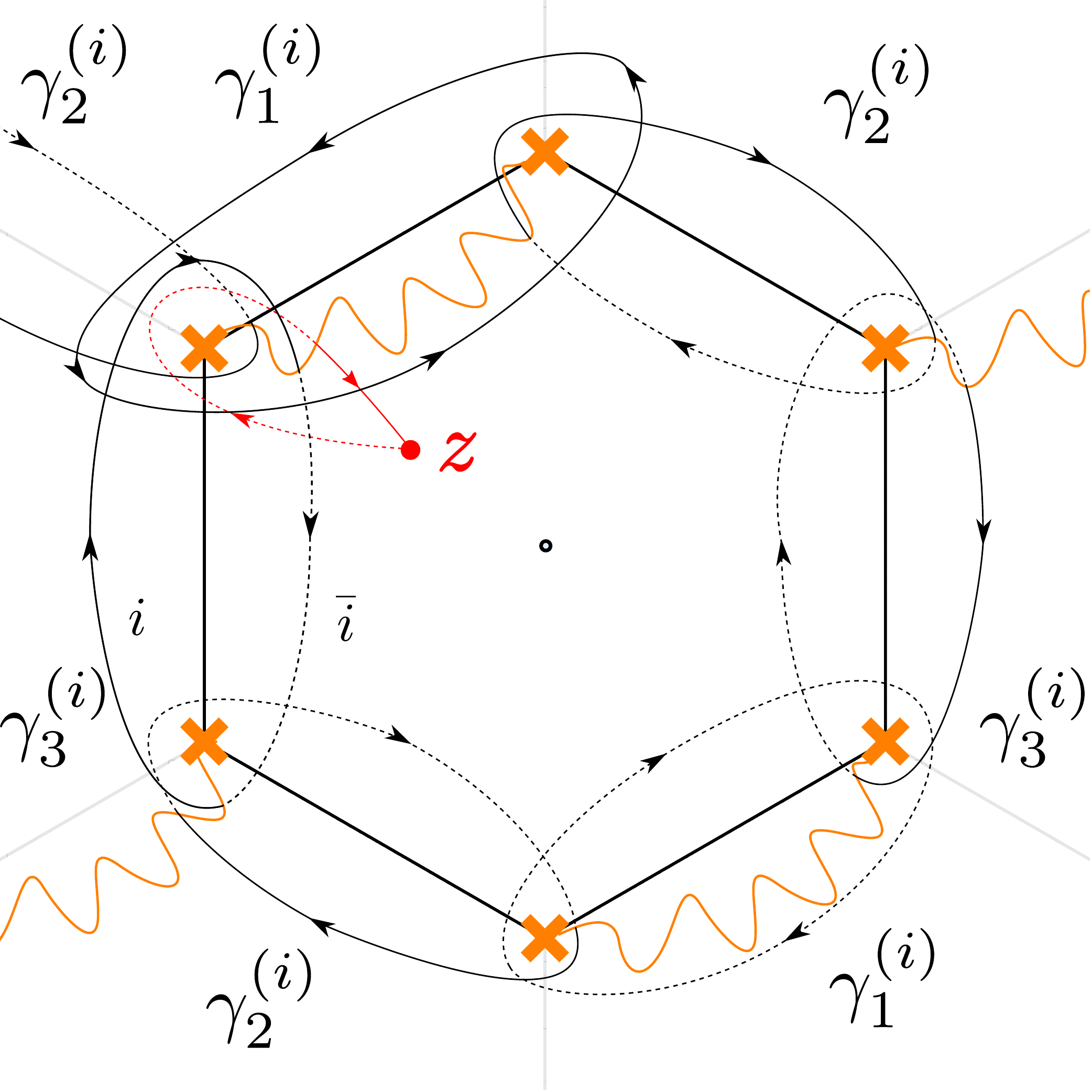}
\caption{Charges of 2d-4d and 4d BPS states for a surface defect of the $\CN=2^*$ theory. Paths on sheet $i$ are denoted by solid lines, while paths on sheet $\bar\imath$ by dashed lines.}
\label{fig:hexagon-marked}
\end{center}
\end{figure}

As before, to count 2d-4d particles we keep track of $\CS$-walls that sweep across $z$ while we increase the phase. To fix conventions, let $a_{\bar \imath i}$ be the soliton charge and $\gamma_i$ the three generators of  the 4d charge lattice depicted in Figure~\ref{fig:hexagon-marked}. Running through the phases of sector $I$ (\emph{i.e.} for $\vartheta_c-\pi/2 \leq \vartheta <\vartheta_c$) we observe the following soliton charges, in order of increasing phase
\beaa\label{eq:n2star-2d-4d-spectrum-I}
	&a_{\bar \imath i} \,, 
	&& a_{\bar \imath i}+ \gamma^{(i)}_3\,,
	&& a_{\bar \imath i}+ \gamma^{(i)}_2+\gamma^{(i)}_3 \,, \\
	& a_{\bar \imath i}+ \gamma^{(i)}_1 + \gamma^{(i)}_2+\gamma^{(i)}_3\,,
	&& a_{\bar \imath i} + \gamma^{(i)}_1 + \gamma^{(i)}_2+2 \gamma^{(i)}_3\,, 
	&& a_{\bar \imath i} + \gamma^{(i)}_1 + 2\gamma^{(i)}_2+2\gamma^{(i)}_3\,, \\
	&a_{\bar \imath i}+ 2\gamma^{(i)}_1 + 2\gamma^{(i)}_2+2\gamma^{(i)}_3\,, && \dots &&
\eeaa
Therefore, the Stokes factor for the first phase sector will be\footnote{Recall that we omit the factors $e_{\bar \imath i}$, as their role is captured by the algebra of formal variables associated with open paths (\ref{eq:q-torus-alg-oo}).}
\be
\begin{split}
	\bS_I^{(2d-4d)} 
	 \ = \ & 
	\mathbbm{1} + \sum_{k\geq 0} \sum_{i=1}^{\lfloor N/2 \rfloor} \bigg[
	X_{a_{\bar \imath i}  - k (\gamma_1^{(i)}  + \gamma_2^{(i)} + \gamma_3^{(i)})} 
	+ X_{a_{\bar \imath i}  +  \gamma_1^{(i)}  + k (\gamma_1^{(i)}  + \gamma_2^{(i)} + \gamma_3^{(i)})} \\
	&\qquad\qquad\qquad
	+ X_{a_{\bar \imath i}  + \gamma_1^{(i)} +  \gamma_2^{(i)} + k (\gamma_1^{(i)}  + \gamma_2^{(i)} + \gamma_3^{(i)})} \bigg]
	\\
	\ = \ &
	\mathbbm{1} 
	+ \sum_{i=1}^{\lfloor N/2 \rfloor} 
	\frac{
	1 + q^{\frac{1}{2} \kappa_i}  X_{\gamma_3^{(i)}}  +  X_{\gamma_2^{(i)}+\gamma_3^{(i)}}   
	}{
	1 - q^{\frac{1}{2} \kappa_i}X_{  \gamma_1^{(i)}  + \gamma_2^{(i)} + \gamma_3^{(i)}}   
	}
	X_{a_{\bar \imath i} }   \,.
	\\
\end{split}
\ee
Here, akin to~(\ref{eq:stokes-sector-I}), we used the quantum torus algebra. However, now, the intersection pairings (as read off from Figure~\ref{fig:hexagon-marked}) are given by
\beaa\label{eq:n2star-pairings}
	 \langle\gamma^{(i)}_k,\gamma^{(j)}_{k+1}\rangle & \ = \  -2 \delta_{ij} \,, \qquad k \ \in \ \{ 1,2,3 \} \,,
	\\
	 \langle a_{\bar \imath i} , \gamma_1^{(j)} \rangle &  \ = \   - \langle a_{\bar \imath i} , \gamma_2^{(j)} \rangle  \ = \   \langle a_{\bar \imath i} , \gamma_3^{(j)} \rangle  \ = \  \delta_{ij}\,.
\eeaa

Next, let us compute the Stokes factor for sector $IV$ (\emph{i.e.} phases $\vartheta_c+\pi < \vartheta\leq \vartheta_c+3\pi/2$). In this sector, the charges of 2d-4d solitons are given as follows
\beaa
	&{\cdots} && {\cdots} && {\cdots} \\
	& a_{\bar \imath i}- 2\gamma^{(i)}_1 - 2\gamma^{(i)}_2 - 2\gamma^{(i)}_3 \,,
	&& a_{\bar \imath i} - 2\gamma^{(i)}_1 -2 \gamma^{(i)}_2- \gamma^{(i)}_3 \,,
	&& a_{\bar \imath i} - 2\gamma^{(i)}_1 - \gamma^{(i)}_2- \gamma^{(i)}_3 \,, \\
	& a_{\bar \imath i} - \gamma^{(i)}_1 - \gamma^{(i)}_2- \gamma^{(i)}_3 \,,
	&& a_{\bar \imath i}- \gamma^{(i)}_1-\gamma^{(i)}_2 \,,
	&& a_{\bar \imath i}- \gamma^{(i)}_1 \,,
\eeaa
in order of increasing phase. Therefore, the Stokes factor for the fourth phase sector reads
\be
\begin{split}
	\bS_{IV}^{(2d-4d)} 
	\ = \ &
	\mathbbm{1} 
	+ \sum_{i=1}^{\lfloor N/2 \rfloor} 
	\frac{
		 q^{-\frac{1}{2} \kappa_i}  X_{-\gamma_1^{(i)}}   
		 +  X_{-\gamma_1^{(i)}-\gamma_2^{(i)}} 
		 + q^{-\frac{1}{2} \kappa_i} X_{ - \gamma_1^{(i)}  - \gamma_2^{(i)} - \gamma_3^{(i)}} 
	}{
		1 - q^{-\frac{1}{2} \kappa_i}X_{ - \gamma_1^{(i)}  - \gamma_2^{(i)} - \gamma_3^{(i)}}   
	}
	X_{a_{\bar \imath i} }   
	\\
	\ = \ &
	\mathbbm{1} 
		- \sum_{i=1}^{\lfloor N/2 \rfloor} 
		\, \frac{
		 1
		+ q^{\frac{1}{2} \kappa_i} X_{\gamma_3^{(i)}} 
		+  X_{\gamma_2^{(i)}+\gamma_3^{(i)}}   
	}{
	1 - q^{\frac{1}{2} \kappa_i}X_{  \gamma_1^{(i)}  + \gamma_2^{(i)} + \gamma_3^{(i)}}   
	}
	X_{a_{\bar \imath i} }   \,.
	\\
\end{split}
\ee
Clearly, once again we arrive at
\be
	\bS^{(2d-4d)} _{IV}\,\cdot \,\bS^{(2d-4d)} _I  \ = \ \mathbbm{1}\,.
\ee
An analogous computation shows that $\bS^{(2d-4d)}_{II}$ and $\bS^{(2d-4d)}_{III}$ also annihilate. Thus, just like in the case analyzed in Section~\ref{sec:2d-4d-macdonald}, it turns out that 2d-4d BPS solitons \emph{do not contribute} to the trace of the 2d-4d BPS spectrum generator.

The next task is to study the remaining contributions, namely the spectrum of 2d BPS particles. There are two equivalent ways to compute them (the alternative approach is mentioned at the end of this subsection). They can be computed from spectral networks, as we observed in Subsection~\ref{sec:2d-particle}. In fact, we can borrow them directly from the computation in~\cite[eq. (4.49)]{Longhi:2016wtv} (see footnote~\ref{foot:2d-particles} for more details), which implies that there is a particle for each vacuum, with charge
\be\label{eq:n2star-2d-particles}
	\gamma_{\sigma_{i}}  \ = \  \gamma_{\sigma_{\bar\imath}}  \ = \  2\gamma^{(i)}_1+2\gamma^{(i)}_2+2\gamma^{(i)}_3\,.
\ee
Note that these are pure-flavor charges from the viewpoint of the 4d theory (recall footnote~\ref{foot:flavor}).
Indeed, in this case we have a relation between these charges and flavor fugacities (analogous to~(\ref{eq:X-gamma-a-rel})),
\be
	X_{2\gamma^{(i)}_1+2\gamma^{(i)}_2+2\gamma^{(i)}_3}
	  \ = \  
	a^{2(N-2 i +1)} \,,
\ee
where again $a = e^{m/2}$, and the Higgs field has a simple pole of the form~(\ref{eq:simple-pole}).

Following again the argument of~\cite[p.126]{Gaiotto:2011tf}, we take $\omega_{a_{i  \bar \imath},\gamma_{\sigma_i},j_{\sigma_i}} = \langle\gamma_{\sigma_i},a_{i \bar  \imath}\rangle$ to compute
\be
	\omega_{a_{i \bar \imath}, \gamma_1^{(i)}+\gamma_2^{(i)}+\gamma_3^{(i)}, j_{i}}  \ = \  \omega_{ -a_{\bar \imath i},  \gamma_1^{(i)}+\gamma_2^{(i)}+\gamma_3^{(i)}, j_{i} }  \ = \  2 \,,
\ee
where we made use of equation~(\ref{eq:n2star-pairings}).
As before, we expect by symmetry
\be\label{eq:omega-vacua}
	\omega_{i, \gamma_{\sigma_i}, j_{\sigma_i}}  \ = \ -1\,, \quad \text{and}  \quad 
	\omega_{\bar \imath , \gamma_{\sigma_{\bar \imath}}, j_{\sigma_{\bar \imath}}} \ = \ 1\,.
\ee

Regarding the spin of the 2d particles, we can again run the same argument that led to~(\ref{eq:particle-spin-N-eq-2}) for the $N=2$ case. It again leads to $j^{(N=2)}_{\sigma_1} = -1$, based on the dictionary proposed in~\cite{Galakhov:2014xba}, which identifies spin with self-intersections of soliton paths. This result lifts to $N>2$ by the arguments from the Subsection~\ref{sec:spins}, where we established that the $q$-powers that count (self-)intersections get corrected by a quadratic coefficient $(N-2k+1)^2$ on the $k$-th sheet. Therefore, we are led to the following ansatz for generic $N$
\be\label{eq:spin-proposal1}
	j^{(N)}_{\sigma_k}  \ = \  (N-2k+1)^2 j^{(N=2)}_{\sigma_1}  \ = \  -(N-2k+1)^2 \,,
\ee
where $k$ is restricted to run over the first ``half'' of vacua $1\leq k\leq \lfloor (N+1)/2 \rfloor$. For the second half of vacua we expect $j^{(N)}_{\sigma_{\bar k}} = -j^{(N)}_{\sigma_{k}}$, since for $N=2$ the two spins have opposite sign (recall~(\ref{eq:particle-spin-N-eq-2})).

As mentioned above, there is an alternative way to compute 2d particles, leading to the same result. We briefly explain it here because it will be useful for quickly deriving 2d particle spectra in the case of higher-genus surfaces with one puncture (see Subsection~\ref{sec:one-puncture-hg}).\footnote{The spectral network technique invoked above can also be adapted to higher genus without any problems. However, the slick argument we are about to give is much more convenient.}
The key is to note that they are solutions to the BPS equations with identical boundary conditions at both spatial endpoints: beginning in vacuum $i$ and ending in the same vacuum. The BPS equations of the 2d field theory (in a 2d-4d system, in this case) translate geometrically into the flow equations of the spectral network.
Therefore, 2d particles are captured by closed geodesics in the foliation induced by the quadratic differential, that begin and end at the same point $z^{(i)}$, the lift of $z$ to a certain sheet $i$ labeling the vacuum where the particle lives. From Figure~\ref{fig:N2-star-hexagon-networks}, we see that closed geodesics appear at $\vartheta_c$, and they come in homology classes corresponding to $2 \gamma^{(i)}_1+2\gamma^{(i)}_2+2\gamma^{(i)}_3$. This agrees indeed with the previous statement~(\ref{eq:n2star-2d-particles}).

We conclude that all the key features encountered in the computation of the 2d-4d BPS spectrum generator of Section~\ref{sec:2d-4d-macdonald} are still present in the case of the torus with one puncture. The 2d-4d cancellation mechanism still works, and the 2d BPS particle spectrum still features particles with the same spin and charges. It follows immediately that the trace of the 2d-4d spectrum generator eventually leads once again to the difference operators acting on the 4d Schur index as obtained in~(\ref{eq:general-operator-formula}).

\subsection{Higher-genus surfaces with one puncture}\label{sec:one-puncture-hg}

The $\CN=2^*$ theory, analyzed in Section~\ref{sec:N2star}, is only the first in a series of theories whose BPS graphs cannot be brought into the form shown in Figure~\ref{fig:A1-puncture-behavior}. In fact, for the sequence of theories defined by a genus-$g$ Riemann surface with a single puncture, $C \equiv C_{g,1}$, the BPS graph exists but it has a special form. As we argue in Appendix~\ref{sec:more-punctures}, when the number of punctures is greater than 1, it is always possible to bring the BPS graph into the form shown in Figure~\ref{fig:A1-puncture-behavior} around one of the punctures. This fact relies on the possibility of performing Whitehead moves.

When there is only one puncture, the BPS graph on $C_{g,1}$ viewed as a ribbon graph has a single boundary component, which circles the puncture.
The BPS graph arises as the critical graph of a Strebel differential associated to a pair-of-pants decomposition of $C_{g,1}$, where the heights of all cylinders has been taken to zero.
According to theorems from the theory of quadratic differentials this is always possible~\cite{Strebel,Liu}. From a physics perspective, the fact that these theories are \emph{complete} in the sense of~\cite{Alim:2011ae} also guarantees the existence of the BPS graph. Then, by the fact that punctures of $C_{g,n}$ correspond to boundary components of the BPS graph, it follows that the graph has exactly one boundary component when $n=1$.
Therefore, cutting the Riemann surface along the BPS graph $\cG$, we obtain a presentation of the surface as a polygon, with opposite edges identified with opposite orientations.

A pants decomposition for $C_{g,1}$ involves $2g-1$ trinions glued together at all punctures except for one. A trinion is the Riemann surface defining the $T_2$ theory, whose BPS graph contains two branch points, as we have seen in Figure~\ref{fig:T2-BPSg}.
Therefore, there will be $V=4g-2$ branch points in the BPS graph of the theory on $C_{g,1}$.
Since each branch point has three slots for edges, and each edge must end on two branch points, there must be $E=6g-3$ number of edges. The number of punctures is by definition $F=1$. Then, if we fill in the puncture, it follows that the Ribbon graph built from the BPS graph has the correct Euler characteristic $V-E+F = 2-2g$ of a compact genus $g$ surface, which is the one it lives in.

Therefore, we can represent the BPS graph, and the Riemann surface itself, as a polygon with $12g-6$ edges glued together. 
Viewing $\cG$ as a ribbon graph, each edge contributes twice to be boundary, once with its ``left'' side, and once with its ``right'' side.
With respect to the cyclic ordering of boundary components, the two contributions for each edge of $\cG$ must come at opposite sides in the polygon, and must clearly have opposite orientations. For the same reason, each branch point must appear three times as a vertex. Indeed, $3V=2 E$ is both the number of vertices of the polygon and the number of its edges. 
For example, when $g=1$ we obtain a hexagon. Indeed, we have already seen in Figure~\ref{fig:N2-star-hexagon-networks} (in the middle frame)  that this is an alternative way of thinking about the more familiar $\CN=2^*$ BPS graph from Figure~\ref{fig:N2star-BPSg}. 

The foliation of $C_{g,1}$ induced by the quadratic differential is very simple for $\vartheta=\vartheta_c$. It consists of closed geodesics circling around the puncture, and asymptoting to the polygon boundary, \emph{i.e.} the BPS graph $\cG$.
We can infer the topology of the foliation at other phases, based on simple considerations. 
For one thing, the flow induced by quadratic differentials is always single-valued (up to orientation). Another basic observation is that topological jumps of the foliation can only occur at $\vartheta_c +  k\pi$, with $k\in \mathbb{Z}$, since we are fixing the moduli to be at the Roman locus. The third important clue is the local behavior of the foliation close to the puncture. Parametrizing the Higgs field as in~(\ref{eq:simple-pole}), the sheets of $\Sigma$ are~(\ref{eq:sheets-near-puncture}). Therefore, the differential equation for the flow (which coincides with the $\CS$-wall equation given in~(\ref{eq:S-wall})) becomes
\be
	(\partial_t , \lambda_{\bar\imath} - \lambda_i)  \ \sim \ \frac{m}{z} \frac{dz}{dt} \ \sim \ e^{\ii\vartheta} \,,
\ee
where $\sim$ denotes proportionality up to a real positive constant. The solution is 
\be
	z(t)  \ = \  z_0 \exp\((t - t_0)\, \frac{e^{\ii\vartheta}}{m} \)\,.
\ee
The generic flow line is a spiral if $\e^{\ii\vartheta} / m \notin \IR \cup \ii\IR$, it is a closed circle around the puncture when $\e^{\ii\vartheta} / m \in \ii\IR$, and it is a radial straight line if $\e^{\ii\vartheta} / m \in \IR$.

Based on these observations we conclude that the topology of the spectral network on these Riemann surfaces at the Roman locus must take qualitatively the same form as in Figure~\ref{fig:N2-star-hexagon-networks}. 
The main difference is the number of $\CS$-walls ending on the puncture, which will be $12g-6$ corresponding to one $\CS$-wall from each vertex of the polygon. 

\begin{figure}[h!]
\begin{center}
\includegraphics[width=0.45\textwidth]{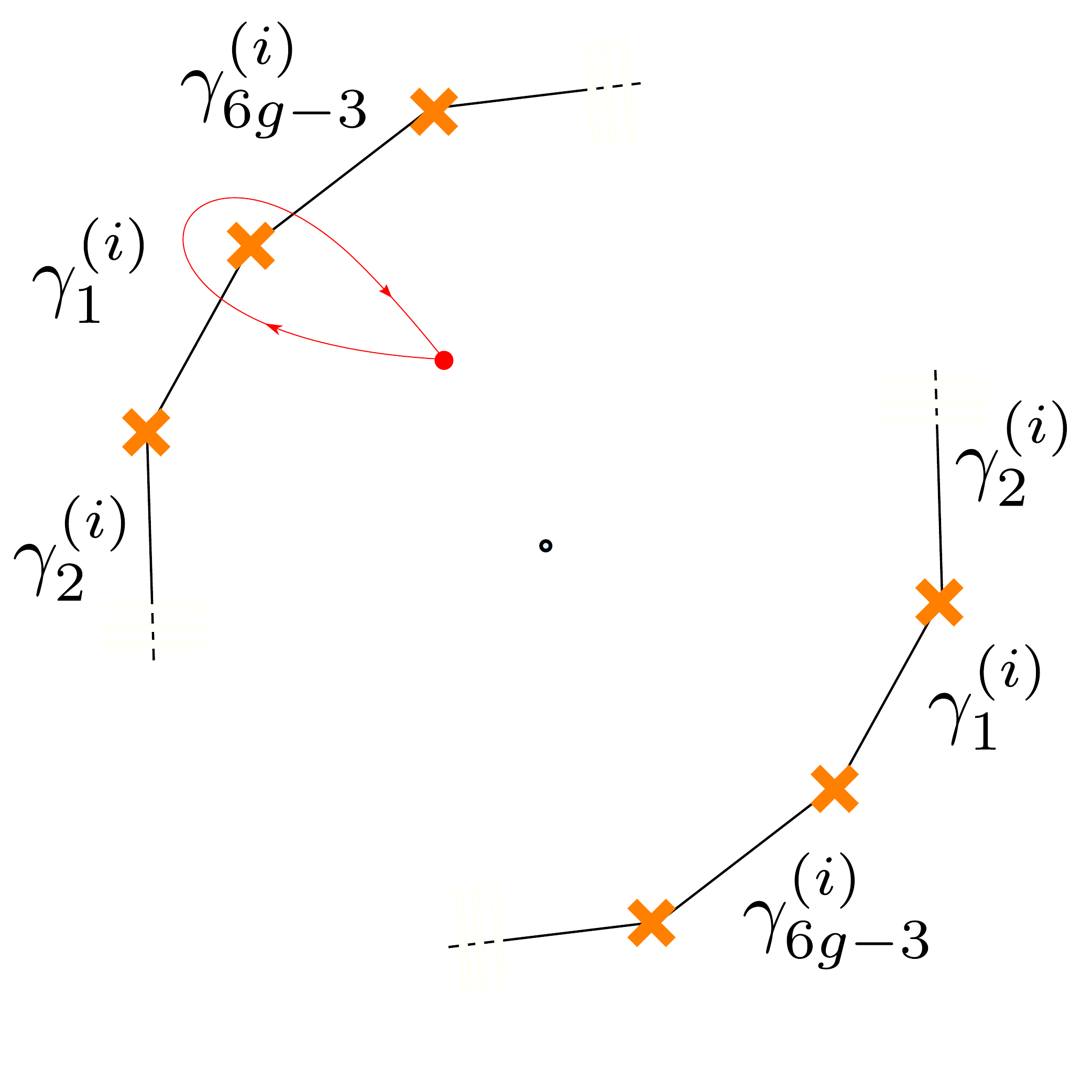}
\caption{BPS graph of the class $\CS$ theory of type $A_1$ on a genus-$g$ Riemann surface $C$ with one regular puncture. Here $C$ is presented as a $(12g-4)$-gon with a puncture in the middle, and where opposite sides are identified.}
\label{fig:polygon-marked}
\end{center}
\end{figure}

A surface defect placed near the puncture will have a spectrum of 2d-4d BPS solitons closely analogous to~(\ref{eq:n2star-2d-4d-spectrum-I}). 
The 4d charge lattice is generated by  charges $\gamma_1^{(i)}, \dots ,\gamma_{6g-3}^{(i)}$, we choose to assign labels as shown in Figure~\ref{fig:polygon-marked}.
The 2d-4d spectrum for phases in sector $I$ will be 
\beaa
	{}& a_{\bar \imath i} \,, 
	&& a_{\bar \imath i}+ \gamma^{(i)}_1 \,, 
	&& a_{\bar \imath i}+ \gamma^{(i)}_1+\gamma^{(i)}_2 \,, 
	&& \cdots  \\
	{}& a_{\bar \imath i}+ \gamma^{(i)}_1+\cdots+\gamma^{(i)}_{6g-4} \,,
	&& a_{\bar \imath i}+ \gamma^{(i)}_1+\cdots+\gamma^{(i)}_{6g-3} \,, 
	&& a_{\bar \imath i}+ 2 \gamma^{(i)}_1+\gamma^{(i)}_2+\cdots+\gamma^{(i)}_{6g-3} \,, 
	&& \cdots \,.
\eeaa
The corresponding Stokes factor is
\be
\begin{split}
	\bS_I^{(2d-4d)} 
	&  \ = \  
	\mathbbm{1} 
	+ \sum_{k\geq 0} \sum_{i=1}^{\lfloor N/2 \rfloor}  \sum_{\ell=0}^{6g-4}
	X_{a_{\bar \imath i}  + ( \gamma_0^{(i)} + \cdots + \gamma_\ell^{(i)}) +  k (\gamma_1^{(i)}  +\cdots+ \gamma_{6g-3}^{(i)})} 
	\\
	&  \ = \  
	\mathbbm{1} 
	+ \sum_{i=1}^{\lfloor N/2 \rfloor} 
	\frac{
	1 +  \sum_{\ell=1}^{6g-4} 
	q^{\frac{1}{2} \langle\langle a_{\bar \imath i} , \gamma_{1}^{(i)} + \cdots + \gamma_{\ell}^{(i)}  \rangle\rangle}     
	X_{\gamma_{1}^{(i)} + \cdots + \gamma_{\ell}^{(i)}}
	}{
	1 - q^{\frac{1}{2}  \langle\langle a_{\bar \imath i} , \gamma_{1}^{(i)} + \cdots + \gamma_{6g-3}^{(i)}  \rangle\rangle }X_{  \gamma_1^{(i)}  + \cdots + \gamma_{6g-3}^{(i)}}   
	}
	X_{a_{\bar \imath i} }   \,,
	\\
\end{split}
\ee
where $\gamma_0^{(i)} \equiv 0$ has been introduced in the first line, to write the formula more compactly.

By the same reasoning, throughout phases of sector $IV$, the 2d-4d BPS states on the surface defect will be
\beaa
	{}& 
	&& 
	&& \cdots
	&& a_{\bar \imath i} - 2 \gamma^{(i)}_{6g-3}-\gamma^{(i)}_{6g-4} \cdots  -\gamma^{(i)}_1 , &\\
	{}& a_{\bar \imath i} - \gamma^{(i)}_{6g-3} \cdots  -\gamma^{(i)}_1 \,,
	&& \cdots \,,
	&& a_{\bar \imath i}-  \gamma^{(i)}_{6g-3}- \gamma^{(i)}_{6g-4} \,,
	&& a_{\bar \imath i}- \gamma^{(i)}_{6g-3} \,, &
\eeaa
resulting in the following Stokes factor
\be
\begin{split}
	\bS_{IV}^{(2d-4d)} 
	&  \ = \  
	\mathbbm{1} 
	+ \sum_{k\geq 0} \sum_{i=1}^{\lfloor N/2 \rfloor}  \sum_{\ell=0}^{6g-4}
	X_{a_{\bar \imath i}  - ( \gamma_{6g-3}^{(i)} + \cdots + \gamma_{6g-3-\ell}^{(i)}) -  k (\gamma_1^{(i)}  +\cdots+ \gamma_{6g-3}^{(i)})} 
	\\
	&  \ = \ 
	\mathbbm{1} 
	+ \sum_{i=1}^{\lfloor N/2 \rfloor} 
	\frac{
	\sum_{\ell=0}^{6g-4} 
	q^{-\frac{1}{2} \langle\langle a_{\bar \imath i} , \gamma_{6g-3}^{(i)} + \cdots + \gamma_{6g-3-\ell}^{(i)}  \rangle\rangle}
	X_{-\gamma_{6g-3}^{(i)} - \cdots - \gamma_{6g-3-\ell}^{(i)}  }     
	}{
	1 - q^{-\frac{1}{2}  \langle\langle a_{\bar \imath i} , \gamma_{1}^{(i)} + \cdots + \gamma_{6g-3}^{(i)}  \rangle\rangle }X_{  -\gamma_1^{(i)}  - \cdots - \gamma_{6g-3}^{(i)}}   
	}
	X_{a_{\bar \imath i} }   
	\\
	&  \ = \  
	\mathbbm{1} 
	- \sum_{i=1}^{\lfloor N/2 \rfloor} 
	\frac{
	1 + \sum_{\ell'=1}^{6g-4} 
	q^{\frac{1}{2} \langle\langle a_{\bar \imath i} , \gamma_{1}^{(i)} + \cdots + \gamma_{\ell'}^{(i)}  \rangle\rangle}
	X_{\gamma_{1}^{(i)} + \cdots + \gamma_{\ell'}^{(i)}  }     
	}{
	1 - q^{\frac{1}{2}  \langle\langle a_{\bar \imath i} , \gamma_{1}^{(i)} + \cdots + \gamma_{6g-3}^{(i)}  \rangle\rangle }X_{  \gamma_1^{(i)}  + \cdots + \gamma_{6g-3}^{(i)}}   
	}
	X_{a_{\bar \imath i} }   \,,
\end{split}
\ee
where we used that $X_{\gamma_0}\equiv 1$ and substituted $\ell' = 6g-4-\ell$. The Stokes factor $\bS_{IV}$ is manifestly the inverse of $\bS_I$, and therefore the 2d-4d soliton cancellation mechanism again works as it did previously.

The description of the 2d particle spectrum is identical to the $g=1$ case, which we treated in Subsection~\ref{sec:N2star}. This should be expected on physical grounds as well, since the 2d particle spectrum is independent of the 4d theory, it only depends on the 2d theory (hence on $N$). We conclude once again that all the key features encountered in the computation of the 2d-4d BPS spectrum generator of Section~\ref{sec:2d-4d-macdonald} are still present in the case of genus-$g$ surfaces with one puncture. The 2d-4d cancellation mechanism still works, and the 2d BPS particle spectrum still features particles with the same spin and charges. It follows immediately that the trace of the 2d-4d spectrum generator eventually leads once again to difference operators acting on the 4d Schur index given in~(\ref{eq:general-operator-formula}).

\section{Including irregular punctures}\label{sec:irregular}

Since our derivation of the 2d-4d BPS monodromy only requires the presence of a single regular puncture and knowledge of the BPS graph, we can easily extend our discussion to class $\CS$ theories of type $A_1$ engineered by a sphere with one regular and one irregular puncture. 
For instance, one class of such theories is known as the Argyres-Douglas $(A_1, D_M)$ series~\cite{Argyres:1995jj,Argyres:1995xn}, and its class $\CS$ description was studied in~\cite{Bonelli:2011aa,Xie:2012hs,Wang:2015mra}.

The quadratic differential that parametrizes the Coulomb branch has the following form
\be
	\phi_2  \ = \  \frac{P_M(z)}{z^2}\, \mathrm{d} z^2\,,
\ee
where
\be
	P_M(z)  \ = \  z^M + u_{1} z^{M-1} + \cdots + u_{M-1} + m^2
\ee
is a polynomial of degree $M$ whose coefficients include $M-1$ complex Coulomb moduli $u_i$, and a UV mass $m$.\footnote{Here, we fixed translation invariance by placing the regular puncture (corresponding to the double pole) at $z=0$, and therefore $u_{M-1}$ cannot be eliminated by a shift in $z$.}
There is a regular puncture at $z=0$ and an irregular one at $z=\infty$. We shall denote this particular type of irregular puncture at infinity by $I_{2,M}$ following the nomenclature in~\cite{Xie:2012hs,Xie:2013jc}.

\begin{figure}
\begin{center}
\includegraphics[width=0.22\textwidth]{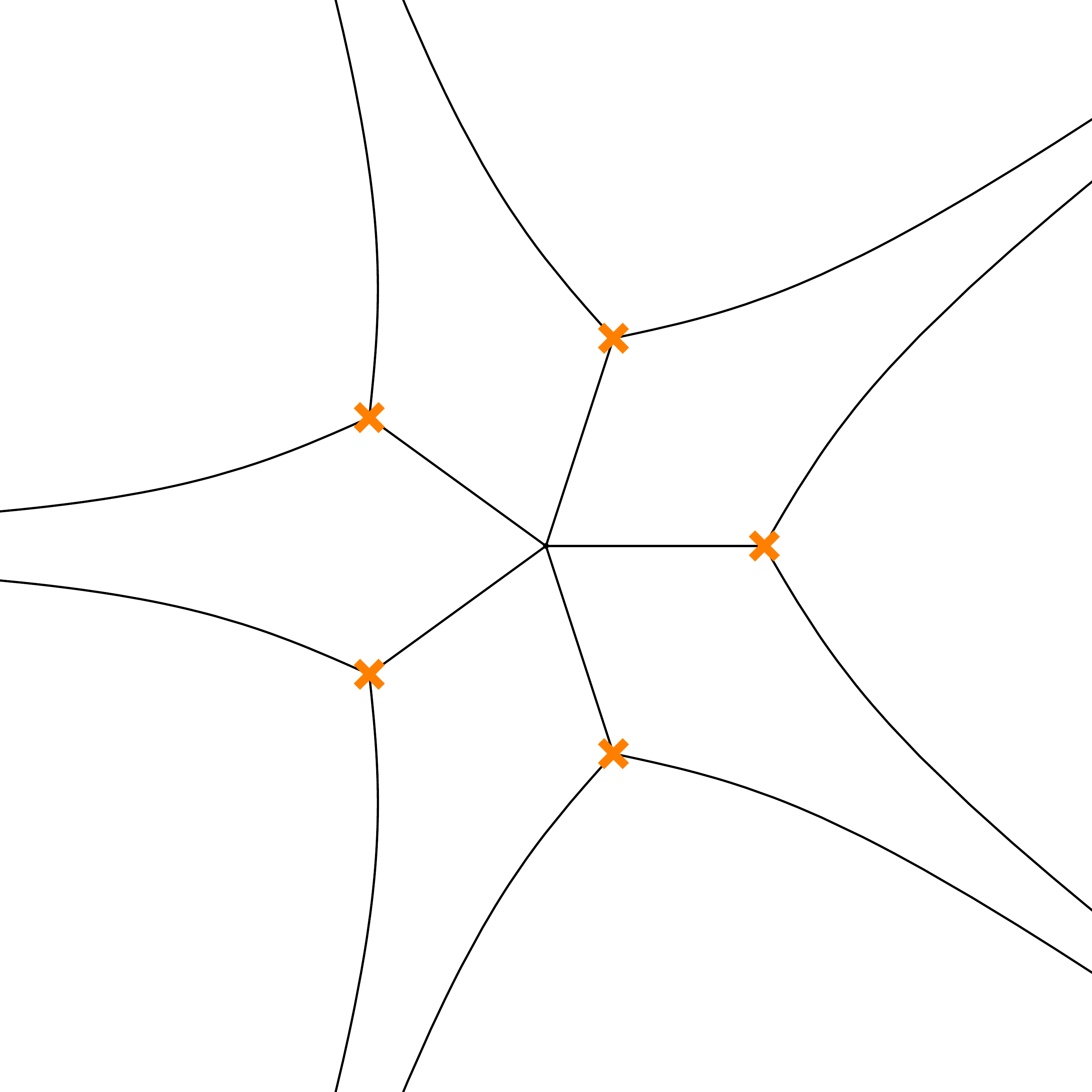}\hfill
\includegraphics[width=0.22\textwidth]{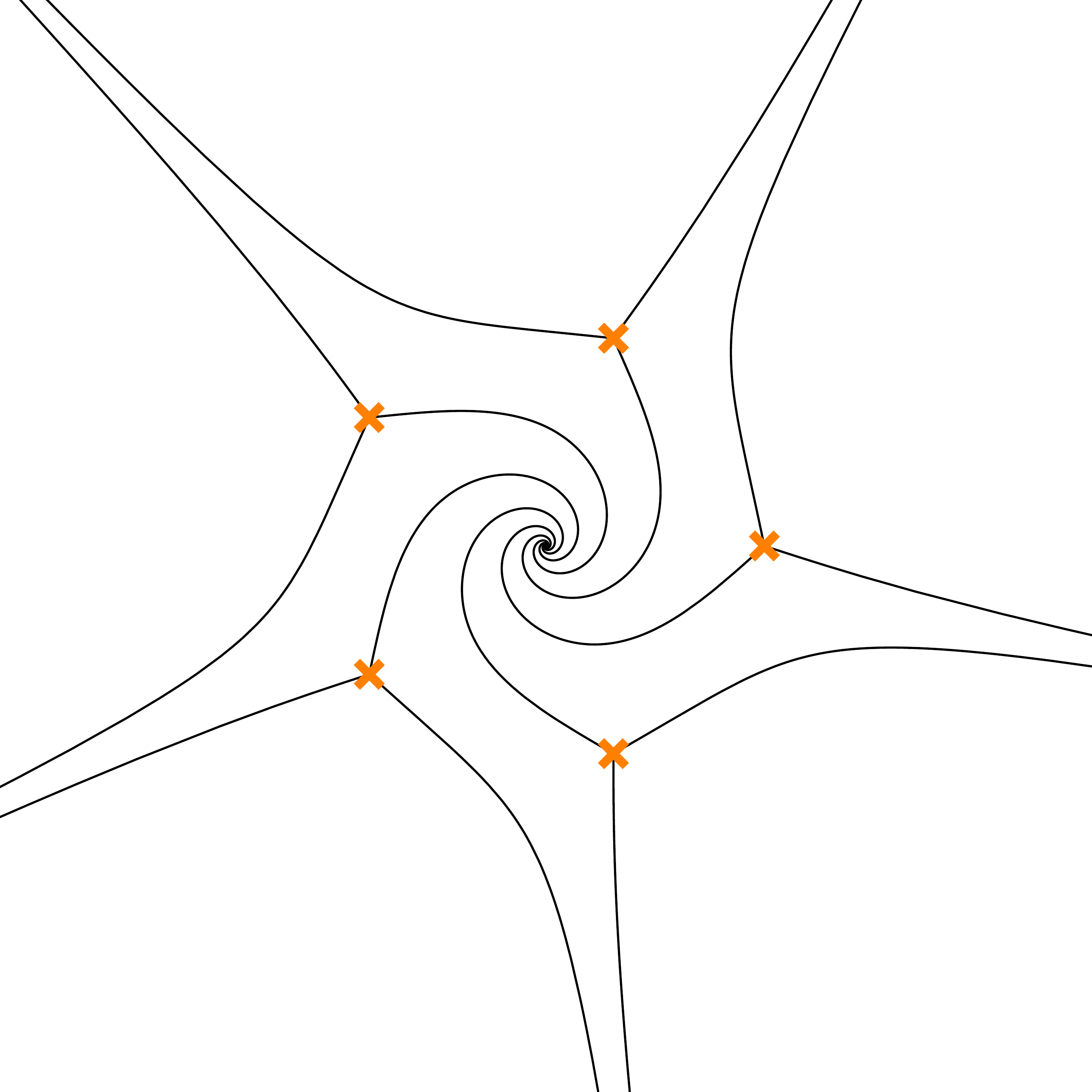}\hfill
\includegraphics[width=0.22\textwidth]{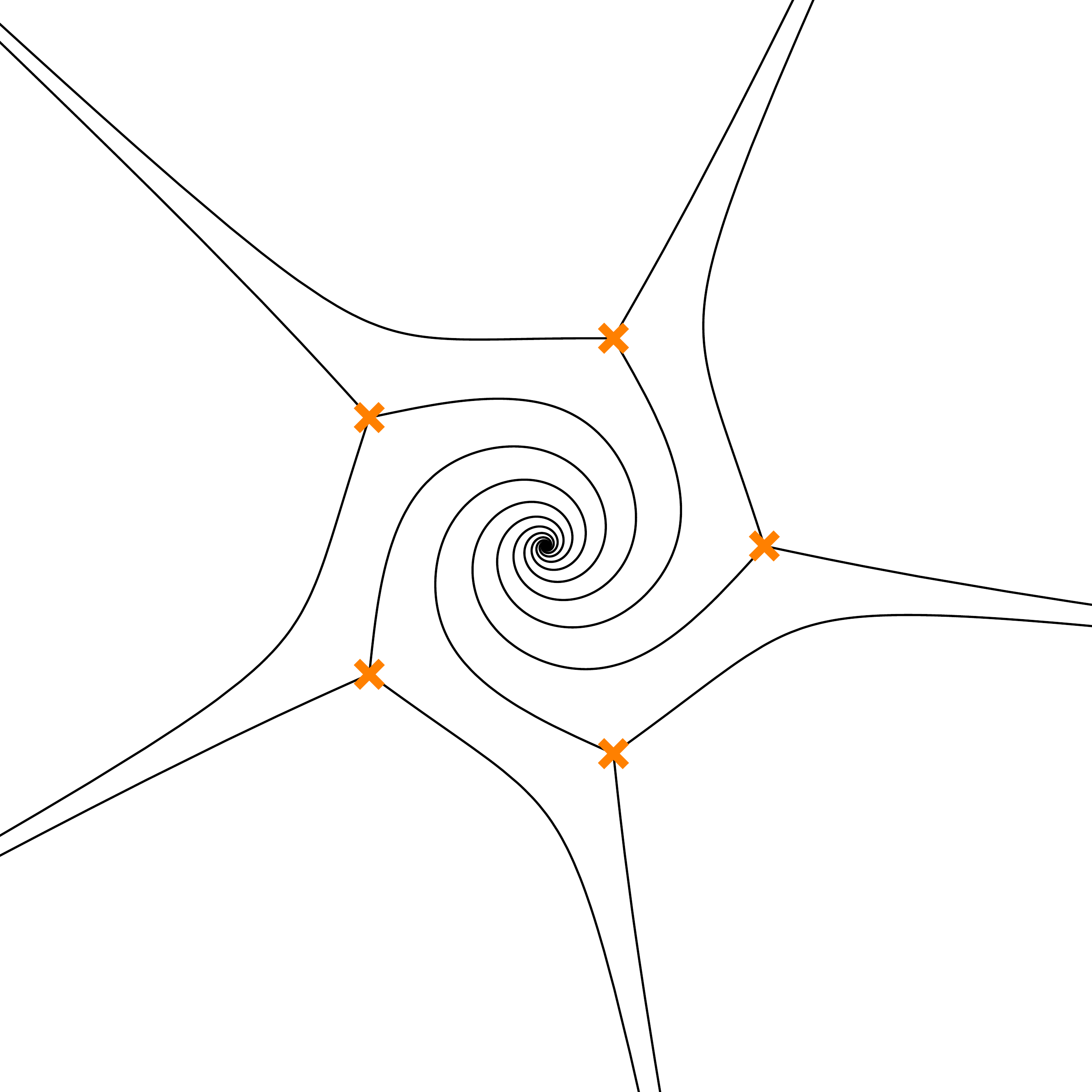}\hfill
\includegraphics[width=0.22\textwidth]{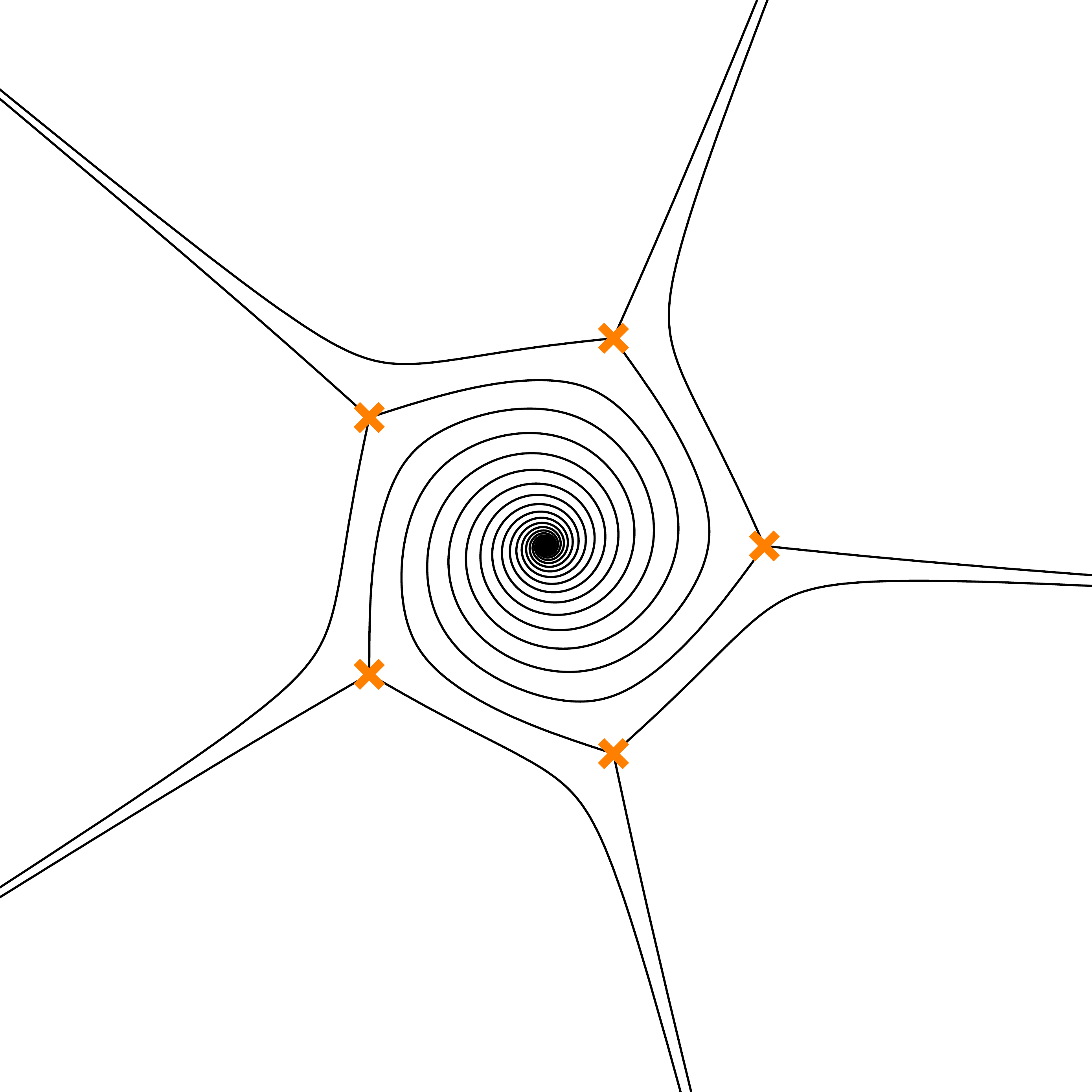}\\[5pt]
\includegraphics[width=0.22\textwidth]{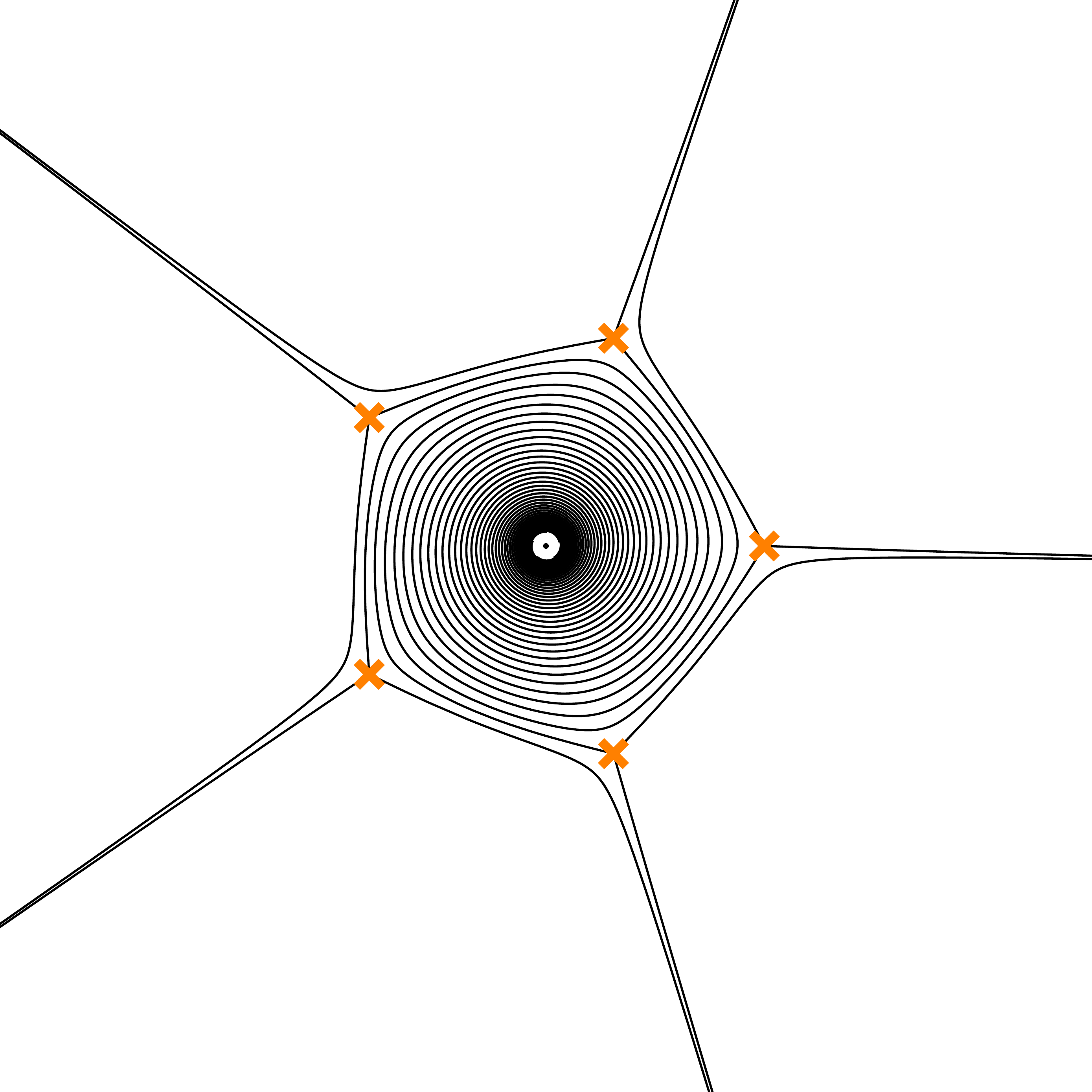}
\hfill\includegraphics[width=0.06\textwidth]{figures/dots.pdf}\hfill
\includegraphics[width=0.22\textwidth]{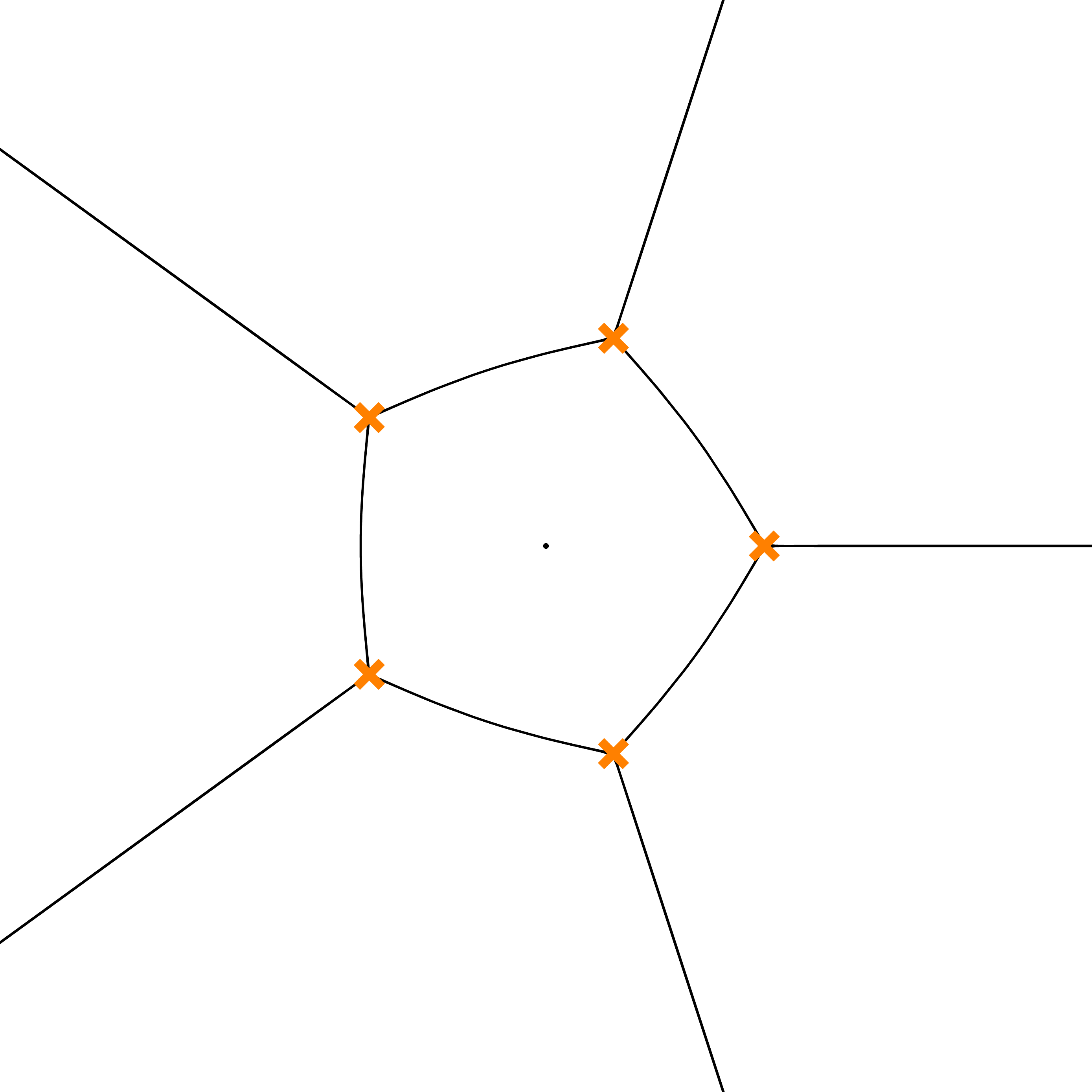}
\hfill\includegraphics[width=0.06\textwidth]{figures/dots.pdf}\hfill
\includegraphics[width=0.22\textwidth]{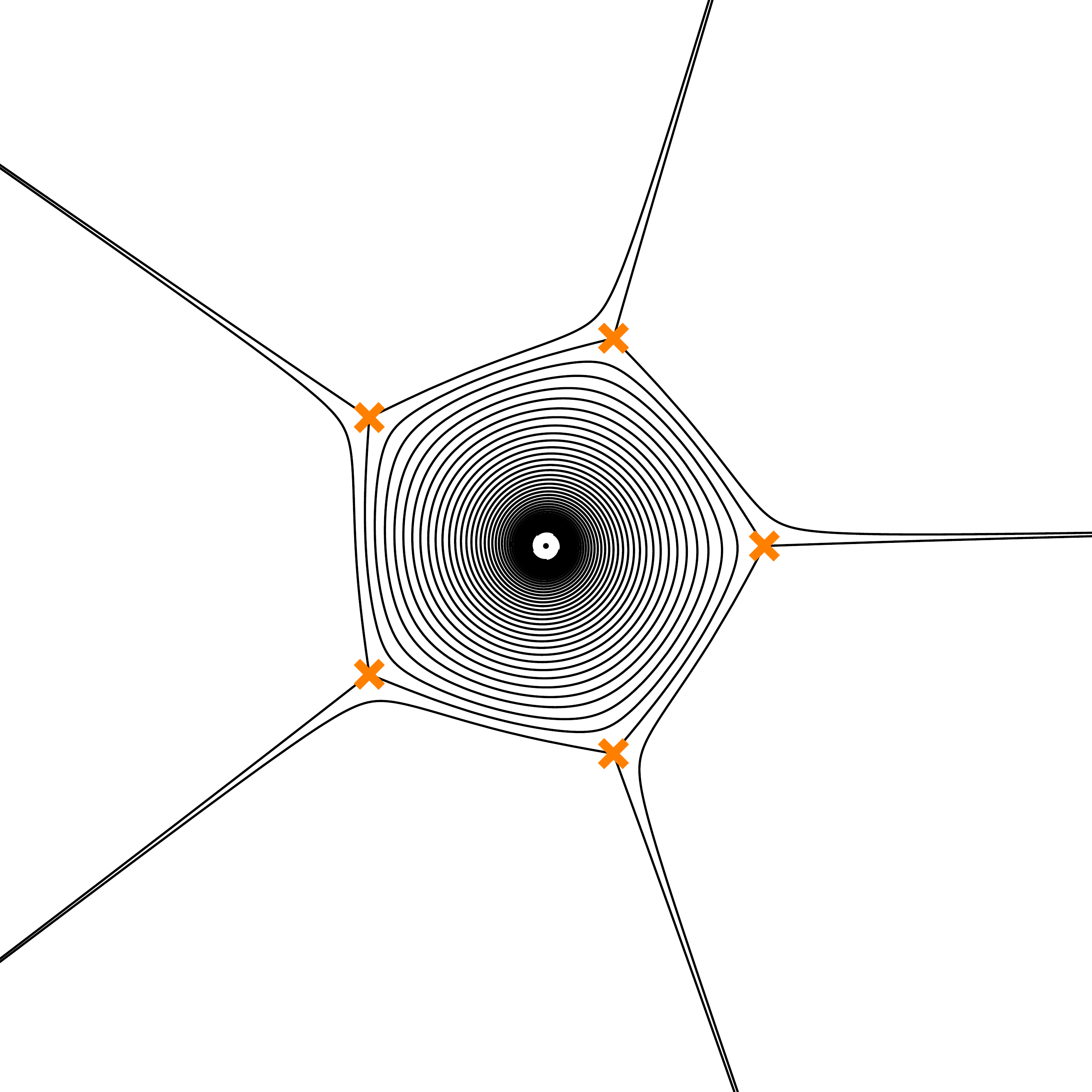}\\[5pt]
\includegraphics[width=0.22\textwidth]{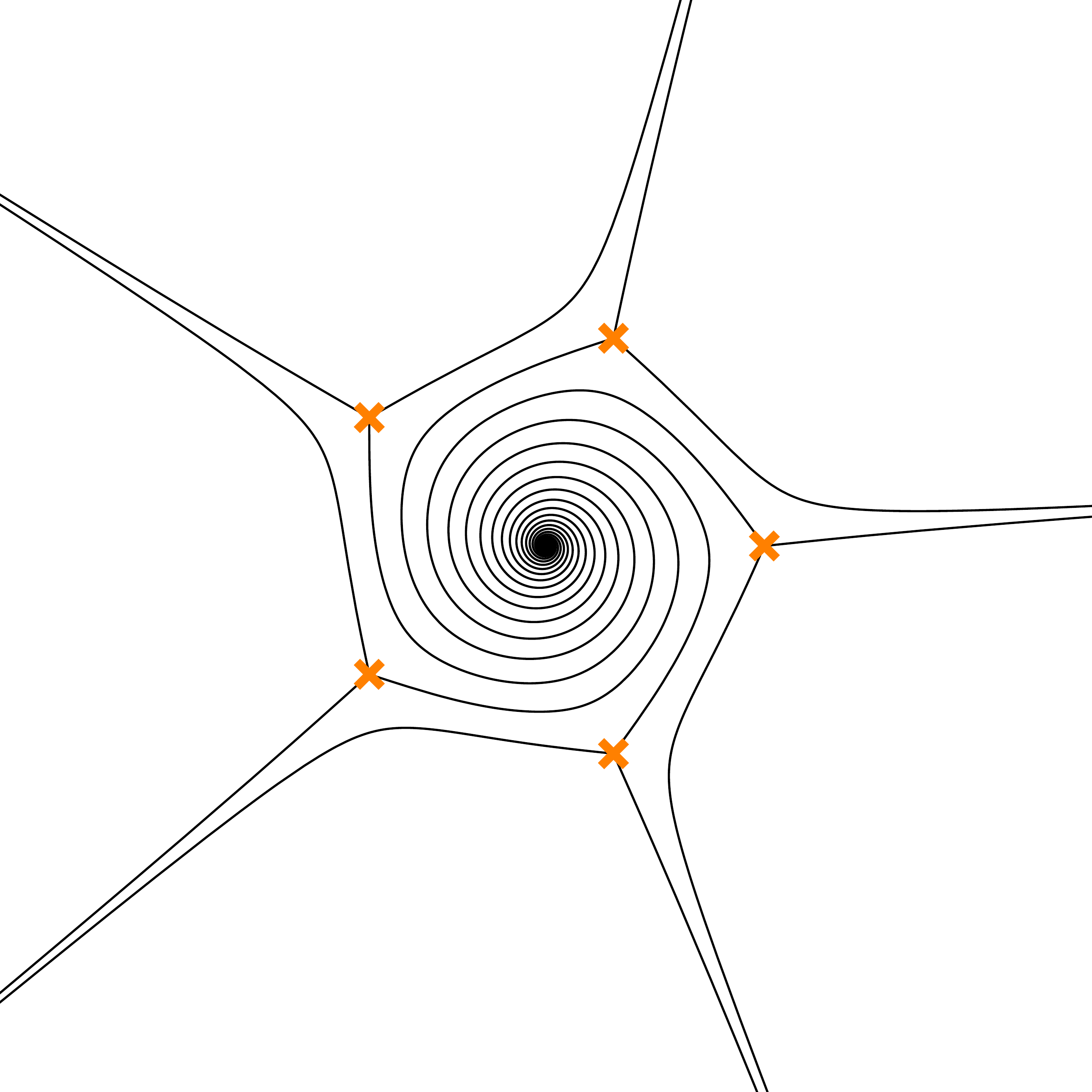}\hfill
\includegraphics[width=0.22\textwidth]{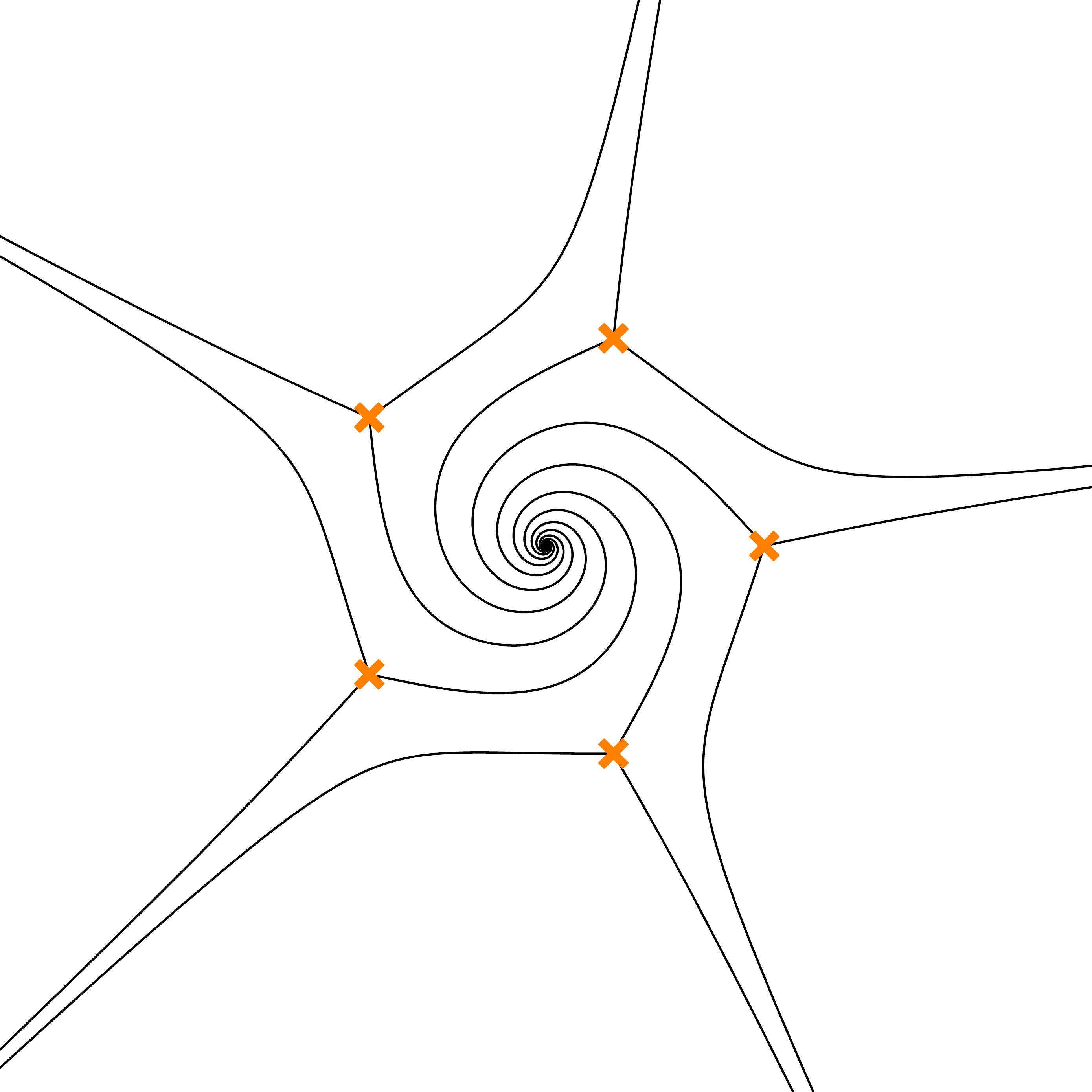}\hfill
\includegraphics[width=0.22\textwidth]{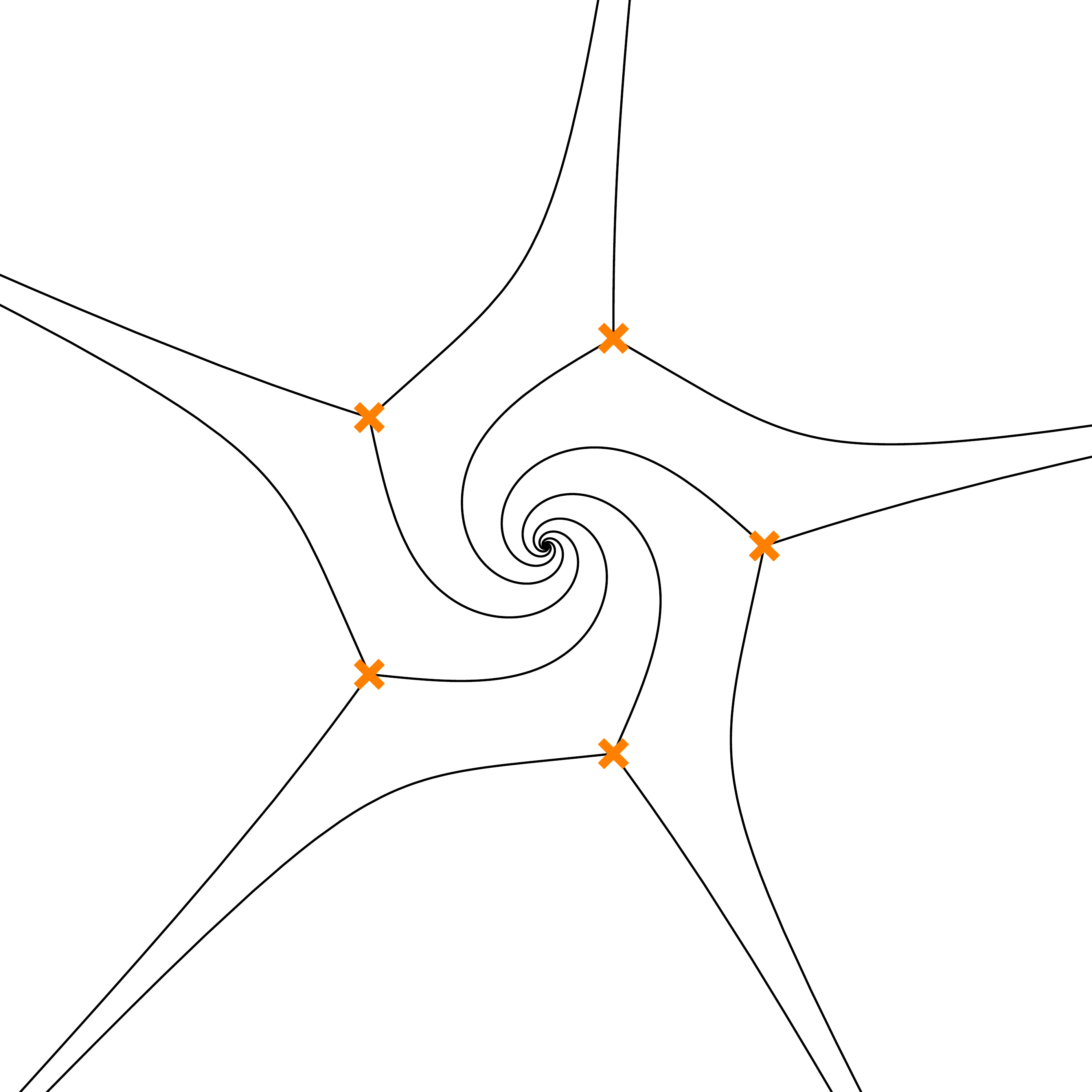}\hfill
\includegraphics[width=0.22\textwidth]{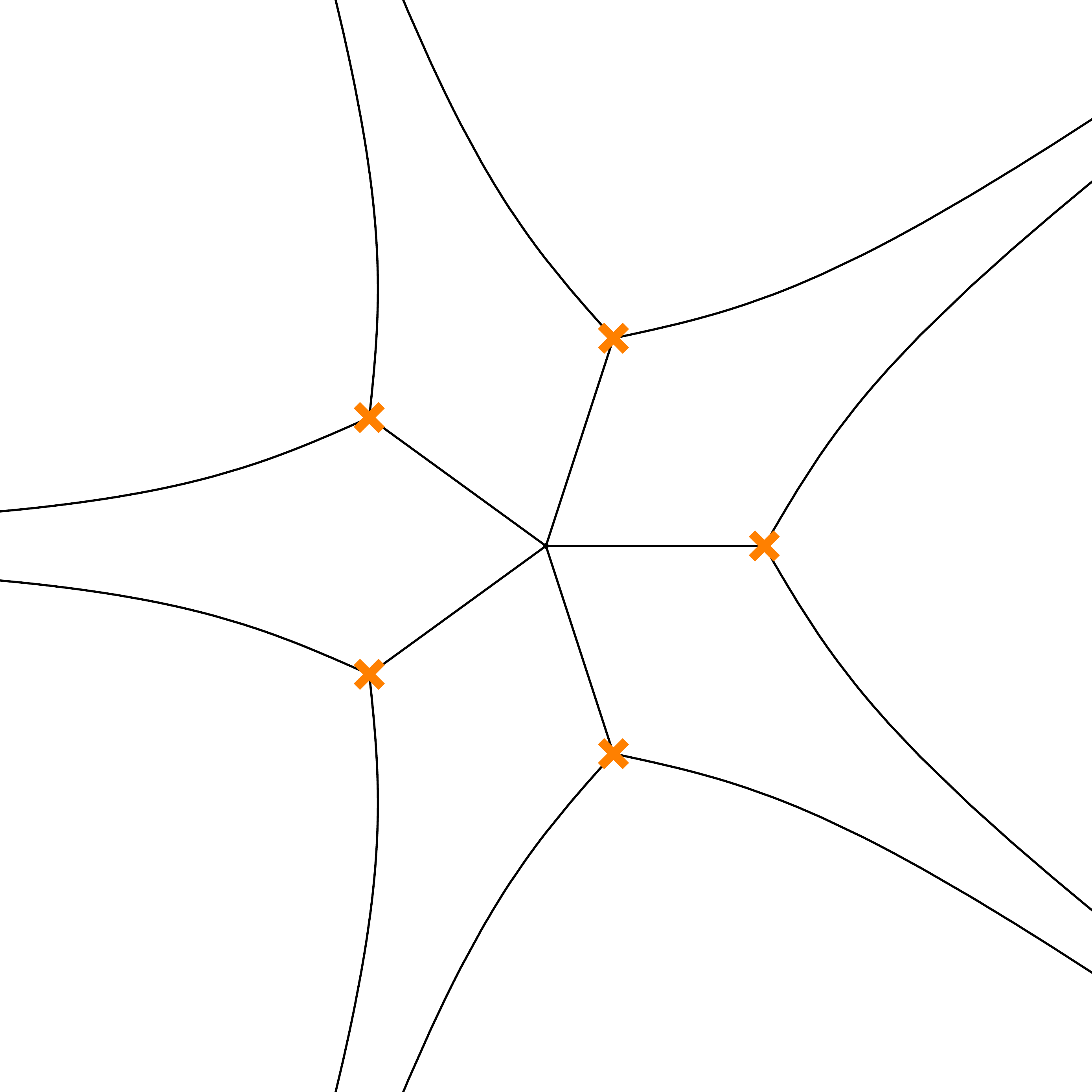}\\
\end{center}
\caption{Sequence of spectral networks of the $(A_1,D_5)$ Argyres-Douglas theory.}
\label{fig:A1-DN-network}
\end{figure}

For any $M\geq 3$, we can choose a point on the Roman locus by taking 
\be
	P_M  \ = \  \prod_{k=0}^{M-1} \(z - \zeta^k\) \,,
\ee
with $\zeta = e^{2\pi \ii k / M}$ the $M$-th root of unity. Since $m^2=-1$, the mass is purely imaginary, and the BPS graph appears at $\vartheta_c = \pi/2$. The evolution of the spectral network for various phases within sectors $I$ and $II$ is shown in Figure~\ref{fig:A1-DN-network} for the example with $M=5$.

\begin{figure}[h!]
\begin{center}
\includegraphics[width=0.45\textwidth]{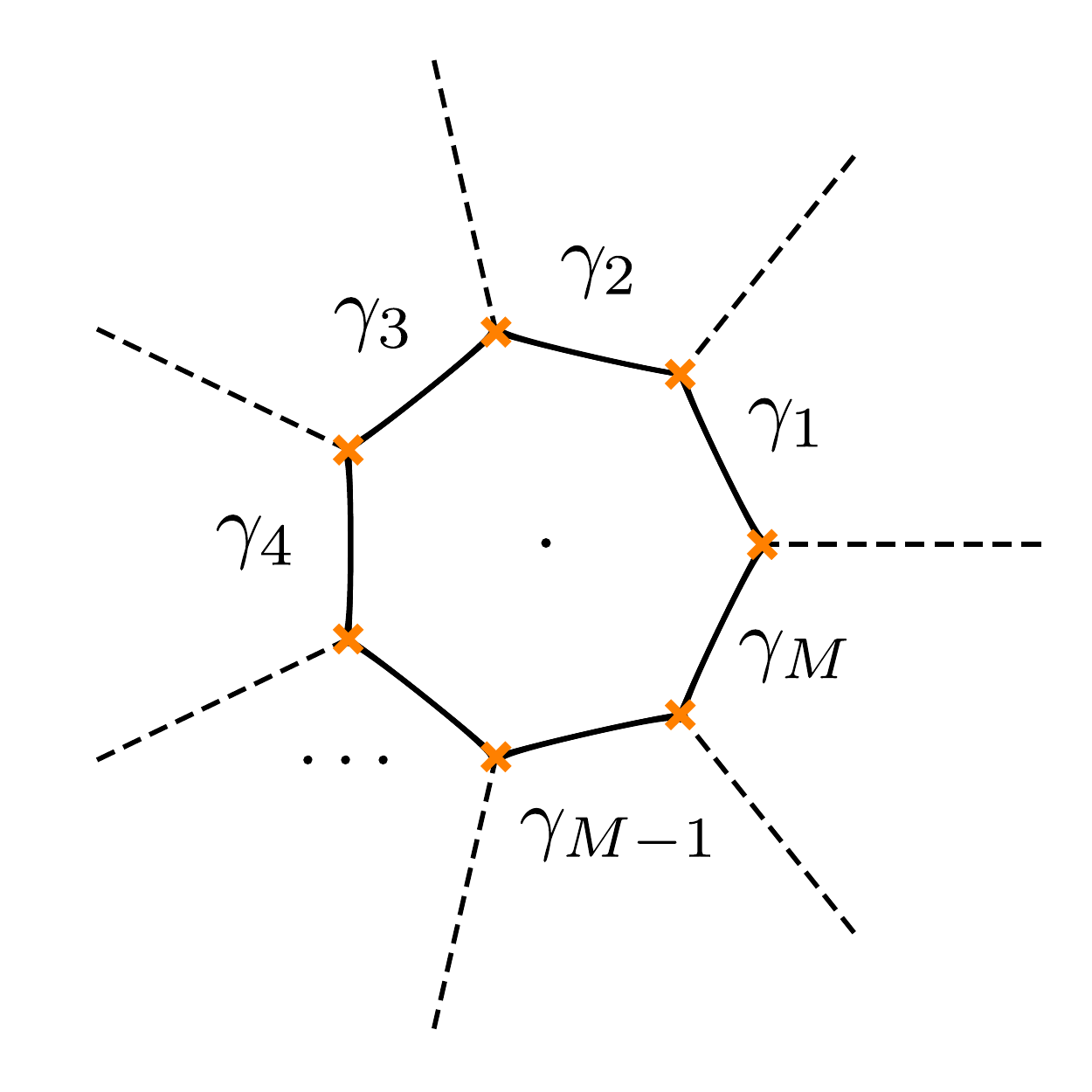}
\caption{BPS graph of the $(A_1,D_7)$ Argyres-Douglas theory.}
\label{fig:A1-DM-graph}
\end{center}
\end{figure}

The BPS graph of this class of theories is shown in Figure~\ref{fig:A1-DM-graph} for $M=7$.
It is a polygon made of $M$ double-walls. Each edge corresponds to a generator of the charge lattice, whose dimension is $2r+f = M$, where $f$ the rank of the flavor lattice, while $r$ is the complex dimension of the Coulomb branch.\footnote{There are in addition $M$ single (or ``one-way'') $\CS$-walls that emanate from branch points, but end up on the puncture at infinity. They are denoted by dashed lines. By definition, these are not part of the BPS graph.}
Let $\gamma_1, \ldots , \gamma_M$ be the homology cycles corresponding to lifts of the double walls, ordered counterclockwise. When $M$ is odd, the theory has an $SU(2)$ flavor symmetry, and hence $f=1$ and $r=(M-1)/2$. The homology class corresponding to the flavor charge is simply $\gamma_f = \sum_{i=1}^M\gamma_i$.
For $M$ even, the theory has an $SU(2)\times U(1)$ flavor symmetry (enhanced to $SU(3)$ for M=4), and thus $f=2$ and $r=(M-2)/2$. The homology class corresponding to the $SU(2)$ flavor charge is again $\gamma_f^{(SU(2))} = \sum_{i=1}^M\gamma_i $, while the $U(1)$ generator can be taken to be $\gamma_f^{(U(1))} = \sum_{i \ \text{odd}}\gamma_i$.

The theory has two types of surface defects: a ``regular'' defect parameterized  by the region near the regular puncture (inside the chamber delimited by double-walls), and an ``irregular'' one corresponding to the region near the irregular puncture (outside the chamber delimited by double-walls). The irregular defect corresponds to the defect considered in~\cite{Cordova:2017ohl, Cordova:2017mhb}; its BPS spectrum can be studied by a straightforward extension of the techniques adopted in the present paper, and it is relevant for the connection with chiral algebras~\cite{Beem:2013sza}. We will instead focus on the regular defect, since our general techniques can be directly applied for that case. The description of the BPS spectrum is in fact very similar to that of higher-genus surfaces studied above.

For example, the 2d-4d spectrum for phases in sector $I$ are
\beaa
	{} &a_{\bar \imath i} \,,
	&& a_{\bar \imath i}+ \gamma^{(i)}_1 \,,
	&& a_{\bar \imath i}+ \gamma^{(i)}_1+\gamma^{(i)}_2 \,, 
	&&  \quad\cdots\quad \\
	{} &a_{\bar \imath i}+ \gamma^{(i)}_1+\cdots+\gamma^{(i)}_{M-1} \,,
	&&a_{\bar \imath i}+ \gamma^{(i)}_1+\cdots+\gamma^{(i)}_{M} \,, 
	&& a_{\bar \imath i}+ 2 \gamma^{(i)}_1+\gamma^{(i)}_2+\cdots+\gamma^{(i)}_{M} \,,
	&& \quad\cdots\quad \,.
\eeaa
Therefore, by a computation essentially identical to the one for higher-genus surfaces with one puncture, the  corresponding Stokes factor is
\be
\begin{split}
	\bS_I^{(2d-4d)} 
	&  \ = \  
	\mathbbm{1} 
	+ \sum_{i=1}^{\lfloor N/2 \rfloor} 
	\frac{
	1 +  \sum_{\ell=1}^{M-1} 
	q^{\frac{1}{2} \langle\langle a_{\bar\imath i} , \gamma_{1}^{(i)} + \cdots + \gamma_{\ell}^{(i)}  \rangle\rangle}     
	X_{\gamma_{1}^{(i)} + \cdots + \gamma_{\ell}^{(i)}}
	}{
	1 - q^{\frac{1}{2}  \langle\langle a_{\bar\imath i} , \gamma_{1}^{(i)} + \cdots + \gamma_{M}^{(i)}  \rangle\rangle }X_{  \gamma_1^{(i)}  + \cdots + \gamma_{M}^{(i)}}   
	}
	X_{a_{\bar \imath i} }   \,.
	\\
\end{split}
\ee
Similarly, throughout phases of sector $IV$, the 2d-4d BPS states on the surface defect will be
\beaa
	{} &  
	&& \cdots
	&& a_{\bar \imath i} - 2 \gamma^{(i)}_{M}-\gamma^{(i)}_{M-1} \cdots  -\gamma^{(i)}_1 \,,
	&& a_{\bar \imath i} - \gamma^{(i)}_{M} \cdots  -\gamma^{(i)}_1\,, \quad \cdots \\
	{} & 
	&& \cdots \quad
	&& a_{\bar \imath i} -  \gamma^{(i)}_{M}- \gamma^{(i)}_{M-1}\,, 
	&& a_{\bar \imath i}- \gamma^{(i)}_{M} \,, &
\eeaa
resulting in the following Stokes factor
\be
\begin{split}
	\bS_{IV}^{(2d-4d)} 
	&  \ = \ 
	\mathbbm{1} 
	- \sum_{i=1}^{\lfloor N/2 \rfloor} 
	\frac{
	1 + \sum_{\ell'=1}^{M-1} 
	q^{\frac{1}{2} \langle\langle a_{\bar\imath i} , \gamma_{1}^{(i)} + \cdots + \gamma_{\ell'}^{(i)}  \rangle\rangle}
	X_{\gamma_{1}^{(i)} + \cdots + \gamma_{\ell'}^{(i)}  }     
	}{
	1 - q^{\frac{1}{2}  \langle\langle a_{\bar\imath i} , \gamma_{1}^{(i)} + \cdots + \gamma_{M}^{(i)}  \rangle\rangle }X_{  \gamma_1^{(i)}  + \cdots + \gamma_{M}^{(i)}}   
	}
	X_{a_{\bar\imath i} }   \,.
\end{split}
\ee
The Stokes factor $\bS_{IV}$ is manifestly the inverse of $\bS_I$, and therefore the 2d-4d soliton cancellation mechanism again works as before.

The description of the 2d particle spectrum is identical to the case of higher-genus Riemann surfaces with a single regular puncture.\footnote{This would differ however for the case of an ``irregular defect''. In fact, while the 2d-4d BPS spectrum and 4d BPS spectrum are related by invariance of the 4d and 2d-4d BPS spectrum generators, this constraint does not apply \emph{a priori} to the spectrum of 2d particles, leaving open the possibility for interesting 2d wall-crossing behavior from the ``irregular'' regime to the ``regular'' one. We leave this for future investigation.} We conclude once again that all the key features encountered in the computation of the 2d-4d BPS spectrum generator of Section~\ref{sec:2d-4d-macdonald} are still present in the case of a Riemann surface with one regular and one irregular puncture. The 2d-4d cancellation mechanism still works, and the 2d BPS particle spectrum still features particles with the same spin and charges.\footnote{One small difference from the case of higher-genus surfaces with one regular puncture, is that the flavor charge in that case was \emph{twice} the sum of all double-walls. This was due to the global properties of the embedding of the BPS graph. In the case of $(A_1,D_M)$ Argyres-Douglas theories each edge contributes once to the flavor charge associated with the $SU(2)$ regular puncture.} 
It follows immediately that the trace of the 2d-4d spectrum generator eventually once again leads to the difference operators acting on the 4d Schur index that we obtained in~\eqref{eq:general-operator-formula}. The relevant fugacity for the action of these operators is that of the $SU(2)$-flavor symmetry associated with the regular puncture.

\section{``Bootstrapping'' the IR formula}\label{sec:bootstrap}

We now switch gears and turn toward ``bootstrapping'' the IR formula for class $\CS$ theories of rank one (including type $I_{2,M}$ irregular punctures) by using the IR surface defects introduced and computed in the above sections. 

Let us first recall some of the key results of our (IR) surface defect. We stress (and shall repeatedly do so throughout) that these properties were derived \emph{solely} from an IR perspective.\footnote{Having said that, and given the fact that we obtain the same surface defects as in~\cite{Gaiotto:2012xa}, these properties were observed in this reference as well. In particular, they were used to ``bootstrap'' the (full) $\CN=2$ superconformal index. Notice however, that there is a subtle difference in that our ``IR surface defects'' can be defined for \emph{arbitrary} (not necessarily conformal) 4d $\CN=2$ theories, while in the latter reference, it is (with some caveats) assumed that we are dealing with a superconformal field theory.} To emphasize that quantities/properties arise from IR formalism as opposed to the superconformal index, we shall in some instances add the specifier ``IR''. We have shown that the following properties hold for rank-one class $\CS$ theories associated with a genus-$g$ Riemann surface with $n$ regular punctures and possibly one irregular puncture of type $I_{2,M}$ (the latter case was discussed in Section~\ref{sec:irregular}).  We believe (and know from the superconformal index perspective) that analogous arguments to the ones used in the previous sections will lead to similar properties for other types of irregular punctures as well as at higher rank; we will defer a treatment for those cases to the future.
\begin{enumerate}
\item[(i)] 
The IR Schur index only depends on the topological data of the Riemann surface, \emph{e.g.} for rank-one class $\CS$ theories with only regular punctures, it solely depends on the genus $g$ and the number of punctures, $n$. The reason is that this is the only data required to define the 4d BPS spectrum generator. This became clear from its construction in terms of BPS graphs~\cite{Longhi:2016wtv}. In fact, regular punctures are interchangeable and indistinguishable; this property can be interpreted as S-duality of the 4d UV theory~\cite{Gaiotto:2009we}. The action of the latter on BPS graphs was studied in~\cite{Gang:2017ojg}, and was shown to act by conjugation on the spectrum generator, which leaves its trace (and thus the IR Schur index) invariant.
Finally, the index is invariant under inversion of the $SU(2)$ fugacities assigned to a given puncture $a \to a^{-1}$. 
\item[(ii)] The IR surface defects act locally on the punctures of the UV curve $C_{g,n}$, with $n>0$. We argued that the IR surface defects act ``locally'' and independently of the remainder of the theory on a given puncture. The argument initially relies on the fact that we have two or more punctures (see Section~\ref{sec:BPS-spectrum} and Appendix~\ref{sec:more-punctures}), and in Section~\ref{sec:one-puncture} we treat the single-punctured case separately, and further extend it to the case with a single irregular puncture of type $I_{2,M}$ in~Section~\ref{sec:irregular}. 
\item[(iii)] In terms of the IR Schur index,~$\cI_{g,n} \( a_1, \ldots, a_n \)$ -- \emph{i.e.} upon taking the trace of the quantum monodromy operator -- of an arbitrary rank-one class $\CS$ theory associated to the UV curve $C_{g,n}$ of genus $g$ and with $n>0$ (regular) punctures, the IR surface defects are shown to be given in terms of a particular set of difference operators. Explicitly, we have shown that on general grounds they act as follows (see equation~\eqref{eq:general-operator-formula} and Section~\ref{sec:one-puncture})
\be\label{IRSDs}
	\mathfrak{S}_{v,a_j}^{IR} \circ \ \cI_{g,n} \( a_1, \ldots a_n \) \ = \sum_{k=0}^{v}
	q^{-\frac{1}{2}(v-2k)^2}  a_j^{-2(v-2k)} \CI_{g,n}\( a_1, \ldots, a_j q^{\frac{1}{2} \(v-2k\)} , \ldots, a_n\)\,.
\ee
We remark that these operators are related to the Macdonald operators, by taking the ``degeneration'' limit $t=q$, and upon conjugation by the square root of the Schur index of a free $SU(2)$ vector multiplet $\cI_{V}(a)$, \emph{i.e.}, schematically (up to some overall factors and some redefinitions)~\cite{Gaiotto:2012xa,Alday:2013kda}
\be\label{SvsD}
	\mathfrak{S}^{IR}_{v=1,a} \ \sim \ \CI_{V}^{-1/2}(q;a) D_{1,a}^{(2)} \CI_{V}^{1/2}(q;a)\big|_{t\to q} \,,
\ee
where $D_{1,a}^{(2)}$ is the Macdonald $q$-difference operators associated to the $A_1$ root system acting on the fugacity $a$~\cite{MR1354144}. We recall that the set of eigenfunctions of the Macdonald operators, $D_{v}^{(2)}$, is given by the (symmetric) Macdonald polynomials, which reduce to the Schur polynomials in~\eqref{Schurpoly} upon taking the ``degeneration'' limit $t\to q$.

\item[(iv)] The IR Schur index for the $T_{2}$-theory -- the class $\CS$ theory associated to the three-punctured sphere $C_{0,3}$, or equivalently the free bifundamental hypermultiplet -- is given by (see Section~\ref{sec:T2} and in particular equation~\eqref{eq:T2-4d-index})\footnote{Notice, that the form in equation~\eqref{Igs} can be obtained from the one in~\eqref{eq:T2-4d-index} by comparing poles~\cite{Gadde:2011uv}.}
\be\label{eqn:I03}
	\cI_{0,3} (a_1,a_2,a_3) \ = \ \sum_{\lambda=1}^{\infty} \( \frac{ \dim_{q} \lambda}{\left( q^{2} ; q\right)_{\infty}} \)^{2} \prod_{j=1}^{3} \Psi_{\lambda}\left( q;a_i \) \,,
\ee
where $\Psi_{\lambda}$ is given as follows
\be\label{Psi}
	\Psi_{\lambda} \(a ;q \) \ = \  \frac{(-1)^{\lambda+1}}{(1-q) \dim_{q} \lambda} \frac{\chi_{\lambda} (a)}{ \( q a^{2};q \)_{\infty}\left( q a^{-2};q \)_{\infty}}\,,
\ee
where $\chi_\lambda$ are the Schur polynomials, $\dim_q \lambda$ the quantum dimension and $(a; q)_\infty$ the Pochhammer symbols. We refer to Appendix~\ref{app:Schur} for relevant definitions. Notice that the $a$-dependent piece in the denominator of~\eqref{Psi} is proportional to the square-root of the Schur index of a free $SU(2)$ vector multiplet, $\cI_{V}^{1/2}(a)$. This is of course closely related to the conjugation of the difference operators in~\eqref{SvsD}.
\item[(v)] Similarly, the Schur index for the $\cN=2$ SQCD theory -- the class $\CS$ theory associated to the sphere with four punctures $C_{0,4}$ -- can be easily computed from the IR formula (see~\cite{Cordova:2016uwk} as well as the details in Section~\ref{sec:examples}), and shown to be given by
\be\label{eqn:I11}
	\cI_{0,4} (q) \ = \ \sum_{\lambda=1}^{\infty} \( \frac{ \mathrm{dim}_{q} \lambda}{\left( q^{2} ; q\right)_{\infty}} \)^{2} \prod_{j=1}^{4} \Psi_{\lambda}\left( q;a_i \) \,,
\ee
with $\Psi_{\lambda}$ given in~\eqref{Psi}.
\end{enumerate}

Let us now outline our strategy: We employ our results (i)~--~(v), which we explicitly derived from the IR formalism, of the IR surface defects and Schur indices, and use them to ``bootstrap'' the form of the resulting IR indices in full generality for rank-one theories of class $\CS$ associated to the UV curve $C_{g,n}$, plus possibly an additional irregular puncture of type $I_{2,M}$. We find that the resulting IR Schur indices precisely agree with the known Schur indices for theories of class $\CS$ associated to the same UV curve (see Appendix~\ref{app:Schur}). Thus, we show by consistency that the IR formalism, and in particular the IR formula for Schur indices, is equivalent to the Schur limit of the $\CN=2$ superconformal index (and surface defects) for this infinite class of theories. 

Notice that similar arguments and results as in (i)~--~(v), relying on surface defects, have been used to make general statements for the (full) superconformal index in~\cite{Gaiotto:2012xa}. Here, we have derived the statements (i)~--~(v) from the IR formalism, and their agreement with the corresponding well-known ``UV statements'' basically imply the equivalence of the (UV) Schur index with one derived from the IR formalism. Nevertheless, let us now provide more detailed arguments, allowing us to fully recover the general formula for rank-one theories of class $\CS$ as given in~\eqref{Igs} and~\eqref{Igs-irreg} from the IR properties (i)~--~(v).

Let us for now fix the number of regular punctures $n>0$. Then, a combination of (i)~--(iii) implies that the IR defects act \emph{locally} and \emph{irrespective} of the puncture and consequently we deduce that the index factorizes in terms of eigenfunctions $\Xi_{\lambda}^{(g)} (q;a)$ of the difference operators $\mathfrak{S}^{IR}_{v,a}$ in~\eqref{IRSDs}. A priori, it is not clear whether such functions exist and if their spectrum $\{\lambda\}$ is degenerate. However, we can use the connection to the Macdonald operators in~\eqref{SvsD}, and use known results for them. For instance, there exists creation operators, $\tilde{B}_{1}^{+}$~\cite{lapointe1997creation}, whose repeated application on the ``trivial'' eigenfunction $\Xi_{1}^{(g)} (q;a) \sim \CI_{V}^{-1/2}(q;a)$ (recall we have to account for the conjugation in~\eqref{SvsD}) yields a whole discrete tower of eigenfunctions labeled by $\lambda \in \mathbb{Z}_{>0}$. It is straightforward to check that up to conjugation by $\CI_V$ these eigenfunctions are proportional to the Schur polynomials, \emph{i.e.}
\be\label{XiSchur}
	\Xi_{\lambda}^{(g)} (q;a) \ \propto \ \CI_{V}^{-1/2} (q;a) \chi_{\lambda}(a) \,.
\ee
Notice, that we have used here that the eigenfunctions are symmetric under $a \to a^{-1}$, as discussed in (i).

Thus, we conclude that in general the IR Schur index is factorized and can be written as an expansion in terms of the eigenfunctions $\Xi_{\lambda}^{(g)}$, 
\be
	\left( \mathfrak{S}^{IR}_{v,a} \circ \Xi_{\lambda}^{(g)} \) (q;a) \  = \ E_{v, \lambda}^{(g)} (q) \Xi_{\lambda}^{(g)} (q;a) \,,
\ee
and which are up to normalization given by~\eqref{XiSchur}, \emph{i.e.}
\be\label{eqn:igsfac}
	\cI_{g,n} \ = \  \sum_{\lambda} \Phi_{\lambda}^{(g)} (q) \prod_{i=1}^{n} \Xi_{\lambda}^{(g)}(q;a_i) \,.
\ee
Here, the functions $\Phi_{\lambda}^{(g)} (q)$ are arbitrary functions of $q$, independent on the puncture-data. In the case in which the irregular puncture has an associated flavor symmetry that piece is contained in $\Phi_{\lambda}^{(g)} (q)$, and, for ease of notation, we leave the dependence on the additional fugacity implicit. 

A couple of remarks are in order: 
Firstly, notice that we assumed that the eigenfunctions are non-degenerate. We justified this by recalling that the IR defects are conjugate to a ``degeneration limit'' of the Macdonald operators whose (unique) non-degenerate orthonormal eigenfunctions are given by the Macdonald polynomials~\cite{MR1354144}. On general grounds, there could be ``more'' eigenfunctions. However, we know that Schur polynomial constitute a basis for symmetric polynomials, and thus, viewed as an expansion, at each order we get a complete basis. Along the same lines, we know that these functions have some orthogonality properties with an appropriate measure given by a product of $\cI_{V}$ and the $SU(2)$ Haar measure.
Secondly, in equation~\eqref{eqn:igsfac} we have picked a particular normalization of the functions $\Xi$ and $\Phi$; we imposed the normalization such that the puncture-dependence is fully contained in the eigenfunctions $\Xi$, while $\Phi$ is only dependent on the genus of the UV curve. There remains a $q$-dependent normalization, which we shall fix in the next step.

Thus far, we have assumed that the eigenfunctions $\Xi_{\lambda}^{(g)}$ are genus dependent. However, since the IR surface defects act locally and thus do not care about the other pieces of the Riemann surface, the eigenvalues and eigenfunctions have to be independent of the genus. Of course,  there is a remaining choice of $q$-dependent normalization of the eigenfunctions, and we shall fix the remaining freedom such that
\be
	\Xi_{\lambda} (q;a) \ \equiv \ \Xi_{\lambda}^{(g)}(q;a) \,.
\ee
To show this more explicitly, we define a gluing functional (``gluing the $i$-th and $j$-th punctures'') as follows\footnote{Of course, from the superconformal index perspective, we know that this ``gluing'' functional corresponds to gauging. However, this is a priori not a notion that is easily accessible from the IR perspective, and we just abstractly define it in the following. }
\bea\label{eq:gluing}
\mathfrak{G}_{ij}^{(g)} : &&\Big( \cI_{g,n} (\ldots, a_i, \ldots ), \cI_{\tilde{g},\tilde{n}}  (\ldots, b_j, \ldots ) \Big) \nn\\
&&\hspace{.1 cm}\mapsto \ \Big\langle \cI_{g,n} (\ldots, a_i, \ldots ), \cI_{\tilde{g},\tilde{n}}  (\ldots, b_j, \ldots ) \Big\rangle \ = \ \cI_{g+\tilde{g}, n+\tilde{n}-2}(\ldots, \overline{a_i},  \ldots , \overline{b_j}, \ldots )\,,
\eea
where we remove the ``barred'' variables. Since the surface defects act locally on punctures, regardless of the rest of the theories, we realize that as long as they do not act on the $i$-th and the $j$-th puncture, they must commute with gluing. Thus, the eigenfunctions and eigenvalues of the punctures are not affected by the gluing and remain the same (upon proper normalization by $q$-dependent pieces). Therefore, the difference operators have to be self-adjoint under the pairing (\ref{eq:gluing}). Given their relation to the Macdonald $q$-difference operators, we conclude that the gluing functional is (up to an $\CI_{V}$-factor) given by the degeneration of the Macdonald integral measure, which is simply the $SU(2)$ Haar measure. Thus, this functional is linear and acts at each level $\lambda$ separately on the eigenfunctions $\Xi_{\lambda}$. 
Now, we may glue a given Riemann surface $C_{g,n}$ with a trinion $C_{0,3}$, to get the Riemann surface $C_{g,n+1}$. Since we know from the IR formalism that the only defining ingredients are $g$ and $n$, see property (i), we ought to end up with the IR Schur index for $C_{g,n+1}$. 
Thus, upon gluing the $n$-th puncture with the first puncture of the trinion $C_{0,3}$, we get
\begin{align}
	\cI_{g,n+1} \ = \ &  
	\left\langle \cI_{g,n} , \cI_{0,3} \right\rangle\nn\\
	 \ = \ & 
	\sum_{\lambda, \lambda^{\prime}} \prod_{i=1}^{n-1} \Xi_{\lambda}(q;a_i)\left\langle\Phi_{\lambda}^{(g)} (q) \Xi_{\lambda}(q;a_n),\Phi^{(0)}_{\lambda^{\prime}} (q)  		\Xi_{\lambda^{\prime}}(q;b_1) \right\rangle \Xi_{\lambda^{\prime}}(q;b_2)\Xi_{\lambda^{\prime}}(q;b_3)\,,
\end{align}
and therefore, by consistency, we require
\be
	\Phi_{\lambda}^{(g)} (q) \Xi_{\lambda}(q;b_2)\Xi_{\lambda}(q;b_3) \ = \ 
	\sum_{\lambda^{\prime}}\left\langle\Phi_{\lambda}^{(g)} (q) \Xi_{\lambda}(q;a_n),\Phi_{\lambda^{\prime}}^{(0)} (q)  \Xi_{\lambda^{\prime}}(q;b_1) \right				\rangle\Xi_{\lambda^{\prime}}(q;b_2)\Xi_{\lambda^{\prime}}(q;b_3)\,,
\ee
for all $\lambda \in \IZ_{>0}$. However, we recall that under the appropriate measure the functions $\Xi_{\lambda}(q;a)$ are orthogonal, and thus applying the corresponding inner product on both sides, we conclude
\be
	\Phi_{\lambda}^{(g)} (q) \delta_{\lambda \lambda^{\prime}} \ = \ 
	\Phi_{\lambda}^{(g)} (q) \Phi_{\lambda^{\prime}}^{(0)} (q)  \left\langle\Xi_{\lambda}^{(g)}(q;a_n), \Xi_{\lambda^{\prime}}^{(0)}(q;b_1) \right\rangle\,.
\ee
Thus, it follows immediately that
\be
	\left\langle\Xi_{\lambda} (q;a_n), \Xi_{\lambda}(q;b_1) \right\rangle \ = \ \frac{1}{\Phi_{\lambda}^{(0)} (q)} \,.
\ee

Now, we may similarly glue a cylinder to a given theory connecting two punctures of the first theory, thus reducing the punctures by two and increasing the genus by one, \emph{i.e.} $(g,n) \to (g+1,n-2)$. A similar computation as above gives
\be
\begin{split}
	\Phi_{\lambda}^{(g+1)} (q)  \ \equiv \ & 
	\Phi_{\lambda}^{(g)} (q)  \left\langle\Xi_{\lambda}(q;a), \Xi_{\lambda} (q;b) \right\rangle  \\
	 \ = \ & 
	\Phi_{\lambda}^{(g)} (q) \frac{1}{\Phi_{\lambda}^{(0)} (q)} \,.
\end{split}
\ee
Hence, we conclude that
\be
	\Phi_{\lambda}^{(g)} (q) \ = \ \(  \frac{1}{\Phi_{\lambda}^{(0)} (q)} \)^{g-1} \,.
\ee
Notice that by employing the gluing functional to two punctures of $C_{g,2}$, we also get the result for the case without any punctures by consistency, \emph{i.e.}
\be
	\cI_{g,0} \ = \ \sum_{\lambda} \( {\Phi_{\lambda}^{(0)} (q)} \)^{1-g} \,.
\ee 

Given this analysis, it remains to fix $\Phi_{\lambda}^{(0)}$ and the correctly normalized eigenfunctions $\Xi_{\lambda}$. We can do this by looking at two examples, which we pick to be the theories associated to the UV curves $C_{0,3}$ and $C_{0,4}$. The IR Schur indices for these theories are given in equations~\eqref{eqn:I03} and~\eqref{eqn:I11}, and we can readily read off $\Xi_{\lambda}$ and $\Phi_{\lambda}$ as follows
\begin{align}
	\Xi_{\lambda}(q;z) \ = \ & \Psi_{\lambda} \(z ;q \) \nn\\
	\ = \ & \frac{(-1)^{\lambda+1}}{(1-q) \dim_{q} \lambda} \frac{\chi_{\lambda} (z)}{ \( q z^{2};q \)_{\infty}\left( q z^{-2};q \)_{\infty}}\,,\\
	\Phi_{\lambda}^{(g)} (q)  \ = \ &  \( \frac{\left( q^{2} ; q\right)_{\infty}}{\dim_{q} \lambda} \)^{2g-2} \,.
\end{align}
As expected, the resulting IR Schur index indeed reproduces the known result in equation~\eqref{Igs}, and we conclude the IR formalism is consistent and the IR Schur index indeed coincides with the Schur limit of the superconformal index.

As the structure of the IR surface defects is not affected by treating class $\CS$ theories associated to UV curves with an irregular puncture (see Section~\ref{sec:irregular}), our arguments straightforwardly extend to this case, and we get the Schur indices for theories associated to a Riemann surface $\tilde C_{g,n}$ of genus $g$, $n$ regular punctures and one irregular puncture of type $I_{2,M}$. This can be done by understanding the structure of the Schur index for the base-case given by the Argyres-Douglas theories of type $(A_1,D_{M})$, \emph{i.e.} the UV curve with a single regular and a single irregular puncture, and then extending inductively as before. By the same logic we end up with the result in~\eqref{Igs-irreg}.

\section{Discussion}\label{sec:disc}

In this paper, we set out to study a large class of~``vortex surface defects'' from the IR Coulomb branch perspective. 
We engineered these defects in class $\CS$ theories of type $A_1$ by considering Hitchin spectral curves in higher symmetric representation, and identifying vorticity with the dimension of the representation (up to a constant shift). We then used spectral networks, taken at a Roman locus, to compute the spectrum of 2d-4d BPS states, 2d BPS particles and 4d wall-crossing invariants from the BPS graph. Rather surprisingly, we observed that 2d-4d BPS states never contribute to the trace of the 2d-4d BPS monodromy, despite their abundant presence, for vortex defects placed near regular punctures (we do not expect this to hold otherwise).

By explicit evaluation of the 2d-4d BPS monodromy, we have shown that its trace takes the form of a $q$-difference operator $\mathfrak{S}^{IR}_v$ acting on (the trace of) the BPS monodromy of the original 4d $\CN=2$ theory. 
The explicit form of these operators reveals that they coincide (up to conjugation) with a specialization of Macdonald $q$-difference operators, that were employed in the bootstrap of the superconformal index in~\cite{Gaiotto:2012xa}.
We further argued that the operators  $\mathfrak{S}_v^{IR}$, together with certain intrinsic properties of (the trace of) the BPS monodromy, establish a bootstrap characterization for the latter. 
This parallels the UV bootstrap of the Schur index and provides the strongest support yet for the validity of the IR formula conjectured in~\cite{Cordova:2015nma}.

Let us now briefly mention some interesting questions and extensions for the future.

A first straightforward generalization consists of extending our arguments to theories with other ``irregular'' punctures. In this paper, we have mostly focused on regular punctures, except for irregular punctures of type $I_{2,M}$, which appear in $(A_1,D_M)$ Argyres-Douglas theories. However, we believe that the IR surface defects we constructed will behave the same way in the presence of other types and/or multiple of irregular punctures. From the TQFT perspective to the superconformal index, we expect that there exists a wavefunction for such irregular punctures and by gluing together Riemann surfaces, this would allow for computing the index of class $\CS$ theories with such irregular punctures. One hindrance may be that we require the IR surface defects to act universally and locally, but showing this from the IR perspective might require some care. 

Another obvious generalization includes dealing with higher-rank theories, or more generally class $\CS$ theories of types ADE. We expect that our logic and arguments follow through in this case as well, although some technicalities (such as the cancellation of the 2d-4d states) are highly non-trivial and deserve further study. Similarly, from the superconformal index perspective, we know that vortex defects act locally on the respective punctures; showing the analogous statement from the IR perspective for higher rank theories may prove difficult. In the rank-one case we relied heavily on powerful theorems, which are not available at higher rank. Nonetheless, given our results, we are confident that this should be true, and we expect that it will shed light on how to deal with surface defects in higher rank theories.

Along similar lines, it would be interesting to understand surface defects near non-full and/or irregular punctures. For instance for higher rank class $\CS$ theories, we could consider surface defects near non-full punctures. Similarly, as already mentioned in the above, for theories with irregular punctures, the vortex defects near the regular punctures differ from the ones near the irregular punctures (\emph{e.g.} compare our result for the vortex defects in Argyres-Douglas theories of type $(A_1,D_M)$ with the corresponding results in~\cite{Cordova:2017ohl, Cordova:2017mhb}). Although the contribution from 4d and 2d-4d BPS states are manifestly wall-crossing invariant, the 2d particles might allow for interesting wall-crossing behavior of such defects. From the superconformal index perspective the corresponding two types of defects seem to be arising from different constructions (\emph{i.e.} they can be viewed as arising from Higgsing seemingly different UV theories). It would also be interesting to understand the IR description of yet another type of surface defect (associated with the R-symmetry) discussed in~\cite{Dedushenko:2019yiw} for Lagrangian theories. In this reference, these ``R-symmetry defects'' were shown to be closely related to the choice of spin structure (and thus modularity properties) of the Schur index, and their IR constructions might shed light on corresponding properties of BPS monodromies.

Thus far, we have solely argued at the level of the trace of the BPS monodromy. Our arguments showed (from the IR) that upon taking the trace it factorizes into eigenfunctions of the vortex surface defects. However, by taking the trace we obviously lose some (important) information about the IR spectrum of the theory. It might prove interesting to study the (in some sense) ``categorified'' version of the Schur index, namely the BPS monodromies themselves. 
Some indications that BPS monodromies behave nicely under gluing, comes from a direct study of the construction in~\cite{Longhi:2016wtv} upon gluing BPS graphs~\cite{GL:unpublished}. 
We also believe that the factorization of the Schur index may be mirrored for the monodromies. 
If this is indeed the case, it might lead to novel ways to think about, and possibly compute, IR data. 
We leave a careful study for the future.

Another interesting direction may be to pursue the 5d uplift of the IR formula, proposed, and supported by evidence, in~\cite{Papageorgakis:2016cej}. This would likely involve a 5d version of spectral networks  developed recently~\cite{Eager:2016yxd, Banerjee:2018syt}, and may further lead to interesting connections to enumerative geometry~\cite{Iqbal:2012xm}.

\section*{Acknowledgments}
We would like to thank Subhojoy Gupta and Shu-Heng Shao for insightful discussions and correspondence. 
PL is grateful for hospitality from the Simons Center for Geometry and Physics summer workshop, Caltech, and the Kavli IPMU at Tokyo University where part of this work was carried out.
The work of MF is supported by the JSPS Grant-In-Aid for Scientific Research Wakate(A) 17H04837, the WPI Initiative, MEXT, Japan at IPMU, the University of Tokyo, the David and Ellen Lee Postdoctoral Scholarship, and the U.S. Department of Energy, Office of Science, Office of High Energy Physics, under Award Number de-sc0011632.
The work of PL is supported by a grant from the Swiss National Science foundation. He also acknowledges the support of the NCCR SwissMAP that is also funded by the Swiss National Science foundation. 
PL was also supported by the grants ``Geometry and Physics'' and ``Exact Results in
Gauge and String Theories'' from the Knut and Alice Wallenberg foundation during part of this work.

\noindent 

\appendix

\section{The Schur index and vortex surface defects from Higgsing}\label{app:Schur}

In this section, we review some aspects of the 4d Schur index (see~\cite{Rastelli:2014jja} for a nice review). We first introduce the Schur index as a limit of the 4d superconformal index, and then proceed to give the \emph{full} formula for rank-one theories of class $\CS$ (with regular punctures). Then, we extend it to the case where one puncture is of irregular type $I_{2,M}$. Finally, we turn towards the known results of the Schur index in the presence of surface defects obtained from the infinite-tension limit of vortices (or from ``Higgsing'')~\cite{Gaiotto:2012xa}.\footnote{See also \emph{e.g}~\cite{Gadde:2013ftv,Alday:2013kda,Razamat:2013jxa,Bullimore:2014nla,Razamat:2014pta,Chen:2014rca,Gaiotto:2015usa,Maruyoshi:2016caf,Ito:2016fpl,Yagi:2017hmj,Cordova:2017ohl,Cordova:2017mhb,Nazzal:2018brc,Razamat:2018zel,Nishinaka:2018zwq} for some generalizations and other examples of this construction for superconformal indices in the presence of vortex defects.}

\subsection{The Schur index}

From an operator-counting perspective, the Schur index can be viewed as a particular limit -- the ``Schur limit'' -- of the 4d $\cN=2$ superconformal index~\cite{Gadde:2011ik,Gadde:2011uv}.\footnote{We follow the conventions of~\cite{Cordova:2015nma,Cordova:2016uwk}, in which $(-1)^{F} = e^{2\pi \ii R}$. This convention slightly modifies the free hypermultiplet Schur index as compared with~\cite{Gadde:2011ik,Gadde:2011uv}, and the corresponding modifications are taken into account in the following formulae.} It is defined as follows
\be
	\cI(q) \ = \ \mathrm{Tr}_{\cH_{S^{3}}} e^{2\pi \ii R} q^{E - R} \prod_{f} a^{f} \,,
\ee
where the trace is taken over the states of the Hilbert space of operators on $S^{3}$ (or equivalently over the space of local operators), $E$ is the conformal Hamiltonian, or dilatation parameter of the respective states, and $R$ is their $SU(2)_{R}$-charge. Finally, we introduced a set of ``flavor fugacities'', $a^{f}$, which commute with the supercharge used to define the index.  

This particular limit of the superconformal index was studied in~\cite{Gadde:2011ik,Gadde:2011uv}, where it was found to be related to a 2d TQFT (q-deformed Yang-Mills theory)~\cite{Gadde:2009kb,Kawano:2012up,Fukuda:2012jr,Mekareeya:2012tn} by an ``AGT-type'' 2d/4d relation (see \emph{e.g.}~\cite{Tachikawa:2016kfc} for a nice review). More recently, the Schur index was found to be given by characters of particular 2d vertex operator algebras~\cite{Beem:2013sza}, and in the theme of this paper it was related to infrared data on the Coulomb branch of 4d (not necessarily conformal) supersymmetric field theory in~\cite{Cordova:2015nma}.\footnote{In~\cite{Dumitrescu:20xx}, the authors suggest that localizing on a special $S^{3}\times S^{1}$-background lets us define the $\cN=2$ Schur index away from the superconformal fixed points.}

Using the relation to TQFT one can associate a wavefunction to boundaries of the Riemann surface (UV curve) $C$ of a class $\CS$ construction. Since this wavefunction is known for a variety of different boundaries (or punctures)~\cite{Gadde:2011ik,Gadde:2011uv,Buican:2015ina,Buican:2015tda,Song:2015wta}, general TQFT arguments allow writing down the TQFT partition function and thus the Schur index for a large class of examples. In particular, it was found that the Schur index, $\cI_{g,n} (q) $, for general 4d rank-one class $\CS$ theories associated to a UV curve $C=C_{g,n}$ of genus $g$ and with $n$ punctures is given by
\be\label{Igs}
	\cI_{g,n} (q) \ = \ \sum_{\lambda=1}^{\infty} \( \frac{\left( q^{2} ; q\right)_{\infty}}{ \mathrm{dim}_{q} \lambda} \)^{2g-2} \prod_{j=1}^{n} \left[ \frac{(-1)^{\lambda+1}}{(1-q) \dim_{q} \lambda} \frac{\chi_{\lambda} (a_j)}{ \( q a_{i}^{\pm 2};q \)_{\infty}}\right] \,.
\ee
Here, the sum is taken over positive integers $\lambda \in \mathbb{Z}_{>0}$, 
\be
	\left( z;q \)_{\infty} \ = \ \prod_{i=0}^{\infty} \( 1- z q^{i} \)
\ee
is the Pochhammer symbol, $\dim_{q}\lambda$ is the quantum dimension, which is explicitly given by
\be
	\dim_{q}\lambda \ \coloneqq \ \frac{q^{-\lambda/2}- q^{\lambda/2}}{q^{-1/2} - q^{1/2}} \,, \qquad \lambda\in \mathbb{Z}_{>0} \,,
\ee
and finally $\chi_{\lambda} (a_i)$ is the Schur polynomial
\be\label{Schurpoly}
	\chi_{\lambda} (a_i) \ \coloneqq \ \frac{a_i^{\lambda} - a_i^{-\lambda}}{a_i-a_i^{-1}} \,, \qquad \lambda\in \mathbb{Z}_{>0} \,,
\ee
where $a_i$ corresponds to the flavor fugacity associated to the $i$-th puncture of $C_{g,n}$. Notice, in~\eqref{Igs} we suggestively write the index in terms of ``puncture data'' and ``genus data'', \emph{i.e.} the factors in the product correspond to the wavefunctions associated with the punctures of the UV curve $C_{g,n}$,
\be\label{AppPsi}
	\Psi_{\lambda} \(a_j;q \) \ = \  \frac{(-1)^{\lambda+1}}{(1-q) \dim_{q} \lambda} \frac{\chi_{\lambda} (a_j)}{ \( q a_{i}^{\pm 2};q \)_{\infty}} \,,
\ee
whereas the first factor depends on the genus. Of course this splitting involves a choice of normalization of the wavefunctions.

\subsection{Schur index for class $\CS$ theories with irregular punctures of type $I_{2,M}$}

One can also define the Schur index for class $\CS$ theories with irregular punctures. This can for example be done by leveraging the TQFT structure of the index and assigning a wavefunction for the particular irregular punctures~\cite{Buican:2015ina,Song:2015wta,Buican:2017uka}.\footnote{This can of course also be derived from the IR Schur formula~\cite{Cordova:2015nma}. Alternatively, in~\cite{Maruyoshi:2016tqk,Maruyoshi:2016aim,Agarwal:2016pjo} the same answer for the Schur index was provided from a UV $\CN=1$ theory which is argued to flow to the Argyres-Douglas theories in question with enhanced $\CN=2$ supersymmetry in the IR.} Although we expect our arguments to go through for other types of irregular punctures, we shall (in the present context) focus on the irregular punctures appearing in $(A_1,D_{M+2})$ Argyres-Douglas theories with integer $M$, and which we label by $I_{2,M}$. In our conventions, the corresponding wavefunctions for odd $M=2\ell-1$ read
\be
	\tilde \Psi_{\lambda}^{I_{2,2\ell-1}} \( q \) \ = \ \left\{\begin{array}{ll} 
	\left( -1 \)^{(\lambda-1)/2} q^{\frac{(2\ell+1) (\lambda^{2}-1)}{8}} \frac{ ( q^{2};q )_{\infty}}{\dim_{q} \lambda} \,, \quad & \lambda \text{ odd} \,, \\ 
	0 \,, & \lambda \text{ even} \,.
\end{array} \right.
\ee
Similarly, for even $M=2\ell$, the wavefunctions are given by
\be
	\tilde \Psi_{\lambda}^{I_{2,2\ell}} ( q ) \ = \ 
	 \frac{q^{\frac{(\ell+1)(\lambda^{2}-1)}{4}}}{(1-q)\dim_{q}\lambda} \, \Tr_{R_{\lambda-1}} \( z^{2J_{3}} q^{-(\ell+1) J_3^{2}}\right) \,,
\ee
where $R_{\lambda}$ is the spin-$\frac{\lambda}{2}$ representation and $J_3$ the Cartan generator of $\mathfrak{su}(2)$.\footnote{Notice that the Schur polynomials can be written as $\chi_{\lambda}(a) = \Tr_{R_{\lambda-1}} a^{2J_3}$.} Additionally, we associate the $U(1)$ flavor fugacity $z$ with the this particular irregular puncture.

Then, the TQFT structure predicts the Schur index of a theory associated with a genus-$g$ Riemann surface with a single irregular $I_{2,M}$ puncture, $n$ regular punctures is given by
\be\label{Igs-irreg}
	\sum_{\lambda=1}^{\infty} \( \frac{\left( q^{2} ; q\right)_{\infty}}{ \mathrm{dim}_{q} \lambda} \)^{2g-2} 
	\tilde \Psi_{\lambda}^{I_{2,M}}( q ) \,
	\prod_{j=1}^{n} \left[ \frac{(-1)^{\lambda+1}}{(1-q) \dim_{q} \lambda} \frac{\chi_{\lambda} (a_j)}{ \( q a_{i}^{\pm 2};q \)_{\infty}}\right]\,.
\ee
Of course, one can play this game with other types of irregular punctures as well as if we have theories with several of them.

\subsection{Vortex surface defects from Higgsing}\label{App:Higgsing}

We now briefly review the construction of surface defects arising from Higgsing the 4d $\cN=2$ superconformal index~\cite{Gaiotto:2012xa}. We start with an arbitrary theory $\cT$ which contains an $SU(2)_{F}$ flavor symmetry. Then, we couple $\cT$ to a bifundamental hypermultiplet -- with flavor symmetry $SU(2)_{1}\times SU(2)_2\times SU(2)_3$ -- by gauging the diagonal combination of the $SU(2)_{F}\times SU(2)_{1}$ and we obtain the gauged theory $\cT_{\rm UV}$. The latter theory then contains an $SU(2)_{2}\times SU(2)_{3}$ flavor symmetry factor, with the latter factor corresponding to the enhanced baryonic symmetry, in which the complex scalars rotate as a doublet. Then, giving a VEV to the baryonic combination of the complex scalars fully ``Higgses'' the gauged diagonal $SU(2)$ symmetry, and, upon flowing to the IR, we end up at the original theory $\cT$.

However, we may turn on a position dependent VEV to the baryonic operator $\langle B \rangle \sim z^{v} $, where $(z,w) \in \mathbb{C}^{2}$ are complex coordinates of $\mathbb{R}^{4}$. Upon flowing to the IR, some nontrivial degrees of freedom survive at the origin $z=0$, which can be interpreted as a surface defect~\cite{Gaiotto:2012xa}. This is confirmed by considering Lagrangian theories coupled to two-dimensional $\cN=(2,2)$ theories in~\cite{Gadde:2013ftv} (see also~\cite{Cordova:2017ohl,Cordova:2017mhb,Nishinaka:2018zwq}).

In practice, the index in the presence of such types of surface defects can be evaluated by taking residues at appropriate poles for the index of the UV theory $\cT_{\rm UV}$. It can be shown that they act as difference operators and in the Schur limit of interest in this paper they are related (via conjugation and upon taking a limit) to the well-known Macdonald difference operators~\cite{MR1354144} (see also~\cite{Gaiotto:2012xa,Bullimore:2014nla}). We shall here simply present the general result for rank-one theories. The defects are labeled by representations of $SU(2)$, which we shall take to be the integer $v \in \mathbb{N}$ (corresponding to the vortex number). Then, the defects act on the index $\cI_\cT$ of the original theory with (at least) one $SU(2)$ flavor fugacity $a$ as follows
\be\label{HiggsSDs}
	\mathfrak{S}^{UV}_{v,a} \circ \ \cI_\cT \( a, b, \cdots \) \ = \ \sum_{m_1+m_2 = v} a^{2 (m_1-m_2)} q^{v+ 2 m_1 m_2}\cI_\cT \( q^{\frac{v}{2} - m_1} a, b, \cdots \) \,,
\ee
where the sum is taken over pairs $(m_1,m_2) \in \mathbb{Z}_{\geq 0} \times \mathbb{Z}_{\geq 0}$ satisfying $m_1+m_2=v$.

Finally, since the surface operators as given in~\eqref{HiggsSDs} are intimately related to Macdonald operators, it is no surprise that their eigenfunctions are given by some normalization of Schur operators. In particular, the wavefunctions $\Psi_{\lambda} \(a;q \)$ are precisely eigenfunctions
\be
	\mathfrak{S}_{v,a}^{UV} \circ \Psi_{\lambda} \(a;q \) \ = \ E_{v,\lambda} \Psi_{\lambda} \(a;q \)
\ee
 with eigenvalues $E_{v,\lambda}$ given by
\be
	E_{v,\lambda} \ = \ (-1)^{v} q^{\frac{v (v+2)}{2}} \frac{S_{\lambda,v}}{S_{\lambda,0}}\,,
\ee
where the (analytic continuation of the) modular S-matrix of the $SU(2)$ WZW model (away from integer level) $S_{\lambda,v}$ can be explicitly written as follows
\be
	S_{\lambda , v} \ = \ \frac{q^{-\frac{\lambda(1+v) }{2}-\frac{v}{2}}-q^{\frac{\lambda (1+v) }{2}+\frac{v}{2}+1}}{1-q}\,.
\ee

\section{Surfaces with multiple punctures, Whitehead moves and bigons}\label{sec:more-punctures}

In this appendix, we explain why, for any class $\CS$ theory of type $A_1$ with $C$ having at least two punctures, there is a choice of moduli for which the BPS graph locally (around a puncture) looks like in Figure \ref{fig:A1-puncture-behavior}.

Recall that both for the sphere with three and four punctures we found more than one BPS graph (see Figures~\ref{fig:T2-BPSg} and~\ref{fig:nf4}).
In both cases, the different BPS graphs are related by a ``Whitehead move'' -- or ``flip'' in terminology from~\cite{Gabella:2017hpz}. 
Physically, a Whitehead move corresponds to varying moduli while moving along the Roman locus (hence preserving alignment of central charges), and going through a singularity where a cycle of the Seiberg-Witten curve shrinks.\footnote{It is natural to think of singularities as complex codimension-one loci in the Coulomb branch of a theory. However, here we are varying the UV moduli such as masses and couplings as well, and therefore, these singularities should be understood as loci in the enlarged moduli space.} 
The shrinking cycle is precisely the one obtained by lifting the edge affected by the Whitehead move.

$A_1$ theories of class $\CS$ are \emph{complete theories}, meaning that there exists a point in their moduli space, such that any configuration of central charges is realized~\cite{Cecotti:2011rv}.
Geometrically, this means that any Whitehead move on the BPS graph is realized by a path in the moduli space of the quadratic differential -- since the move corresponds to taking a period of the Seiberg-Witten curve $Z_\gamma$ to zero along the ray $e^{\ii\vartheta_c}\IR_{\geq 0} $ -- and then continuing along the opposite ray $e^{\ii\vartheta_c+\pi}\IR_{\geq 0}$.\footnote{
In fact, the existence of a sequence of differentials that implements a Whitehead move can also be argued through the theory of half-translation surfaces, for background see~\cite{2015arXiv150502939G, 2016arXiv160706931G}.
One way of specifying a holomorphic quadratic differential on a surface is to describe charts to $\IC$ that differ by half-translations. 
The canonical quadratic differential $\diff z^2$ on these charts then induces a holomorphic quadratic differential on the entire surface.  
In the case of BPS graphs, one is gluing a collection of semi-infinite Euclidean cylinders, each carrying a canonical quadratic differential with a pole of order two at infinity.
Since boundary identifications are half-translations, one obtains a holomorphic quadratic differential on the resulting  punctured surface.  
Whitehead moves just correspond to changing the pattern of identifications. We thank S.~Gupta for explaining this point.
}
Therefore, given the BPS graph in the left frame of Figure~\ref{fig:nf4}, we could have \emph{predicted} the existence of the one on the right, since they belong to the same Whitehead equivalence class. The same applies to the graphs in Figure~\ref{fig:T2-BPSg}.

More generally, the BPS graph of an $A_1$ theory of class $\CS$ is always dual to an ideal triangulation~\cite{Gabella:2017hpz}. A Whitehead move on the BPS graph is dual to a flip of the triangulation. Since any two ideal triangulations are related by flips, one can get the BPS graph for free by simply \emph{choosing} an ideal triangulation and dualizing it. The completeness property of $A_1$ theories then guarantees that such a BPS graph exists.

\begin{figure}[h!]
\begin{center}
\includegraphics[width=0.5\textwidth]{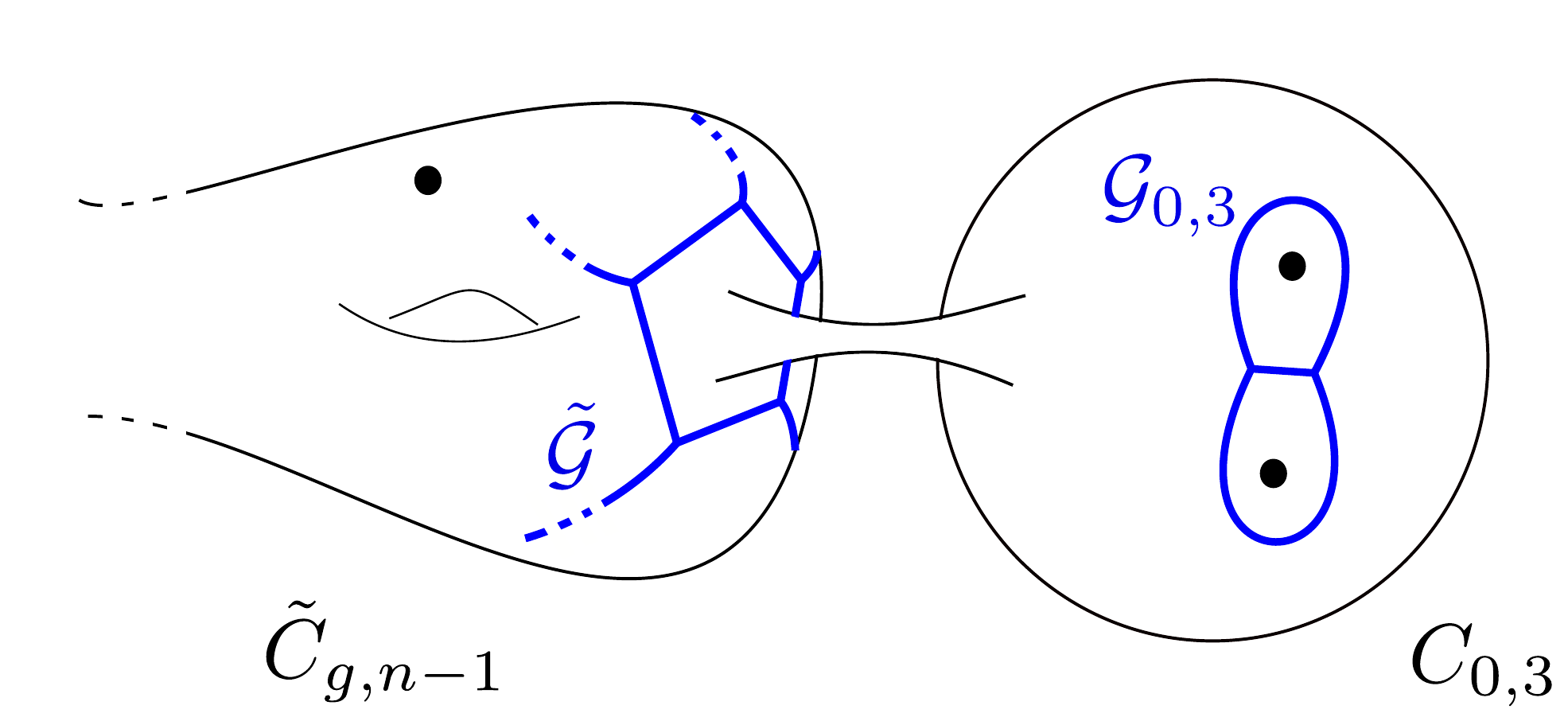}
\caption{When $C$ has two punctures, we can always choose a pairs-of-pants decomposition featuring a trinion glued to the rest of $C$ along a single puncture. The critical graph of the Strebel differential respecting the pants decomposition is shown in blue.}
\label{fig:two-punctures}
\end{center}
\end{figure}

\begin{figure}[h!]
\begin{center}
\raisebox{-0.5\height}{\includegraphics[width=0.35\textwidth]{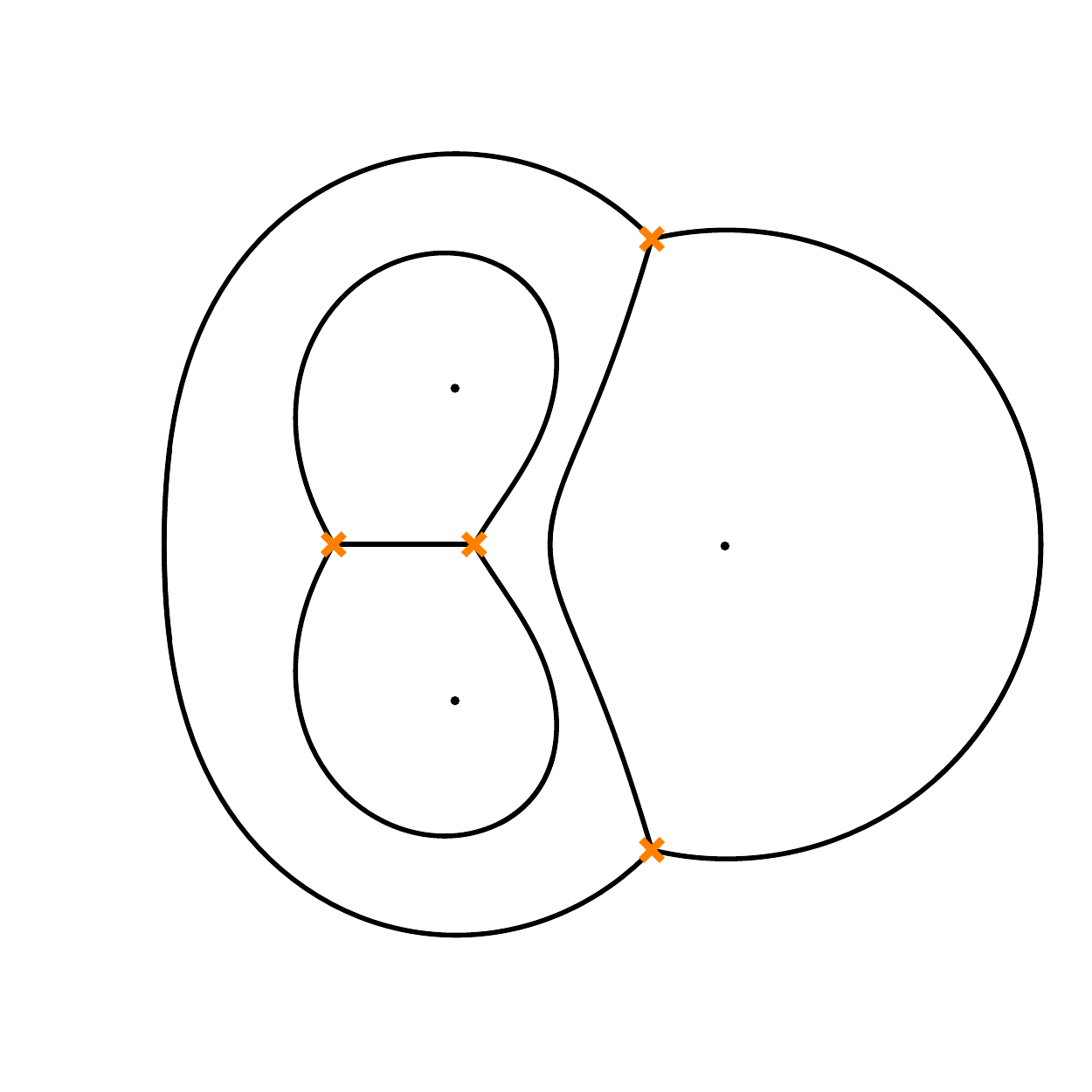}}
\ \ $\longrightarrow$\ \ 
\raisebox{-0.5\height}{\includegraphics[width=0.35\textwidth]{figures/tetrahedron.pdf}}

\caption{Left: the $N_f=4$ Strebel graph obtained at $\vartheta=0$ from $M_{a,b}=3, M_{b,c}=1.2$, and $u=4.3$. One can clearly view it as a $T_2$ where we glued another $T_2$ in place of one puncture. Tuning moduli eventually brings it into the form shown on the right (which is the BPS graph in the left frame in Figure~\ref{fig:nf4}).}
\label{fig:nf4-uncontracted}
\end{center}
\end{figure}

If the UV curve $C$ has \emph{at least two} (regular) punctures, it is always possible to choose the BPS graph such that there is a bigon around one of the punctures. The choice of puncture is arbitrary.
The argument uses a combination of the results of Strebel and Liu, and the completeness of class $\CS$ theories.
First of all, given a Riemann surface with at least two punctures we can choose a pair-of-pants decomposition such that it features a long thin tube separating the two distinguished punctures from the rest of the surface. 
The thin neck, together with the three punctures, are the three holes of the three-punctured sphere glued to the rest of the surface (see Figure~\ref{fig:two-punctures}).
Then, the Riemann surface decomposes as the connected sum
\be
	C_{g,n}  \ = \  \tilde C_{g,n-1} \ \#\  C_{0,3} \,.
\ee
By the completeness property of $A_1$ theories, there must exist a BPS graph on $\tilde C_{g,n-1}$, let us call this graph $\tilde \cG$. We also know that there is always a BPS graph on $C_{0,3}$, let us denote this by $\cG_{0,3}$.
On the connected sum, these statements imply the existence of a Strebel differential whose critical graph is the disjoint union 
\be
	\tilde \cG \ \sqcup\  \cG_{0,3}\,.
\ee
A sketch of this idea is shown in Figure~\ref{fig:two-punctures}, while an actual example is shown in Figure~\ref{fig:nf4-uncontracted}.

By definition, the foliation induced by the Strebel differential features closed loops around the tube. The differential induces a transverse measure along this tube, which computes the distance between its boundaries on $\tilde \cG$ and $\cG_{0,3}$. We shall call this $h$, the \emph{height} of the tube.
According to~\cite{Strebel, Liu}, there is a Strebel differential for any $h \geq 0$. In particular, we can take $h\to 0$ and get the BPS graph of the theory defined by $C_{g,n}$.

\begin{figure}[h!]
\begin{center}
\includegraphics[width=0.95\textwidth]{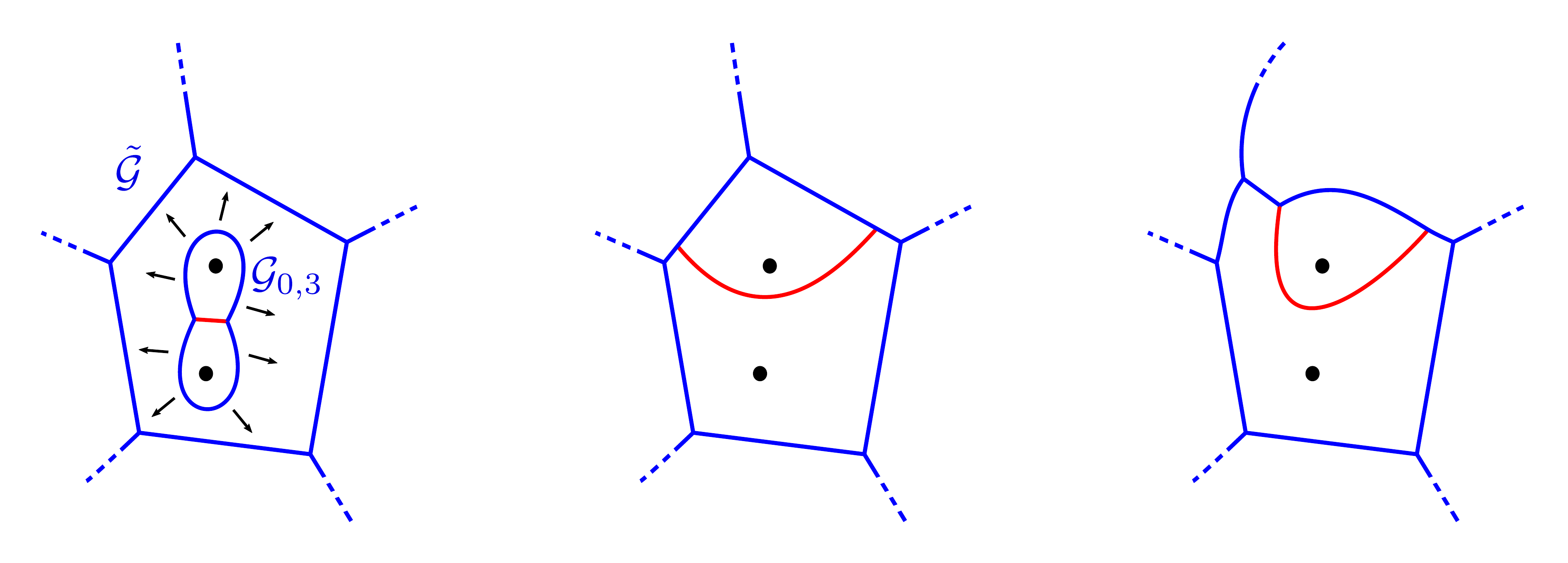}
\caption{The contraction of $\cG_{0,3}$ with $\tilde \cG$ is determined by the cyclic position of the vertices of $\cG_{0,3}$ along the polygonal boundary of $\tilde\cG$. A sequence of Whitehead moves always produces a bigon.}
\label{fig:secant}
\end{center}
\end{figure}

The BPS graph of this theory is obtained by \emph{contraction} of $\tilde \cG$ and $\cG_{0,3}$ as discussed in~\cite{Hollands:2013qza, Gabella:2017hpz}.
There are several ways to contract two BPS graphs, they are classified by the cyclic ordering of the branch points of $\tilde\cG$ and $\cG_{0,3}$ along the collapsing boundaries of the tube.
Let $\tilde P$ be the polygonal boundary of the tube on the $\tilde\cG$ side.
After contraction of $\tilde \cG$ with $\cG_{0,3}$, the two branch points of the latter will end up on  $\tilde P$. 
Therefore, the new BPS graph is determined by a choice of \emph{secant} for $\tilde P$ (see Figure~\ref{fig:secant}).
By a sequence of Whitehead moves, it is always possible to modify the secant such that it forms a bigon around one of the punctures. 
To give a real-world example, consider the BPS graph of the $N_f=4$ theory obtained by contraction as in Figure~\ref{fig:nf4-uncontracted}. We showed in Figure ~\ref{fig:nf4}, that there exists a choice of moduli that implements a Whitehead move so that the new graph will feature a bigon around at least one puncture.

Once the BPS graph features a bigon, we claim that the spectral network behaves qualitatively as described in Section~\ref{sec:near-punctures}, even away from the critical phase $\vartheta\neq \vartheta_c$.
The reason is that the basin of attraction of the puncture is entirely enclosed by the bigon boundaries made of degenerate $\CS$-walls, and two branch points.
An infinitesimal perturbation of the phase to $\vartheta_c\pm \epsilon$ yields a network with exactly two walls ending on the puncture: each of these must be sourced at the branch points corresponding to bigon vertices.
For larger phases away from $\vartheta_c$, there cannot be any other walls ending on the puncture, because that would require a topological change in the network. This is because the moduli of the 4d theory are fixed at the Roman locus, therefore by construction the only jump of the network occurs at $\vartheta_c$, barring this possibility.
Therefore, only the two $\CS$-walls sourced by the nearest branch points will end on the puncture, at all phases.

Moreover, the two $\CS$-walls that spiral into the puncture will cross the surface defect in alternating order. 
They are bound to spiral slower and slower around the puncture as the phase increases (or decreases) away from $\vartheta_c$. 
The alternating walls cannot be broken, because that would require one of the walls to cross the other, which is impossible by single-valuedness of the flow of the foliation, and the absence of topological jumps of the network except for $\vartheta_c$. 
Thus, we conclude that the behavior of the spectral network near a puncture surrounded by a bigon (at the critical phase) is universally described by the analysis of Section~\ref{sec:near-punctures}.

\cleardoublepage

\bibliography{refs}

\end{document}